\documentclass[aps,rmp,reprint,amsmath,amssymb,longbibliography]{revtex4-1}

\usepackage[english]{babel} % Specify a different language here - english by default
\usepackage{graphicx,xspace}
\usepackage{tabularx}
\usepackage{bm}
\usepackage{overpic}
\usepackage{upgreek}
\usepackage{multirow}
\usepackage[shortlabels]{enumitem}
\usepackage{threeparttable}
\usepackage{pgf, etoolbox, needspace}
\usepackage{mathtools, slashed}

\usepackage{fancyhdr}
\input{babarsym.tex}

\newcommand{\bfacs}{$B$~factories\xspace}

\newcommand{\FRA}{\texttt{FR}}
\newcommand{\FEI}{\texttt{FEI}}
\newcommand{\SER}{\texttt{SER}}

\newcommand{\Br}{{\cal B}}
\newcommand{\decay}[2]{\ensuremath{#1\!\to #2}\xspace}         % {\Pa}{\Pb \Pc}

\def\Dssm    {{\ensuremath{D^{*-}_{s}}}\xspace}
\def\BdorBs  {{\ensuremath{\B^0_{(s)}}}\xspace}

\newcommand{\Dx}{\ensuremath{D^{(*)}}\xspace}
\newcommand{\Dxbar}{\kern 0.18em\overline{\kern -0.18em D}^{(*)}{}\xspace}
\newcommand{\Btag}{\ensuremath{B_{\rm tag}}\xspace}
\newcommand{\Bsig}{\ensuremath{B_{\rm sig}}\xspace}

\def\nn{\nonumber}

\newcommand{\MatrixElem}[3]{\big\langle #1 \big| #2 \big| #3 \big\rangle}

\def\epem{\ensuremath{e^+e^-}\xspace}
\def\BB{\ensuremath{B\Bbar}\xspace}
\def\bb{\ensuremath{b\bbar}\xspace}
\def\pp{\ensuremath{pp}\xspace}
\def\lqcd{\ensuremath{\Lambda_{\text{QCD}}}\xspace}
\def\pipipi{\ensuremath{\pi\pi\pi}\xspace}

\def\Lb{\ensuremath{\Lambda_b}\xspace}
\def\Bs{\ensuremath{B_s}\xspace}
\def\Bc{\ensuremath{B_c}\xspace}
\def\Bp{\ensuremath{B^+}\xspace}
\def\Bz{\ensuremath{B^0}\xspace}
\def\Xib{\ensuremath{\Xi_b}\xspace}
\def\Omegab{\ensuremath{\Omega_b}\xspace}
\def\Bsstar2{\ensuremath{B_{s2}^{**}}\xspace}

\def\tauon{\ensuremath{\tau}\xspace}
\def\neu{\ensuremath{\nu}\xspace}

\def\aone{\ensuremath{a_1}\xspace}

\def\minpipi{\ensuremath{\text{min}(m_{\pip\pim})}\xspace}
\def\maxpipi{\ensuremath{\text{max}(m_{\pip\pim})}\xspace}
\def\mpipi{\ensuremath{m_{\pip\pip}}\xspace}
\def\mthreepi{\ensuremath{m_{\pip\pim\pip}}\xspace}
\def\ttau{\ensuremath{t_{\tau}}\xspace}
\def\Dplus{\ensuremath{D^+}\xspace}
\def\Dsm{\ensuremath{D_s^-}\xspace}

\def\Dsstar{\ensuremath{D_s^*}\xspace}
\def\Dsstarstar{\ensuremath{D_s^{**}}\xspace}
\def\Ddouble{\ensuremath{D^{**}}\xspace}

\def\Jpsi{\ensuremath{J/\psi}\xspace}
\def\Lc{\ensuremath{\Lambda_c}\xspace}
\def\Lcstar{\ensuremath{\Lambda_c^*}\xspace}

\def\taupipipi{\ensuremath{\tau \rightarrow \pi^- \pi^+ \pi^- \nu}\xspace}
\def\tauellnu{\ensuremath{\tau \rightarrow \ell \nu \nub}\xspace}
\def\taumunu{\ensuremath{\tau \rightarrow \mu \nu \nub}\xspace}
\def\tauenu{\ensuremath{\tau \rightarrow e \nu \nub}\xspace}
\def\taupinu{\ensuremath{\tau \rightarrow \pi \nu}\xspace}
\def\taurhonu{\ensuremath{\tau \rightarrow \rho \nu}\xspace}

\def\Bpiellnu{\ensuremath{B \rightarrow \pi \ell\nu}\xspace}
\def\Bpitaunu{\ensuremath{B \rightarrow \pi \tau\nu}\xspace}

\def\BDtaunu{\ensuremath{B \rightarrow D \tau\nu}\xspace}
\def\BDstaunu{\ensuremath{B\rightarrow D^* \tau\nu}\xspace}
\def\BDxtaunu{\ensuremath{B \rightarrow D^{(*)} \tau \nu}\xspace}
\def\BDsstaunu{\ensuremath{B \rightarrow D^{**} \tau\nu}\xspace}
\def\BsDssstaunu{\ensuremath{B_s \rightarrow D^{**}_s \tau\nu}\xspace}

\def\BDellnu{\ensuremath{B \rightarrow D \ell \nu}\xspace}
\def\BDsellnu{\ensuremath{B \rightarrow D^* \ell \nu}\xspace}
\def\BDxellnu{\ensuremath{B \rightarrow D^{(*)} \ell \nu}\xspace}
\def\BDssellnu{\ensuremath{B \rightarrow D^{**} \ell \nu}\xspace}
\def\BsDsssellnu{\ensuremath{B_s \rightarrow D^{**}_s \ell \nu}\xspace}

\def\BDlnu{\ensuremath{B \rightarrow D l \nu}\xspace}

\def\BDslnu{\ensuremath{B \rightarrow D^* l \nu}\xspace}
\def\BDsltnu{\ensuremath{B \rightarrow D^{*} l \nu}\xspace}
\def\BDxlnu{\ensuremath{B \rightarrow \Dx l \nu}\xspace}
\def\BDxltnu{\ensuremath{B \rightarrow D^{(*)} l \nu}\xspace}
\def\BDssltnu{\ensuremath{B \rightarrow D^{**} l \nu}\xspace}
\def\BDsslnu{\ensuremath{B \rightarrow D^{**} l \nu}\xspace}
\def\BsDssslnu{\ensuremath{B_s \rightarrow D^{**}_s l \nu}\xspace}

\def\BDsmunu{\ensuremath{B \rightarrow D^* \mu \nu}\xspace}
\def\HbHctaunu{\ensuremath{H_b \rightarrow H_{c} \tau \nu}\xspace}
\def\HbHcellnu{\ensuremath{H_b \rightarrow H_{c} \ell \nu}\xspace}
\def\HbHcutaunu{\ensuremath{H_b \rightarrow H_{c,u} \tau \nu}\xspace}

\def\BDsptaunu{\ensuremath{\Bbar^0\rightarrow D^{*+} \tau^-\nutb}\xspace}
\def\BDsztaunu{\ensuremath{B^-\rightarrow D^{*0} \tau^-\nutb}\xspace}

\newcommand{\BJtaunu}{\ensuremath{B_c \rightarrow \jpsi \tau\nu}\xspace}
\newcommand{\BJlnu}{\ensuremath{B_c \rightarrow \jpsi l\nu}\xspace}

\newcommand{\BJmunu}{\ensuremath{B_c \rightarrow \jpsi \mu\nu}\xspace}

\def\BXclnu{\ensuremath{B \rightarrow X_c\ell\nu}\xspace}

\def\bctaunu{\ensuremath{b \rightarrow c\tau\nu}\xspace}
\def\butaunu{\ensuremath{b \rightarrow u\tau\nu}\xspace}

\def\RDz{\ensuremath{\calR(D^0)}\xspace}
\def\RDp{\ensuremath{\calR(D^+)}\xspace}
\def\RDsz{\ensuremath{\calR(D^{*0})}\xspace}
\def\RDsp{\ensuremath{\calR(D^{*+})}\xspace}
\def\RD{\ensuremath{\calR(D)}\xspace}
\def\RDx{\ensuremath{\calR(\Dx)}\xspace}
\def\RDs{\ensuremath{\calR(D^*)}\xspace}
\def\RDdouble{\ensuremath{\calR(D^{**})}\xspace}

\def\RJ{\ensuremath{\calR(\jpsi)}\xspace}
\def\RLc{\ensuremath{{\cal R}(\Lambda_{c}^{+})}\xspace}

\def\RKx{\ensuremath{\calR_{K^{(*)}}}\xspace}

\def\Esl{\ensuremath{E^*_\ell}\xspace}

\def\mmiss{\ensuremath{m_{\rm miss}^2}\xspace}

\def\ECL{\ensuremath{E_{\rm ECL}}\xspace}
\def\pmiss{\ensuremath{p_{\rm miss}}\xspace}

\def\mES{\ensuremath{m_{\rm ES}}\xspace}

\def\cosHel{\ensuremath{\cos\theta_{v}}\xspace}

\usepackage{xcolor}

\definecolor{niceblue}{rgb}{0.1,0.1,0.5}
\definecolor{nicegreen}{rgb}{0.1,0.5,0.1}
\usepackage[bookmarks=false]{hyperref}
\hypersetup{colorlinks,citecolor= nicegreen,linkcolor= niceblue}

\makeatletter
\g@addto@macro\bfseries{\boldmath}
\makeatother

 \usepackage{lineno}
 \newcommand*\patchAmsMathEnvironmentForLineno[1]{%
   \expandafter\let\csname old#1\expandafter\endcsname\csname #1\endcsname
   \expandafter\let\csname oldend#1\expandafter\endcsname\csname end#1\endcsname
   \renewenvironment{#1}%
      {\linenomath\csname old#1\endcsname}%
      {\csname oldend#1\endcsname\endlinenomath}}% 
 \newcommand*\patchBothAmsMathEnvironmentsForLineno[1]{%
   \patchAmsMathEnvironmentForLineno{#1}%
   \patchAmsMathEnvironmentForLineno{#1*}}%
 \AtBeginDocument{%
 \patchBothAmsMathEnvironmentsForLineno{equation}%
 \patchBothAmsMathEnvironmentsForLineno{align}%
 \patchBothAmsMathEnvironmentsForLineno{flalign}%
 \patchBothAmsMathEnvironmentsForLineno{alignat}%
 \patchBothAmsMathEnvironmentsForLineno{gather}%
 \patchBothAmsMathEnvironmentsForLineno{multline}%
 }

\allowdisplaybreaks

\begin{document}

\title{Semitauonic $b$-hadron decays: A lepton flavor universality laboratory}

\author{Florian U. Bernlochner}
\email{florian.bernlochner@uni-bonn.de}
\affiliation{\mbox{Physikalisches Institut der Rheinischen Friedrich-Wilhelms-Universit\"at Bonn, 53115 Bonn, Germany}}
\author{Manuel Franco Sevilla}
\email{manuelf@umd.edu}
\affiliation{\mbox{University of Maryland, College Park, MD, USA}}
\author{Dean J. Robinson}
\email{drobinson@lbl.gov}
\affiliation{\mbox{Ernest Orlando Lawrence Berkeley National Laboratory, University of California, Berkeley, CA, USA}}
\author{Guy Wormser}
\email{wormser@lal.in2p3.fr}
\affiliation{\mbox{Laboratoire Ir\`ene Joliot-Curie, Universit\'e Paris-Saclay, CNRS/IN2P3, Orsay, France}}

\date{\today{}}

\begin{abstract}
The study of lepton flavor universality violation (LFUV) in semitauonic $b$-hadron decays has become
increasingly important in light of long-standing anomalies in their measured branching fractions, and
the large datasets anticipated from the LHC experiments and Belle~II. In this review, a comprehensive
survey of the experimental environments and methodologies for semitauonic LFUV measurements at
the B factories and LHCb is undertaken, along with an overview of the theoretical foundations and
predictions for a wide range of semileptonic decay observables. The future prospects of controlling
systematic uncertainties down to the percent level, matching the precision of standard model (SM)
predictions, are examined. Furthermore, new perspectives and caveats on combinations of the LFUV data 
are discussed and the world averages for the ${\cal R}(D^{(*)})$ ratios are revisited. Here it is demonstrated
that different treatments for the correlations of uncertainties from $D^{**}$ excited states 
can vary the current $3\sigma$ tension with the SM within a $1\sigma$ range. 
Prior experimental overestimates of $D^{**}\tau\nu$ contributions may further exacerbate this. 
The precision of future measurements is also estimated; their power to exploit full differential information, 
 and solutions to the inherent difficulties in self-consistent new physics interpretations of LFUV observables, are explored.
% \begin{center}
% To be submitted to {\it Reviews of Modern Physics}.
% \end{center}
\end{abstract}

\maketitle

\tableofcontents{}

\makeatletter
	\let\t@section\section
	\renewcommand{\section}[1]{
		\needspace{8\baselineskip}
		\t@section{#1}
		\addtocontents{toc}{\protect\nopagebreak}
		\gdef\sectiontitle{#1}
	}
\makeatother
	
\section{Introduction}
\label{sec:intro}

Over the past decade, collider experiments have provided ever-more precise measurements of Standard Model (SM) parameters, 
while direct collider searches for new interactions or particles have yielded ever-more stringent bounds on New Physics (NP) beyond the SM.
This, in turn, has brought renewed attention to the NP discovery potential of indirect searches:
measurements that compare the interactions of different species of elementary SM particles
to SM expectations.

A key feature of the Standard Model is the universality of the electroweak gauge coupling to the three known fermion generations or families.
In the lepton sector, this universality results in an accidental lepton flavor symmetry,
that is broken in the SM (without neutrino mass terms) only by Higgs Yukawa interactions responsible for generating the charged lepton masses.
A key prediction, then, of the Standard Model is that physical processes involving charged leptons should feature a \emph{lepton flavor universality}: 
an approximate lepton flavor symmetry among physical observables, such as decay rates or scattering cross-sections,
that is broken in the SM only by charged lepton mass terms in the amplitude and phase space.
(Effects of additional Dirac or Majorana neutrino mass terms in extensions of the SM are negligible in all contexts we consider.)
In the common parlance of the literature, testing for lepton flavor universality violation (LFUV) in any particular process thus refers 
to measuring deviations in the size of lepton flavor symmetry breaking versus SM predictions.

An observation of LFUV would clearly establish the presence of physics beyond the Standard Model, 
and could thus provide an indirect window into resolutions of the nature of dark matter, 
the origins of the matter-antimatter asymmetry, or the dynamics of the electroweak scale itself.
Decades of LFUV measurements have yielded results predominantly in agreement with SM predictions.
Various strong constraints have been obtained from (semi)leptonic decays of light hadrons, gauge bosons, or leptonic $\tau$ decays (see~\cite{Zyla:2020zbs}),
among many other measurements. 
A notable recent addition is the measurement of $\BR(W \to \tau\nu)/(W \to \mu\nu)$~\cite{Aad:2020ayz}, 
resolving a long-standing LFUV anomaly from LEP that deviated from the SM prediction at $2.7\sigma$.
Moreover, sources of LFUV that implicate NP interactions with the first two quark generations 
are typically strongly constrained by, e.g., precision $K$-$\Kbar$ and $D$-$\Dbar$ mixing measurements.
Such LFUV bounds involving third generation quarks, however, are typically much weaker~\cite{Cerri:2018ypt}.

This review focuses on the rich experimental landscape for testing LFUV in semileptonic $b$-hadron decays. 
Not only do these decays provide a high statistics laboratory to measure LFUV that is relatively theoretically clean, 
but results from the last decade of measurements have indicated anomalously high rates for various semitauonic $b \to c \tau \nu$ decays compared to precision SM predictions.
In particular, the ratios 
\begin{equation}
	\RDx = \frac{\BR(\BDxtaunu)}{\BR(\BDxellnu)}\,, \qquad \ell =e\,,~\mu\,, 
\end{equation}	
where \Dx refers to both $D$ and $D^*$ mesons, deviate from SM predictions at the $3\sigma$ level when taken together~\cite{Amhis:2019ckw}.
(We revisit later the construction of these world averages and their degree of tension with the SM.)
Apart from these results, there are additional measurements for various other $b \to c \tau \nu$ decays and other observables, 
including $\RJ$, the $\tau$ polarization and $D^*$ longitudinal fractions (see Sec.~\ref{sec:measurements}).
Some of these measurements presently agree with SM predictions only at the $1.6-1.8\sigma$ level, 
and when combined with $\RDx$ can mildly increase the degree of tension with the SM.
Some tensions also currently exist in several $b \to s ee$ versus $b \to s \mu\mu$ transitions, each at the $2.5\sigma$ level~\cite{Aaij:2017vbb,Aaij:2019wad}.
See \cite{Ciezarek:2017yzh,Bifani:2018zmi} for prior experimental reviews that consider aspects of LFUV in semileptonic decays.

Upcoming runs of the LHC, the high-luminosity (HL)-LHC, and Belle II will yield large new datasets for a wide range of $b \to c \tau \nu$ and $b \to u \tau \nu$ processes. 
Given this expected deluge of data, it is important to review and synthesize our understanding of the various strategies and channels through 
which LFUV might be discovered.
To this end, we undertake this review along two different threads.
First, in Sec.~\ref{sec:theory} we provide a compact yet comprehensive overview of the current theoretical state of the art for the SM (and NP) description of semitauonic decays.
This includes not only a survey of SM predictions in the literature, but also several novel results first calculated for this review.

Second, we provide a substantial review of the various experimental methods and strategies used to measure LFUV.
This includes an assessment of the various experimental methods in Sec.~\ref{sec:experiments}, 
and a summary of the LFUV measurements published to date in Sec.~\ref{sec:measurements}.
An effort has been made to synthesize all of the available information from current measurements and,
when possible, to make direct comparisons across experiments that provide further context. 
For instance, we present the various approaches towards reconstructing the momentum of the parent $b$-hadron in Sec.~\ref{sec:bframe} 
and provide a comparison between the two hadronic $B$ tag measurements of \RDx by \babar and Belle in Sec.~\ref{sec:bfactories_hadtag_rdx}.

These two threads of the review are woven in Secs.~\ref{sec:systematics} and~\ref{sec:interpretation} 
into discussions of the main challenges arising from systematic uncertainties, 
and into discussions of current interpretations and combinations of the data, respectively.
In particular, in Sec.~\ref{sec:systematics} we provide an extended analysis of the main sources of systematic uncertainty in the LFUV measurements, 
and the prospects to control them in the future down to the percent level.
This will be essential for establishing a conclusive tension with the Standard Model.
We examine key challenges in computation, the modeling of $b$-hadron semileptonic decays in signal and background modes, and estimations of other important backgrounds.
We also point out the potential sensitivity of $\RDx$ analyses to the assumptions used for the \BDsstaunu branching fractions (Sec.~\ref{sec:syst:dss_bf}), 
which are presently overestimated compared to SM predictions. 

Section~\ref{sec:interpretation} begins by examining the $\RDx$ results and other SM tensions for different light-lepton normalization modes or isospin channels,
before turning to revisit entirely the world average combinations of the $\RDx$ ratios. 
We specifically analyze the sensitivity of these combinations to the treatment of the correlation structure 
assigned to the uncertainties from \BDssellnu decays across different measurements,
and show they may vary the degree of their current $\sim3\sigma$ tension with the SM over approximately a $1\sigma$ range.
As an illustration, incorporating such correlations as a free fit parameter in the combination, 
we show that the resulting $\RDx$ world averages 
would feature a tension of $3.6$ standard deviations with respect to the SM. 
This is $0.5\sigma$ higher than the current world average~\cite{Amhis:2019ckw}. 
We further explore a comparison of inclusive versus exclusive measurements;
caveats and challenges in establishing NP interpretations of the current $\RDx$ anomalies;
and possible connections to anomalies in neutral current rare $B$ decays.

Beyond the current state of the art, in Sec.~\ref{sec:outlook} we proceed to explore the power of 
future LFUV ratio measurements for a variety of hadronic states,
taking into account the discussed prospects for the evolution of the systematic uncertainties
and the data samples that LHCb and Belle~II are expected to collect over the next two decades (Sec.~\ref{sec:outlook:ratios}). 
The power of future analyses to exploit full differential information is briefly explored (Sec.~\ref{sec:outlook:distributions}),
as well as the role of proposed future colliders (Sec.~\ref{sec:outlook:future_colliders}).

\def\rDs{r^*}
\def\rD{r}
\def\SR{S\!R}
\def\SL{S\!L}
\def\VR{V\!R}
\def\VL{V\!L}

\section{Theory of Semileptonic Decays}
\label{sec:theory}
In this section we introduce the foundational theoretical concepts required to describe $b \to c l \nu$ semileptonic decays.
Throughout this review, we adopt the notation
\begin{equation}
	l = \tau\,,~\mu\,,~e\,, \qquad
	\ell  = \mu\,,~e\,.
\end{equation}
While our focus is the SM description of $b \to c l \nu$, in some contexts we present a model-independent discussion, 
in order to accommodate discussion of beyond Standard Model (BSM) physics.
We discuss first \BDxlnu decays, since they are of predominant experimental importance in current measurements, before turning to processes involving excited states, 
charm-strange mesons, charmonia, baryons, as well as $b \to u l \nu$ and inclusive processes.
The LFUV observables (anticipating their definitions in later parts of the review) for which predictions are discussed, and their respective sections, comprise:
\begin{align*}
	\RDx: \quad & \text{Sec.~\ref{subsec:SM_pred}}  			& F_{L}(\Dstar),\,P_{\tau}(\Dx)\!: \quad & \text{Sec.~\ref{sec:th:longpol}} \\
	\RDdouble: \quad & \text{Sec.~\ref{sec:th:other_states}} 		& \calR(D_s^{(*)}): \quad & \text{Sec.~\ref{sec:th:other_states}} \\
	\RJ: \quad &  \text{Sec.~\ref{sec:th:other_states}} 			& \calR(\Lambda^{(*)}_c): \quad & \text{Sec.~\ref{sec:th:other_states}} \\
	\calR(\pi): \quad & \text{Sec.~\ref{sec:th:vub} } 				& \calR(\rho),\, \calR(\omega)\!: \quad & \text{Sec.~\ref{sec:th:vub} } \\
	\calR(X_c): \quad & \text{Sec.~\ref{sec:th:incl}}\,. 	&& \\
\end{align*}

%%%%%%%%%%%%%%%%%%%%%%%%%%%%%%%%%%%%%%%%%
\subsection{SM operator and amplitudes}
In the SM, $b \to c l \nu$ processes are mediated by the weak charged current, generating the usual $V-A$ four-Fermi operator
\begin{equation}
	\label{eqn:smop}
	\mathcal{O}_{\text{SM}} = 2\sqrt{2} G_F V_{cb} \big( \cbar \g^\mu P_L b\big) \big( \bar{l} \g_\mu P_L \nu_l)\,,
\end{equation}
at leading electroweak order. Here we use the projectors $P_{L,R} = (1\mp \g^5)/2$, and $G_F^{-1} = 8 m_W^2 / (\sqrt{2} g_2^2) = \sqrt{2}v^2$, with $v \simeq 246.22$\,GeV~\cite{Zyla:2020zbs}.
Further, $g_2$ denotes the $SU(2)$ weak coupling constant, and $V_{cb}$ is the quark-mixing Cabibbo-Kobayashi-Masakawa (CKM) matrix element.
The corresponding amplitude for this charged-current process has the diagrammatic form
\begin{equation}
	\mathcal{A}_{\textrm{SM}} \quad = \quad \begin{gathered} \includegraphics[width = 3.25cm]{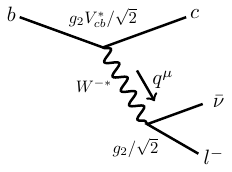} \end{gathered}\,,
\end{equation}
in which the quarks may be ``dressed'' into various different hadrons. 
It is conventional to define the momentum $q = p - p' = p_l + p_\nu$
where $p$ ($p'$) is the beauty (charm) hadron momentum.

\begin{figure}[t]
	\includegraphics[width=0.36\linewidth]{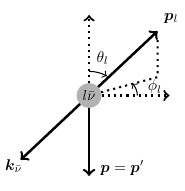}\hspace{-0.4cm}
	\includegraphics[width=0.36\linewidth]{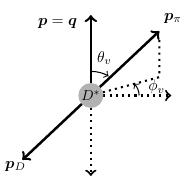}\hspace{-0.5cm}
	\includegraphics[width=0.36\linewidth]{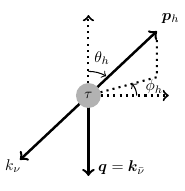}
	\caption{Left: Definition of the $\theta_l$ and $\phi_l$ helicity angles in the lepton pair rest frame. 
	Center: Definition of the $\theta_v$ and $\phi_v$ helicity angles in the $D^*$ rest frame. 
	Right: Definition of the $\theta_h$ and $\phi_h$ helicity angles in the $\tau$ rest frame frame, for $B \to \Dx (\tau \to h\nu) \bar\nu$ decay.}
	\label{fig:polar_def}
\end{figure}

The leptonic amplitude $W \to l \nu$ always take the form of a Wigner-$D$ function $D^j_{m1,m2}(\theta_l, \phi_l)$, with $j = 0$ or $1$, and $|m_{1,2}| \le j$.
The helicity angles $\theta_l$ and $\phi_l$ are defined herein as in Fig.~\ref{fig:polar_def}. 
We show also in Fig.~\ref{fig:polar_def} the definition of helicity angles for subsequent $D^* \to D \pi$ or $\tau \to h\nu$ decays, for example, where $h$ is any hadronic system or $\ell\nu$.
The helicity angle definition also applies for the case of $D^* \to D \gamma$, though with a different fully differential rate.
Some literature uses the definition $\theta_l \to \pi- \theta_l$, such that caution must be used in adapting fits to fully differential measurements from one convention to the other. 
The phase $\phi_l$ is unphysical unless defined with reference to spin-polarizers of the charm or beauty hadronic system or the lepton,
such as the subsequent decay kinematics of the $\tau$ or charm hadron, or the spin of the initial $b$-hadron. 
For example, in $B \to (D^* \to D \pi) \ell \nu$, the only physical phase is $\chi \equiv \phi_l - \phi_v$.

%%%%%%%%%%%%%%%%%%%%%%%%%%%%%%%%%%%%%%%%%
\subsection{Hadronic matrix elements and form factors}
The predominant theory uncertainty in $\BDxlnu$ arises in the description of the hadronic matrix elements $\langle \Dx | \cbar\,\Gamma\,b | \Bbar \rangle$,\footnote{
All definitions and sign conventions hereafter apply to $b \to c$ transitions; they may be extended to $\bbar \to \cbar$ with appropriate sign changes. 
To emphasize this, while we do not typically distinguish between $\Bbar \to \Dx$ and $B \to \Dxbar$ in this discussion, 
we do retain such notation in the explicit definition of matrix elements or where charge assignments of other particles have been made explicit.
Throughtout the review, inclusion of charge-conjugate decay modes is implied, unless stated otherwise.}
where (anticipating a discussion of New Physics (NP) below) $\Gamma$ is any Dirac operator.
More generally, one seeks a theoretical framework to describe the matrix elements $\big\langle {}^{2s_c+1}(L^c)_{J_c} \big| \cbar\,\Gamma\,b \big| {}^{2s_b+1}(L^b)_{J_b} \big\rangle$, 
using here the spectroscopic notation to describe the hadron in terms of its quark constituents' total spin $s$, 
their orbital angular momentum $L = S$, $P$, $D$, \ldots, and the total angular momentum of the hadron $J$.
We focus first on the description for $B \to \Dx$, i.e. ${}^1S_0 \to {}^1S_0$ or ${}^3S_1$: The ground state charmed mesons.

Hadronic matrix elements incorporate non-perturbative QCD and cannot be computed from first principles.
However, the transition matrix element between hadrons of definite spin and parity, mediated by any particular operator, 
can be described by a finite set of amplitudes involving partial waves of definite orbital angular momentum.
Each such amplitude can be represented by a tensor product of the external momenta, polarizations and spins, 
multiplied by an unknown hadronic function: a form factor.
One may represent the matrix element by different linear combinations of these tensor products, 
defining a basis for the form factors.

For $B \to \Dx$ SM transitions, the matrix elements are represented by two (four) independent form factors. 
In terms of two (three) common form factor bases,
\begin{subequations}
\label{eqn:FFdefs}
\begin{align}
\MatrixElem{D}{\cbar \g^\mu b}{\Bbar} 
	& = f_+(p + p')^\mu \nn\\
	& \quad + (f_0 - f_+) q^\mu (m_B^2 - m_D^2)/q^2  \\
	& = \sqrt{m_B m_D} \big[ h_+(v+v')^\mu + h_-(v-v')^\mu\big] \nn \\
\MatrixElem{D^*}{\cbar \g^\mu b}{\Bbar} 
	& = 2 i \widetilde{g}\, \varepsilon^{\mu\nu\alpha\beta}\,\epsilon^*_{\nu}p'_\alpha p_\beta \\
	& = i\sqrt{m_B m_{D^*}}\, h_V\, \varepsilon^{\mu\nu\alpha\beta}\,\epsilon^*_{\nu}v'_\alpha v_\beta \nn \\
	& = 2 i V (m_B + m_{D^*})^{-1}\varepsilon^{\mu\nu\alpha\beta}\,\epsilon^*_{\nu}p'_\alpha p_\beta \nn \\
\MatrixElem{D^*}{\cbar \g^\mu \g^5 b}{\Bbar}  
	& = f \epsilon^{*\mu} + a_+ \epsilon^* \!\cdot\! p\, (p + p')^\mu+ a_- (\epsilon^* \!\cdot\!  p) q^\mu \nn \\
	& = \sqrt{m_B m_{D^*}}\, \big[h_{A_1} (w+1)\epsilon^{*\mu} \\
 	& \quad - h_{A_2}(\epsilon^* \!\cdot\!  v)v^\mu  - h_{A_3}(\epsilon^* \!\cdot\! v)v'^\mu \big]\,, \nn \\
	& = A_1(m_B + m_{D^*})\epsilon^{*\mu} -A_2  \frac{\epsilon^* \!\cdot\! p(p + p')^\mu}{m_B + m_{D^*}} \nn \\
	& \quad + 2m_{D^*}q^\mu(A_0 - A_3)( \epsilon^* \!\cdot\! p)/q^2 \,, \nn
\end{align}
\end{subequations}
noting $\MatrixElem{D}{\cbar \g^\mu \g^5 b}{\Bbar} = 0$ because of angular momentum and parity conservation.
Here we have used the spectroscopic basis $\{f_+, f_0, f, \widetilde{g}, a_\pm\}$ (cf.~\cite{PhysRevD.39.799});\footnote{
The form factor $\widetilde{g}$ is often written as $g$, but should not be confused with $g=2\widetilde{g}$ in the helicity basis defined in Eq.~\eqref{eqn:SHrel}.} 
the heavy quark symmetry (HQS) basis $\{h_\pm, h_V, h_{A_{1,2,3}}\}$ (e.g.~\cite{Neubert:1993mb}); 
and the basis $\{V, A_{0,1,2,3}\}$ \cite{Wirbel:1985ji}, in which 
$2m_{D^*}A_3 = A_1(m_B + m_{D^*}) - A_2(m_B - m_{D^*})$.
Furthermore, the velocities $v = p/m_B$ and $v' = p'/m_{\Dx}$, $\epsilon^*$ is the $D^*$ polarization vector, and the recoil parameter 
\begin{equation}
	w = v \cdot v' = \frac{m_B^2 + m_{\Dx}^2 - q^2}{2 m_B m_{\Dx}}\,.
\end{equation} 
The form factors are functions of $q^2$ or equivalently $w$. 
Their explicit forms may also involve the scheme-dependent parameters $m_b/m_c$ and $\alpha_s$, though any such scheme dependency must vanish in physical quantities.
In the HQS basis, $h_{A_1}$ and the three form factor ratios
\begin{align}
 R_1(w) & = \frac{h_V}{h_{A_1}}\,, \qquad  R_2(w) = \frac{h_{A_3} + \rDs\, h_{A_2}}{h_{A_1}}\,,~\text{and}\\
 R_0(w) & = \frac{(w+1)h_{A_1} - (w -\rDs)h_{A_3} - (1 - w\rDs)h_{A_2}}{(1+\rDs)\, h_{A_1}}\,, \nn
\end{align}
where $r^{(*)} = m_{D^{(*)}}/m_B$, fully describe the $B \to D^*$ transition. Note $R_0$ enters only in terms proportional to $m_l$.

Particular care must be taken with sign conventions in Eqs.~\eqref{eqn:FFdefs}:
For $B \to \Dx$, the conventional choice in the literature, and here, is such that $\text{Tr}[\g^\mu\g^\nu\g^\rho\g^\sigma\g^5] = +4i \varepsilon^{\mu\nu\rho\sigma}$, 
equivalent to fixing the identity $\sigma^{\mu\nu} \g^5 \equiv -(i/2)\, \varepsilon^{\mu\nu \rho \sigma} \sigma_{\rho \sigma}$, with $\sigma_{\mu\nu} = (i/2)[\g^\mu, \g^\nu]$.
One may further choose either $\varepsilon^{0123} = +1$ or~$-1$. 
In $B \to D^{**}$ literature, as well as $\Lambda_b \to \Lambda_c$, typically the choice is instead $\text{Tr}[\g^\mu\g^\nu\g^\rho\g^\sigma\g^5] = -4i \varepsilon^{\mu\nu\rho\sigma}$,
equivalent to $\sigma^{\mu\nu} \g^5 \equiv +(i/2)\, \varepsilon^{\mu\nu \rho \sigma} \sigma_{\rho \sigma}$.
These sign choices affect the sign of $R_1$,
but leave physical quantities unchanged provided they are used consistently both in the form factor definitions \emph{and} in the calculation of the amplitudes. 
Care must be taken in adapting form factor fit results obtained in one convention to expressions defined in the other.
In our sign conventions, the form factor ratio $R_1 > 0$.

An additional common choice for $B \to D^*$ decays is the helicity basis (cf.~\cite{Boyd:1995sq,Boyd:1997kz}), 
with form factors $\{g, f, F_1, P_1\}$, that are particularly convenient for expressing the $B \to D^*$ helicity amplitudes.
Explicit relations between the HQS and helicity bases are
\begin{subequations}
\label{eqn:SHrel}
\begin{align}
	& h_{A_1}   = \frac{f}{m_B \sqrt{\rDs}(w+1)}\,, \qquad  h_{V}  =  g m_B \sqrt{\rDs} \\
	&h_{A_1} \big(w-\rDs -(w-1)R_2\big) = \frac{F_1}{m_B^2 \sqrt{\rDs}(w+1)}\,, \label{eqn:fF1rel}\\
	& h_{A_1} R_0 = P_1\,.
\end{align} 
\end{subequations}
The SM differential rate can then be written compactly in terms of Legendre polynomials of $\cos\theta_l$,
\begin{align}
	& \frac{d^2\Gamma}{dw\,d\cos\theta_\ell}  
	 = 2\Gamma_0 \sqrt{w^2-1} \, {\rDs}^3 \bigg[\frac{\bar{q}^2 \!- \!  r_l^2}{\bar{q}^2}\bigg]^2\Bigg\{ \label{eqn:fulldiff} \\
	& \quad \bigg(1 + \frac{r_l^2}{2 \bar{q}^2}\bigg)\big(\mathcal{H}_+ + 2\bar{q}^2 \mathcal{H}_1\big) + \frac{ 3 r_l^2}{2\bar{q}^2} \mathcal{H}_0 \nn \\
	& \quad + \cos\theta_l \mathcal{H}_{+0} + \frac{3\cos^2\!\theta_l \!-\!1}{2}\bigg[\frac{\bar{q}^2 \!- \!  r_l^2}{\bar{q}^2}\bigg]\big(\bar{q}^2 \mathcal{H}_1 - \mathcal{H}_+ \big)\Bigg\}\,, \nn
\end{align}
in which $\Gamma_0 \equiv G_F^2 \eta_{\text{ew}}^2 |V_{cb}|^2/(192\pi^3)$, $r_l = m_l/m_B$, $\bar{q}^2 = q^2/m_B^2 = 1 - 2\rDs w + {\rDs}^2$, 
$\eta_{\text{EW}} \simeq 1 + \alpha/\pi\log(m_Z/m_B) \simeq1.0066$ is an electroweak correction~\cite{Sirlin:1981ie}, and
\begin{subequations}
\label{eqn:Hexpr}
\begin{align}
	\mathcal{H}_1 & =  \frac{f^2}{\rDs m_B^2} + g^2 \rDs m_B^2 (w^2 \!-\!1) \,,\\
	\mathcal{H}_+ & = \frac{F_1^2}{\rDs m_B^4}\,,\\
	\mathcal{H}_0 & =  P_1^2  (\rDs \!+\! 1)^2(w^2 \!-\! 1)\,,  \\ 
	\mathcal{H}_{+0} & = 6\bar{q}^2 f g \sqrt{w^2-1} - \frac{3 r_l^2}{\bar{q}^2} \sqrt{\mathcal{H}_+ \mathcal{H}_0}\,.
\end{align}
\end{subequations}
The $\theta_l$-independent term in Eq.~\eqref{eqn:fulldiff} is simply $1/2~d\Gamma/dw$. The overall sign of the $\cos\theta_l$ term, 
and the relative sign of the $f g$ term in $\mathcal{H}_{+0}$, are sensitive to sign conventions. 
In the massless lepton limit, it is common to express the differential rate $d\Gamma/dw$ in terms of the single form factor combination
\begin{equation}
	\mathcal{F}^2(w) = \frac{\mathcal{H}_+ + 2\bar{q}^2 \mathcal{H}_1}{(1-\rDs)^2(w+1)^2+  4w(w+1)\bar{q}^2}\,,
\end{equation}
normalized such that $\mathcal{F}(1) = h_{A_1}(1)$.

The $B \to D$ rate may be expressed similarly. In the form factor basis $\{\mathcal{G} \equiv V_1, S_1\}$,\footnote{Some literature uses the notation $V_1$, while others $\mathcal{G}$.} defined via
\begin{subequations}
\begin{align}
	\mathcal{G} \equiv V_1 & = h_+ - \frac{1-\rD}{1+\rD}h_-\,, \\
	S_1 & = h_+ - \frac{1+\rD}{1-\rD}\frac{w-1}{w+1} h_-\,,
\end{align}
\end{subequations}
the SM differential rate has the same form as Eq.~\eqref{eqn:fulldiff} and Eqs.~\eqref{eqn:Hexpr}, but with $\rDs \to \rD$, 
\begin{subequations}
\begin{align}
	\mathcal{H}_+ & = V_1^2(1+\rD)^2(w-1)^2\,, \\
	\mathcal{H}_0 & = S_1^2(1-\rD)^2(w+1)^2\,, 
\end{align}
\end{subequations}
and, by definition, no $f$ or $g$ terms, i.e $\mathcal{H}_1 = 0$ and $\mathcal{H}_{+0} = -3r_l^2/\bar{q}^2 \sqrt{\mathcal{H}_+ \mathcal{H}_0}$.

Note that the expressions of this section apply similarly to any other ${}^1S_0 \to {}^1S_0$ or ${}^3S_0$ transition, such as $B \to \pi l\nu$ or $B \to \rho l \nu$ (with the additional replacement of $V_{cb} \to V_{ub}$).

%%%%%%%%%%%%%%%%%%%%%%%%%%%%%%%%%%%%%%%%%
\subsection{Theoretical frameworks}
\label{sec:th:frames}

Various theoretical approaches exist to parametrize the $B \to \Dx$ or other exclusive decay form factors. Broadly, these fall into four overlapping categories:
\begin{enumerate}
	\item Use of the functional properties of the hadronic matrix elements---analyticity, unitarity, and dispersion relations---to constrain the form factor structure;
	\item Use of heavy quark effective theory (HQET) to generate order-by-order relations in $1/m_{c,b}$ and $\alpha_s$ between form factors; 
	\item Various quark models, including those that may approximately compute the form factors (in various regimes), such as QCD sum rule (QCDSR) and light cone sum rule (LCSR) approaches; and
	\item Lattice QCD (LQCD) calculations, presently available only for a limited subset of form-factors and kinematic regimes.
\end{enumerate}

The details of the various approaches to the form factor parametrization are particularly important for measurements that are sensitive to the differential shape of exclusive semileptonic decays, 
such as the extraction of the CKM matrix element $|V_{cb}|$.
Hadronic uncertainties, however, mostly factor out of observables that consider ratios of $|V_{cb}|$-dependent quantities,
including measurements that probe lepton universality relations between the $\BDxellnu$ and $\BDxtaunu$ decays or other exclusive processes.
Instead, in the latter context the main role and importance of form factor parametrizations lies in their ability to generate predictions for lepton universality relations, and the precision thereof.

%%%%%%%%%%%%%%%%%%%%%%%%%%%%%%%%%%%%%%%%%
\subsubsection{Dispersive bounds}
\label{sec:th:disp}
A dispersion relations-based approach does not alone generate lepton universality relations between the $\BDxlnu$ rates or other exclusive processes,
but does provide crucial underlying theoretical inputs to approaches that do.
The dispersive approach \cite{Boyd:1995sq,Boyd:1997kz} begins with the observation that the matrix element $\langle H_c | J | H_b\rangle$ for a hadronic transition $H_b \to H_c$,
mediated by current $J = \cbar\,\Gamma\,b$, may be analytically continued beyond the physical regime $q^2 < (m_{H_b} - m_{H_c})^2 \equiv q_-^2$ into the complex $q^2$ plane. 
For $q^2 > (m_{H^0_b} + m_{H^0_c})^2 \equiv q_+^2$, where $H^0_{c,b}$ denote the lightest pair of hadrons that couple to $J$, 
the matrix element features a branch cut from the crossed process $H^0_b H_c^{0\dagger}$ pair production.
For $B \to \Dstar$ processes, it is typical to take $q_+ \equiv (m_B + m_{\Dstar})^2$ for both vector and axial vector currents. 
For e.g. $B_c \to \Jpsi$, the branch points are taken as $(m_{B} + m_{D})^2$ and $(m_{B^{*}} + m_{D})^2$ for vector  and axial vector currents, respectively.
A $bc$ bound state that is created by $J$ but with mass $m^2 < q_+^2$, is a ``subthreshold'' resonance.

The conformal transformation 
\newcommand{\sqrtb}[1]{\sqrt{\smash[b]{#1}}\vphantom{b}}
\begin{equation}
	z(q^2, q_0^2) = \frac{\sqrtb{q_+^2 - q^2}- \sqrtb{q_+^2 - q_0^2}}{\sqrtb{q_+^2 - q^2} + \sqrtb{q_+^2 - q_0^2}}
\end{equation}
maps $q^2 > q_+^2$ to the boundary of the unit circle $|z| = 1$, centered at $q^2 = q_0^2$. 
Two common choices of $q_0^2$ are $q_-^2$, in which case $z(w=1) = 0$, or $q_+^2 (1 - [1- q_-^2/q_+^2]^{1/2}) \equiv q_{\text{opt}}^2$, which minimizes $|z(q^2 = 0)|$.
This allows the matrix element to be written as an analytic function of $z$ on the unit disc $|z| \le 1$, up to simple poles that are expected at each `sub-threshold'  resonance. 
These poles must fall on the interval $q_-^2 \le q^2 \le q_+^2 \Leftrightarrow (0 \ge) z_- \ge z \ge -1$.

The second ingredient is the vacuum polarization $\Pi_J = i\int d^4 x e^{iqx} \langle 0 | TJ^\dagger(x) J(0) | 0 \rangle$,
which obeys a once-subtracted dispersion relation
\begin{equation}
	\label{eqn:dispbnd}
	\chi_J(q^2) \equiv \frac{\partial \Pi_J}{\partial q^2} = \frac{1}{\pi}\int \frac{dt}{(t-q^2)^2} \text{Im}\Pi_J\,.
\end{equation}
The QCD correlator $\chi_J$ can be computed at one-loop in perturbative QCD for $q^2 > q_+^2$, and then analytically continued to $q^2 <  q_-^2$.
$\text{Im}\Pi_J$ may be reexpressed as a phase-space-integrated sum over a complete set of $b$- and $c$-hadronic states 
$\sim \sum_{X=H_b H_c^\dagger,\ldots} |\langle 0 | J | X \rangle|^2$ 
with appropriate parity and spin. For $J = \cbar \gamma^\mu b$, one may have $H_{b} H_c^\dagger = BD^\dagger$, $BD^{*\dagger}$ and so on.
The positivity of each summand allows the dispersion relation to provide an upper bound---a so-called `weak' unitarity bound---for any given hadron pair $H_{b} H_c^\dagger$.
(A ``strong'' unitarity bound would, by contrast, impose the upper bound on a finite sum of hadron pairs coupling to $J$.)
Crossing symmetry permits these bounds to be applied to the transition matrix elements $\langle H_c | J | H_b\rangle$ of interest here.

Making use of the conformal transformation, the unitarity bound can be expressed in the form
\begin{equation}
	\label{eqn:utbnd}
	\int_{|z| = 1} \frac{dz}{2\pi i z} \sum_i | P^J_i(z) \phi^J_i(z) F^J_i(z)|^2 \le 1\,.
\end{equation}
in which $F^J_i$ is a basis of form factors and the ``outer'' functions $\phi^J_i$ are analytic weight functions 
that encode both their $q^2$-dependent prefactors arising in $\langle H_c | J | H_b\rangle$, as well as incorporating the $1/\sqrt{\pi \chi_J}$ prefactor. 
The additional Blaschke factors $P^J_i$ satisfy $|P^J_i(|z| = 1)| = 1$ by construction, and do not affect the integrand on the $|z| =1$ contour. 
However, the choice $P^J_i = \prod_\alpha(z-z_{\alpha,i})/(1- z z_{\alpha,i})$ explicitly cancels the known poles at $z= z_{\alpha,i}$ on the negative real axis.
Each term in the sum must then be analytic, i.e., $P^J_i(z) \phi^J_i(z) F^J_i(z) = \sum_{n=0}^\infty a^{Ji}_{n} z^n$, 
so that Eq.~\eqref{eqn:utbnd} requires the $a^{Ji}_{n}$ coefficients to satisfy a unitarity bound $\sum_{i,n}  |a^{Ji}_{n}|^2 \le 1$.

The Boyd-Grinstein-Lebed (BGL) parametrization~\cite{Boyd:1995sq,Boyd:1997kz} uses this approach to express the $f$, $g$, $F_1$ and $P_1$ form factors in terms of an analytic expansion in $z = z(q^2, q_-^2)$.
In particular for the light lepton modes, with $F_A = f, F_1$, 
\begin{align}
	g(z) & = \frac{1}{P_V(z) \phi_g(z)} \sum_n a^g_n z^n\,,&  \sum_{n}  |a^{g}_{n}|^2 & \le 1\,, \nn \\
	F_A(z) & = \frac{1}{P_A(z) \phi_{F_A}(z)} \sum_n a^{F_A}_n z^n\,,&  \sum_{F_A,n}  |a^{F_A}_{n}|^2 & \le 1\,, \nn
\end{align}
further noting that $F_1(q_-^2)/\phi_{F_1}(q_-^2)= f(q_-^2) m_B(1-\rDs)/\phi_{f}(q_-^2)$ from Eq.~\eqref{eqn:fF1rel}. 
This relatively unconstrained parameterization provides a hadronic model-independent approach to measuring $|V_{cb}|$ from light leptonic $\BDsellnu$ modes, 
but does not relate $\BDstaunu$ to $\BDsellnu$: 
E.g. a fit to light lepton data, taking $m_\ell \to 0$, to determine $f$, $g$, $F_1$ provides no \emph{prediction} for $P_1$, and hence no prediction for the $\BDstaunu$ rate.
(The general SM expectation remains, however, that the unitarity bound for $P_1$ should not be violated in a direct fit to $\BDstaunu$ data.)
Instead, additional theoretical inputs are required.

%%%%%%%%%%%%%%%%%%%%%%%%%%%%%%%%%%%%%%%%%
\subsubsection{Heavy quark effective theory}
\label{sec:th:hqet}
HQET inputs may be combined with the BGL approach, in order to generate SM (or NP) predictions for lepton universality observables.
A ``heavy'' hadron is defined as containing one heavy valence quark---i.e. the heavy quark mass $m_Q \gg \mathcal{O}(\lqcd)$, the QCD scale---dressed 
by light quark and gluon degrees of freedom---so-called brown muck---in a particular spin and parity state.
An HQET~\cite{Isgur:1989vq, Isgur:1989ed, Eichten:1989zv, Georgi:1990um} (for a review, see e.g.~\cite{Neubert:1993mb})
is an effective field theory of the brown muck, in which interactions with the heavy quark enter at higher orders in $1/m_Q$.
An apt analogy arises in atomic physics in which the electronic states are insensitive to the nuclear spin state, up to hyperfine corrections.
This provides a hadronic model-independent parametrization of not only the spectroscopy of heavy
hadrons but also order-by-order in $1/m_Q$ relations between their transition matrix elements.
The form factors of $\BDxellnu$ are then related to those of $\BDxtaunu$, allowing for lepton universality predictions.

In this language, the spectroscopic $^1S_0$ and $^3S_1$ states---e.g. the $D$ and $D^*$ or $B$ and $B^*$---may instead be considered to belong to 
a heavy quark (HQ) spin symmetry doublet of a pseudoscalar (P) and vector (V) meson, 
formed by the tensor product of the light degrees of freedom in a spin-parity $s^P_\ell = 1/2^-$ state, 
combined with the heavy quark spin: $\mathbf{\frac{1}{2}_{\text{HQ}}} \otimes \mathbf{\frac{1}{2}_{\text{light}}} = \mathbf{0} \oplus \mathbf{1}$.
Their masses can be expressed as
\begin{equation}
	m_{P,V} = m_Q + \bar\Lambda - \frac{\lambda_1}{2m_Q} \mp \frac{(2J_{V,P}+1)\lambda_2}{2m_Q} + \ldots\,,
\end{equation}
where $\bar\Lambda =  \mathcal{O}(\lqcd)$ is the brown muck kinetic energy for $m_{Q} \to \infty$, and $\lambda_{1,2} = \mathcal{O}(\lqcd^2)$.
Furthermore, one expects that in the limit that $m_Q \to \infty$ (and, $\alpha_s \to 0$)---the heavy quark limit---the physics of heavy 
hadron flavor-changing transitions such as $B \to \Dx$ 
should be insensitive to---and therefore preserve---the spin of the underlying heavy quarks, while being sensitive to the change in heavy quark velocity.

Following this intuition, the QCD kinetic term \mbox{$\bar{Q} (i\slashed{D} - m_Q) Q$} may itself be reorganized into an effective theory of brown muck---i.e. an HQET---parametrized 
by the heavy quark velocity $v = p_Q/m_Q$.
This effective theory features a $1/m_Q$ expansion in which the leading order terms conserve heavy quark spin, while higher order terms in $1/m_Q$ do not. 
A heavy quark flavor violating interaction like $J = \cbar\,\Gamma\,b$ can be similarly reorganized, 
such that at leading order, the transition is sensitive only to the difference of the incoming and outgoing heavy hadron velocities $v$ and $v'$, respectively.
It is then natural to express the matrix elements as in Eq.~\eqref{eqn:FFdefs}, with the natural form factor basis in the SM being $h_\pm, h_V, h_{A_{1,2,3}}$.

When organized in this way, the key result is that any $B \to \Dx$ matrix element can be written as a spin-trace
\begin{equation}
\label{eqn:TraceLM}
\frac{\langle \Dx |\, \cbar\, \Gamma\, b\, | \Bbar \rangle}{\sqrt{m_{\Dx} m_B}}
  = - \xi(w)\, \text{Tr} \big[ \bar H^{(c)}_{v'}\, \Gamma\,H^{(b)}_v \big]  + \mathcal{O}(\varepsilon_c, \varepsilon_b, \hat\alpha_s)\,,
\end{equation}
where $H^{(c,b)}$ are HQET representations of the HQ doublet and $\xi(w)$ is a leading \emph{Isgur-Wise function}. 
Higher order terms in $\varepsilon_{c,b} = \bar\Lambda/(2m_{c,b})$,
can be similarly systematically constructed in terms of universal \emph{subleading} Isgur-Wise functions, 
while radiative corrections in $\hat\alpha_s = \alpha_s/\pi$ can be incorporated at arbitrary fixed order.
Heavy quark flavor symmetry implies that $\xi(1) = 1$, preserved at order $\varepsilon_{c,b}$ by Luke's theorem.

The CLN parametrization~\cite{Caprini:1997mu} applies dispersive bounds to the $B \to D$ form factor $V_1$, expanded up to cubic order as
\begin{equation}
	\frac{V_1(w)}{V_1(1)} = 1-\rho^2_1(w-1)+c_1(w-1)^2 +d_1(w-1)^3  +\ldots.
\end{equation}
It thus extracts approximate relations between the parameters $\rho^2$, $c_1$ and $d_1$, 
by saturating the dispersive bounds at (the then) $1\sigma$ uncertainty in the QCD correlators $\chi_J$. 
The parametrization then makes use of heavy quark symmetry to relate this form factor to all other form factors in the $B \to \Dx$ system, 
incorporating additional, quark model inputs from QCD sum rules (QCDSR, see below), to constrain the $1/m_{c,b}$ terms.
In particular, predictions are obtained for a $z$ expansion of $h_{A_1}$, with coefficients dependent only on $\rho^2_1$, 
plus predictions for $R_{1,2,0}(w)$ up to a fixed order in $(w-1)$: 
$R_i(w) = R_i(1) + R^{\prime}_i(1)(w-1) + 1/2R^{\prime\prime}_i(1)(w-1)^2 + \ldots$.

The intercepts $R_{i}(1)$ are theoretically correlated order-by-order in the HQ expansion with the slope and gradients $R_{i}^{(\prime,\prime\prime)}(1)$, 
and therefore must be determined simultaneously when measured.
A common experimental fitting practice of floating $R_{1,2}(1)$ while keeping $R_{1,2}^{(\prime,\prime\prime)}(1)$ fixed to their QCDSR predictions is
inconsistent with HQET at subleading order, when fits are performed to recent higher precision unfolded datasets, 
such as the 2017 Belle tagged analysis~\cite{Abdesselam:2017kjf}.
The Bernlochner-Ligeti-Papucci-Robinson (BLPR) parametrization~\cite{Bernlochner:2017jka} removes this inconsistency, 
and exploits higher precision data-driven fits to the subleading IW functions to obviate the need for QCDSR inputs. 
It furthermore consistently incorporates the $1/m_{c,b}$ terms for NP currents, important for NP predictions of $\BDxtaunu$. 

There has been long-standing debate about the size of the $1/m_{c}^2$ corrections, 
partly because quark-model-based calculations predicted them to have coefficients somewhat larger than unity.
Recent data-driven fits, however, in the baryonic $\Lambda_b \to \Lambda_c$ system provide good evidence 
that the $1/m_c^2$ corrections obey power counting expectations \cite{Bernlochner:2018kxh}; 
see also \cite{Bordone:2019guc} with regard to $B_{(s)} \to D_{(s)}^*$.
%%%%%%%%%%%%%%%%%%%%%%%%%%%%%%%%%%%%%%%%%
\subsubsection{Quark models}

Beyond dispersive bounds and HQET, 
quark model-based approaches have historically played an important role in descriptions of the form factors, 
and have provided useful constraints in generating lepton universality predictions.
The Isgur-Scora-Grinstein-Wise updated model (ISGW2) parametrization~\cite{Isgur:1988gb,Scora:1995ty} 
implements a non-relativistic constituent quark model, providing estimates of the form factors
by expressing the transition matrix elements for each spectroscopic combination of hadrons in terms of wave-function overlap integrals. 
In addition, it incorporates leading order and $\mathcal{O}(1/m_{c,b})$ constraints from heavy quark symmetry and higher-order hyperfine corrections.

The ISGW2 parametrization of the form factors is treated as fully predictive, being typically implemented without any undetermined parameters.
This amounts to fixed choices for, e.g., the heavy and light quark masses or the brown muck kinetic energy $\bar\Lambda$. 
It therefore is not considered to provide state-of-the-art form factors, compared to data-driven fits.
Non-relativistic quark models may, however, be useful choices for double heavy hadron transitions such as $B_c \to J/\psi$ or $\eta_c$ 
(for a very recent example see e.g. \cite{Penalva:2020ftd}), 
where heavy-quark symmetry cannot be applied.

\subsubsection{Sum rules}

QCDSRs exploit the analytic properties of three-point correlators constructed by sandwiching an operator of interest with appropriate interpolating hadronic currents. 
This allows the expression of an Isgur-Wise function in terms of the Borel transform of the correlator, 
the latter of which can be computed in perturbation theory via an operator product expansion (OPE).
One must further assume quark-hadron duality to estimate the spectral densities of relevant excited states. 
Renormalization improved results for the $1/m_{c,b}$ Isgur-Wise functions and their gradients at zero recoil are 
known~\cite{Ligeti:1993hw,Neubert:1992pn,Neubert:1992wq,Neubert:1993mb}.
While theoretical uncertainties associated with the perturbative calculations are well understood, 
there is no systematic approach to assessing uncertainties arising from quark-hadron duality and scale variations. 
Rough estimates of the uncertainties are large compared to the precision obtained by more recent data-driven methods.

LCSRs operate in a similar spirit to QCDSR,
reorganizing the OPE such that one expands in the ``transverse distance'' of partons from the light cone.
The resulting sum rules are valid for the regime in which the outgoing hadron kinetic energy is large.
LCSR have broad application in exclusive heavy-light quark transitions, such as for $b \to u$ transitions including $B \to \rho$, $\omega$, or $\pi$, 
in which the valence parton is highly boosted compared to the spectator.

%%%%%%%%%%%%%%%%%%%%%%%%%%%%%%%%%%%%%%%%%
\subsubsection{Lattice calculations}
\label{sec:th:LQCD}
Lattice QCD (LQCD) results are available for the SM form factors at zero recoil for both $B_{(s)} \to D_{(s)} $ and $B_{(s)} \to D^*_{(s)} $, with the most precise $B \to \Dx$ results~\cite{Aoki:2019cca}
\begin{align}
	\mathcal{G}(1)  \equiv V_1(1) & = 1.054(4)_{\text{stat}}(8)_{\text{sys}}\,,\nn\\ 
	\mathcal{F}(1) & = 0.906(4)_{\text{stat}}(12)_{\text{sys}}\,.
\end{align}	
LQCD results for the both the $B_{(s)}  \to D_{(s)} $ form factors $f^{(s)}_{+,0}$ are available beyond zero recoil, 
with respect to the optimized expansion in $z = z(q^2, q^2_{\text{opt}})$.
Further, preliminary results for the $B_s \to D_s^*$~\cite{Harrison:2021tol} and $B \to D^*$~\cite{Bazavov:2021bax}
SM form factors beyond zero recoil have recently become available.

The $B \to D$ LQCD data allows for lattice predictions for the differential rate of $\BDtaunu$, 
and when combined with HQET relations plus QCDSR predictions, 
may also predict $\BDstaunu$, but with slightly poorer precision compared to data-driven approaches~\cite{Bernlochner:2017jka}.
Beyond zero-recoil LQCD results are also available for $B_c \to \Jpsi l\nu$~\cite{Harrison:2020gvo} (see Sec.~\ref{sec:th:other_states}),
as well as for the baryonic $\Lambda_b \to \Lambda_c l \nu$~\cite{Detmold:2015aaa} decays including NP matrix elements.

%%%%%%%%%%%%%%%%%%%%%%%%%%%%%%%%%%%%%%%%%
\subsection{Ground state observables and predictions}
\subsubsection{Lepton universality ratios} \label{subsec:SM_pred}
Lepton universality in $b \to c l \nu$ may be probed by comparing the ratios of total rates for $l = e$, $\mu$ and $\tau$, 
in particular the ratio of the semitauonic to light semileptonic exclusive decays
\begin{equation}
	\label{eqn:defRHc}
	\calR(H_c) = \frac{\Gamma[H_b \to H_c \tau \nu]}{\Gamma[H_b \to H_c \ell \nu]}\,,\qquad \ell = e,\mu\,,
\end{equation}
where $H_{c,b}$ are any allowed pair of $c$- and $b$-hadrons. 
(The ratios of the electron and muon modes are in agreement with SM predictions, i.e. near unity; see Sec.~\ref{sec:analysis_rdx}.
One may also consider ratios $\calR(H_u)$ for $H_b \to H_u \tau \nu$ decays, in which the valence charm quark is replaced by a $u$ quark.)
The ratios $R(H_c)$ should differ from unity not only from the reduced phase space as $m_\tau \gg m_{e,\mu}$, 
but also because of the mass-dependent coupling to the longitudinal $W$ mode.
The theory uncertainties entering into the SM predictions for this quantity are then dominated by uncertainties in the form factor contributions coupling exclusively to the lepton mass, 
such as the form factor ratios $S_1/V_1$ and $R_0(w)$ in $B \to D$ and $\Dstar$, respectively.

In Table~\ref{tab:th:RDDs} we show a summary of various predictions as collated by the Heavy Flavor Averaging Group (HFLAV)~\cite{Amhis:2019ckw}. 
Before 2017, $\RDx$ predictions based on experimental data used the CLN parametrization, since this was the only experimentally implemented form factor parametrization.
An unfolded analysis by Belle~\cite{Abdesselam:2017kjf} has since allowed the use of other parameterizations, 
with the different (and more consistent) theoretical inputs as described in Table~\ref{tab:th:RDDs}.
At present, given the different theoretical inputs and correlations in the results of these analyses, 
the HFLAV SM prediction is a na\"ive arithmetic average of the $\RD$ and $\RDs$ predictions and uncertainties, for each mode independently.
A subsequent Belle 2018 analysis of $B \to \Dstar \ell \nu$ \cite{Waheed:2018djm} provided response functions and efficiencies, 
into which different parametrizations may be folded to generate predictions for bin yields in various different marginal distributions. 
For example, \cite{Gambino:2019sif} finds $\RDs = 0.254^{+0.007}_{-0.006}$ and \cite{Jaiswal:2020wer} finds $0.251^{+0.004}_{-0.005}$, 
with and without LCSR inputs, respectively.
Finally, preliminary lattice results for $B \to D^*$ beyond zero recoil predict $\RDs = 0.266(14)$~\cite{Bazavov:2021bax}.

\begin{table}[t]
\caption{$\RDx$ predictions as currently collated and arithmetically averaged by HFLAV. Predictions shown below the HFLAV line are not included in the arithmetic average.}
\scalebox{0.84}{
\parbox{1.15\linewidth}{
\renewcommand*{\arraystretch}{1.75}
%\newcolumntype{C}{ >{\centering\arraybackslash } m{4cm} <{}}
\label{tab:th:RDDs}
\begin{tabular}{ccccc}
	\hline\hline
	Inputs & & $\qquad \RD \qquad $ & $\qquad \RDs \qquad $ & corr. \\
	\hline
	\ \shortstack{\\LQCD \\  \!+\!  Belle/\babar Data} & \footnote{\cite{Bigi:2016mdz}} & $0.299 \pm 0.003$ & --- & --- \\
	 \shortstack{LQCD \!+\!  HQET $\mathcal{O}(\alpha_s, 1/m_{c,b})$ \\ \!+\! Belle 2017 analysis\footnote{\cite{Abdesselam:2017kjf}}} & \footnote{The `BLPR' parametrization~\cite{Bernlochner:2017jka}}  & $0.299 \pm 0.003$ & $0.257 \pm 0.003$  & $0.44$ \\
	 \shortstack{BGL \!+\! BLPR \!+\!  $\sim 1/m_c^2$ \\ \!+\!  Belle 2017 analysis} & \footnote{Includes estimations of $1/m_c^2$ uncertainties~\cite{Bigi:2017jbd}.\\
	 See also \cite{Gambino:2019sif}.} & --- & $0.260 \pm 0.008$ & --- \\
	 \shortstack{BGL \!+\!  BLPR \!+\!  $\sim 1/m_c^2$ \\ \!+\!  Belle 2017 analysis} & \footnote{Fits nuisance parameters for $1/m_c^2$ terms~\cite{Jaiswal:2017rve}.\\
	 See also \cite{Jaiswal:2020wer}.
	 } & $0.299 \pm 0.004$ &  $0.257 \pm 0.005$ & $0.1$ \\
	\hline
	HFLAV arithmetic averages & & $0.299 \pm 0.003$ &  $0.258 \pm 0.005$ & --- \\
	\hline
	LQCD & \footnote{World average~\cite{Aoki:2019cca}} & $0.300 \pm 0.008$ & --- & ---\\
	\shortstack{CLN \\ \!+\! Belle Data} & \footnote{\cite{Fajfer:2012vx}} & --- & $0.252 \pm 0.003$ & --- \\ 
	\hline\hline
\end{tabular}
}}
\renewcommand*{\arraystretch}{1}
\end{table}

On occasion, the phase-space constrained ratio
\begin{equation}
	\widetilde\calR(H_c) = \frac{\int_{m_\tau^2}^{Q_-^2} dq^2~\frac{d\Gamma[H_b \to H_c \tau \nu]}{d q^2}}{\int_{m_\tau^2}^{Q_-^2} dq^2~\frac{d\Gamma[H_b \to H_c \ell \nu]}{d q^2}}\,,\quad \ell = e,\mu\,,
\end{equation}
is also considered, in which the relative phase-space suppression for the tauonic mode is factored out. 
For instance, the SM predictions are, using the fit results of~\cite{Bernlochner:2017jka}
\begin{equation}
	\widetilde\calR(D)  = 0.576(3)\,, \quad
	\widetilde\calR(D^*) = 0.342(2)\,,
\end{equation}
with a correlation coefficient of $0.53$.

%%%%%%%%%%%%%%%%%%%%%%%%%%%%%%%%%%%%%%%%%
\subsubsection{Longitudinal and polarization fractions}
\label{sec:th:longpol}
In the helicity basis for the $D^*$ polarization, the $D^* \to D\pi$ decay amplitudes within $B \to (D^* \to D\pi)l\nu$ decays
are simply $L=1$ spherical harmonics $e^{i\lambda\phi_v}Y_{1,\lambda}(\theta_v)$, with respect to the helicity angles defined in Fig.~\ref{fig:polar_def}. 
That is, the $B \to (D^* \to D\pi)l\nu$ amplitudes may be expressed in the schematic form $\sum_\lambda A_\lambda[B \to D^* l \nu](\theta_l, \phi_l -\phi_v) \times Y_{1,\lambda}(\theta_v)$.
The $D^*$ longitudinal polarization fraction\footnote{Another common notation is $F_{L,\tau}(D^*) =F_L^{D^*}$.} 
\begin{equation}
	F_{L,l}(D^*) = \frac{\Gamma_{\lambda = 0}[B \to D^* l \nu]}{\Gamma[B \to D^* l \nu]}\,,
\end{equation}
thus arises as a physical quantity in $B \to (D^* \to D\pi)l\nu$ decays, via the marginal differential rate
\begin{equation}
	\frac{1}{\Gamma}\frac{d\Gamma_{B \to (D^* \to D\pi)l\nu}}{d \cos \theta_v} = \frac{3}{2}\Big[ F_{L,l}\cos^2\theta_v + (1-F_{L,l})\frac{\sin^2\theta_v}{2}\Big]\,.
\end{equation}
The interference terms between amplitudes with different $\lambda$ vanish under integration over $\phi_l - \phi_v$.
Similar to $\RDx$, theory uncertainties in $|V_{cb}|$ are factored out of $F_{L,l}$. 
Some recent (and new) SM predictions for $F_{L,\tau}(D^*)$ are provided in Table~\ref{tab:th:FLPt}, using a variety of theoretical inputs. 
We also include an SM prediction for $F_{L,\ell}(D^*)$.

\begin{table}[t]
%\newcolumntype{C}{ >{\centering\arraybackslash } m{4cm} <{}}
\caption{SM predictions for the $D^*$ longitudinal fraction and the $\tau$ polarization in $B \to \Dx$. 
We also show simple arithmetic averages of the predictions and uncertainties. The CLN-based prediction shown below the line is not included in the arithmetic average.}
\label{tab:th:FLPt}
\scalebox{0.9}{
\parbox{1.1\linewidth}{
\renewcommand*{\arraystretch}{1.75}
\begin{tabular}{cccccc}
	\hline\hline
	Inputs 	& & $~~F_{L,\tau}(D^*)~~$ 	& $~~F_{L,\ell}(D^*)~~$ 	& $~~P_\tau(D^*)~~$ & $~~ P_\tau(D)~~$\\
	\hline
	\shortstack{BLPR, $\sim 1/m_c^2$, \\ LCSR} 	& \footnote{\cite{Huang:2018nnq}, using the fit of~\cite{Jung:2018lfu}}&  $0.441(6)$ & ---   & $-0.508(4)$ & $0.325(3)$ \\
	\shortstack{BGL, BLPR, \\ $\sim 1/m_c^2$, LCSR}&\footnote{\cite{Bordone:2019vic}, with Belle 2019 data~\cite{Waheed:2018djm}}&  $0.464(10)$ 	& --- 	& $-0.496(15)$ & $0.321(3)$ \\
	\shortstack{BGL, BLPR, \\ $\sim 1/m_c^2$}&\footnote{\cite{Jaiswal:2020wer}, with Belle 2019 data~\cite{Waheed:2018djm}}&  $0.469(10)$ 		& --- 	& $-0.492(25)$ & --- \\
	BLPR 	& \footnote{Using the fit of~\cite{Bernlochner:2017jka}. 
	The correlation between  $P_\tau(D^*)$ and $P_\tau(D)$ is $\rho = 0.33$ }	& $0.455(3)$	& $0.517(5)$ 	& $-0.504(4)$ & $0.323(2)$ \\
	\hline 
	\shortstack{\\Arithmetic \\ averages}  & & $0.455(6)$ & $0.517(5)$ & $-0.501(11)$ & $0.324(3)$ \\
	\hline
	CLN		& \footnote{\cite{Alok:2016qyh}} &  $0.46(4)$ 	 &   ---   & 	--- & ---	 \\	
	\hline\hline
\end{tabular}
}}
\renewcommand*{\arraystretch}{1}
\end{table}

A similar analysis may be applied to $\tau \to h\nu$ decay amplitudes within $B \to \Dx (\tau \to h \nu) \bar\nu$.
For example, in the helicity basis for the $\tau$, the $\tau \to \pi \nu$ amplitudes 
are the $j =\frac{1}{2}$ Wigner-$D$ functions $e^{i\phi_h/2}\sin(\theta_h/2)$ or $e^{-i\phi_hi/2}\cos(\theta_h/2)$,
for $\lambda_\tau = \mp$, respectively, where the helicity angles $\theta_h$ and $\phi_h$ are defined in Fig.~\ref{fig:polar_def}.
The $\tau$ polarization
\begin{equation}
	P_\tau(\Dx) = \frac{\big(\Gamma_{\lambda_\tau = +}- \Gamma_{\lambda_\tau = -}\big)[B \to \Dx \tau \nu]}{\Gamma[B \to \Dx \tau \nu]}\,,
\end{equation}
is a physical quantity in $B \to \Dx (\tau \to \pi \nu) \bar\nu$ decays, via the marginal differential rate
\begin{equation}
	\frac{1}{\Gamma}\frac{d\Gamma_{B \to \Dx (\tau \to \pi\nu) \bar\nu}}{d \cos \theta_h} = \frac{1}{2}\Big[1 + P_\tau(\Dx) \cos\theta_h \Big]\,.
\end{equation}
The interference terms between amplitudes with different $\lambda_\tau$ vanish under integration over $\phi_\tau - \phi_h$.
This generalizes to other final states, such as $h = \rho$, $3\pi$ as
\begin{equation}
	\label{eqn:th:ptanz}
	\frac{1}{\Gamma}\frac{d\Gamma_{B \to \Dx (\tau \to h\nu) \bar\nu}}{d \cos \theta_h} = \frac{1}{2}\Big[1 + \alpha_h P_\tau(\Dx) \cos\theta_h \Big]\,,
\end{equation}
in which $\alpha_h$ is the analyzing power, that depends on the final state $h$. 
In particular the pion is a perfect polarizer, $\alpha_\pi = 1$, while $\alpha_\rho = (1- 2m_\rho^2/m_\tau^2)/(1+2m_\rho^2/m_\tau^2)$.
Just as for $F_{L,\tau}(D^*)$, some recent (and new) SM predictions for $P_\tau(\Dx)$ are provided in Table~\ref{tab:th:FLPt}, 
using a variety of different theoretical inputs. 
The missing energy in the $\tau$ decay means that $\theta_h$ is reconstructible only up to $2$-fold ambiguities in present experimental frameworks.

%%%%%%%%%%%%%%%%%%%%%%%%%%%%%%%%%%%%%%%%%
\subsection{Excited and other states}
\label{sec:th:other_states}
Thus far we have discussed mainly the ground state meson transitions $\BDxlnu$. 
However, much of the above discussion can be extended to excited charm states, baryons, charm-strange hadrons, or double heavy hadrons.
Several of these processes exhibit fewer HQ symmetry constraints or greater theoretical cleanliness compared to the ground states.
This may be exploited to gain higher sensitivity to NP effects or better insight or control over theoretical uncertainties, such as $1/m_c^2$ contributions.

Four orbitally-excited charm mesons, collectively labelled as the $D^{**}$, comprise in spectroscopic notation, 
the states $D_0^* \sim {}^3P_0$, $D_1^\prime \sim {}^3P_1$, $D_2^* \sim {}^3P_2$ and the $D_1 \sim {}^1P_1$.\footnote{The $D_1^\prime$ is also often denoted by $D_1^*$.}
In the language of HQ symmetry, the $D_0^*$ and $D_1^\prime$ ($D_1$ and $D_2^*$) furnish a heavy quark doublet whose dynamics 
is described by the $s^P_\ell = 1/2^+$ ($s^P_\ell = 3/2^+$) HQET. 
The $1/2^+$ doublet is quite broad, with widths $\sim 0.2$ and $0.4$~GeV, while the $3/2^+$ states are an order of magnitude narrower.
The $B \to D^{**}l\nu$ decays produce important feed-down backgrounds to $\BDxlnu$ (see Sec.~\ref{sec:measurements} and~\ref{sec:syst:dssbg}).

Several of the $B \to D^{**}$ form factors vanish at leading order in the heavy-quark limit at zero recoil, 
so that the higher-order $\mathcal{O}(1/m_{c,b})$ corrections become important, as included in the Leibovich-Ligeti-Stewart-Wise (LLSW) parametrization
~\cite{Leibovich:1997tu,Leibovich:1997em}.
This can lead to higher sensitivities to various NP currents compared to the ground states~\cite{Biancofiore:2013ki,Bernlochner:2017jxt}.
These decays must be therefore incorporated consistently, especially for LFUV analyses with NP contributions.
The current SM predictions for all four modes, from fits to Belle data including higher-order HQET contributions at $\mathcal{O}(\alpha_s,1/m_{c,b})$, 
are~\cite{Bernlochner:2016bci,Bernlochner:2017jxt}
\begin{align}
  \label{eqn:th:RDss}
  \calR(D_0^*) & = 0.08(3), &  \calR(D_1^\prime) &= 0.05(2), \nn\\
  \calR(D_1) & =  0.10(2), & \calR(D_2^*) &= 0.07(1)\,.
\end{align}
These are smaller than $\RDx$ because of the smaller phase space and reduced $w$ range.
An additional useful quantity is the ratio for the sum of the four $D^{**}$ states~\cite{Bernlochner:2016bci,Bernlochner:2017jxt}
\begin{equation}
	\label{eqn:th:RbarDss}	
	\calR(D^{**}) = \frac{\sum_{X \in D^{**}} \Gamma[B \to X \tau \bar\nu]}{\sum_{X \in D^{**}} \Gamma[B \to X \ell \bar\nu]} = 0.08(1)\,,
\end{equation}
taking into account correlations in the SM predictions. 

An identical discussion proceeds for $B_{s} \to D^{(*,**)}_s l\nu$ decays, with the light spectator quark replaced by a strange quark. 
The typical size of flavor $SU(3)$ breaking, seen in e.g. $f_K/f_\pi$, suggests $\sim 20\%$ corrections compared to the predictions for $B \to D^{(*,**)}$. 
Lattice studies are available for $B_s \to D_s$~\cite{McLean:2019qcx} beyond zero-recoil as are preliminary results for $B_s \to D_s^*$~\cite{Harrison:2021tol}, 
with the respective predictions
\begin{equation}
	\calR(D_s) = 0.2987(46)\,,\quad \calR(D_s^*) = 0.2442(79)(35)\,,
\end{equation}	
and there is some evidence of relative insensitivity of the matrix elements to the (light) spectator quark~\cite{McLean:2019sds}, despite the expectations from $SU(3)$ breaking.
A recent analysis for $B_{(s)} \to D^{(*)}_{(s)}$~\cite{Bordone:2019guc} combines model-dependent QCDSR inputs with LCSR inputs extrapolated from beyond the physical recoil limit. 
This analysis predicts
\begin{align}
  \label{eqn:th:RDstrange}
  \RD & = 0.298(3)\,, &  \calR(D_s) &= 0.297(3) \nn \\
 \RDs & =  0.250(3)\,, & \calR(D^*_s) &= 0.247(8)\,.
\end{align}
The resulting $\RDs$ prediction agrees with the prior predictions in Table~\ref{tab:th:RDDs} at the $1$-$2\sigma$ level. 
At the LHC, or on the $Z$ peak, non-negligible feed-downs to $\RDs$ arise from $\Bs \to D^{\prime}_{s1}\tau\nu$ decays, because of their subsequent decay to $\Dx\tau\nu X$, 
that must be taken into account. 
Likewise $\Bs \to D^{*}_{s2}\tau\nu$ decays may feed-down to $\RD$: see Sec.~\ref{sec:lhcb_untagged}.

The light degrees of freedom in the ground state baryons $\Lambda_{b,c}$ have spin-parity $s^P_\ell = 0^+$, 
corresponding to the simplest, and therefore most constrained, HQET. 
In particular, the $\Lambda_b \to \Lambda_c$ form factors receive hadronic corrections to the leading order IW function only at $1/m_{c,b}^2$.
Beyond zero-recoil lattice data is available for both SM and NP form factors~\cite{Detmold:2015aaa}.
Predictions for $\Lambda_b \to \Lambda_c \tau \nu$, however, 
are at present more precise when LQCD results are combined with data-driven fits for $\Lambda_b \to \Lambda_c \ell\nu$ plus HQET relations.
In particular, a data-driven HQET-based form factor parametrization, when combined with the lattice data, provides the currently most precise prediction~\cite{Bernlochner:2018kxh}
\begin{equation}
	\label{eqn:th:RLcpred}
	\calR(\Lambda_c) = 0.324(4)\,, 
\end{equation}
as well as the ability to directly extract or constrain the $1/m_{c}^2$ corrections. The latter are found to be consistent with HQ symmetry power counting expectations. 
Similar techniques will be applicable to the two $\Lambda_c^*$ excited states with $s^P_\ell = 1^-$~\cite{Leibovich:1997az, Boer:2018vpx}, once data is available.
At present, predictions for $\calR(\Lambda_c^*)$ may be derived using a constituent quark model approach~\cite{Pervin:2005ve} similar to ISGW2, 
yielding $\calR(\Lambda_c^*(2595)) \simeq 0.16$ and $\calR(\Lambda_c^*(2625)) \simeq 0.11$.

Finally, the semileptonic decay $B_c \to \Jpsi(\to \ell \ell)l\nu$ provides an extremely clean signature to test LFUV.
The aforementioned HQ symmetry arguments, however, cannot be applied to double heavy quark mesons such as $B_c$ and $\Jpsi$ (or the pseudoscalar $\eta_c$): 
They cannot be thought of as a single heavy quark dressed by brown muck.
Rather, large kinetic energy terms break the heavy quark flavor symmetry,  
leaving an approximate residual heavy quark spin symmetry~\cite{Jenkins:1992nb}.
Hence an HQET description is not used for these modes.
A variety of quark-model-based analyses and predictions have been conducted, with wide-ranging predictions for $\RJ \sim 0.2$--$0.4$.
A recent model-independent combined analysis for $B_{(s)} \to \Dx_{(s)}$ and $B_c \to \Jpsi$ and $\eta_c$, making use of a combination of dispersive bounds, 
lattice results and HQET where applicable, provided a prediction $\calR(\Jpsi) = 0.25(3)$~\cite{Cohen:2019zev}.
A subsequent LQCD result provides the high-precision prediction~\cite{Harrison:2020nrv}
\begin{equation}
        \label{eqn:th:RJPsi}
	\calR(\Jpsi) = 0.2582(38)\,.
\end{equation}
Preliminary lattice results for the $B_c \to \eta_c$ form factors beyond zero recoil are also available~\cite{Colquhoun:2016osw}.

%%%%%%%%%%%%%%%%%%%%%%%%%%%%%%%%%%%%%%%%%
\subsection{$b \to u l \nu$ processes}
\label{sec:th:vub}
The dispersive analysis used in Sec.~\ref{sec:th:disp} to parametrize the form factors for $B \to \Dx$ may also be employed for the light hadron $b \to u l\nu$ processes. 
For $B \to \pi l \nu$ in particular, significant simplifications arise because there is only a single possible subthreshold resonance---the $B^*$---for the $f_+$ form factor, 
and no subthreshold resonance for $f_0$. 
Combining this with general analyticity properties of the $B \to \pi$ matrix element, leads to the Bourrely-Caprini-Lellouch (BCL) parametrization~\cite{Bourrely:2008za}. 
Expanding in $z = z(q^2, q_{\text{opt}}^2)$
\begin{align}
	f_+(q^2) & = \frac{1}{1- q^2/m_{B^*}^2} \sum_{j=0}^N b^+_j \Big[z^j - (-1)^{j-N}\frac{j}{N} z^N\Big]\,,\nn\\
	f_0(q^2) & = \sum_{j=0}^N b^0_j z^j\,,
\end{align}
where $N$ is the truncation order.
Lattice results beyond zero recoil are available for all $B \to \pi$ form factors~\cite{Lattice:2015tia,Bailey:2015nbd}, 
that can be incorporated into global fits to available experimental data. The 
SM prediction is~\cite{Bernlochner:2015mya}
\begin{equation}
	\calR(\pi) = 0.641(16)\,.
\end{equation}

Higher-twist LCSR results are available for the $B \to \rho$ and $B \to \omega$ SM and NP form factors, 
parametrized by the optimized $z = z(q^2, q_{\text{opt}}^2)$ expansion~\cite{Straub:2015ica}.
These results may be applied to obtain a correlated, beyond zero recoil fit between 
the SM and NP form factors and the measured $q^2$ spectra of the corresponding light-lepton modes. 
The SM predictions from this fit are~\cite{Bernlochner:2021rel}
\begin{equation}
	\calR(\rho) = 0.535(9)\,, \qquad \calR(\omega) = 0.543(15)\,.
\end{equation}

Quark model approaches have also been applied to the double-heavy to heavy-light decays $B_c \to \Dx l \nu$ (see e.g.~\cite{Ivanov:2006ni,Leljak:2019fqa});
lattice results are, however, soon anticipated for these decays.

%%%%%%%%%%%%%%%%%%%%%%%%%%%%%%%%%%%%%%%%%
\subsection{Inclusive processes}
\label{sec:th:incl}
The inclusive process $B \to X_c l \nu$, where $X_c$ is a single-charm (multi)hadron final state of any invariant mass,
admits a different, cleaner theoretical description than the exclusive processes. For instance,
in the limit $m_{b} \to \infty$, the inclusive process is described simply by the underlying $b \to c l \nu$ free quark decay, rather than in terms of an unknown Isgur-Wise function.

The square of the inclusive matrix element $|\langle X_c | J | \Bbar\rangle|^2$ can be reexpressed 
in terms of the time-ordered forward matrix element $\langle \Bbar | T(J^\dagger J) | \Bbar \rangle$. 
The latter can be computed via an OPE order-by-order in $1/m_b$ and $\alpha_s$, yielding theoretically clean predictions.
State-of-the-art predictions include $1/m_b^2$ terms~\cite{Ligeti:2014kia} and two-loop QCD corrections~\cite{Biswas:2009rb}, 
that may be combined to generate the precision prediction~\cite{Freytsis:2015qca}
\begin{equation}
	\label{eqn:th:Rincl}
	\calR(X_c) = 0.223(4)\,,
\end{equation}
as well as precision predictions for the dilepton invariant mass and lepton energy distributions.
Because the theoretical uncertainties in $B \to X_c l \nu$ are of a different origin to the exclusive modes, 
the measurement of $B \to X_c \tau \nu$ would provide a hadronic-model-independent cross-check of lepton flavor universality (see Sec.~\ref{sec:incl_excl.pdf}).
The inclusive baryonic decays $\Lambda_b \to X_c l \nu$ may be similarly considered, see e.g.~\cite{Balk:1997fg,Colangelo:2020vhu}.

%%%%%%%%%%%%%%%%%%%%%%%%%%%%%%%%%%%%%%%%%
\subsection{New Physics operators}
NP may enter the $b \to c \tau\nu$ processes via a heavy mediator, such that the semileptonic decay is generated by four-Fermi operators of the form
\begin{equation}
	\label{eqn:npop}
	\mathcal{O}_{XY} = \frac{c_{XY}}{\Lambda_\text{eff}^2} \big( \cbar\,\Gamma_{X}\,b\big) \big( \bar\tau\,\Gamma_{Y} \,\nut)\,,
\end{equation}
where $\Gamma_{X(Y)}$ is any Dirac matrix with $X$ ($Y$) labeling the chiral structure of the quark (lepton) current, and $c_{XY}$ is a Wilson coefficient defined at scale $\mu \sim m_{c,b}$. 
The Wilson coefficient is normalized against the SM such that $\Lambda_\text{eff} = (2\sqrt{2} G_F V_{cb})^{-1/2} \simeq 870$\,GeV. 
If we denote by $M$ the characteristic scale of an ultraviolet (UV) completion that matches onto the effective NP operators in Eq.~\eqref{eqn:npop},
then order $10$--$20$\% variations in $\RDx$ or other observables from SM predictions typically probe $M \sim \Lambda_\text{eff}/\sqrt{c_{XY}} \sim \text{few}$~TeV.
This is tantalizingly in range of direct collider measurements and nearby the natural scale for UV completions of electroweak dynamics.

A common basis choice for $\Gamma_{X}$ is the set of chiral scalar, vector and tensor currents: $P_{R,L}$, $\g^\mu P_{R,L}$, and $\sigma^{\mu\nu} P_{R,L}$, respectively.
Assuming only SM left-handed neutrinos, the lepton current is always left-handed, and the tensor current may only be left-handed.
It is common to write the five remaining Wilson Coefficients as $c_{XY} = c_{\SR}$, $c_{\SL}$, $c_{\VR}$, $c_{\VL}$ and $c_T$.
 We use this notation for the Wilson coefficients hereafter.
As for the SM, the NP leptonic amplitude still takes the form $D^j_{m1,m2}(\theta_l, \phi_l)$, with $j = 0$ or $1$, and $|m_{1,2}| \le j$,
and the structure of the differential decay rate resembles Eq.~\eqref{eqn:fulldiff}, but with additional dependencies on NP Wilson coefficients, $w$, and $r$.

The (pseudo)scalar and tensor operators run under the Renormalization Group (RG) evolution of QCD, 
while the vector and axial vector operators correspond to conserved currents and do not 
(for this reason the normalization of Eq.~\eqref{eqn:smop} is well-defined).
At one-loop order in the leading-log approximation, the running of $c_{\SR,\SL,T}$ is dominated by contributions below the top quark mass $m_t$, 
and only weakly affected by variations in $M \sim \Lambda_{\text{eff}}$.
Electroweak interactions, however, may induce mixing between $c_{\SR,\SL,T}$, that can become non-negligible for RG evolution above the weak scale~\cite{Gonzalez-Alonso:2017iyc}.
RG evolution from $M \simeq \Lambda_{\text{eff}} > m_t$ to $\mu \simeq \sqrt{m_c m_b}$ generates at leading-log order
\begin{equation}
	\label{eqn:rgerhos}
	c_{\SR,\SL}(\mu)/c_{\SR,\SL}(M) \simeq 1.7\,, \quad c_{T}(\mu)/c_{T}(M) \simeq 0.84\,.
\end{equation}
These running effects are particularly important in translating the low scale effective field theory (EFT) implications of $b \to c\tau\nu$ measurements to collider measurements at high scales.

%%%%%%%%%%%%%%%%%%%%%%%%%%%%%%%%%%%%%%%%%
\subsection{Connection to other processes}
\label{sec:th:conn}
LFUV in $b \to c l \nu$ necessarily implies violation in the crossed process $B_c \to l \nu$.
The latter decays are extremely theoretically clean: 
Their tauonic versus leptonic LFUV ratios are simply the ratios of chiral suppression and 2-body phase space factors, 
i.e. $m_\tau^2 (1 - r_\tau^2)^2/ m_{\ell}^2(1 - r_\ell^2)^2$, in which $r_l =  m_l/m_{B_c}$. These ratios are precisely known.

In the SM, the branching ratio
\begin{equation}
	\BR[B_c \to l \nu] = \tau_{B_c} G_F^2 |V_{cb}|^2 m_{B_c}^3 f_{B_c}^2 r_l^2\big(1 - r_l^2\big)^2/8\pi\,,
\end{equation}
in which the decay constant $ f_{B_c} \simeq 0.434(15)$\,GeV from lattice data~\cite{Colquhoun:2015oha},
and the $B_c$ lifetime, $\tau_{B_c} = 0.510(9)\times 10^{-12}$\,s is well measured~\cite{Zyla:2020zbs}. 
In particular, in the SM one predicts $\BR[B_c \to \tau \nu] \simeq 2.2(2)\% \times (|V_{cb}|/0.04)^2$.

In the presence of NP, the NP Wilson coefficients generate an additional factor
\begin{equation}
	\BR[B_c \to \tau \nu] = \BR_{SM}\bigg|1 + c_{\VL} - c_{\VR} + \frac{m_{B_c}^2(c_{\SR} - c_{\SL})}{m_\tau(\overline{m}_b + \overline{m}_c)} \bigg|^2\,,
\end{equation}
where $\overline{m}_{c,b}$ are quark masses in the rescaled minimal subtraction ($\overline{\text{MS}}$) renormalization scheme at scale $\mu$, entering via equations of motion. 
Because the NP pseudoscalar current induces a chiral flip there is no chiral suppression in the pseudoscalar term.
As a result this term is enhanced by a factor of $m_{B_c}/m_\tau \sim 3.5$ versus the $V-A$ current contribution. 
This leads to large tauonic branching ratio enhancements, 
that may then be in tension with naive expectations that the $B_c$ hadronic branching ratios $\sim 70$--$90\%$~\cite{Li:2016vvp,Alonso:2016oyd,Akeroyd:2017mhr,Bardhan:2019ljo}.
A corollary is that a future measurement or bounds of $\BR[B_c \to \tau \nu]$ alone would tightly constrain the NP pseudoscalar contributions.

In the absence of any NP below the electroweak scale, the NP effective operators in Eq.~\eqref{eqn:npop} 
must match onto an electroweak-consistent EFT constructed from SM quark and lepton doublets and singlets under $SU(2)_L \times U(1)_Y$. 
In particular, because the SM neutrino belongs to an electroweak lepton doublet, $L_L$, 
then electroweak symmetry requires the presence of at least two electroweak doublets in any operator that generates the $b \to c\tau\nu$ decay.
(An exception applies if right-handed sterile neutrinos are present.)
In any given NP scenario, this may generate relations between $b \to c\tau\nu$ and other processes, 
that arise when at least one of the four fermions is replaced by its electroweak partner.
For example, various minimal NP models, depending on their flavor structure, 
may be subject to tight bounds from the rare $b \to s\nu\nu$ or $b \to s \tau\tau$ decays or bounds on $Z \to \tau\tau$ or $W \to \tau \nu$ branching ratios~\cite{Freytsis:2015qca,Sakaki:2013bfa}, 
or the high-$p_T$ scattering $pp \to b\tau\nu$~\cite{Altmannshofer:2017poe} and $pp \to \tau\tau$ or $\tau\nu$~\cite{Greljo:2017vvb,Greljo:2018tzh}.
Ultraviolet completions with nontrivial flavor structures may further generate relations to charm decay processes, or $b \to s \ell\ell$. 
The latter is particularly intriguing, because of an indication for \emph{light} lepton universality violation in the ratios~\cite{Aaij:2017vbb,Aaij:2019wad}
\begin{equation}
	\label{eqn:th:RK}
	\RKx \equiv \frac{\Gamma[B \to K^{(*)} \mu\mu]}{\Gamma[B \to K^{(*)} ee]}\,,
\end{equation}
at the $2-3\sigma$ level in each mode (see Sec.~\ref{sec:fcnc}). 
Extensive literature has considered possible common origins of LFUV in semitauonic processes with LFUV in these rare decays.
See~\cite{Bhattacharya:2014wla, Calibbi:2015kma, Buttazzo:2017ixm, Kumar:2018kmr}, among many others, 
for extensive discussions of combined explanations for semileptonic and rare decay LFUV anomalies.

\begin{table*}
 \renewcommand*{\arraystretch}{1.25}
  \centering
  \caption{Approximate number of $b$-hadrons produced and expected at the \bfacs~\cite{Bevan_2014,Kou:2018nap} and at the LHCb
    experiment~\cite{Albrecht:2653011}, including some of the latest developments~\cite{hl_lhc:2020tdr}. The LHCb numbers take into account an average
    geometrical acceptance of about 15\%. Note that the overall $B$ reconstruction efficiencies at LHCb are usually significantly lower than those at the \bfacs (see text).
    The two values of integrated luminosities and center-of-mass energies shown
    for Belle and Belle II correspond to data taking at the
    $\Upsilon(4S)$ and $\Upsilon(5S)$ resonances, respectively. The $B$-factory experiments also
    recorded data sets at lower center-of-mass energies (below the open beauty threshold) that are not included in this table.}

  \label{tab:production_comp}
    \begin{tabular*}{\textwidth}{@{\extracolsep{\fill}} l ccc cccc}
      \hline\hline
       \multirow{2}{*}{Experiment}      & \multirow{2}{*}{\babar} & \multirow{2}{*}{Belle} & \multirow{2}{*}{Belle II} & \multicolumn{4}{c}{LHCb} \\
      \cline{5-8}
                                        &       &             &              & Run 1             & Run 2  & Runs 3--4        & Runs 5--6         \\
      \hline
      Completion date                   & 2008  & 2010        & 2031         & 2012              & 2018   & 2031            & 2041             \\
      Center-of-mass energy       & 10.58~GeV & 10.58/10.87~GeV & 10.58/10.87~GeV  & 7/8~TeV      & 13~TeV & 14~TeV          & 14~TeV           \\
      \bbbar cross section [nb]         & 1.05 & 1.05/0.34    & 1.05/0.34               & (3.0/3.4)$\times 10^5$ & $5.6\times10^5$ & $6.0\times10^5$ & $6.0\times 10^5$ \\
      Integrated luminosity [fb$^{-1}$] & 424  & 711/121 & $(40/4) \times 10^3$ & 3   & 6   & 40    & 300    \\
      \hline
      \Bz mesons [10$^9$]               & 0.47 & 0.77    & 40                   & 100 & 350 & 2,500 & 19,000 \\
      \Bp mesons [10$^9$]               & 0.47 & 0.77    & 40                   & 100 & 350 & 2,500 & 19,000 \\
      \Bs mesons [10$^9$]               & -    & 0.01    & 0.5                  & 24  & 84 & 610 & 4,600  \\
      \Lb baryons [10$^9$]              & -    & -       & -                    & 51  & 180 & 1,300 & 9,800 \\
      \Bc mesons [10$^9$]               & -    & -       & -                    & 0.8 & 4.4 & 19    & 150    \\
      \hline\hline
    \end{tabular*}
\end{table*}

\section{Experimental Methods}
\label{sec:experiments}

%%%%%%%%%%%%%%%%%%%%%%%%%%%%%%%%%%%%%%%%%%%%%%%%%%%%%%%%%%
%%%%%%%%%%%%%%%% PRODUCTION/DETECTION %%%%%%%%%%%%%%%%%%%%
\subsection{Production and detection of $b$-hadrons}
\label{sec:production}

Since the discovery of the $b$ quark in 1977~\cite{Herb:1977ek}, large samples of $b$-hadrons have been produced at
colliders such as CESR, LEP, or Tevatron. However, it was not until the advent of the \bfacs and the LHC, with
their even larger samples and specialized detectors, that the study of third generation LFUV in $B$ mesons
became feasible. This is because of the stringent analysis selections that are required to achieve adequate
signal purity when reconstructing final states that include multiple unreconstructed neutrinos.  The
\bfacs~\cite{Bevan_2014}, KEKB in Japan and PEP-II in the United States, took data from 1999 to 2010. Their detectors,
Belle ~\cite{Abashian:2000cg} and \babar ~\cite{TheBABAR:2013jta},
recorded over a billion of \BB events originating from clean \epem collisions.  The LHCb
detector \cite{Alves:2008zz,lhcbdet2015} at the CERN LHC, which started taking data in 2010, has recorded an unprecedented trillion \bb
pairs as of 2020, which allows it to compensate for the more challenging environment of $pp$ collisions.
The recently commissioned Belle~II experiment and the LHCb detector, to be upgraded in 2019--21 and 2031, are expected to continue taking data over the next decade and
a half, surpassing the current data samples by more than an order of magnitude.
In the following, we describe how $b$-hadrons are produced and detected at these facilities.\footnote{
Other current experiments might also be able to make contributions to semitauonic LFUV measurements in the future.
For instance, the CMS experiment at the LHC recorded in 2018 a large (parked) sample of unbiased $b$-hadron decays, with the primary goal of measuring the $\calR_{K^{(*)}}$ ratios.
This sample could conceivably also be used to measure semitauonic decays if, e.g., the challenges arising from the multiple neutrinos in the final state can be overcome.}
Table~\ref{tab:production_comp} summarizes the number of $b$-hadrons produced and expected at the \bfacs and at the LHCb
experiment.

%%%%%%%%%%%% THE B-FACTORIES DETECTORS %%%%%%%%%%%%%%%%
\subsubsection{The \bfacs}
\label{sec:bfactories}
%%%%%%%%%%%% THE B-FACTORIES DETECTORS %%%%%%%%%%%%%%%%

%% Belle detector and event display
\begin{figure*}
  \includegraphics[width=0.9\textwidth]{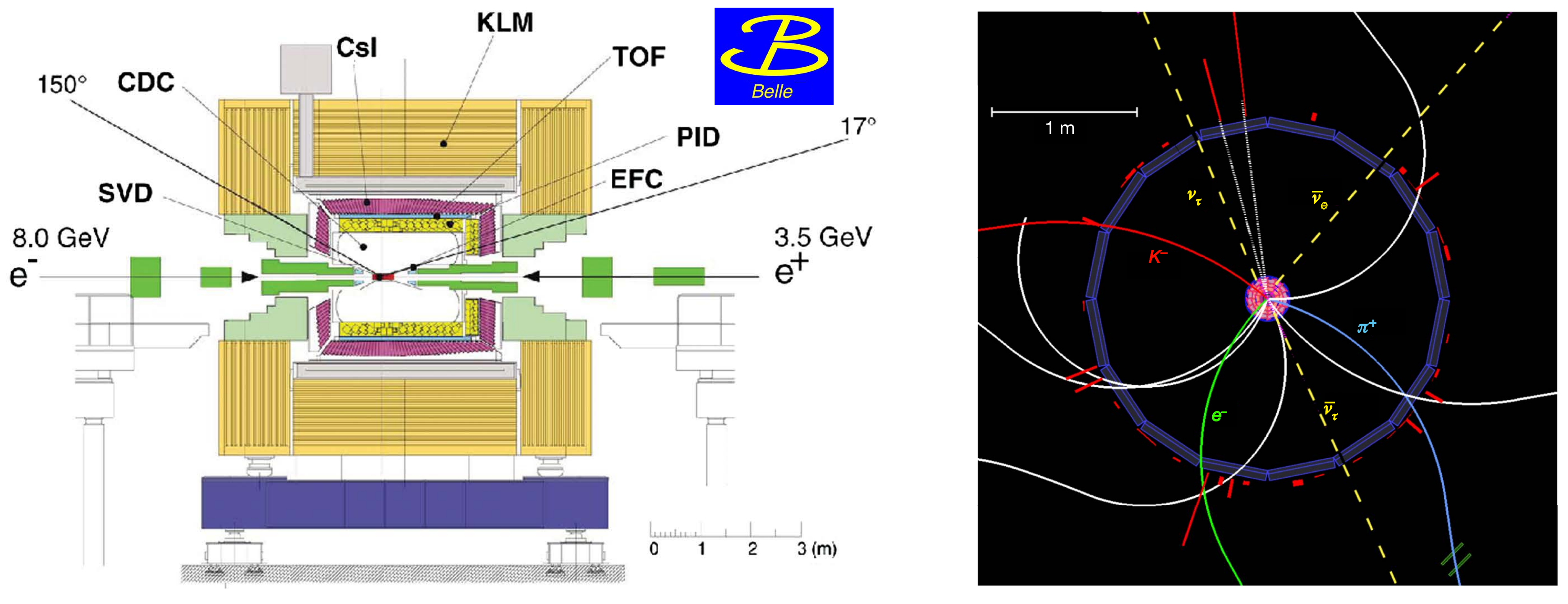} \caption{Left: Side view of the Belle
 detector. See \cite{Abashian:2000cg} for further detail on the subdetectors and their acronyms. The \babar detector has a
 similar configuration. Right: view perpendicular to the beam axis. The displayed event is reconstructed as a
 $\Upsilon(4S)\to B^+B^-$ candidate, with $B^-\to D^0\tau^-\overline{\nu}_\tau$, $D^0\to K^-\pi^+$ and $\tau^-\to
 e^-\nu_\tau \overline{\nu}_e$, and the $B^+$ decaying to five charged particles (white solid lines) and two
 photons. The  directions of undetected neutrinos are indicated as dashed lines. From \cite{Abashian:2000cg,Ciezarek:2017yzh}.}
\label{fig:detector_event_belle.pdf}
\end{figure*}

KEKB and PEP-II produced $B$ mesons by colliding electron and positron beams at a center-of-mass energy of $10.579$~GeV. 
At this energy, $e^+$ and $e^-$ annihilation produces $\Y4S$ mesons in about 24\% of the hadronic collision processes,
with the production of \ccbar and other light quark pairs accounting for the remaining 76\%.
Together with other processes producing pairs of fermions, the latter form the so-called continuum background.

The $\Y4S$ meson is a $\bb$ bound state which, as a result of having a
mass only about 20 MeV above the $\BB$ production threshold, decays almost exclusively to $\BpBm$ or $\BzBzb$
pairs. Some limited running away from the $\Y4S$ resonance was performed in order to study the continuum
background and the properties of the bottomonium resonances $\Y1S-\Y5S$. The largest dataset produced by KEKB was used 
to study $\Bs$ mesons obtained from $\Y5S$ decays. However, the resulting
$B_s^{(*)}\Bbar_s^{(*)}$ data sample was small, about $3\%$ of the total $\BB$ sample as shown in Table~\ref{tab:production_comp}.

On the one hand, compared to hadron colliders, the $\bbbar$ production cross section in lepton colliders such as the
\bfacs is much smaller: even at the (so far) highest instantaneous luminosity of
$2.4 \times10^{34}$~cm$^{-2}$s$^{-1}$ achieved by SuperKEKB in the Summer of 2020, $\BB$ pairs were produced only at a rate of
about $25$\,Hz. On the other hand, one of the significant advantages of colliding fundamental particles like
electrons and positrons is that the initial state is fully known; i.e., nearly $100\%$ of the $\epem$ energy is transferred
to the $\BB$ pair. This feature can be exploited by tagging techniques (Sec.~\ref{sec:tagging}) that
 reconstruct the full collision event and can determine the momenta of missing particles such as neutrinos, so long as the detectors are capable of reliably
reconstructing all of the visible particles. The \babar and Belle detectors managed to cover close to $90\%$ of
the total solid angle by placing a series of cylindrical subdetectors around the interaction point and
complementing them by endcaps, that reconstructed the particles that were ejected almost parallel to the beam
pipe. This is sketched in Fig.~\ref{fig:detector_event_belle.pdf}.

The specific technologies employed in both $B$-factory detectors have been described in detail in \cite{Bevan_2014}.
Four or five layers of precision silicon sensors placed close to the interaction point reconstruct the decay
vertices of long-lived particles, as well as the first $\approx10$~cm of the tracks left by charged
particles. Forty to fifty layers of low-material drift chambers measure the trajectories and ionization
energy loss as a function of distance ($dE/dx$) of charged particles. Time-of-flight and Cherenkov systems
provide particle identification (PID) that allow kaon/pion discrimination. Crystal calorimeters measure
the electromagnetic showers created by electrons and photons. A solenoid magnet generates the $1.5$\,T magnetic
field parallel to the beam pipe that bends the trajectories of charged particles, to allow for determination of their momenta. 
A series of steel layers instrumented with muon chambers guide the return of the magnetic
flux and provide muon and $\KL$ PID.

Between 1998 and 2008--10, the \babar and Belle detectors recorded a total of 471 and 772 million $\BB$ pairs,
respectively. These large samples, which are still being analyzed, allowed for the first
measurement of CP violation in the $B$ system, the observation of $B$ mixing, as well as many other novel
results~\cite{Bevan_2014}. These further included the first observations of $\BDxtaunu$ decays (see Sec.~\ref{sec:measurements}), which in turn began the study of third
generation LFUV: the focus of this review.  The success of the \bfacs has led to the
upgrade of the accelerator facilities at KEKB, so called SuperKEKB~\cite{Akai:2018mbz}, such that it will be capable of delivering instantaneous luminosities 30 times higher than before. 
The upgraded Belle detector, Belle II~\cite{abe2010belle}, started taking data in 2018 with the aim of recording a total of over $40$ billion $\BB$ pairs. 
The LFUV prospects for Belle II are discussed in Sec.~\ref{sec:out:belle}.

%%%%%%%%%%%% THE LHCB DETECTOR %%%%%%%%%%%%%%%%
\subsubsection{The LHCb experiment}
\label{sec:lhcb}
%%%%%%%%%%%% THE LHCB DETECTOR %%%%%%%%%%%%%%%%

%% LHCb detector and event display
\begin{figure*}
  \includegraphics[width=0.9\textwidth]{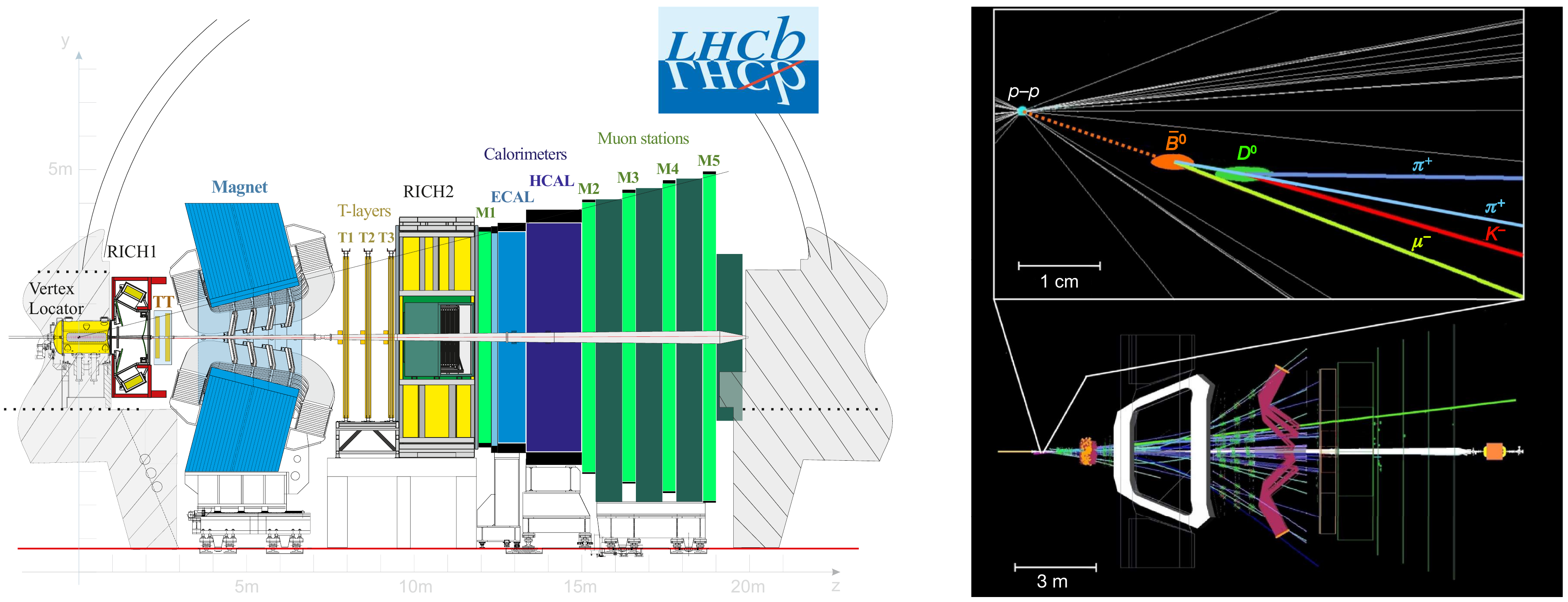}

  \caption{Left: Side view of the LHCb detector. See \cite{LHCb_Collaboration_2008,lhcbdet2015} for further
    details on the subdetectors and their acronyms. Right: side view of an event display for a $B^0\to
    D^{*+}\tau^-\overline{\nu}_\tau$ decay. The area around the interaction point is enlarged in the inset at
    the top. The trajectory of the $B^0$ meson is indicated with the thick dotted line, and the trajectories of
    the particles from the subsequent $D^{*+} \to D^0\pi^+$, $D^0 \to K^-\pi^+$, and $\tau^- \to \mu^- \numb
    \nut$ decays are illustrated with thick solid lines. Adapted from \cite{lhcbdet2015,Ciezarek:2017yzh}.}
\label{fig:detector_event_lhcb.pdf}
\end{figure*}

At hadron colliders such as the LHC, $b$ quarks are predominantly pair-produced in $\pp$ collisions via the gluon fusion process $gg \to \bbbar$ plus subleading quark fusion contributions,
with an approximate production cross-section $\sigma(\bb) \sim 560\,\mu$b at $\sqrt{s} = 13$\,TeV, scaling approximately linearly in $\sqrt{s}$~\cite{Aaij_2017}.
Electroweak production cross-sections for single or pairs of $b$ quarks via Drell-Yan processes, Higgs or top quark decays are five or more orders of magnitude smaller, 
with the largest such cross-section $\sigma(Z \to \bbbar) \sim 10$\,nb. 
As a result, $b$ quarks are effectively always accompanied in LHC collisions by a companion $\bbar$ quark. 
This feature is extremely important for unbiased trigger strategies enabling the study of one $b$-hadron decay while triggering on the other.

At leading order, the hadronization of a $b$ quark at the LHC is quite similar to the one observed in detail by the LEP experiments. 
For instance, the momentum distribution of the non $b$-hadron fragments, which is relevant for same-side tagging studies,
is well described by LEP-inspired Monte Carlo simulations~\cite{Sjostrand_2015}. 
More important is the relative production of the various $b$-hadron species: the main features---dominant production of
$\Bz$ and $\Bp$ mesons, and sizeable production fraction of $\Bs$ and $\Lb$---are the same, 
except that a much larger \Lb production fraction is observed for $p_{T}$ (momentum transverse to the beam axis) below 10~GeV~\cite{PhysRevD.100.031102}.
LHCb can also study the decays of $B_c$ mesons, in spite of its very low production rate, approximately $0.6\%$ of the \Bp production cross-section~\cite{bcprod}.
As discussed in Secs.~\ref{sec:th:other_states} and~\ref{sec:th:conn}, $B_c$ mesons provide a very interesting laboratory for testing LFUV in $\BJtaunu$ or $B_c \to \tau\nu$ decays.

The parton center-of-mass energy required to produce a $b$-hadron pair at threshold is far smaller than the total available collision energy in the $\pp$ system,
leading to the production of a significant fraction of $\bbbar$ pairs with very large forward or backward boosts.
This characteristic is the basis of the LHCb experimental concept~\cite{Alves:2008zz,lhcbdet2015}, 
which studies the $\bb$ pairs produced within a $400$\,mrad cone covering the forward  region, corresponding to a pseudorapidity $2 \le \eta \le 5$.
Despite this very small solid angle, the LHCb detector captures $\sim 15\%$ of the full $\bb$ cross-section~\cite{Aaij:2017deq}.

Within this acceptance, the $b$-hadrons have a typical transverse momentum, $p_{T}$, of 10~GeV, corresponding to an overall energy of $\sim 200$~GeV.	
This in turn corresponds to a typical boost factor of about 50, resulting in a mean flight distance of over $2$~cm for each electroweakly-decaying ground-state $b$-hadron: 
namely $B^{0,+}$, $B_{s}$, $B_{c}$, or $\Lb$. 
The sophisticated silicon trackers used in the LHCb detector provide a typical position resolution of 300~$\mu$m for the $B$ vertex along its flight direction,
resulting in flight distance significances between the $b$-hadron decay vertex and its primary vertex (PV) of over 100$\sigma$. This precision
leads to extremely clean signals even for high-multiplicity decay channels where the combinatorial background is potentially very important~\cite{Aaij:2017deq}, 
provided the primary production vertex can be identified. 

The LHCb luminosity was kept low enough~\cite{lhcbdet2015} so that the mean number of primary vertices per event until 2018 was between 1 and 2. This number
is expected to rise to about 5 after the 2019-2021 upgrade~\cite{Bediaga:2012uyd} and possibly 50 after the 2031 upgrade~\cite{Aaij:2244311}.
The longitudinal size of the LHCb luminous region is 20\,cm, so that with only a handful of $\pp$ interactions in a given event,
the primary vertex misconstruction is kept to a very low level. 
The \mbox{ATLAS} and CMS experiments typically accumulate $50$ primary vertices in a given event (rising to $200$ after 2027) and therefore face a different challenge. 
Nevertheless, they are capable of cleanly reconstructing low multiplicity $b$-hadron decays thanks to their large coverage and high-granularity subdetectors.
It should be stressed, however, 
that for semitauonic $b$-hadron decays the goal is not just to isolate a decay vertex from a primary vertex,
but rather to identify a chain of vertices comprising the PV, the $b$-hadron decay, and, in the case of hadronic-$\tau$ measurements, the $\tau$ decay. 
At the LHC, this is currently only feasible at LHCb.

As is the case in the \bfacs, PID capabilities are critical to properly identify $b$-hadron decays.  For instance,
at a hadron collider, misidentifying a pion as a kaon could lead to confusing a $\Bs$ meson for a $\Bz$ meson,
and identifying a pion as a proton can lead to a $\Lb$ baryon impersonating a $\Bz$ meson. PID information is
provided by the two Ring Imaging Cherenkov (RICH) detectors shown in the left panel of
Fig.~\ref{fig:detector_event_lhcb.pdf}.
	
Table~\ref{tab:production_comp} lists the known production rates for all ground-state $b$-hadron species, at
both LHCb and the \bfacs.  While the geometrical acceptance is included for the LHCb values, the average trigger
and analysis requirements must be taken into account as well in order to compare LHCb with the \bfacs.  These
requirements limit the LHCb useful yield at LHCb to about $0.1\%$ or less of the available sample.
As an example, for their respective measurements of ${\cal R}_{K^+}$, LHCb~\cite{Aaij:2021vac} and
Belle~\cite{Abdesselam:2019lab} reconstructed 3850 and 42.3 $B^+ \to K^+ \mu^+ \mu^-$ signal candidates.
These correspond to $8.6\times10^{-9}$ and $54.9\times10^{-9}$ candidates per $B^+$ meson in
Table~\ref{tab:production_comp}, respectively, which translates to an overall signal reconstruction efficiency for this particular decay
this is about six times lower for LHCb than for Belle.

Another feature of LHCb physics is the large production rate of excited $b$-hadron states: $B^{**}$, $B_s^{**}$, $\Lambda_b^{**}$
can be studied in detail, as well as baryons containing both $b$ and $s$ quarks, such as $\Xib$, $\Omegab$, and their
excited states. These can be useful to study semitauonic decays because, as described in Sec.~\ref{sec:bframe:rfa},
the decay $\Bsstar2 \to B K$ can provide access to kinematic variables in the $B$ center-of-mass frame via $B$ tagging.

%%%%%%%%%%%%%%%%%%%%%%%%%%%%%%%%%%%%%%%%%%%%%%%%%%%%%%%%%%
%%%%%%%%%%%%%%% PARTICLE RECONSTRUCTION %%%%%%%%%%%%%%%%%%
\subsection{Particle reconstruction}
\label{sec:reco}
%%%%%%%%%%%% PARTICLE RECONSTRUCTION %%%%%%%%%%%%%%%%

Ground state $b$-hadrons---i.e. hadrons decaying only through flavor-changing electroweak currents---have
lifetimes of the order of one picosecond. Thus, they decay fast enough that they
must all be reconstructed from their more stable decay products. 
At the same time, they live and fly long enough so that their decay vertices can be separated from the vertex of the primary collision (\epem in the
case of the \bfacs and \pp in the case of LHCb). 
The reconstruction of these stable decay products proceeds in a similar fashion for the \bfacs and the LHCb experiment, 
with some key differences.

%% BaBar particle reco
\begin{figure*}
  \includegraphics[width=0.8\textwidth]{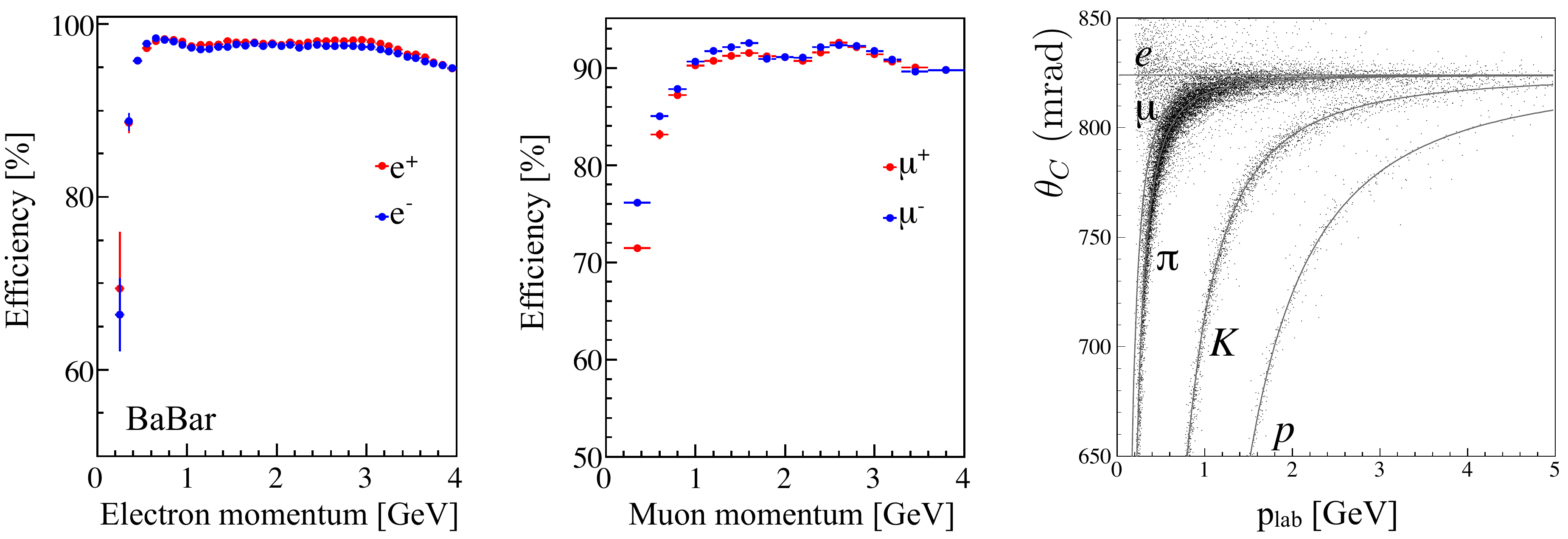} 
 \caption{ Examples of particle reconstruction performance for the \babar detector; the performance for the
   Belle detector is similar. Left: electron reconstruction efficiency. Middle: muon reconstruction
   efficiency. Right: Cherenkov angle measurement for different particles species at \babar's Detector of
   internally reflected Cherenkov light (DIRC). Adapted from~\cite{TheBABAR:2013jta,FrancoSevilla:2012thesis}.  }
\label{fig:particle_reco_babar.pdf}
\end{figure*}

%% LHCb particle reco
\begin{figure*}
  \includegraphics[width=\textwidth]{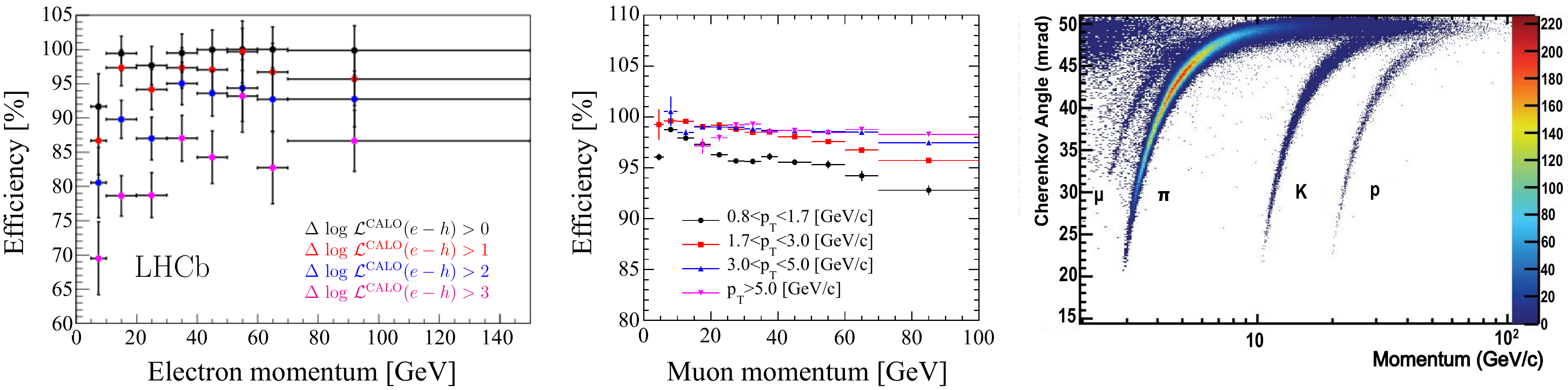} 
 \caption{ Examples of particle reconstruction performance for the LHCb detector. Left: electron
   reconstruction efficiency. Middle: muon reconstruction efficiency. Right: Cherenkov angle measurement for
   different particle species at LHCb's Ring Imaging Cherenkov detector 1
   (RICH1). Adapted from~\cite{lhcbdet2015}. }
\label{fig:particle_reco_lhcb.pdf}
\end{figure*}

%% Charged particles
\subsubsection{Charged particle reconstruction}
The trajectories of charged particles---``tracks''---are reconstructed based on the energy deposits left in the
trackers---``hits''. The momenta of these particles are determined based on the bending of these trajectories
induced by the magnetic fields in each detector. As shown in Figs.~\ref{fig:detector_event_belle.pdf} and \ref{fig:detector_event_lhcb.pdf}, charged particles follow helical
trajectories in the \bfacs due to their solenoidal magnetic fields, while in LHCb the particles are
simply deflected by the dipole magnet. In either case, charged track reconstruction proceeds with efficiencies
of over 95\%---for $p>300$~MeV at the \bfacs~\cite{Bevan_2014} and $p>5$~GeV at LHCb~\cite{lhcbdet2015}---and the momentum
determination is achieved with a typical resolution of $0.5$-$1$\%.

%% Vertex reconstruction
The reconstruction of the $b$-hadron secondary vertices is of primary importance to distinguish
signal from background decays, especially in LHCb. In the \bfacs~\cite{Bevan_2014}, the decay vertices of the short-lived 
$B$ and $D$ mesons were reconstructed with a resolution of $60-100~\mu$m when they decayed inside
the vertex trackers (about 80\% of the time), and $100-400~\mu$m when decaying outside. LHCb
reconstructs the impact parameter of the tracks, that is, their distance to the primary vertex in the
plane transverse to the beam line, with an impressive resolution of 45~$\mu$m for $\pt=1$~GeV, and down to
15~$\mu$m for very high momenta tracks. As discussed in Sec.~\ref{sec:lhcb}, the vertex resolution along
the beam line is of the order of 250~$\mu$m, which, given the large boost of most particles
at LHCb, is sufficient to suppress prompt background processes by multiple orders of magnitude
(Sec.~\ref{sec:lhcb_hadronic_rds}).

%% Leptons
For both the \bfacs and LHCb, charged leptons have generically clean signatures that can be differentiated
from other types of particles with high efficiency. Electrons are reconstructed from tracks that match a cluster in the electromagnetic
calorimeter with the appropriate shape and energy; muons are generally identified as tracks that leave hits in
the outer muon detectors, with some additional inputs from the other subdetectors. However, the performance of the two kinds of experiments
diverges substantially in the details. 

At the \bfacs, both electrons and muons are reconstructed with efficiencies over $90\%$ and with low mis-identification rates, though
the performance is generally better for electrons; see Fig.~\ref{fig:particle_reco_babar.pdf} and \cite{TheBABAR:2013jta,FrancoSevilla:2012thesis}.
For instance, a typical 2~GeV electron is reconstructed
with $96\%$ efficiency and $0.3\%$ pion misidentification probability, whereas a 2~GeV muon would have $92\%$ efficiency 
and $2.5\%$ pion misidentification probability. In contrast, at LHCb the electron reconstruction is much
more challenging  because of the lower granularity of the electromagnetic calorimeter and the larger amount
of material before it, compared to the \bfacs. A 20~GeV electron is reconstructed with about $90\%$
efficiency for a misidentification rate of $2.5\%$, while a muon with the same momentum would be reconstructed
with $98\%$ efficiency for a $1\%$ misidentification rate (Fig.~\ref{fig:particle_reco_lhcb.pdf} and \cite{lhcbdet2015}). Additionally, the first level of the LHCb trigger
during 2010--18 was implemented on hardware and did not use information from the trackers, resulting in trigger efficiencies much
lower for electrons than muons. This limitation will be overcome during the 2019--21 upgrade by a software-only trigger.

%% Charged hadrons
Finally, charged (light) hadrons are identified primarily by their signatures in the Cherenkov detectors, as well as
the energy deposition in the drift chamber for low momentum particles in the \bfacs. The right panels of Figs.~\ref{fig:particle_reco_babar.pdf} and \ref{fig:particle_reco_lhcb.pdf}
show the separation achieved for several species of charged hadrons in some of the Cherenkov detectors for
\babar and LHCb, respectively.

%% Neutral particles
\subsubsection{Neutral particle reconstruction}
\label{sec:reco:neutral}
Another key difference between \bfacs and LHCb lies in the ability to efficiently reconstruct neutral
particles: primarily photons in the case of LFUV measurements. The low material in front of the
$B$-factory calorimeters, as well as their good resolution and granularities, allows them to fully
reconstruct final states that contain $\piz$ mesons decaying to two photons---present, for instance, via the copious $D^0\to
K^- \pi^+ \piz$ decay---as well as photons, such as those coming from $D^{*0}\to D^0 \gamma$ decays.  At LHCb,
the granularity and detector material challenges discussed above, as well as the high number of $b$-hadrons,
have thus far led its LFUV measurements to avoid the reconstruction of final states with $\piz$ mesons or photons.

%%%%%%%%%%%%%%%%%%%%%%%%%%%%%%%%%%%%%%%%%%%%%%%%%%%%%%%%%%
%%%%%%%%%%%%%%%%% BACKGROUND REJECTION %%%%%%%%%%%%%%%%%%%
\subsection{Kinematic reconstruction: The $b$-hadron momentum}
\label{sec:bframe}

%%%%%%%%%%%% B FRAME RECONSTRUCTION %%%%%%%%%%%%%%%%

One of the major challenges in the reconstruction of semitauonic $\HbHctaunu$ decays is the determination of
the parent $b$-hadron momentum. This momentum is necessary to measure important kinematic variables such
as the momentum transfer $q^2=\big(p_{H_b}-p_{H_{c}}\big)^2 \equiv \big(p_{\tau}+p_{\nu}\big)^2$, 
which is not directly accessible because of the undetected neutrinos in the final state. In measurements involving the \tauellnu decay, 
the momentum of the parent $b$-hadron is further employed
to reconstruct other invariants, such as the invariant mass of the unreconstructed particles
\begin{equation}
	\label{eqn:kin:mmiss}
	\mmiss = \big(p_{H_b}-p_{H_{c}}-p_\ell\big)^2\,,
\end{equation}
or the energy of the charged lepton in the $H_b$ rest frame, 
\begin{equation}
	\label{eqn:kin:Eell}
	\Esl = (p_\ell \cdot p_{H_b})/m_{H_b}\,.
\end{equation}
In these leptonic-$\tau$ measurements, the
signal and normalization modes (\HbHctaunu and \HbHcellnu, respectively) are reconstructed in the same exact final state,
differing only in the number of undetected neutrinos. Since normalization events only have one neutrino,
their reconstructed $\mmiss$ distribution is sharply peaked at zero, in contrast to the broad \mmiss distribution of signal
events. Additionally, charged leptons in the signal events are generated in the secondary $\tau$ decay and thus have
a lower maximum $\Esl$ than those arising from normalization \HbHcellnu decays.

In Sec.~\ref{sec:tagging} we describe how the \bfacs take advantage of their precisely known $\epem$ beam energies
to determine the momentum of the signal $B$ in a $B\Bbar$ event by reconstructing the accompanying tag $\Bbar$. 
This procedure is not available in the busier hadronic environment of $pp$ collisions. Instead, LHCb
employs the untagged methods detailed in Secs.~\ref{sec:bframe:tauvertex} and \ref{sec:bframe:rfa}. 
These methods have much higher efficiency than $B$ tagging, but at the cost of significantly worse
$p_{H_b}$ resolution.

%%%%%%%%%%%% TAGGING AT B-FACTORIES
\subsubsection{$B$ tagging at the \bfacs}
\label{sec:tagging}

As described in Sec.~\ref{sec:bfactories}, the \bfacs produce $B$ mesons via $\epem \to \Upsilon(4S) \to \BB$ decays.
Since the momenta of the colliding electron-positron beams are known with high precision, the
complete reconstruction of one of the two $B$ mesons (the tag $B$ or \Btag) can be used to fully determine the momentum
of the other $B$ meson (the signal $B$ or \Bsig), simply via $p_{\Bsig} = p_{\epem}-p_{\Btag}$.

\begin{table}
  \centering
  \caption{Reconstruction efficiencies of some of the $B$ tagging algorithms employed by the
  \bfacs. \FEI\ stands for ``Full event interpretation'', \FRA\ for ``Full reconstruction'',
  and \SER\ for ``Semi-exclusive reconstruction''. The numbers are extracted from \cite{Keck:2018lcd,Lees:2013uzd}}
  
  \label{tab:tagging_perf}
  \vspace{1ex}
  \begin{tabular}{lllrr}
    \hline\hline
     $B$ tagging                  & Experiment & Algorithm              & $B^\pm$  & $B^0$    \\ \hline
    \multirow{4}{*}{Hadronic}     & Belle II   & \FEI                   & $0.76$\% & $0.46$\% \\ 
                                  & Belle II   & \FEI\ (\FRA\ channels) & $0.53$\% & $0.33$\% \\
                                  & Belle      & \FRA                   & $0.28$\% & $0.18$\% \\
                                  & \babar     & \SER                   & $0.4$\%  & $0.2$\%  \\ \hline		
    \multirow{3}{*}{Semileptonic} & Belle II   & \FEI                   & $1.80$\% & $2.04$\% \\  
                                  & Belle      & \FRA                   & $0.31$\% & $0.34$\% \\
                                  & \babar     & \SER                   & $0.3$\%  & $0.6$\%  \\ 
    \hline\hline		
  \end{tabular}
\end{table}

\begin{figure}
  \centering \includegraphics[width=0.49\textwidth]{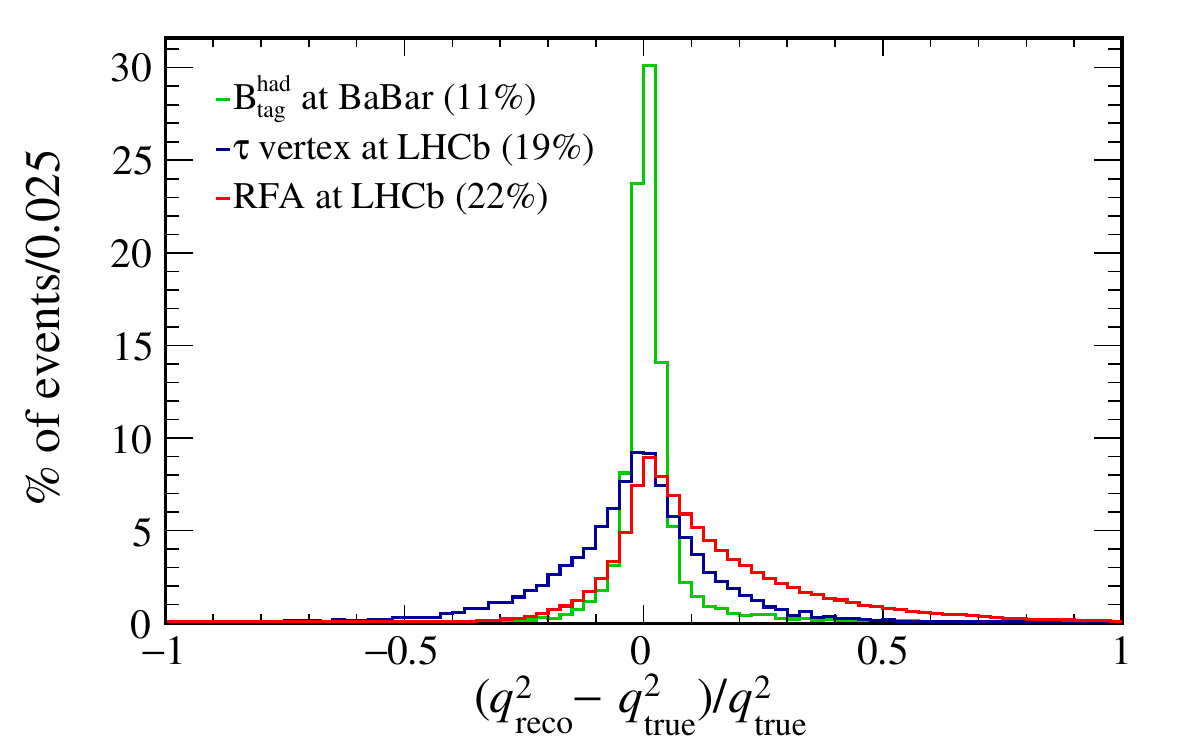} \caption{Resolution on the \qsq
  reconstruction in simulated \BDstaunu decays for the different methods of estimating the $p_{\Bsig}$
  momentum. The $\tau$ vertex and Rest-Frame Approximation~(RFA) methods used at LHCb are described in
  Secs.~\ref{sec:bframe:tauvertex} and \ref{sec:bframe:rfa}, respectively. The values in parentheses
  correspond to the RMS of each distribution. The various curves are extracted
  from \cite{Lees:2013uzd,Aaij:2017deq,Aaij:2015yra}.}
\label{fig:q2_resolution.pdf}
\end{figure}

This ``tagging'' has been implemented by the \bfacs~\cite{Bevan_2014} in the following ways:
\begin{itemize}
  \item \emph{Hadronic $B$ tagging}: the \Btag is fully reconstructed in final states that contain a charm
  hadron plus a number of pions and kaons. The full reconstruction of the decay results in the best possible
  $p_{\Bsig}$ resolution (11\% as shown in Fig.~\ref{fig:q2_resolution.pdf}) at the price of a lower 0.2--0.8\%
  efficiency (Table~\ref{tab:tagging_perf}).
  
  \item \emph{Semileptonic $B$ tagging}: the \Btag is reconstructed in its $\Bbar_\text{tag} \rightarrow
  D^{(*)} \ell \nub$ decays. This leads to efficiencies as high as 2\% thanks to the large values of the  semileptonic branching fractions.
  The presence of an unreconstructed neutrino, however, results in a poor resolution of  $p_{\Bsig}$. To mitigate this effect,
  analyses employing this technique exploit the full reconstruction of the collision event and require that no unassigned charged or neutral 
  particles should be present. They further avoid the direct use of $p_{\Bsig}$. 
  
  \item \emph{Inclusive $B$ tagging}: no attempt is made to explicitly reconstruct the $B$ decay chain. Instead, a 
  specific \Bsig candidate is first reconstructed. The tag side is then reconstructed using all remaining charged and neutral particles. 
  This leads to a high efficiency, but also poor resolution of the tag-side momentum. 
  
\end{itemize}

%% q2 resolution

Table~\ref{tab:tagging_perf} summarizes the performance of the most efficient algorithms employed by \babar, Belle, and Belle II. The Belle II numbers are based on simulations.

The hadronic $B$ tagging algorithm of \babar is based on the semi-exclusive reconstruction (\SER) of a charmed seed state of a $B \to H_c \, X$ cascade. 
Here $H_c$ can either be a charmed meson or a $J/\psi$ particle and $X$ is a number of charged
and neutral pions or a single kaon. Combinations of seed mesons with different $X$ constituents are selected based on the purity obtained from simulated samples. 
 
\begin{figure}
 \centering 
 \includegraphics[width=0.43\textwidth]{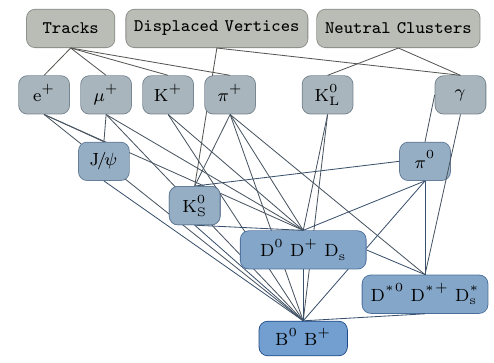} \hfil
 \caption{Schematic illustration of the \FEI\ algorithm. From \cite{Keck:2018lcd}. }
\label{fig:FEI_illustration.pdf}
\end{figure}

Belle uses a similar Ansatz, but relies on multivariate methods (either neural networks or boosted decision trees) 
to distinguish correctly-reconstructed versus wrongly-reconstructed tag candidates in a staged approach.
Figure~\ref{fig:FEI_illustration.pdf} illustrates this procedure for the Full Event Interpretation (\FEI) algorithm described in~\cite{Keck:2018lcd}.  
This algorithm reconstructs one of the $B$ mesons produced in the collision event using either hadronic or semileptonic decay channels. 
Instead of attempting to reconstruct as many $B$ meson decay cascades as possible, the \FEI\ algorithm employs a hierarchical reconstruction Ansatz in several stages.
At the initial stage, boosted decision trees are trained to identify charged tracks and neutral energy depositions as detector stable particles ($e^+$, $\mu^+$, $K^+$, $\pi^+$, $K_L^0$, $\gamma$). 
At the following stages, these candidate particles are combined into composite particles ($\pi^0$, $K_S^0$) and later heavier meson candidates ($J/\psi$, $D^0$, $D^{+}$, $D_s$). 
For each target final state, a boosted decision tree (BDT) is trained to identify probable candidates. 
The input features are the classifier outputs of the previous stages, vertex fit probabilities, and the four-momenta. 
Candidates for $D^{*0}$, $D^{*+}$, and $D_s^*$ mesons are formed similarly. 
At the final stage, all the information of the previous stages is combined to assess the viability of a \Btag candidate. 
The Full Reconstruction (\FRA) algorithm uses a very similar approach but is based on neural networks instead of BDTs. A more detailed description can be found in \cite{Feindt:2011mr}. 
The performance of the \FEI\ algorithm on early Belle~II data is discussed in~\cite{Abudinen:2020dla}.

In the future deep learning or graph-based network approaches might allow further increases 
in the reconstruction efficiency of algorithms like \FEI\ at Belle~II~\cite{boeckh_tobias_2020_21962,keck_thomas_2017_21404}. 

%% Schematic of FEI

%%%%%%%%%%%% TAU VERTEX RECONSTRUCTION AT LHCB
\subsubsection{\taupipipi vertex reconstruction at LHCb}
\label{sec:bframe:tauvertex}

At the LHC, the energies of the partons whose collisions produce the $\bbbar$ pairs are not known, 
so it is not possible to derive the 4-momentum of one $b$-hadron from the reconstruction of the other. 
However, by taking advantage of the excellent vertexing capabilities of LHCb, in the case that the $\tau$ lepton
decays to at least three charged particles, the momentum
of the parent $b$-hadron in $\HbHcutaunu$ events can still be precisely determined up to a discrete ambiguity. 
This procedure was established in 2018 by the hadronic-$\tau$ measurement of $\calR(D^{*+})$ with $\tau \to \pi^+ \pi^+ \pi^- \nu$~\cite{Aaij:2017deq}.
\footnote{The channel  \taupipipi always includes contributions from the $\tau \to \pi^- \pi^+ \pi^- \pi^0 (\pi^0) \nu$ channels, unless specified otherwise.}

\begin{figure}
  \centering \includegraphics[width=0.46\textwidth]{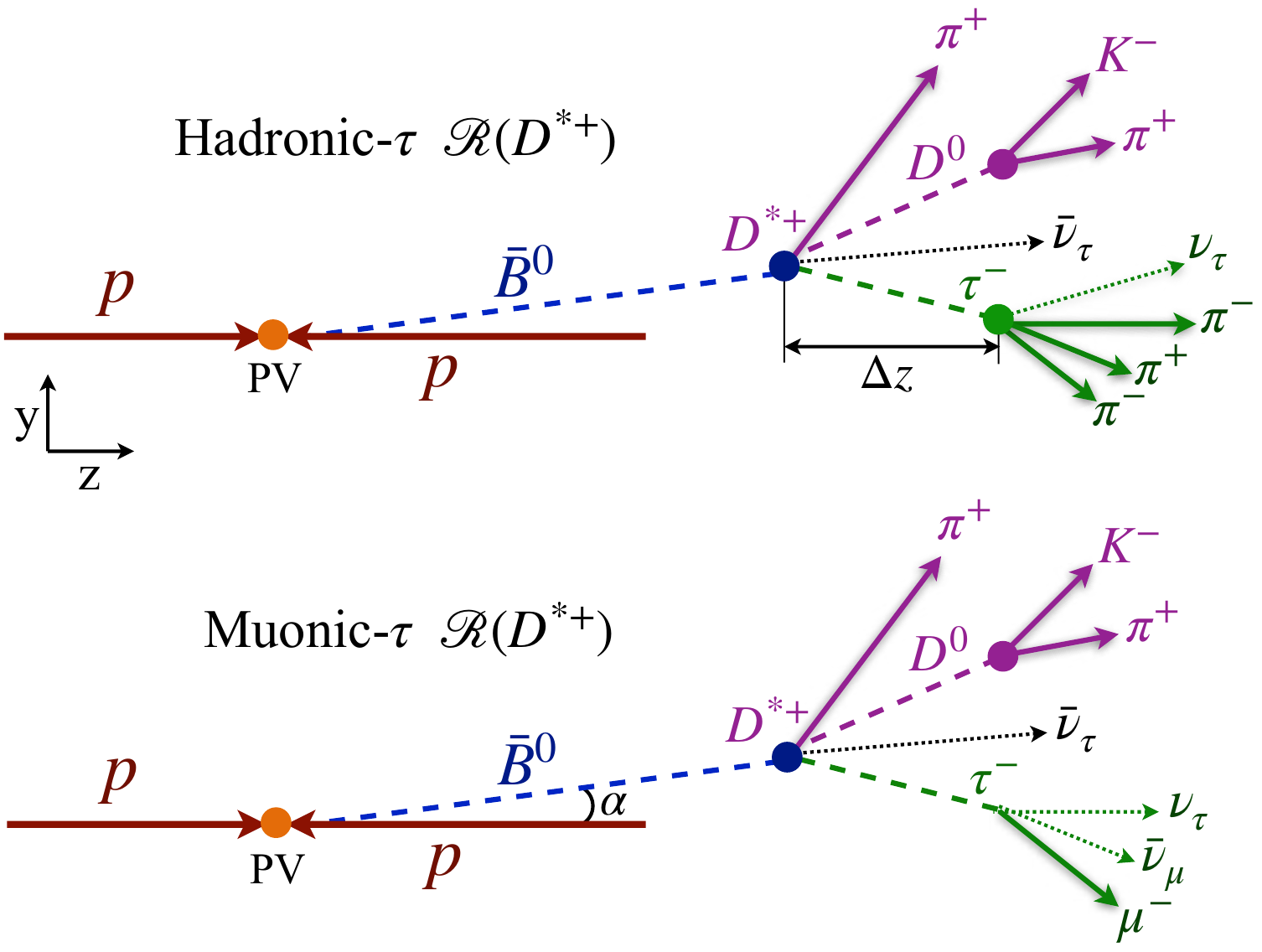}
  
  \caption{Reconstructed topologies for the \BDstaunu decays in the hadronic-$\tau$ (top) and muonic-$\tau$
  (bottom) measurements of $\calR(D^{*+})$ at LHCb~\cite{Aaij:2017deq,Aaij:2015yra}. The filled circles correspond to the reconstructed
  vertices, and solid lines to reconstructed particles. ``PV'' refers to the primary vertex, $\Delta z$ the
  distance in the $z$-direction between the $\Bbar^0$ (or $D^{*+}$) and the $\tau^-$ vertices, and
  $\alpha$ to the angle between the beam axis and the momentum of the $\Bbar^0$ meson.}
\label{fig:topologies}
\end{figure}

In general, about 100 tracks arise from a primary vertex (PV) within a $pp$ collision at LHCb, such that the
location of this vertex can be measured to an excellent precision of around 10~$\mu$m along the beam direction. In
$\Bbar^0 \rightarrow D^{*+} \tau^-\nutb$ events with the $D^{*+}$ meson decaying promptly via the $D^{*+}\to D^0\pi^+$ strong decay, 
the $D^0$ vertex can be reconstructed as the intersection of its kaon and pion daughters
with a 150~$\mu$m precision along the $z$ direction~\cite{Aaij:2017deq} (see Fig.~\ref{fig:topologies} top).
The vertex for the \taupipipi decay can be measured to a 200~$\mu$m precision.
Because of the very small angle between the directions of the bachelor pion produced in the $D^{*+}$ decay  and the reconstructed $D^0$, their intersection
has poor precision and is not used in the determination of the position of the $\Bbar^0$ vertex.
Instead, this position is estimated with a $\sim$1~mm resolution as the intersection of the $D^{*+}$ and \tauon trajectories, 
where the \tauon line of flight is approximated by the $\pi^- \pi^+ \pi^-$ direction. Thanks to the large boost of $b$-hadrons at LHCb,
$\beta\gamma \sim 50$, these three vertices are well separated and determine the directions of flight of the
$\Bbar^0$ meson and $\tau$ lepton momenta---the unit vectors $\hat{\bm{p}}_{B}$ and $\hat{\bm{p}}_{\tau}$, respectively---with fairly good precision.

%% Topologies of LHCb decays

With $\hat{\bm{p}}_{\tau}$ known and the $\pi^- \pi^+ \pi^-$ hadronic state fully reconstructed, the $\tau$ energy can be determined up to a two-fold ambiguity, 
arising from the solution of the quadratic relation $(p_{\tau} - p_{\pi\pi\pi})^2 = 0$. 
This result, when further combined with $\hat{\bm{p}}_{B}$ and the full reconstruction of the $D^{*+}$, 
in turn allows the determination of the $B$ momentum up to a four-fold ambiguity from the quadratic $(p_{B} - p_{D^*} - p_{\tau})^2 = 0$.
The resulting overall $\qsq$ resolution is around $19\%$.

%%%%%%%%%%%% REST FRAME APPROXIMATION AT LHCB
\subsubsection{Rest frame approximation with $\tau \to \mu \nu \overline{\nu}$ at LHCb}
\label{sec:bframe:rfa}

It is not possible to reconstruct the $\tau$ vertex when the $\tau$ lepton is identified by its 1-prong
$\tau \to \mu \nu \overline{\nu}$ decay (Fig.~\ref{fig:topologies} bottom). Thus, semitauonic measurements at
LHCb that make use of this decay mode estimate the momentum of the $b$-hadron via the \emph{rest frame
approximation} (RFA) instead. This procedure assumes that the proper velocity of  the $H_b$ hadron along the $z$-axis---the beam
axis---is the same as that of the reconstructed charm-muon system, $\mu H_{c}$.
This leads to the relationship $(p_{H_b})_z/m_{H_b} =  (p_{\mu H_{c}})_z/m_{\mu H_{c} }$. 
Since the direction of flight of the $b$-hadron can be determined by the displacement of the $H_b$ decay vertex from the primary vertex, 
the $H_b$ momentum can then be estimated via
\begin{equation}
	|p_{H_b}| = \frac{m_{H_b}}{m_{\mu H_{c}}} (p_{\mu H_{c}})_z \sqrt{1+\tan^2\alpha}\,,
\end{equation}
where $\alpha$ is the angle between the $H_b$ direction of flight and the $z$ axis, as shown in Fig.~\ref{fig:topologies}.

In the highly boosted regime of LHCb, the RFA is a fairly good approximation that leads to an adequate
overall $\qsq$ resolution of about $22\%$ (see Fig.~\ref{fig:q2_resolution.pdf}), albeit with a long tail on the positive
side and some bias. It is worth noting that this resolution is highly $\qsq$-dependent, as it varies between
$34\%$ for $q^2<5\text{ GeV}^2$ and $7\%$ at $q^2>9\text{ GeV}^2$.

In general, semitauonic measurements at LHCb that make use of the hadronic-$\tau$ reconstruction will have
better precision for the reconstruction of kinematic distributions than muonic-$\tau$
measurements. In contrast, the latter may have a better ultimate precision in the determination of the ratios
$\calR(H_{c})$ because they do not depend on external branching fractions in the normalization of the
signal \HbHctaunu decays, such as those used in Eq.~\eqref{eq:rd_hadronic} below.

In the future, LHCb may be able to improve the precision on the $b$-hadron momentum reconstruction by taking
advantage of the large samples of $b$-hadrons that will be collected over the next decade and a half. For
instance, the reconstruction of $B^+$ mesons arising from $B_{s2}^*\rightarrow B^+K^-$ decays allows for a
higher-precision determination of the $B^+$ kinematics by constraining the invariant mass of the $B^+K^-$
system to the known $B_{s2}^*$ mass, but it comes at the price of a less than 1\% reconstruction
efficiency. This technique has already been successfully employed to reconstruct $B^- \to
D^{(*,**)0} \mu^- \bar{\nu}_\mu$ decays~\cite{Aaij:2018unp}, and could be applied in the future to semitauonic
decays as well.

\section{Experimental Tests of Lepton Flavor Universality}
\label{sec:measurements}

%%%%%%%%%%%%%%%%%%%%%%%%%%%%%%%%%%%%%%%%%%%%%%%%%%%%%%%%%%%%%%%%%%%%%%
%%%%%%%%%%%%%%%%%%%%%%% OVERVIEW OF MEASUREMENTS %%%%%%%%%%%%%%%%%%%%%
%%%%%%%%%%%%%%%%%%%%%%% OVERVIEW OF MEASUREMENTS %%%%%%%%%%%%%%%%%%%%%

%% Superseded measurements
The decay \BDstaunu was first observed in 2007 by the Belle collaboration~\cite{Matyja:2007kt}, and subsequent
measurements by \babar~\cite{Aubert:2007dsa} and Belle~\cite{Adachi:2009qg,Bozek:2010xy} found evidence
for \BDtaunu decays as well. These measurements all saw values of \RDx that exceeded the SM expectations, but
the significance of these excesses was low due to the large uncertainties involved in these early results:
above 20\% for \RDs and over 30\% for \RD. All of these measurements have now been superseded, so they will
not be further discussed in this review.

%% List of measurements
The first evidence for an excess of \BDxtaunu decays was reported by \babar in 2012~\cite{Lees:2012xj}, a
measurement that also included the first observation of \BDtaunu decays. Similar excesses have been reported
since by the Belle~\cite{Huschle:2015rga, Sato:2016svk, Belle:2019rba, Hirose:2017dxl} and LHCb
experiments~\cite{Aaij:2015yra, Aaij:2017deq}. The persistent
nature of these anomalies has spurred wide interest in semitauonic decays and, as a result, other channels
that proceed via \butaunu or different \bctaunu transitions are being studied. Three such results have been
published so far: Belle's search for \Bpitaunu decays~\cite{Hamer:2015jsa}, and LHCb's measurement
of \RJ~\cite{Aaij:2017tyk} and of \RLc~\cite{Aaij:2022}. The first measurements of the polarization of some of the
decay products have been reported by Belle~\cite{Sato:2016svk, Abdesselam:2019wbt} as well.

%% TOC
In this section we describe the key features of all of these measurements regarding their event selection,
background determination, main uncertainties, and signal extraction. The following subsections \ref{sec:hadronic_tag}--\ref{sec:belle_polarization} group the
various results according to their $b$-hadron \emph{tagging} method which, as we saw in Sec.~\ref{sec:bframe},
can be employed to determining the momentum of the parent $b$-hadron and has a substantial impact on the approach to
determine the signal yields and on the composition of the background contributions. Table~\ref{tab:overview}
shows an overview of the results and the subsections in which they are discussed.  Additionally, the section following
this one (Sec.~\ref{sec:systematics}) offers a deeper dive into the various sources of systematic uncertainty to
which these measurements are subject, as well the prospects for its reduction.
Section~\ref{sec:interpretation} provides combinations of the various \RDx results and comparisons of all the
observables with their respective SM predictions.

\begin{table}[t]
\renewcommand*{\arraystretch}{1.5}
\newcolumntype{D}{ >{\centering\arraybackslash } m{1cm} <{}}
\newcolumntype{C}{ >{\centering\arraybackslash $} m{2.75cm} <{$}}
\newcolumntype{E}{ >{\centering\arraybackslash $} m{3.5cm} <{$}}
\newcolumntype{R}{ >{\centering\arraybackslash \scriptsize\bgroup} c <{\egroup}}
 \caption{Summary of the different results covered by this review, classified by the measured 
 observable and the deployed method. The references for each experiment are given at the bottom of the
 table; the relevant sections of this review are provided below each result.}  \label{tab:overview}
\begin{threeparttable}
  \scalebox{0.82}{
  \parbox{1.1\linewidth}{
    \begin{tabular*}{1.1\linewidth}{@{\extracolsep{\fill}}D C C E}
      \hline\hline
      \textbf{Obs.}                                      & \multicolumn{3}{c}{\textbf{Method}}                                                                                     \\
                                                         & \text{Hadronic tag}  		     & \text{Semilep. tag}	 	      & \text{Untagged}                    \\
      \hline
      \multirow{4}{*}[5pt]{$\RD$}                        & 0.440(58)(42)\tnote{Ba12}                 & 0.307(37)(16)\tnote{B20} 	      &                                    \\[-5pt]
                                                         & \text{\ref{sec:bfactories_hadtag_rdx}}    & \text{\ref{sec:bfactories_leptag_rdx}} &                                    \\
                                                         & 0.375(64)(26)\tnote{B15a}   		     &                                        &                                    \\[-5pt]
                                                         & \text{\ref{sec:bfactories_hadtag_rdx}}    &                                        &                                    \\
      \hline
      \multirow{6}{*}[7.5pt]{$\RDs$}                     & 0.332(24)(18)\tnote{Ba12}  		     & 0.302(30)(11)\tnote{B16b}	      & 0.336(27)(30)\tnote{L15}           \\[-5pt]
                                                         & \text{\ref{sec:bfactories_hadtag_rdx}}    & \text{\ref{sec:bfactories_leptag_rdx}} & \text{\ref{sec:lhcb_muonic_rds}}   \\
                                                         & 0.293(38)(15)\tnote{B15a} 		     & 0.283(18)(14)\tnote{B20} 	      & 0.280(18)(25)(13)\tnote{L18b}      \\[-5pt]
                                                         & \text{\ref{sec:bfactories_hadtag_rdx}}    & \text{\ref{sec:bfactories_leptag_rdx}} & \text{\ref{sec:lhcb_hadronic_rds}} \\
                                                         & 0.270(35)^{(+28)}_{(-25)}\tnote{B17}     &                                        &                                    \\[-5pt]
                                                         & \text{\ref{sec:bfactories_hadtag_rdx}}    &                                        &                                    \\	
      \hline
      \multirow{2}{*}[2.5pt]{$P_{\tau}(\Dstar)$}	 & -0.38(51)^{(21)}_{(16)}\tnote{B17} 	     &                                        &                                    \\[-5pt]
      							 & \text{\ref{sec:bfactories_hadtag_taupol}} &                                        &                                    \\					
      \hline
      \multirow{2}{*}[2.5pt]{$F_{L,\tau}(\Dstar)$\!\!\!} &  & & 0.60(8)(4)\tnote{B19}                                           \\[-5pt]
      							& &  & \text{\ref{sec:exp:Dspol}}                                             \\
      \hline
      \multirow{2}{*}[2.5pt]{$\RJ$	}                &                                           &                                        & 0.71(17)(18)\tnote{L18a}           \\[-5pt]
                                                         &                                           &                                        & \text{\ref{sec:lhcb_rjpsi}}        \\
      \hline
      \multirow{2}{*}[2.5pt]{$\calR(\pi)$}            & 1.05(51)\tnote{B16a}                      &                                        &                                    \\[-5pt]
                                                         & \text{\ref{sec:belle_pitaunu}}            &                                        &                                    \\
      \hline
      \multirow{2}{*}[2.5pt]{$\RLc$}       	& 					    & 					   &   0.242(26)(40)(59)\tnote{L22}           \\[-5pt]
      						       &                                           &                                        & \text{\ref{sec:lhcb_hadronic_rlc}}        \\                 
      \hline\hline
\end{tabular*}
\begin{tablenotes}[normal]
  \item[Ba12] \babar \cite{Lees:2012xj, Lees:2013uzd}, with $\rho = -0.31$. 
  \item[B15a] Belle \cite{Huschle:2015rga}, with $\rho = -0.50$.
  \item[B16a] Belle \cite{Hamer:2015jsa}, when combined with world-averaged $\text{Br}(\Bpiellnu)$.
  \item[L15]  LHCb \cite{Aaij:2015yra}.
  \item[B16b] Belle \cite{Sato:2016svk}.
  \item[B17] Belle \cite{Hirose:2016wfn, Hirose:2017dxl}, with single-prong $\tau$ hadronic decays.
  \item[L18a] LHCb \cite{Aaij:2017tyk}.
  \item[L18b] LHCb \cite{Aaij:2017deq}, with $\tau \to \pi^+\pi^+\pi^- \nu$ updated taking into account
 the latest HFLAV average of $\Br(B^0 \to D^{*+} \ell\nu) = 5.08 \pm 0.02 \pm 0.12)$\%. The third uncertainty is from external branching fractions.
  \item[B19] Belle \cite{Abdesselam:2019wbt}, using inclusive tagging.
  \item[B20] Belle \cite{Belle:2019rba}, with $\rho = -0.52$.
  \item[L22] LHCb \cite{Aaij:2022}, with $\tau \to \pi^+\pi^+\pi^- \nu$. The third uncertainty is from external branching fractions.
\end{tablenotes}
}}
\end{threeparttable}         
\end{table}

There exist, in addition, several measurements of the inclusive $B \to X_c \, \tau \nu$ rate that we will not cover in this section. 
These comprise LEP measurements of $b \to X \tau \nu$~\cite{Barate:2000rc,Abreu:1999xe,Acciarri:1996wt,Acciarri:1994hb,Abbiendi:2001fi},
that require assumptions about the cancellation of 
hadronization effects in order to be interpreted as $B \to X\, \tau \nu$ measurements,
and a recent result~\cite{handle:20.500.11811/7578} that is unpublished. 
A comparison of the predicted and measured rates from inclusive and exclusive semitauonic decays is presented in Sec.~\ref{sec:incl_excl.pdf}.

%%%%%%%%%%%%%%%%%%%%%%%%%%%%%%%%%%%%%%%%%%%%%%%%%%%%%%%%%%%%%%%%%%%%
%%%%%%%%%%%%%%%%%%%%% HADRONIC TAG MEASUREMENTS %%%%%%%%%%%%%%%%%%%%
\subsection{$B$-factory measurements with hadronic tags}
\label{sec:hadronic_tag}

This section describes some of the most recent semitauonic results involving hadronic $B$ tags: the
measurements of \BDxtaunu decays by \babar~\cite{Lees:2012xj, Lees:2013uzd} and Belle~\cite{Huschle:2015rga}
in Sec.~\ref{sec:bfactories_hadtag_rdx}, as well as a 2015 search for $\Bpitaunu$ decays by
Belle~\cite{Hamer:2015jsa} in Sec.~\ref{sec:belle_pitaunu}.  An additional measurement of $\BDxtaunu$ decays by
Belle involving hadronic tags focused on the polarization of the $\tau$ lepton
\cite{Hirose:2016wfn,Hirose:2017dxl}, and is described in Sec.~\ref{sec:belle_polarization}.

%%%%%%%%%%%%%%%%%%%%% B-FACTORIES HADRONIC TAG RD(*) %%%%%%%%%%%%%%%%%%%%
\subsubsection{\RDx with \tauellnu}
\label{sec:bfactories_hadtag_rdx}
%%%%%%%%%%%%%%%%%%%%% B-FACTORIES HADRONIC TAG RD(*) %%%%%%%%%%%%%%%%%%%%

The \babar experiment published the first high-precision measurement of $\RDx$ based on their full dataset of
$471\times10^6$ $\BB$ pairs in 2012~\cite{Lees:2012xj, Lees:2013uzd}. The Belle experiment followed in 2015
with an analysis of their $772\times10^6$ $\BB$ pair dataset~\cite{Huschle:2015rga}, employing a similar
strategy. In both cases, signal $\BDxtaunu$ and normalization $\BDxellnu$ decays were selected using the same
particles in the final state: a $D$ or $D^*$ meson, and a charged light lepton $\ell = e$ or $\mu$. 
In the case of signal events, the light lepton $\ell$ comes from the secondary \tauellnu decay, which leads to two additional neutrinos in the final state and a typically lower lepton
momentum. The $D$ mesons are reconstructed by combinations of $K^+$, $K_S^0$, $\pi^+$, and $\pi^0$ mesons with
invariant masses close to the nominal $D^0$ and $D^+$ masses, covering $25$--$35\%$ of the total $D$ branching
fractions. The heavier $D^*$ mesons are identified by the $D^{*+}\to D^0\pi^+, \,D^+\pi^0$ and $D^{*0}\to
D^0\pi^0,\, D^0\gamma$ decays.

In order to separate signal from normalization decays as well as to reduce background contributions,
the event is also required to have a fully reconstructed hadronic $\Btag$ and no additional tracks; see Sec.~\ref{sec:tagging}. 
As described there, the reconstruction efficiency of the $\Btag$ is only $\approx 0.3\%$, but
it allows these measurements to accurately determine the $4$-momentum of the signal $B$,
which in turn is used to calculate the momentum transfer $q^2=(p_{\Bsig}-p_{\Dx})^2$ and 
the missing momentum of the unreconstructed neutrinos $\pmiss = p_{\Bsig}-p_{\Dx}-p_\ell = p_{\epem}-p_{\Btag}-p_{\Dx}-p_\ell$.
The invariant missing mass $\mmiss=\pmiss^2$ peaks at zero for the one-neutrino normalization events, 
but has a broad distribution at positive values for signal events with three neutrinos in the final state.

A key variable to further reduce background contributions is $\ECL$: the sum of the energy deposits in the
calorimeter that are not associated with the tag or signal $B$ decays. Events involving signal and normalization
decays have all their visible final state particles reconstructed, but background decays to $\Ddouble$ mesons (among
others) can enter the signal selection when their daughter $\piz$ mesons or photons are unassigned. Both \babar and Belle feed
$\ECL$ to multivariate classifiers that are trained to reject these background contributions. In the case of
\babar, the output of the classifier, a boosted decision tree, is required to have a minimum value for the
event to be selected. As we describe below, Belle fits the output distributions of the classifier (from a neural
network) directly. 
Finally, only events with $q^2>4\,\text{GeV}^2$ are selected, a requirement that takes advantage of
the momentum transfer of signal events being kinematically
constrained to lie above $m^2_\tau = 3.16\,\text{GeV}^2$.

The number of signal, normalization, and background events in each of the $D^0\ell$, $D^+\ell$, $D^{*0}\ell$,
and $D^{*+}\ell$ data samples is determined by maximum likelihood fits to the observed data distributions. The
ratios of yields for the isospin-related contributions---e.g., $D^0\ell$ versus $D^+\ell$ or $D^{*0}\ell$
versus $D^{*+}\ell$---are constrained by the known branching fractions and simulated relative efficiencies. 
\babar employs an additional fit
without these constraints that checks the consistency with the expected percent-level degree of isospin
breaking.  The probability distribution functions (PDFs) that describe each of the contributions are taken
from Monte Carlo simulations that make use of the CLN form factor parametrization (Sec.~\ref{sec:th:hqet})
for the signal and normalization modes, the LLSW form factor parametrization~\cite{Leibovich:1997tu} for \BDssltnu decays,\footnote{As a reminder,
throughout this review $l$ stands for $e$, $\mu$, or $\tau$, and $\ell$ for $e$ or $\mu$.} and other
(phase-space based) models augmented with corrections from data control samples for the rest of the background
contributions. Additional assumptions on the \Ddouble branching fractions are described in Sec.~\ref{sec:syst:dss_bf}.

%% Fits of hadronic tag measurements
\begin{figure*}
  \includegraphics[width=\textwidth]{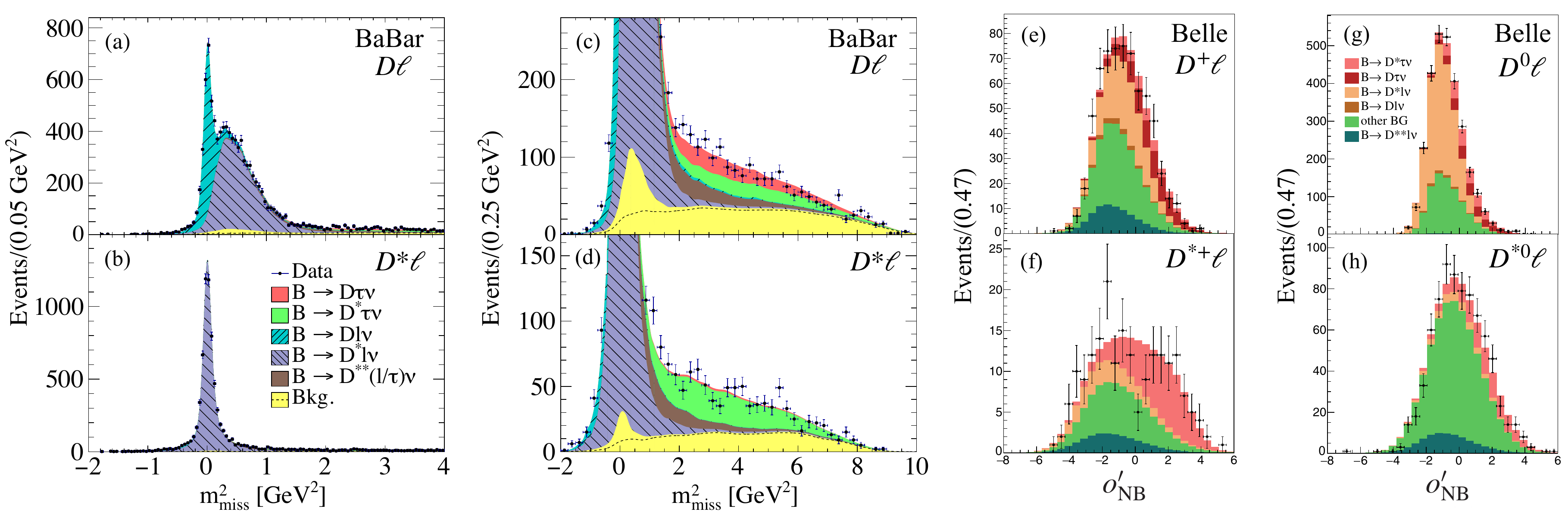}

    \caption{Projections of the signal fits for the \babar~\cite{Lees:2012xj} and Belle~\cite{Huschle:2015rga}
    measurements of \RDx with hadronic tagging. (a-b) Full \mmiss projections of the \babar fit showing the
    normalization components for the $D\ell$ and $D^*\ell$ samples (combination of $D^{(*)0}\ell$ and
    $D^{(*)+}\ell$). (c-d) \mmiss projections of the \babar fit focusing on the signal contributions at
    high \mmiss.  (e-h) Full projections of the fit to the neural network output $o'_\text{NB}$ by Belle in
    the region $\mmiss>0.85\text{ GeV}^2$ for the four $\Dx\ell$ samples.}
\label{fig:had_tag_fits.pdf}
\end{figure*}

\begin{table}
 \renewcommand*{\arraystretch}{1.2}
  \centering
  
  \caption{Comparison of the total yields extracted by the isospin-constrained fits
    from \babar~\cite{Lees:2012xj} and Belle~\cite{Huschle:2015thesis}. The ``$\epsilon$ ratio'' column corresponds to
    the ratio of the Belle fitted yields to the \babar fitted yields normalized by the datasets, 471 million of \BB pairs
    for \babar and 772 million for Belle.}
    
  \label{tab:had_tag_yields}
  \vspace{1ex}
  \begin{tabular}{llrrr}
    \hline\hline
    Sample                     & Contribution & \babar & Belle & $\epsilon$ ratio \\ \hline
    \multirow{4}{*}{$D\ell$}   & \BDtaunu     & 489    & 320   & 0.40  \\
                               & \BDellnu     & 2981   & 3147  & 0.64  \\
                               & \BDssltnu    & 506    & 239   & 0.29  \\
                               & Other bkg.   & 1033   & 2005  & 1.18  \\ \hline
    
    \multirow{4}{*}{$D^*\ell$} & \BDstaunu    & 888    & 503   & 0.35  \\
                               & \BDsellnu    & 11953  & 12045 & 0.61  \\
                               & \BDssltnu    & 261    & 153   & 0.36  \\
                               & Other bkg.   & 404    & 2477  & 3.74  \\ 
    \hline\hline
  \end{tabular}
\end{table}

The \babar analysis employs a two-dimensional fit to the $\mmiss$ and the charged lepton energy in the
$B$ rest frame, $\Esl$, while Belle fits the $\mmiss$ distribution for $\mmiss<0.85\text{ GeV}^2$ and the output of
the classifier at high $\mmiss$. Figure~\ref{fig:had_tag_fits.pdf} shows some of the relevant projections for both
fits. The narrow peaks in Fig.~\ref{fig:had_tag_fits.pdf}(a-b), including that of the \emph{feed-down} $\BDsellnu$
decays reconstructed in the $D\ell$ sample with a broader $\mmiss$ distribution, illustrate the power of
hadronic tagging in discriminating signal from normalization decays. Table~\ref{tab:had_tag_yields} shows a
comparison of their fitted yields. Although the Belle dataset is $64\%$ larger, the signal yields are about
$40\%$ smaller due to the lower reconstruction efficiency. The differences in the background yields
are primarily due to \babar placing a requirement on the multivariate classifier and Belle fitting
its output instead.

The most challenging background contribution arises from $\BDssellnu$ and $\BDsstaunu$ decays. The $\BDssellnu$ processes are
estimated in control samples with the same selection as the signal samples, except for the addition of a $\piz$
meson. In these control samples, decays of the form $\Bbar \rightarrow D^{(*)}\piz \ell^-\nulb$ have values of
$\mmiss$ close to zero, so that their yields are easily determined with fits to this variable. This fit is performed
simultaneously with the fits to the signal samples, and the $\BDssltnu$ contribution to both is linked by
the ratio of expected yields taken from the simulation. Additional backgrounds from continuum and combinatorial $B$
processes are estimated from data control samples and are fixed in the fits.

\begin{table}
  \renewcommand*{\arraystretch}{1.2}
  \centering
  
  \caption{Summary of the relative uncertainties for the \babar~\cite{Lees:2012xj} and
    Belle~\cite{Huschle:2015rga} measurements of \RDx with hadronic tagging.}  
  \label{tab:had_tag_errors}
  \vspace{1ex}
  \begin{tabular}{llccccc}
    \hline\hline
    \multirow{3}{*}{Result} & \multirow{3}{*}{Contribution}    & \multicolumn{4}{c}{Uncertainty [\%]} & \multirow{3}{*}{Ratio} \\
                  && \multicolumn{2}{c}{\babar} & \multicolumn{2}{c}{Belle}  &     \\ 
                  && Sys. & Stat. & Sys. & Stat. & \\ \hline
    \multirow{7}{*}{\RD}  & \BDssltnu               & 5.8 &       & 4.4 &        & 0.76       \\
                          & MC stats                & 5.7  &      & 4.4  &       & 0.78       \\
                          & \BDlnu                  & 2.5  &      & 3.3  &       & 1.30       \\
                          & Other bkg.              & 3.9 &       & 0.7 &        & 0.18       \\
                          & Particle ID             & 0.9 &       & 0.5 &        & 0.54       \\
                          & {\bf Total systematic}  & {\bf 9.6}  && {\bf 7.1}   && {\bf 0.74} \\
                          & {\bf Total statistical} & &{\bf 13.1} && {\bf  17.1} & {\bf 1.31} \\\cline{2-7}
                          & {\bf Total}             & \multicolumn{2}{c}{{\bf 16.2}} & \multicolumn{2}{c}{{\bf 18.5}}  & {\bf 1.14} \\ \hline
    
    \multirow{7}{*}{\RDs} & \BDssltnu               & 3.7 &       & 3.4  &       & 0.90       \\
                          & MC stats                & 2.8  &      & 3.6 &        & 1.31       \\
                          & \BDslnu                 & 1.0  &      & 1.3  &       & 1.31       \\
                          & Other bkg.              & 2.3 &       & 0.7 &        & 0.29       \\
                          & Particle ID             & 0.9  &      & 0.5   &      & 0.54       \\
                          & {\bf Total systematic}  & {\bf 5.6}  && {\bf 5.2}   && {\bf 0.93} \\
                          & {\bf Total statistical} && {\bf 7.1}  && {\bf 13.0}  & {\bf 1.83} \\\cline{2-7}
                          & {\bf Total}             & \multicolumn{2}{c}{{\bf 9.0}}  & \multicolumn{2}{c}{{\bf 14.0}}  & {\bf 1.56} \\ 
    \hline\hline
  \end{tabular}
\end{table}

%% mES and ECL
\begin{figure*}
  \includegraphics[width=\textwidth]{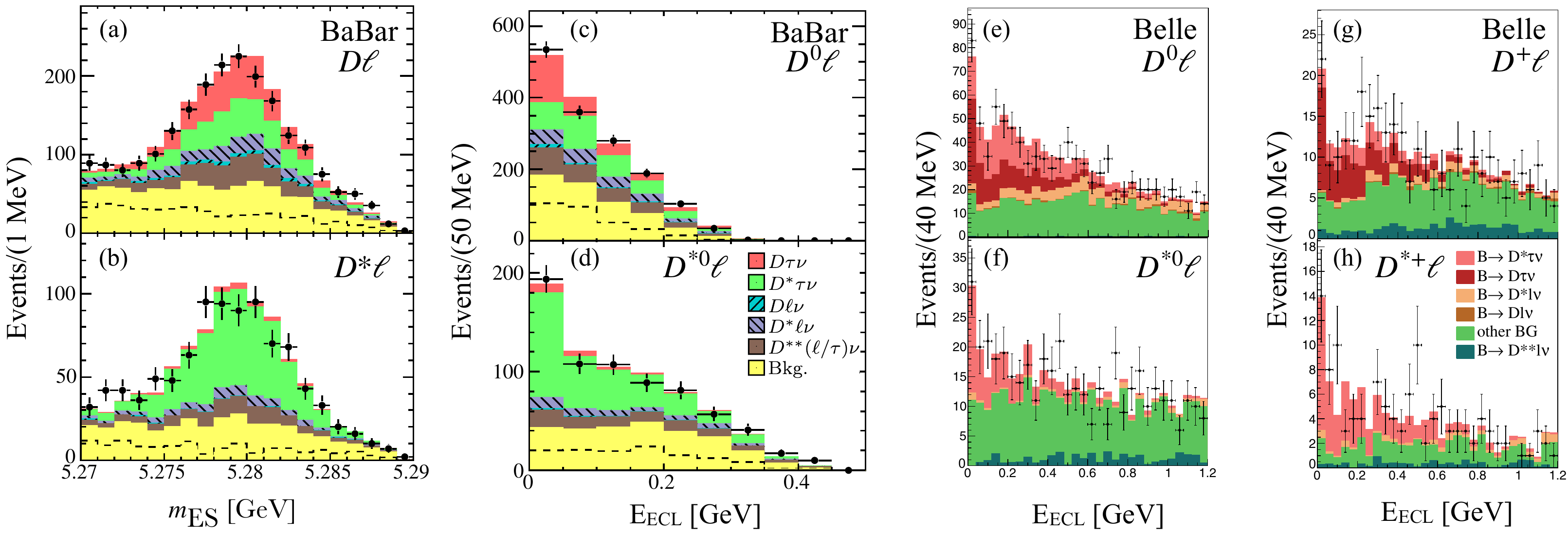}
    \caption{Checks on the kinematic distributions for events in the signal
    enhanced high \mmiss region [$\mmiss>1.5\text{ GeV}^2$ for (a-d) and $\mmiss>2\text{ GeV}^2$ for (e-h)].
    The solid histograms correspond to the simulation scaled to the fit results. Adapted
    from \cite{Lees:2013uzd, Huschle:2015rga}.}
\label{fig:had_tag_checks.pdf}
\end{figure*}

Table~\ref{tab:had_tag_errors} summarizes all the sources of uncertainty in the \RDx ratios measured by both
analyses.  The largest uncertainties come from the \BDssltnu contributions and the limited size of the
simulated samples (``MC stats''). The latter uncertainty affects primarily the PDFs describing the kinematic
distributions of all the components in the fit. The branching fraction ratios are calculated as
\begin{equation}
  \RDx = \frac{N_\text{sig}}{N_\text{norm}} \frac{\epsilon_\text{norm}}{\epsilon_\text{sig}},
\end{equation}
where $N_\text{sig}$ and $N_\text{norm}$ are the number of signal and normalization events determined by the
fit, respectively, and $\epsilon_\text{sig}/\epsilon_\text{norm}$ is the ratio of efficiencies taken from
simulation. Since the signal and normalization decays are reconstructed with the same particles in the final
state, many uncertainties cancel in the ratio leading to a relatively small 2--3\% overall uncertainty for
this quantity.

%% RD(*) results
\begin{table} 
  \renewcommand*{\arraystretch}{1.5}
  \caption{Results of the \babar~\cite{Lees:2012xj} and
  Belle~\cite{Huschle:2015rga} measurements of \RDx with hadronic tagging. The first uncertainty is
  statistical and the second is systematic.}
    
  \label{tab:had_tag_results} \vspace{0.1in}
  \begin{tabular}{l cc} \hline\hline
    Result & \babar                      & Belle                       \\ \hline
    \RD    & $0.440 \pm 0.058 \pm 0.042$ & $0.375 \pm 0.064 \pm 0.026$ \\
    \RDs   & $0.332 \pm 0.024 \pm 0.018$ & $0.293 \pm 0.038 \pm 0.015$ \\
    \hline\hline
  \end{tabular}
\end{table}

%% q2 distributions for SM and tanBeta/mH = 0.45 GeV^2
\begin{figure*}
  \includegraphics[width=0.9\textwidth]{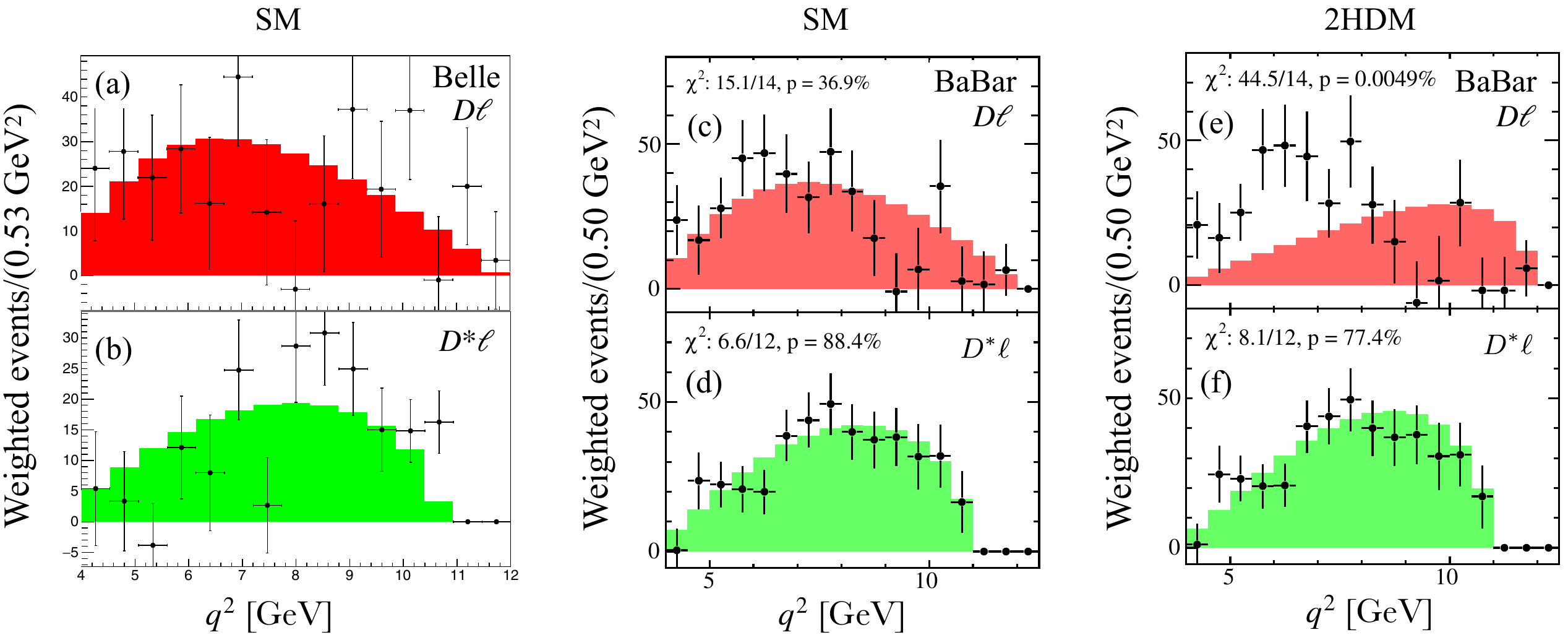} \caption{Efficiency-corrected $q^2$ distributions
    for \BDtaunu (top panels) and \BDstaunu (bottom panels) events with $\mmiss>0.85\text{ GeV}^2$ (a-b) and
    $\mmiss>1.5\text{ GeV}^2$ (c-f). The shaded distributions correspond to the SM expectations in (a-d) 
    and a Type-II 2HDM with $\tan\beta/m_{H^\pm}=0.45\text{ GeV}^{-1}$ in (e-f), the value that reproduces the value of \RD measured by \babar. The $\chi^2$ values are calculated
  based on the statistical uncertainties only. Adapted
    from \cite{Lees:2013uzd, Huschle:2015rga}.}
\label{fig:had_tag_q2.pdf}
\end{figure*}

Table~\ref{tab:had_tag_results} shows the results from the \babar and Belle analyses, which are compatible
within uncertainties. The isospin-unconstrained results from \babar (Table~\ref{tab:iso_unconstrained} in
Sec.~\ref{sec:analysis_rdx}) show good agreement with the expected percent-level degree of isospin breaking.
The total uncertainty on $\RDx$ in
these measurements is dominated by the statistical uncertainty, so the much larger data samples expected to be
collected by Belle II should improve these results significantly.

Thorough checks of the stability of these results were performed, including separate fits to the muon and
electron samples, fits to the various running periods, and fits to samples that modified selection requirements by varying the
signal over background ratio, $S/B$, from $1.27$ to $0.27$. In all cases, the results were compatible with the
nominal result. Additionally, a number of kinematic distributions of signal-enriched samples were compared
with the fitted SM signal plus background model and result in good agreement overall. Figure~\ref{fig:had_tag_checks.pdf}
shows the distributions for the energy substituted mass
$\mES=\sqrt{E^2_\text{beam}-\boldsymbol{p}^2_\text{tag}}$, which peaks at the $B$ mass for correctly
reconstructed events, and \ECL. In both cases, the distributions are consistent with the fitted signal
events to be coming from $B$ mesons with no additional unreconstructed particles in the event.

Finally, Fig.~\ref{fig:had_tag_q2.pdf} shows the measured efficiency-corrected $q^2$ distributions for \BDxtaunu
decays and finds good agreement with the SM expectations. The measured distributions are also compared in
panels (e-f) with the expectations from the Type-II two-Higgs doublet model (2HDM) with
$\tan\beta/m_{H^\pm}=0.45\text{ GeV}^{-1}$, which proceeds primarily via a scalar mediator. The \babar
analysis recalculates the signal PDFs, reweighting the light lepton momentum to approximately account for the
changes in helicity for each value of $\tan\beta/m_{H^\pm}$ and fits the data again, so the data points in
Fig.~\ref{fig:had_tag_q2.pdf} (c-d) are somewhat different than those in panels (e-f) due to the slightly
different background and signal cross-feed subtraction. Including systematic uncertainties, this benchmark
model is excluded at greater than $95\%$ confidence level.

%%%%%%%%%%%%%%%%%%%%%%%%%% BELLE B -> pi tau nu %%%%%%%%%%%%%%%%%%%%%%%%
\subsubsection{Search for \Bpitaunu decays}
\label{sec:belle_pitaunu}
Charmless semitauonic decays offer an interesting, independent probe of LFUV to complement the excesses observed
in various $\RDx$ measurements. Although they involve different four-Fermi operators, and are CKM suppressed, 
they also offer access to third generation semileptonic decays in
an experimental setting with very different background composition. The most promising candidate for a first observation is the
\Bpitaunu channel. Further, even modest precision could already strongly constrain new physics
models involving scalar mediators such as the Type-II 2HDM~\cite{Bernlochner:2015mya}.

A first limit on the branching fraction of this decay was obtained by Belle in 2015~\cite{Hamer:2015jsa},
which followed a similar strategy to that employed by Belle's hadronic tag measurement of \RDx.
For the \Bpitaunu analysis, \Btag mesons are selected only when the best 
candidate is compatible with the decay of a neutral $B$ meson. In order to
boost the reconstructed number of \Bpitaunu signal decays, 
both electronic \tauenu and hadronic one-prong \taupinu and \taurhonu decays were included in
the reconstruction. The signal side is thus required to have at
most two oppositely charged tracks, with one of those tracks having a particle identification 
compatible with an electron in the case of \tauenu decays.
For the $\rho^+ \to \pi^+\pi^0$ reconstruction, neutral pion candidates, which are not used in
the tag-reconstruction, are constructed from neutral energy depositions in the calorimeter. If multiple
$\rho$ candidates exist, the one with a mass closest to the nominal $\rho$ mass is kept. In order to reduce
the background from \BXclnu decays, events with $K_L$ candidates are vetoed. Such candidates are
identified by a cluster in the outer $K_L$-and-muon detector (KLM in Fig.~\ref{fig:detector_event_belle.pdf}) with no energy depositions in the electromagnetic calorimeter near
the flight path of the $K_L$ candidate.

\begin{figure*}  
  \includegraphics[width=0.98\textwidth]{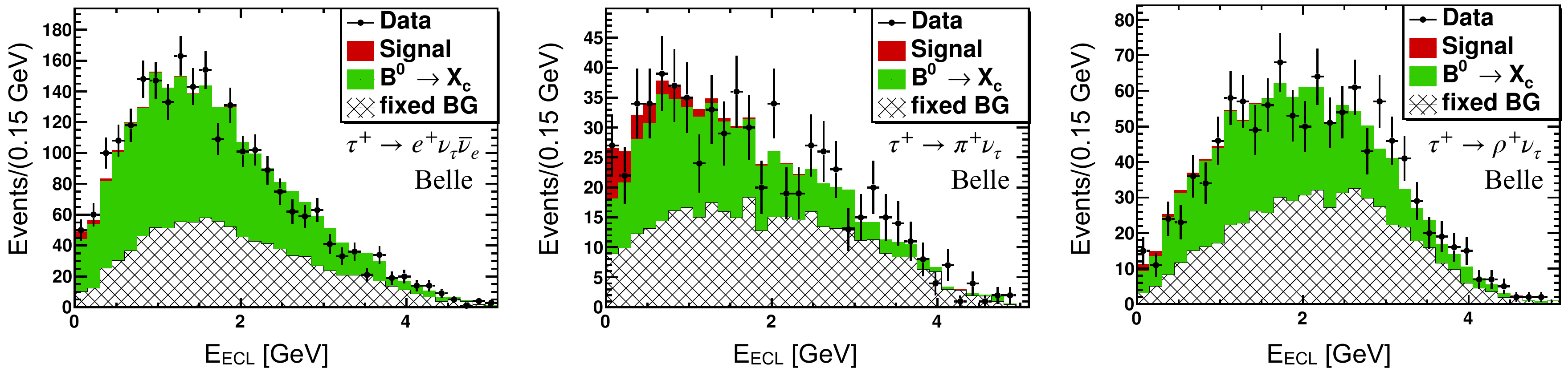}
  \caption{Signal fit for the Belle measurement of \Bpitaunu decays, adapted from~\cite{Hamer:2015jsa}. The
  \ECL distributions for the three reconstructed $\tau$ decay
  modes are shown: (left) \tauenu, (middle) \taupinu, and (right)
  \taurhonu. }
  \label{fig:pitaunu_fit}
\end{figure*}

With all particles assigned to either the tag or signal side, \ECL can be reconstructed from
the remaining neutral clusters in the collision event. To further reduce backgrounds, three boosted decision
trees are trained: one for each probed $\tau$ decay mode. The input variables are:
\begin{itemize}
  \item The four-momenta of all signal particles
  \item $q^2$ as calculated from the tag-side $B$ meson four-momentum and the signal-side pion with the
  highest momentum; for signal decays $q^2 \ge m_\tau^2$, whereas for backgrounds lower values are
  possible.
  \item $m_{\mathrm{miss}}^2$; for signal decays we expect a higher missing mass because of the additional
  neutrinos in the final state. 
\end{itemize}
Requirements on the classifier outputs are chosen to select signal events such that each channel has an optimal
statistical sensitivity. The resulting number of signal events is then extracted via a simultaneous fit of
the respective \ECL distributions. The post-fit distributions are shown in
Fig.~\ref{fig:pitaunu_fit}. The measurement quotes an upper limit of
$\mathcal{B}(\Bpitaunu) < 2.5 \times 10^{-4}$ at 90\% CL. This can be converted to a
value of
\begin{align}
 \calR(\pi) & = 1.05 \pm 0.51 \, ,
\end{align}
which is compatible with the SM expectation of $R(\pi)_{\mathrm{SM}} = 0.641 \pm 0.016$~\cite{Bernlochner:2015mya}.

\begin{table}
  \renewcommand*{\arraystretch}{1.1}
  \newcolumntype{C}{ >{\centering\arraybackslash } m{1.25cm} <{}}
  \centering
  \caption{Summary of the relative uncertainties for the measurement of \Bpitaunu
  decays by Belle~\cite{Hamer:2015jsa}. }
  \label{tab:pitaunu_hadtag_errors}
  \vspace{1ex}
  \begin{tabular}{lCC}
    \hline\hline
    \multirow{2}{*}{Contribution} & \multicolumn{2}{c}{Uncertainty [\%]} \\ 
				  & Sys. & Stat.                         \\ \hline
    \BXclnu                       & 2.2                                  \\
    Signal modeling               & 1.8                                  \\
     Tagging calibration          & 4.6                                  \\
     $K_L$ veto                   & 3.2                                  \\
    Particle ID                   & 2.4                                  \\
    Bkg. modeling                 & 4.4                                  \\
    Other                         & 3.2                                  \\ 
    \vspace{1mm}
    \textbf{Total systematic}     & \textbf{8.3}                         \\
    \textbf{Total statistical}    &      & \textbf{48}                   \\ \hline
    \textbf{Total}                & \multicolumn{2}{c}{\textbf{49}}      \\     
    \hline\hline
  \end{tabular}
\end{table}

Table~\ref{tab:pitaunu_hadtag_errors} shows an overview of the systematic uncertainties of the result. The
largest systematic uncertainties stem from the tagging calibration, as the measurement was not carried out as
a ratio with respect to the light-lepton mode. The $K_L$ veto, used to reduce the background from CKM
favored semileptonic decays, introduces a large uncertainty due to the poorly known $K_L$ reconstruction
efficiency.

%%%%%%%%%%%%%%%%%%%%%%%%%%%%%%%%%%%%%%%%%%%%%%%%%%%%%%%%%%%%%%%%%%%%%%%
%%%%%%%%%%%%%%%%%%%%% BELLE SEMILEPTONIC TAG RD(*) %%%%%%%%%%%%%%%%%%%%
\subsection{Belle measurements with semileptonic tags}
\label{sec:belle_semileptag_rdx}
%%%%%%%%%%%%%%%%%%%%% BELLE SEMILEPTONIC TAG RD(*) %%%%%%%%%%%%%%%%%%%%

\subsubsection{\RDx with \tauellnu}
\label{sec:bfactories_leptag_rdx}

The first measurement of $\RDs$ using semileptonic tagging was performed by Belle~\cite{Sato:2016svk}, a 
result that was subsequently superseded by Belle's combined measurement of $\RD$ and $\RDs$ in 2020~\cite{Belle:2019rba}.
This analysis employs the \texttt{FEI} algorithm (described in Sec.~\ref{sec:tagging}) to efficiently identify semileptonic $B$ meson decays of the second $B$ meson (\Btag) in the event. 
This allows for the full identification of all particles and decay cascades in the collision event and the reliable reconstruction of \ECL, 
the unassigned energy in the calorimeter, as already defined in Sec.~\ref{sec:hadronic_tag}.
\emph{Tag-side} \BDxellnu decays are selected by exploiting the observable
\begin{equation}
	\label{eqn:meas:cosBDxell}
	\cos \theta_{B, D^{(*)} \ell} \equiv \frac{2 E_{\text{beam}} E_{\Dx\ell} - m_B^2 - m_{\Dx\ell}^2 }{2 | \bm{p}_B | | \bm{p}_{\Dx\ell}|}\,,
\end{equation}
in which the energies and momenta, $E$ and $\bm{p}$, are all defined in the centre-of-mass (CM) frame---the $\Upsilon(4S)$ rest frame---of the colliding beams.
In particular, note that $E_{\Dx\ell}$ and $\bm{p}_{\Dx\ell}$ are the energy and momentum of the $\Dx\ell$ system,
respectively, and that in this frame $E_{\text{beam}} = E_B$.
For \BDxellnu decays with a single final state neutrino, which satisfy $(p_B - p_{\Dx\ell})^2 = m_\nu^2 \simeq 0$,
the definition of $\cos \theta_{B, D^{*} \ell}$ corresponds to the cosine of the angle between the tag $B$ meson and $\Dx\ell$ system in the CM frame.
Thus, for correctly reconstructed  tag-side \BDxellnu decays, the right-hand side of Eq.~\eqref{eqn:meas:cosBDxell} 
falls in the physical region such that $-1 \le \cos \theta_{B, D^{*} \ell} \le 1$ 
(with a tail towards negative values due to final state radiation).
However, for incorrectly reconstructed tag-side decays such as $B \to D^{**} \ell \nu$ or semitauonic $B \to D^{(*)} \tau (\to \ell \nu \nu) \nu$ decays,
the right-hand side of Eq.~\eqref{eqn:meas:cosBDxell} will typically produce large negative values
due to the absent term $(p_B - p_{\Dx\ell})^2 /2 | \bm{p}_B | | \bm{p}_{\Dx\ell}| > 0$, which is needed for $\cos \theta_{B, D^{(*)} \ell}$ to represent a physical cosine:
see Fig.~1 of~\cite{Sato:2016svk}. 
Including finite resolution effects, a requirement of $\cos \theta_{B, D^{*} \ell} \in [-2,1]$ thus captures most tag $B \to \Dx \ell \nu$ decays,
while strongly suppressing $B \to D^{**} \ell \nu$ and $B \to \Dx \tau (\to \ell \nu \nub) \nu$ decays.

On the signal side, lepton candidates are combined with $D$ and $D^*$ meson candidates. 
The decay modes used for the $D^0$ and $D^+$ account for about $30\%$ and $22\%$, respectively, of the overall decay branching fractions. 
To further improve the reconstruction, a decay vertex fit of the $D$ daughter particles is carried out. 
The $D^{*\,+}$ is reconstructed using both charged and neutral slow pion candidates, and for the  $D^{*\,0}$ neutral slow pion candidates and photons are used. 
The selection is refined by applying requirements on the masses of these candidates and other variables that are optimized to maximize the statistical significance of the final result. 
In case several tag and signal-side candidates can be reconstructed, the candidate combination with the highest tagging classifier output from the \texttt{FEI},
and on the signal side the combination with the best $D$ vertex fit probability, is selected. Events with additional unassigned charged particles or displaced tracks are rejected. 
At this stage, all signal and tag-side particles are identified and \ECL can be reconstructed. 
Here, only clusters in the barrel, forward region and backward region with energies greater than $50$, $100$, and $150$ MeV, respectively are included. 
For correctly reconstructed normalization and signal decays, one expects no unassigned neutral depositions in the detector and also that \ECL peaks at zero with a tail toward positive values due to reconstruction mistakes on the tag-side, and to a lesser extent due to beam-background depositions and noise in the calorimeter. 
 
\begin{figure*}
  \includegraphics[width=0.85\textwidth]{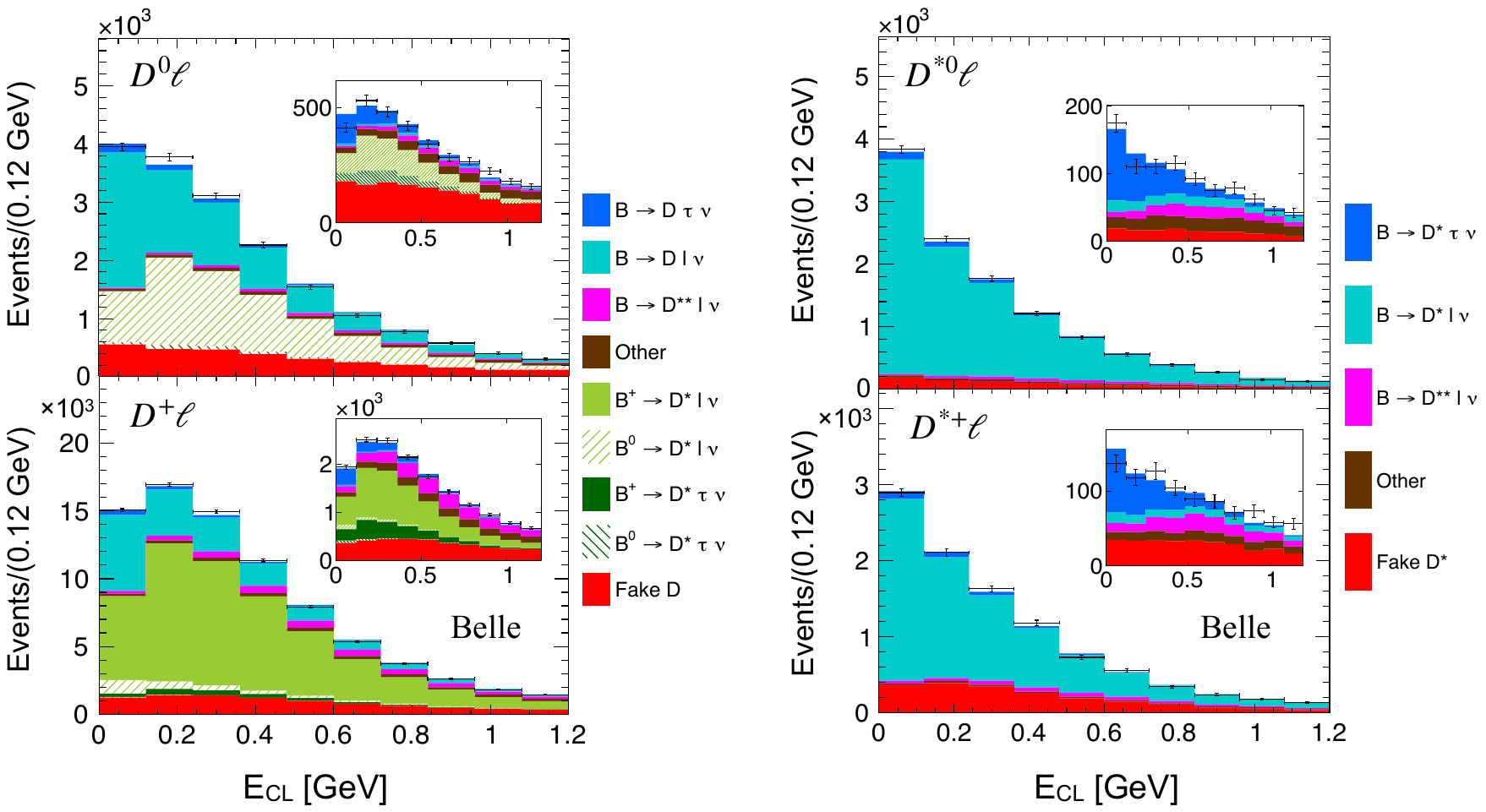}
  
  \caption{Projection of the signal fit for the Belle measurement of \RDx using semileptonic
  tagging, adapted from~\cite{Belle:2019rba}. The four panels correspond to the four
  reconstruction categories: (top left) $D^0 \ell$, (top right) $D^{*0} \ell$, (bottom left) $D^{+} \ell$,
  (bottom right) $D^{*+} \ell$. The signal enriched regions, obtained by a cut on a multivariate
  classifier, are shown in the inset figures. The uncertainties are only statistical. }
  
\label{fig:sl_tag_E_ECL.pdf}
\end{figure*}

To separate signal and normalization mode decays, a boosted decision tree is trained with the following distinguishing features ranked in order of importance: 
\begin{itemize}
 \item Signal side $\cos \theta_{B, D^{*} \ell}$: for normalization mode decays this variable will be in the physical range of $[-1,1]$, whereas for the signal mode large negative values are expected. 
 \item Approximate missing mass squared, $\mmiss$ (more details are given in Sec.~\ref{sec:bframe}): the additional two neutrinos from the $\tau$ decay will produce on average a larger missing invariant mass than the normalization mode. 
 \item The total visible energy $E_{\mathrm{vis}} = \sum_i E_i$ of all reconstructed particles $i$ in the event: the two additional neutrinos from the signal mode also will reduce the visible energy observed in the detector in contrast to the normalization mode. 
\end{itemize}
The classifier output $\mathcal{O}_{\mathrm{sig}}$ is then directly fitted along with the \ECL of the event to disentangle signal, 
normalization, and background contributions. 
This is done by exploiting the isospin relations between the charged and neutral final states for the normalization and signal contributions, i.e. by fixing $\calR(D^{(*)\,0}) = \calR(D^{(*)\, +})$. 
The free parameters of the fit are the yields for the signal, normalization, \BDsslnu, and feed-down from $\Dx\ell$ components. 
The yields of other background contributions from continuum and $B$ meson decays are kept fixed to their expectation values. 

Figure~\ref{fig:sl_tag_E_ECL.pdf} shows the full post-fit projections of \ECL as well as those in the signal-enriched region of $\mathcal{O}_{\mathrm{sig}} > 0.9$. 
The final results are
\begin{align}
 	\RD & = 0.307 \pm 0.037\,\text{(stat)} \pm 0.016\,\text{(syst)} \, , \\
 	\RDs & = 0.283 \pm 0.018\,\text{(stat)} \pm 0.014\,\text{(syst)} \, ,
\end{align}
with the first error being statistical and the second coming from systematic uncertainties, and an anti-correlation of $ \rho = -0.52$ between the two values.
The measurement is the most precise determination of these ratios to date and shows good compatibility with the SM expectation (Table~\ref{tab:th:RDDs}). 

\begin{table}
  \renewcommand*{\arraystretch}{1.1}
  \newcolumntype{C}{ >{\centering\arraybackslash } m{1.25cm} <{}}
  \centering
  \caption{Summary of the relative uncertainties for the Belle measurement of \RDx using semileptonic tagging~\cite{Belle:2019rba}. }
  \label{tab:ls_tag_errors}
  \vspace{1ex}
  \begin{tabular}{llCC}
    \hline\hline
    \multirow{2}{*}{Result} & \multirow{2}{*}{Contribution}              & \multicolumn{2}{c}{Uncertainty [\%]} \\ 
                            &                                            & Sys. & Stat.                         \\ \hline
    \multirow{7}{*}{\RD}    & $B \to D^{**} \ell \bar \nu_\ell$          & 0.8                                  \\
                            & PDF modeling                               & 4.4                                  \\
                            & Other bkg.                                 & 2.0                                  \\
                            & $\epsilon_\text{sig}/\epsilon_\text{norm}$ & 1.9                                  \\ 
                            \vspace{1mm}
                            & \textbf{Total systematic}                  & \textbf{5.2}                         \\
                            & \textbf{Total statistical}                 &      & \textbf{12.1}                 \\\cline{2-4}
                            & \textbf{Total}                             & \multicolumn{2}{c}{\textbf{13.1}}    \\ \hline
    
    \multirow{7}{*}{\RDs}   & $B \to D^{**} \ell \bar \nu_\ell$          & 1.4                                  \\
                            & PDF modeling                               & 2.3                                  \\
                            & Other bkg.                                 & 1.4                                  \\
                            & $\epsilon_\text{sig}/\epsilon_\text{norm}$ & 4.1                                  \\ 
                            \vspace{1mm}
                            & \textbf{Total systematic}                  & \textbf{4.9}                         \\
                            & \textbf{Total statistical}                 &      & \textbf{6.4}                  \\\cline{2-4}
                            & \textbf{Total}                             & \multicolumn{2}{c}{\textbf{8.1 }}    \\ 
    \hline\hline
  \end{tabular}
\end{table}

Table~\ref{tab:ls_tag_errors} summarizes the relative systematic and statistical uncertainties on $\RD$ and $\RDs$. 
The limited size of the simulated sample, used to define the fit templates and to train the multivariate selection, results in the dominant systematic uncertainty. 
Uncertainties from lepton efficiencies and fake rates cancel to only a certain extent in the measured ratios 
because of the large differences in the momentum spectra of the signal and normalization decays. 
This leads to a sizeable uncertainty of the efficiency ratios $\epsilon_\text{sig}/\epsilon_\text{norm}$.
Uncertainties from the \BDsslnu background are less dominant.

%%%%%%%%%%%%%%%%%%%%%%%%%%%%%%%%%%%%%%%%%%%%%%%%%%%%%%%%%%%%%%%%%%%%%%%
%%%%%%%%%%%%%%%%%%%%%%%%%%%% LHCB UNTAGGED %%%%%%%%%%%%%%%%%%%%%%%%%%%%
\subsection{LHCb untagged measurements}
\label{sec:lhcb_untagged}
The measurement of decays with multiple neutrinos in the final state is especially challenging at hadron
colliders given the typically smaller signal-to-background ratios compared to the $B$-factories and the inability to
effectively reconstruct a tag $b$-hadron to constrain the kinematics of the signal decay.  These difficulties
have been overcome by taking advantage of the large data samples of $b$-hadrons produced in high-energy $pp$ collisions
and by cleverly estimating the kinematics of the signal $b$-hadron based on the particles that can be
reconstructed. The measurements described in Secs.~\ref{sec:lhcb_muonic_rds} and \ref{sec:lhcb_rjpsi} make
use of the relatively clean muonic decays of the $\tau$ lepton to limit the background contributions and
estimate the $B$ or $B_c$ kinematics with the so-called rest frame approximation (see Sec.~\ref{sec:bframe:rfa}). The measurement
detailed in Sec.~\ref{sec:lhcb_hadronic_rds} takes advantage of the additional vertex that can be
reconstructed from \taupipipi hadronic decays to not only reduce hadronic backgrounds by four
orders of magnitude, but also to estimate the momentum of the signal $B$ meson relatively precisely; see Sec.~\ref{sec:bframe:tauvertex}.

%%%%%%%%%%%%%%%%%%%%%%%%%%%% LHCB MUONIC RD* %%%%%%%%%%%%%%%%%%%%%%%%%%%%
\subsubsection{\RDsp with \taumunu}
\label{sec:lhcb_muonic_rds}
%%%%%%%%%%%%%%%%%%%%%%%%%%%% LHCB MUONIC RD* %%%%%%%%%%%%%%%%%%%%%%%%%%%%

The LHCb experiment published the first measurement of a \bctaunu transition in a hadron collider environment
in 2015~\cite{Aaij:2015yra}. This result was based on a 3~fb$^{-1}$ sample of $pp$ collision data and measured
\RDsp, which under isospin symmetry has the same value as \RDsz to a very good approximation.
This first analysis chose to focus on \RDs over \RD because the lower \BDtaunu branching fraction,
the lack of the $D^*$ mass constraint, and the larger contributions from feed-down processes make \RD a
significantly more challenging observable to measure at a hadron collider. A combined \RD--\RDs measurement
from LHCb is expected in the near future.

%% Event reconstruction
Signal $\Bzb \to D^{*+} \tau^- \nutb$ and normalization $\Bzb \to D^{*+} \mu^- \numb$ decays are selected
by requiring that the trajectories of a $\mu^-$ and an oppositely charged $D^{*+}$ candidate,
reconstructed exclusively via the decay chain $D^{*+} \to D^0\left( \to K^- \pi^+\right) \pi^+$, are consistent with a
common vertex that is separated from the $pp$ primary vertex (PV). Events with an electron in the final state are not
included because of the trigger and calorimeter limitations described in Sec.~\ref{sec:reco}. Compared to the $B$-factories,
the reduction in signal reconstruction efficiency due to the exclusive use of muons and a single $D^0$ decay chain is
compensated by the far larger production cross-section for $B$ mesons at LHCb.

%% Fit to LHCb muonic control samples
\begin{figure*}
  \includegraphics[width=0.67\textwidth]{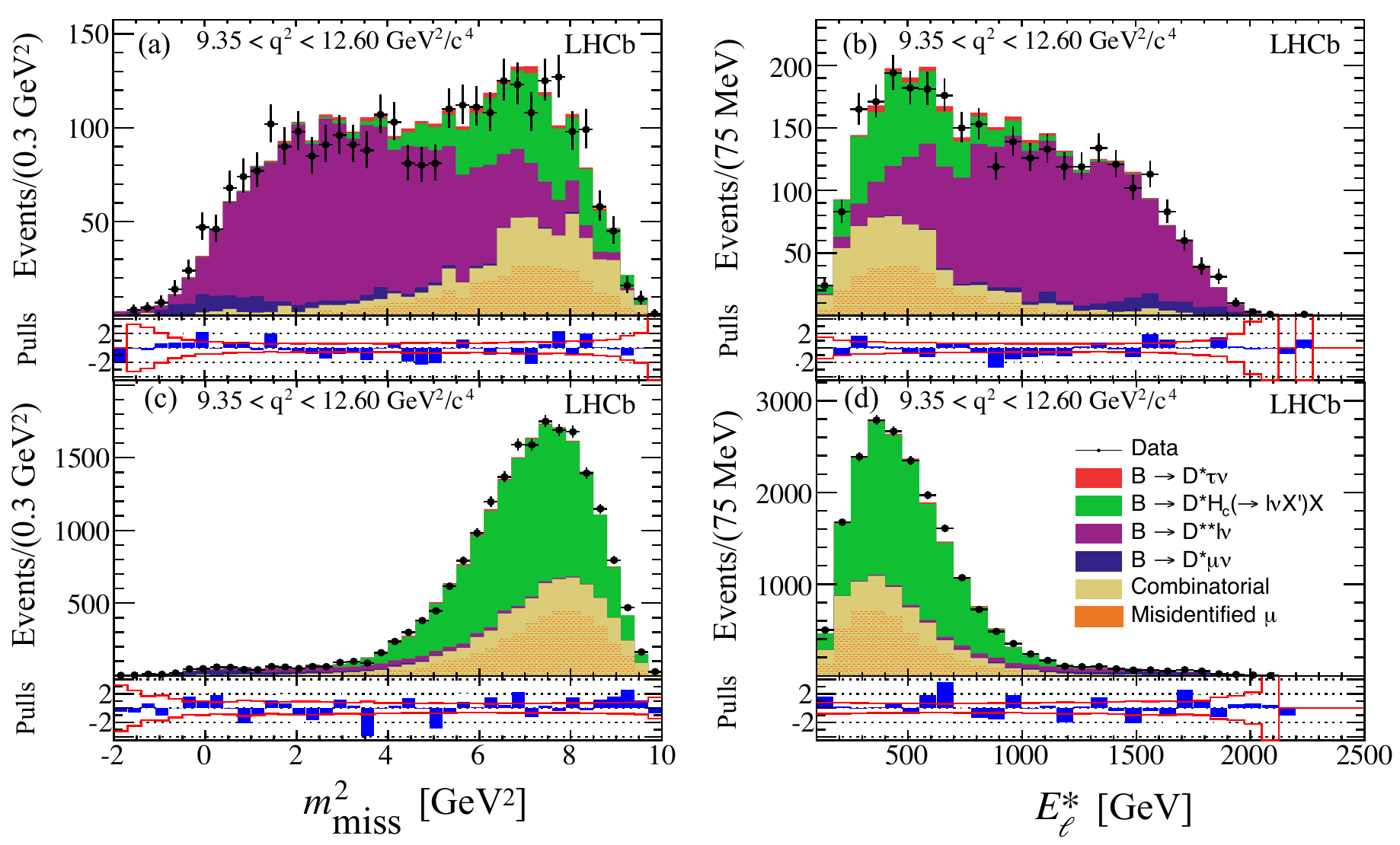}

  \caption{Projections of the control sample fits in the highest $q^2$ bin for the LHCb measurement of \RDsp
  involving muonic $\tau$ decays, adapted from~\cite{Aaij:2015yra}. (a-b) $D^{*+} \mu^-\pi^-$ sample enriched in \BDsslnu decays and (c-d) $D^{*+} \mu^-K^\pm$ sample enriched in
  $\overline{B} \to D^{*+}H_cX$ decays. }
    
\label{fig:lhcb_muonic_cs_fit.pdf}
\end{figure*}

%% Fit to LHCb muonic signal sample
\begin{figure*}
  \includegraphics[width=0.98\textwidth]{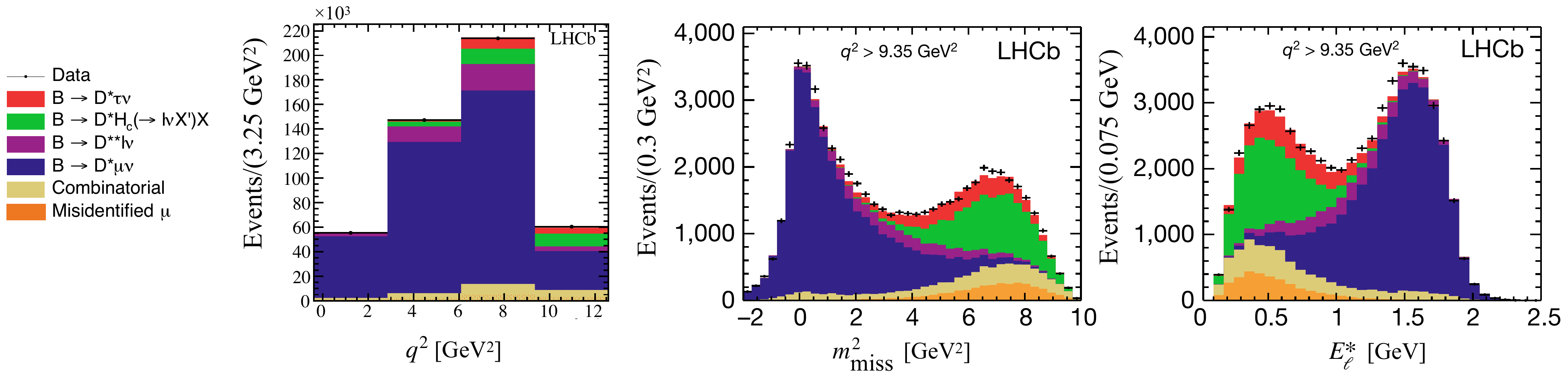}

\caption{Projections of the signal fit for the LHCb measurement of \RDsp
  involving muonic $\tau$ decays~\cite{Aaij:2015yra}. Left: full $q^2$ projection;
    Middle: \mmiss projection in the highest $q^2$ bin; and Right: \Esl projection in the highest $q^2$ bin.}
\label{fig:lhcb_muonic_signal_fit.pdf}
\end{figure*}

%% Isolation
An isolation BDT is trained to reject events arising
from partially reconstructed $B$ decays. For each additional track in the event this algorithm evaluates the
possibility that the track originates from the same vertex as the $D^{*+} \mu^-$ candidate based on quantities
such as the track separation from the decay vertex and the angle between the track and the candidate momentum
vector. The signal sample is made up of events where the $D^{*+} \mu^-$ candidate is found to be isolated from
all other tracks in the event.

%% Backgrounds and control samples
The isolation BDT is employed to further select three data control samples: a $D^{*+} \mu^-K^\pm$ sample that
includes an additional kaon coming from the $D^{*+} \mu^-$ vertex, as well as the $D^{*+} \mu^-\pi^-$ and
$D^{*+} \mu^-\pi^-\pi^+$ samples with an additional pion and pion pair, respectively. The $D^{*+} \mu^-K^\pm$
sample is enriched in double-charm decays of the type $\overline{B} \to D^{*+}H_cX$, where $H_c$ is a charmed hadron that
decays semileptonically and $X$ refers to unreconstructed particles, while the samples with additional pions
are enriched in \BDssltnu decays. Additional data control samples based on wrong charge combinations of the
$D^{*+}$ decay products and muon are used to measure the $D^{*+}$ and $B$ combinatorial backgrounds. The
misidentified muon background is estimated in a $D^{*+}h^\pm$ sample where $h^\pm$ is a track that fails the
muon identification requirements.

%% Rest frame approximation
A three-dimensional binned maximum likelihood fit to the $q^2$, \mmiss (Eq.~\eqref{eqn:kin:mmiss}), and \Esl (Eq.~\eqref{eqn:kin:Eell}) variables is performed to
determine the signal, normalization, and background yields, as well as several parameters describing the shapes
of the different distributions. The momentum of the $B$ meson, necessary to calculate the three fit variables,
is estimated via the rest frame approximation, detailed in Sec.~\ref{sec:bframe:rfa}.

%% Control sample fit
The templates for the combinatorial and misidentified muon backgrounds are taken directly from the data control
samples described above, while the templates for the $\overline{B} \to D^{*+}H_cX$ and \BDssltnu backgrounds
are based on Monte Carlo simulations with corrections extracted from a fit to the $D^{*+} \mu^-K^\pm$ and
$D^{*+} \mu^-\pi^-\left(\pi^+\right)$ samples. Figure~\ref{fig:lhcb_muonic_cs_fit.pdf} shows the excellent
agreement between the data and the resulting background model that is achieved.

%% Signal Fit
The templates for the signal and normalization contributions are parametrized by CLN form factors (Sec.~\ref{sec:th:hqet}) extracted
from the fit to the signal sample. Figure~\ref{fig:lhcb_muonic_signal_fit.pdf} shows the fit projection of the
$q^2$ variable in the full range, as well as the \mmiss and \Esl projections in the $q^2$ bin with the highest
signal-to-background ratio.

\begin{table}
  \renewcommand*{\arraystretch}{1.1}
  \newcolumntype{C}{ >{\centering\arraybackslash } m{1.25cm} <{}}
  \newcolumntype{R}{ >{\raggedright\arraybackslash } m{4cm} <{}}
  \centering
  \caption{Summary of the relative uncertainties for the LHCb measurement of \RDsp
  involving muonic $\tau$ decays~\cite{Aaij:2015yra}.}
  \label{tab:lhcb_muonic_errors}
  \vspace{1ex}
  \begin{tabular}{RCC}
    \hline\hline
\multirow{2}{*}{Contribution} & \multicolumn{2}{c}{Uncertainty [\%]} \\
			      & Sys.         & Stat.                 \\
\hline
Simulated sample size         & 6.2	     &                       \\
Misidentified $\mu$ bkg.      & 4.8          &                       \\
\BDssltnu bkg.                & 2.1          &                       \\
\BDslnu form factors       & 1.9          &                       \\
Hardware trigger              & 1.8          &                       \\
Double-charm bkg.             & 1.5          &                       \\
MC/data correction            & 1.2          &                       \\
Combinatorial bkg.            & 0.9          &                       \\
Particle ID                   & 0.9                                  \\
\textbf{Total systematic}     & \textbf{8.9} &                       \\   
\vspace{1mm}
\textbf{Total statistical}    &              & \textbf{8.0}          \\
\hline
\textbf{Total}                & \multicolumn{2}{c}{\textbf{12.0}}    \\ 
\hline\hline
\end{tabular}
\end{table}

%% Systematics and result
As Table~\ref{tab:lhcb_muonic_errors} shows, the limited size of the simulated samples is the main source of
systematic uncertainty in this analysis, followed by the uncertainty in the background contributions and
\BDsltnu templates. The overall systematic uncertainty is slightly larger than the statistical uncertainty but,
as discussed in Sec.~\ref{sec:systematics}, many of the systematic uncertainties are expected to decrease
commensurately with larger data samples. The result of this measurement is 
\begin{equation}
	\RDsp = 0.336\pm0.027\,\text{(stat)}\pm0.030\,\text{(syst)}\,,
\end{equation}
which is in good agreement with the previous measurements by the $B$-factories.

%%%%%%%%%%%%%%%%%%%%%%%%%%%% LHCB HADRONIC RD* %%%%%%%%%%%%%%%%%%%%%%%%%%%%
\subsubsection{\RDsp with \taupipipi}
\label{sec:lhcb_hadronic_rds}
%%%%%%%%%%%%%%%%%%%%%%%%%%%% LHCB HADRONIC RD* %%%%%%%%%%%%%%%%%%%%%%%%%%%%

Instead of a leptonic $\tau$ decay, the 2018 measurement of \RDsp by LHCb~\cite{Aaij:2017deq} employed the 3-prong
$\tau^- \to \pi^- \pi^+ \pi^- \nu_\tau$ decay. This channel is interesting \emph{a priori} because it
is presently the only $\tau$ decay for which it is practical to reconstruct the $\tau$ decay vertex. This in turn
provides good precision on the reconstruction of the \Bzb momentum as described in
Sec.~\ref{sec:bframe}. Moreover, when aggregated with the $\tau^- \to \pi^- \pi^+ \pi^- \pi^0 \nu_\tau$
channel, the 3-prong decays have a total branching fraction of 13.5\%, comparable to that of the muonic decay
channel, and the pion-triplet dynamics provides very useful discrimination against the largest background contributions.

\begin{figure}
  \centering
  \includegraphics[width=0.43\textwidth]{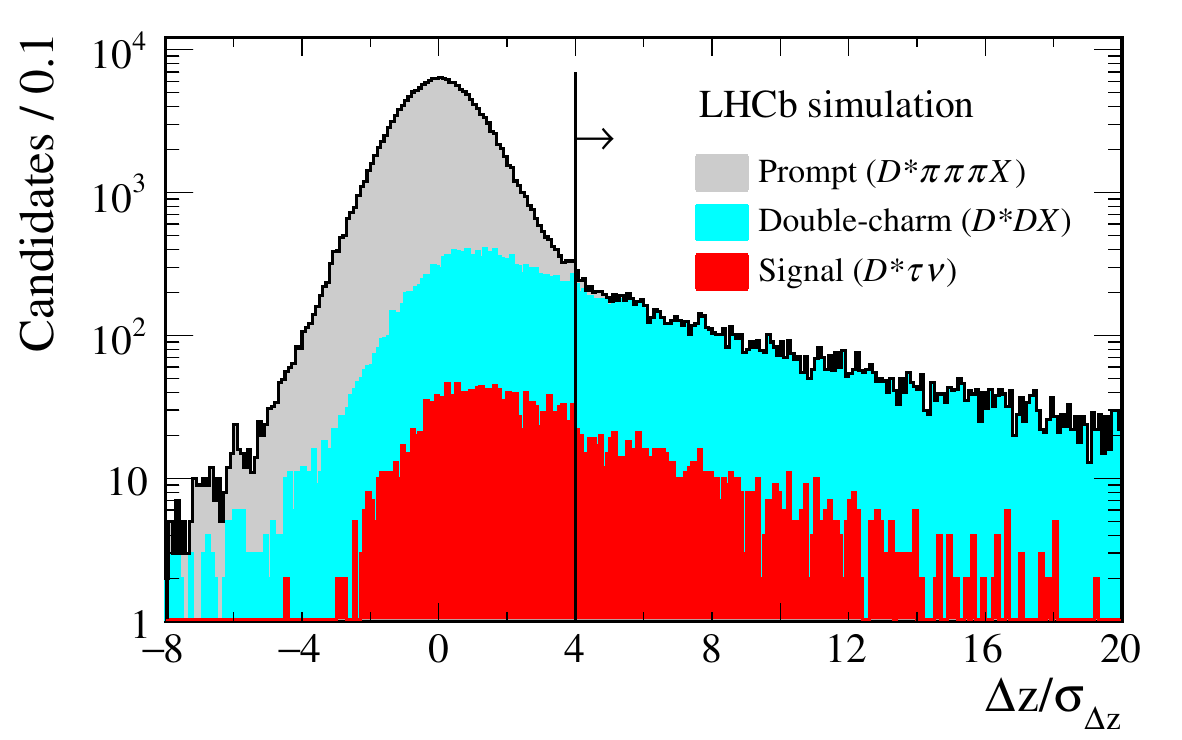}

  \caption{Distribution of the distance between the \Bzb vertex and the $\tau$ vertex along the beam direction
  (Fig.~\ref{fig:topologies} top) divided by its uncertainty in simulated events for the LHCb measurement of \RDsp
  involving \taupipipi decays~\cite{Aaij:2017deq}. The vertical line shows the 4$\sigma$ requirement used in the analysis
  to separate signal events in red (dark gray) from the prompt background component in medium gray. }
\label{fig:deltaz}
\end{figure}

%% Event reconstruction
In this measurement, signal \BDsptaunu decays are selected by requiring that the trajectories of a $\tau^-$
lepton and an oppositely charged $D^{*+}$ candidate, reconstructed exclusively via the decay chain $D^{*+} \to
D^0\left( \to K^- \pi^+\right) \pi^+$, are consistent with a common vertex separated from the PV. The $\tau$
lepton is reconstructed by requiring that the tracks of three pions with the appropriate charges share a
common vertex (Fig.~\ref{fig:topologies} top). Since the final state does not contain any charged lepton,
fully hadronic $\Bbar^0\rightarrow D^{*+} \pi^- \pi^+ \pi^- X$ decays initially dominate the selected event
sample. However, this background contribution may be reduced by four orders of magnitude by taking advantage of the long $\tau$ lifetime: the \pipipi vertex in a signal decay is typically \emph{displaced} downstream of the $B$ vertex. This allows one to distinguish such from the prompt topology of $\Bbar^0\rightarrow D^{*+} \pi^- \pi^+ \pi^- X$ decays, in which the \pipipi and the \Bzb vertices overlap, by requiring that
the distance between the $\tau$ and the $B$ vertex positions along the beam-axis is larger than four times its reconstructed uncertainty (Fig.~\ref{fig:deltaz}). 
Additionally, strict isolation from other charged particles is required to reject charm decays with more than three charged daughters, as well as
fake detached vertices where the \Dstar meson and the three pions come from other $b$-hadrons present in the event.

One of the major challenges in hadronic-$\tau$ measurements is that the normalization $\Bzb \to
D^{*+} \mu^- \numb$ decays are not measured simultaneously with the signal $\Bzb \to D^{*+} \tau^- \nutb$ decays. 
Since absolute branching fraction measurements are exceedingly difficult at LHCb, this analysis
normalizes the signal yield against that of the prompt $\Bbar^0\rightarrow D^{*+} \pi^- \pi^+ \pi^-$ decay, which has the same particle content as the signal,  and then relies on two \emph{external branching fractions} to calculate \RDs via
\begin{equation}
\label{eq:rd_hadronic}
\calR\left(D^*\right) =
\left.\frac{\BR\left(\bar{B}\rightarrow D^*\tau\nu_\tau\right)}
           {\BR\left(\bar{B}\rightarrow D^*\pi\pi\pi\right)}\right|_{\text{fit}} \!\!\!\times
\left.\frac{\BR\left(\bar{B}\rightarrow D^*\pi\pi\pi\right)}
           {\BR\left(\bar{B}\rightarrow D^*\mu\nu_\mu\right)}\right|_{\text{ext}}.
\end{equation}

%% Delta z / Sigma z

%% Fits to control samples
\begin{figure*}
  \begin{center}
  \includegraphics*[width=0.95\textwidth]{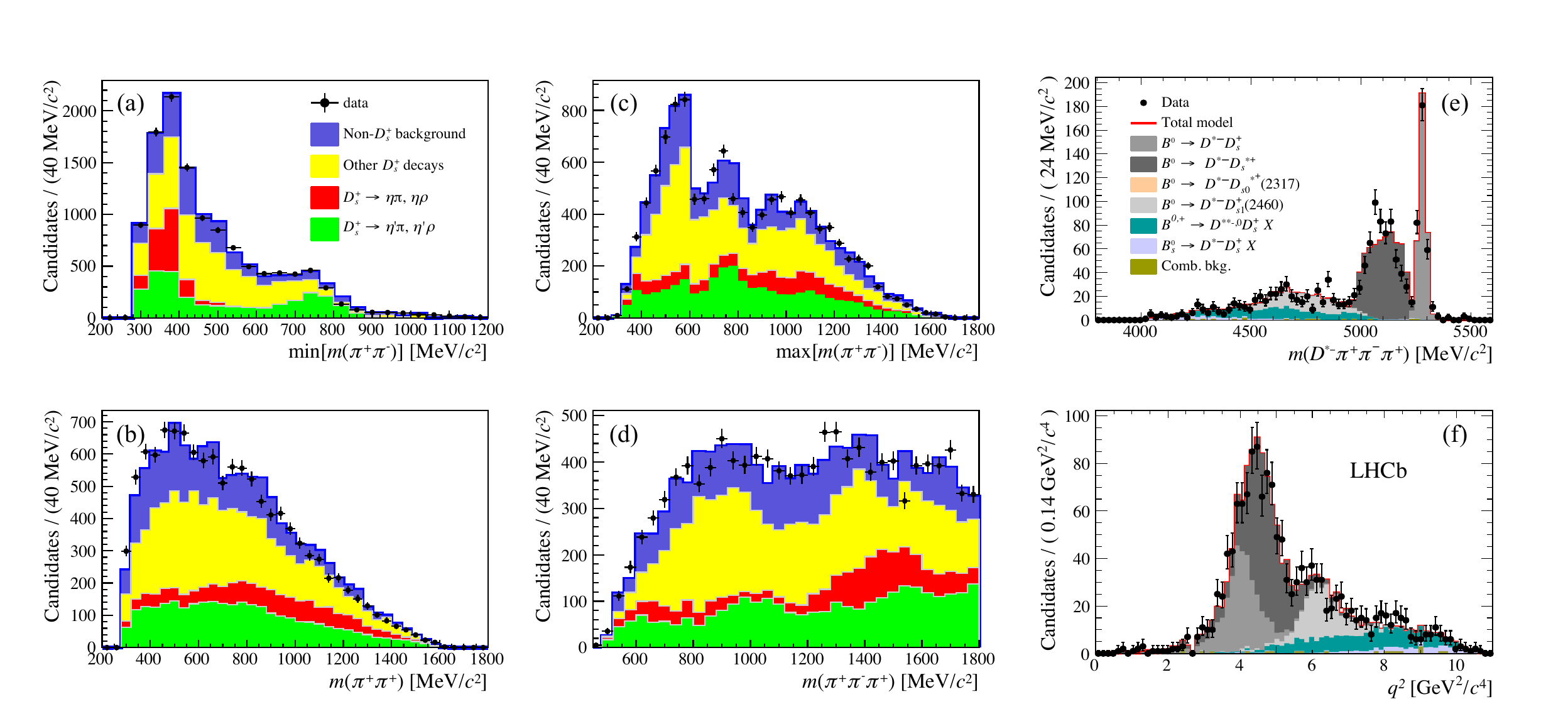}
  \end{center}

  \caption{Control sample fits for the LHCb measurement of \RDsp
  involving \taupipipi decays~\cite{Aaij:2017deq} employed to evaluate the composition of the various double-charm background contributions. (a-d) low-BDT sample
  and (e-f) $B\to D^{*+}D_{s}^-(\to \pim\pip\pim)X$ sample.}  \label{fig:lhcb_hadronic_cs.pdf}
\end{figure*}

%% Double-charm background
After selecting events with large $\tau$ flight significance as described above, the dominant remaining
background contributions consist of double-charm $B\to D^{*+}D_{(s)}^{(*,**)}$ decays. These decays were also
the largest background contributions to the muonic-$\tau$ measurement of \RDsp, but their relative amount in
$D$ and \Ds mesons are very different.  Because of the large inclusive branching fraction of the \Ds meson to
final states with three pions (about 30\%) and the small rate to semileptonic final states, the double-charm
background in the hadronic-$\tau$ sample contains ten times more \Ds mesons than that for the muonic-$\tau$
sample.  Interestingly, the \Ds inclusive three-pion modes proceed mainly from two-body and quasi two-body
decay channels involving $\eta$, $\eta'$, $\omega$, and $\phi$ mesons, which leads to very different three-pion
kinematics with respect to those of the signal. That is, the \taupipipi decay is well-described within resonance chiral theory~\cite{Ecker:1988te,Ecker:1989yg}, 
which features chiral terms as well as single-resonance $\rho$ and double-resonance $a_1 \to \rho$ contributions~\cite{Shekhovtsova:2012ra, Nugent2013}, leading to prominent
$\rho$ peaks in the distribution of both the minimum and maximum masses to the two $\pip\pim$ mass combinations---\minpipi and \maxpipi, respectively.

%% BDT
These kinematic differences are effectively exploited by a BDT that also includes other variables such as the
energy measured in the electromagnetic calorimeter in a cone whose axis is defined by the three-pion momentum. The kinematics of
the three-pion system in background \Dz and \Dplus decays is more similar to that in signal decays because
the inclusive \pipipi final state from these two mesons is dominated by the $K\aone$ channel \cite{Zyla:2020zbs}. Some
discrimination is still possible, however, due to the restricted phase space of this virtual \aone meson.

%% Signal fit plot
\begin{figure}
  \begin{center}
    \includegraphics[width=0.49\textwidth]{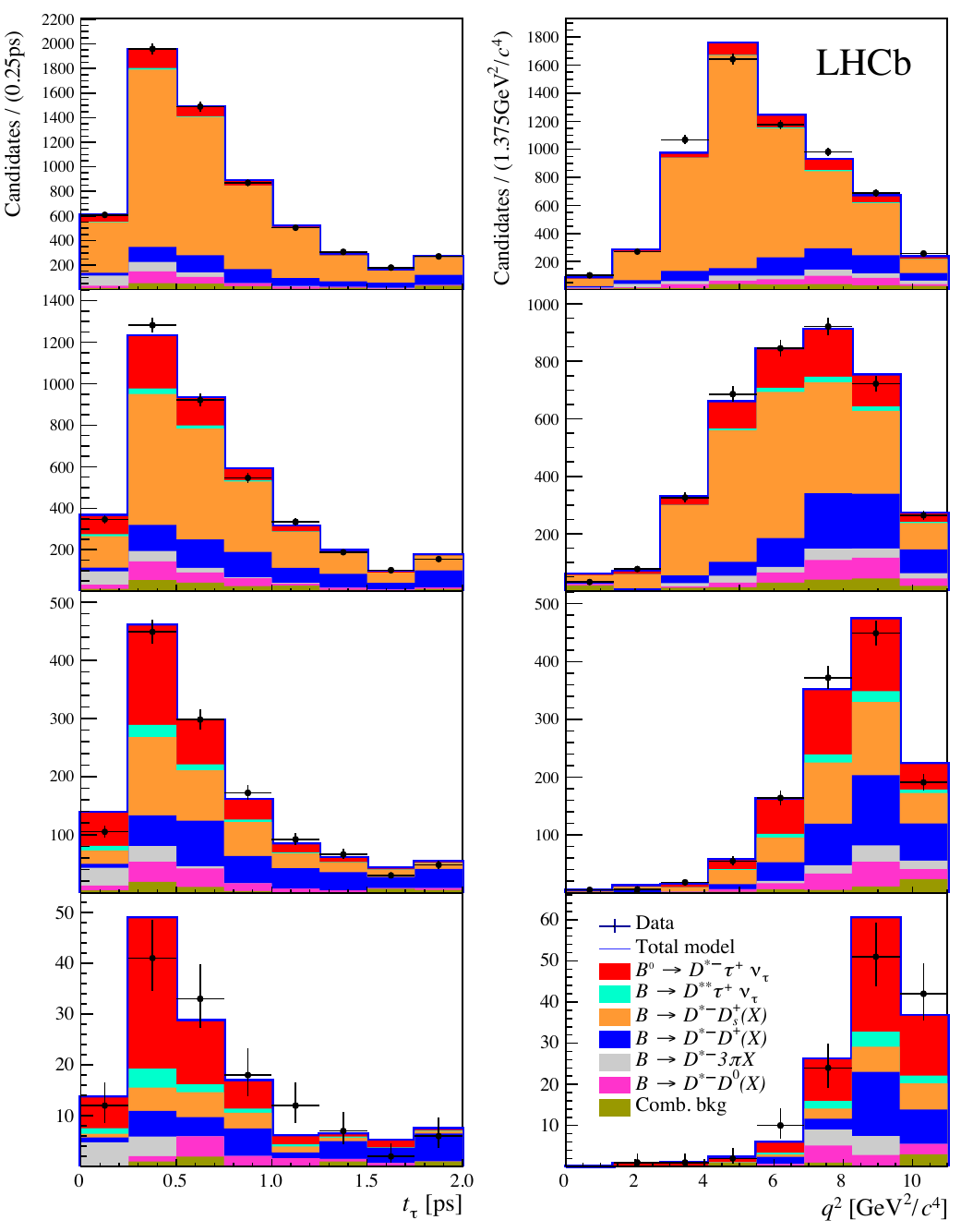}
  \end{center}
  
  \caption{Projections of the signal fit for the LHCb measurement of \RDsp involving \taupipipi
  decays~\cite{Aaij:2017deq}. The four rows correspond to the four BDT bins for increasing values of the BDT
  response.}  \label{fig:signalvsBDT}
\end{figure}

%% Data control samples
Many of the $B$ branching fractions to double-charm final states are known with poor precision or have not
been measured yet. 
The following data control samples are used to reduce the uncertainty due to the
composition of these background contributions:
\begin{itemize}
  \item A low-BDT sample enriched with inclusive \Ds decays constrains the composition of $B\to
  D^{*+}D_{s}^-X$ decays. The simulation is reweighted to match a fit to the \minpipi, \maxpipi,  \mthreepi,
  and \mpipi distributions. These variables capture the combined dynamics of the various inclusive \Ds decay channels to three pions (Fig.~\ref{fig:lhcb_hadronic_cs.pdf} a-d).

  \item A highly pure $B\to D^{*+}D_{s}^-(\to \pim\pip\pim)X$ sample selected by imposing a requirement on \mthreepi
  around the \Ds mass. A template fit to the \mthreepi distribution is used to measure the relative fractions
  of \Ds mesons produced directly and from \Dsstar or \Dsstarstar decays. The shape of the \Dsstar broad peak
  depends on the degree of longitudinal polarization of the \Dsstar and was adjusted in the simulation to
  reproduce the data. These measurements are important since the \qsq distributions of these decays are very
  different from each other, as shown in Fig.~\ref{fig:lhcb_hadronic_cs.pdf} (f).

  \item Clean $B\to D^{*+}\Dz(\to K^-\pip\pim\pip)X$ and $B\to D^{*+}D^-(\to K^-\pip\pim)X$ samples selected
  by explicitly reconstructing the \Dz and $D^-$ mesons. These samples are used to monitor and understand
  the non-\Ds background composition.
  
\end{itemize}

%% Signal fit
A three-dimensional binned maximum likelihood fit to $q^2$, the BDT output, and the decay time of the
reconstructed \tauon is performed to determine the signal and background yields. The calculation of \qsq
relies on the $B$ momentum determination described in Sec.~\ref{sec:bframe:tauvertex}. The decay time of the
reconstructed \tauon, \ttau, is computed from its flight distance and momentum obtained by the partial
kinematic reconstruction. This variable is useful for separating $\tau$ from $D^-$ decays, since the
lifetime of the $D^-$ meson is $3.5$ times longer than that of the \tauon lepton.  The fit results for the LHC
Run 1 data sample, corresponding to a luminosity of 3~\invfb, are displayed in Fig.~\ref{fig:signalvsBDT}.  An
interesting feature of this method compared to the muonic-$\tau$ measurement is that the highest BDT output
bin provides a fairly clean sample of signal decays with a purity of about 40\%. 

%% Uncertainties table
\begin{table}
	\renewcommand*{\arraystretch}{1.1}
 	\newcolumntype{C}{ >{\centering\arraybackslash } m{1cm} <{}}
 	 \newcolumntype{R}{ >{\raggedright\arraybackslash } m{4cm} <{}}	
  \centering 
  \caption{Summary of the relative uncertainties for the LHCb measurement of \RDsp
  involving \taupipipi decays~\cite{Aaij:2017deq}.}
	\label{tab:lhcb_hadronic_errors}
	\begin{tabular}{RCCC}
	\hline\hline
	\multirow{2}{*}{Contribution}	& \multicolumn{3}{c}{Uncertainty [\%]} \\
							& Sys. 	& Ext.  	& Stat. \\
	\hline						
	Double-charm bkg.                	& 5.4    	&&      \\
	Simulated sample size         	& 4.9    	&&     \\
	Corrections to simulation       	& 3.0        	&&  \\
	\BDssltnu bkg.                    	& 2.7        	&&  \\
	Normalization yield              	& 2.2       	&&  \\
	Trigger                          		& 1.6        	&&  \\
	PID                                		& 1.3        	&&  \\
	Signal form factors                     & 1.2        &&  \\
	Combinatorial bkg.            		& 0.7        	&&  \\
	Modeling of $\tau$ decay    	& 0.4        	&&  \\
	{\bf Total systematic}                  & {\bf 9.1} &&   \\
        \vspace{1mm}
	$\BR(B \to D^*\pipipi)$           	&& 3.9  &        \\
	$\BR(B \to D^*\ell\nu)$             && 2.3   &       \\
	${\BR(\taup\to3\pi\nu)}/{\BR(\taup\to3\pi\piz\nu)}$ && 0.7  &        \\
	{\bf Total external}                    	&& {\bf 4.6} &   \\ 
        \vspace{1mm}
	{\bf Total statistical}                    &&& {\bf 6.5}    \\
	\hline
	{\bf Total}                                  &  \multicolumn{3}{c}{{\bf 12.0}}   \\ 
	\hline\hline
	\end{tabular}
\end{table}

%% Uncertainties
As shown in Table~\ref{tab:lhcb_hadronic_errors}, the uncertainties related to the double-charm background and the
limited size of the simulated samples are the dominant systematic uncertainties in this measurement. 
 The uncertainties due to the limited knowledge of external branching fractions in
Eq.~\eqref{eq:rd_hadronic}, currently 4.6\%, are worth mentioning because, unlike many of the other systematic uncertainties,
these will not be reduced with the increasing LHCb data samples that will be collected. Instead, additional measurements from Belle~II will be needed (Sec.~\ref{sec:overview:other_bkg}).

The result of this measurement was reported as $\RDsp = 0.291 \pm 0.019 \pm 0.026\pm 0.013$ in 2018. Taking into account
the latest HFLAV average of $\Br(B^0 \to D^{*+} \ell\nu) = 5.08 \pm 0.02 \pm 0.12)$\%~\cite{Amhis:2019ckw}, the result is 
\begin{equation}
\RDsp = 0.280 \pm 0.018\,\text{(stat)} \pm 0.025\,\text{(syst)}\pm 0.013\,,
\end{equation}
where the third uncertainty is due to the external branching fractions described above.

%%%%%%%%%%%%%%%%%%%%%%%%%%%% LHCB HADRONIC RLc %%%%%%%%%%%%%%%%%%%%%%%%%%%%
\subsubsection{\RLc with $\tau^- \to \pi^- \pi^+ \pi^- \nu_\tau$ [Added 2022] }
\label{sec:lhcb_hadronic_rlc}
%%%%%%%%%%%%%%%%%%%%%%%%%%%% LHCB HADRONIC RLc %%%%%%%%%%%%%%%%%%%%%%%%%%%%

In 2022 LHCb published a measurement of $\RLc$ using the $\tau$ hadronic decay to $3\pi$~\cite{Aaij:2022}. 
This analysis reported the first observation of the semitauonic $\Lb$ decay and provided the first measurement of \RLc. 
As pointed out in Sec.~\ref{sec:th:other_states}, because the $\Lb \to \Lc$ transition 
involves the more constrained $s_\ell^P = 0^+$ HQET and different selection rules than $B \to \Dx$,
it provides a complementary test of LFUV versus $\RDx$. 

The analysis methodology is similar as the one described in Sec.~\ref{sec:lhcb_hadronic_rds}, but with the following  differences: 
a data-driven template is used to describe the background coming from fake \Lc candidates; 
a much stronger vertex detachment requirement is used to further suppress the not-well-known prompt component, 
taking advantage of the short \Lc lifetime;
and a multivariate analysis (MVA)-based isolation criterion is used to improve efficiency. 
This measurement, which is presently only possible at the LHC because there is no $\Lb$ production at $B$ factories, 
was performed using \pp collisions at 7~and 8~TeV, corresponding to an integrated luminosity of $3 \invfb$. 
The $\Lc$ meson is reconstructed using the $pK\pi$ decay channel. 
The final state for signal $\Lb\to\Lc\tau^-\nu_\tau$ 
and normalisation $\Lb\to\Lc\pim\pip\pip$ modes consist of a $\Lc$ candidate and three extra pions. 
The new MVA-based isolation algorithm~\cite{LHCb-PAPER-2015-031}
has an efficiency $20\%$ higher than the cut-based algorithm used in the analysis of Sec.~\ref{sec:lhcb_hadronic_rds}. 
The difference of the positions of the $3\pi$ and the $\Lc$ vertices along the beam direction is required to be at least five times larger than its uncertainty. 
This requirement suppresses the prompt background ten times more than the selection used in the analysis in Sec.~\ref{sec:lhcb_hadronic_rds}, 
reducing this initially dominant background to a negligible level, but at a price of a $50\%$ reduction in the signal efficiency.

The leading contribution to the remaining background originates from double charm events, 
dominated by $\Lb \to \Lc\Dsm(X)$ decays, 
with $\Dsm$ decaying to $3\pi$ and any set of extra particles. 
Such $\Dsm$ decays have a large branching fraction, $\sim$30\%~\cite{Zyla:2020zbs}. 
The BDT classifier described in Sec.~\ref{sec:lhcb_hadronic_rds} is used to separate these events from signal ones, 
based of the resonant structure of the $3\pi$ final state in $\tau$ decays.

\begin{figure}[htb]
 \includegraphics[width=0.40\textwidth]{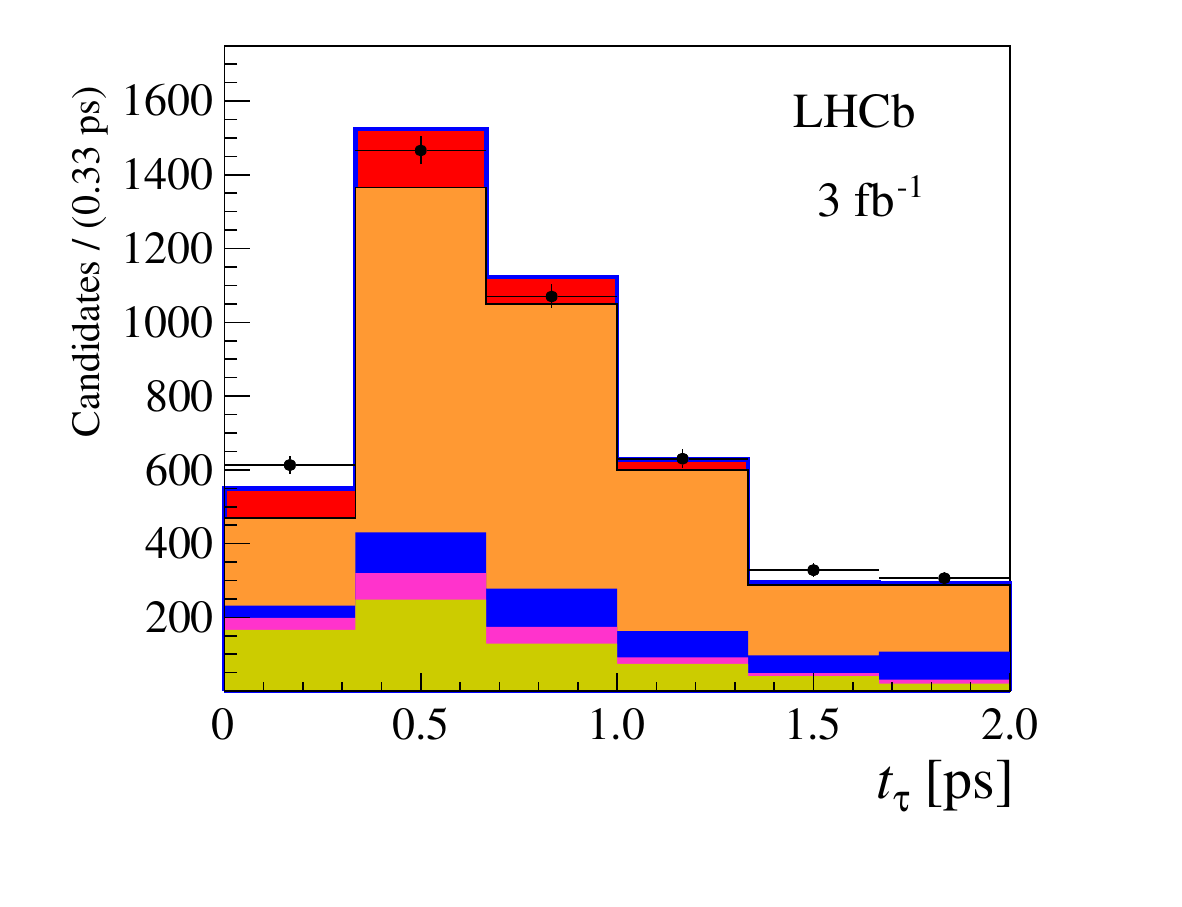}
 \includegraphics[width=0.40\textwidth]{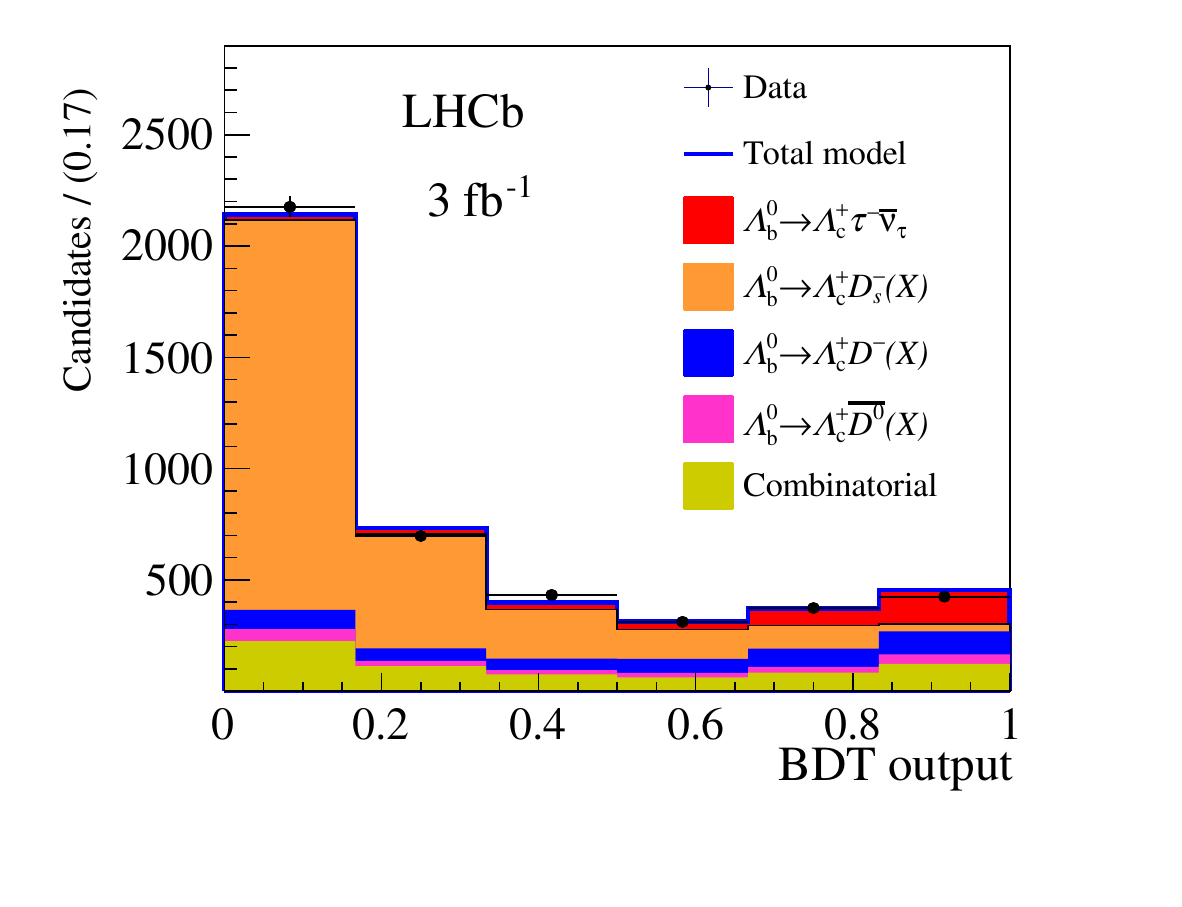}
    \vspace*{-5mm}
   \caption{Distributions of (top) $\tau^-$ decay time and (bottom) BDT output for $\Lb\to\Lc\tau^-\nu_\tau$ candidates. 
   Projections of the three-dimensional fit results are overlaid. The various fit components are described in the legend. 
  }
   \label{fig:fit_results}
 %\captionsetup[figure]{font=small,skip=0pt}
 \end{figure}

\begin{figure}[htb]
 \includegraphics[width=0.40\textwidth]{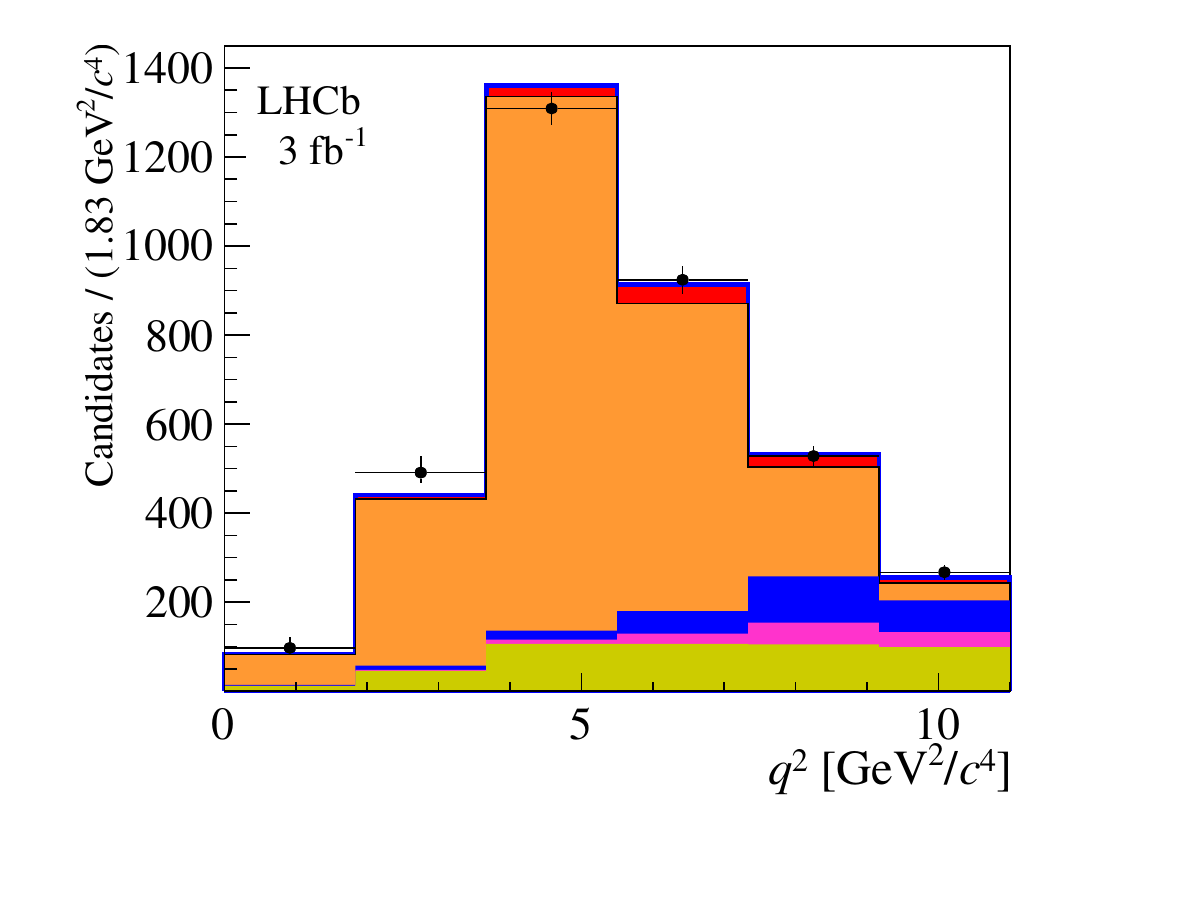}
 \includegraphics[width=0.40\textwidth]{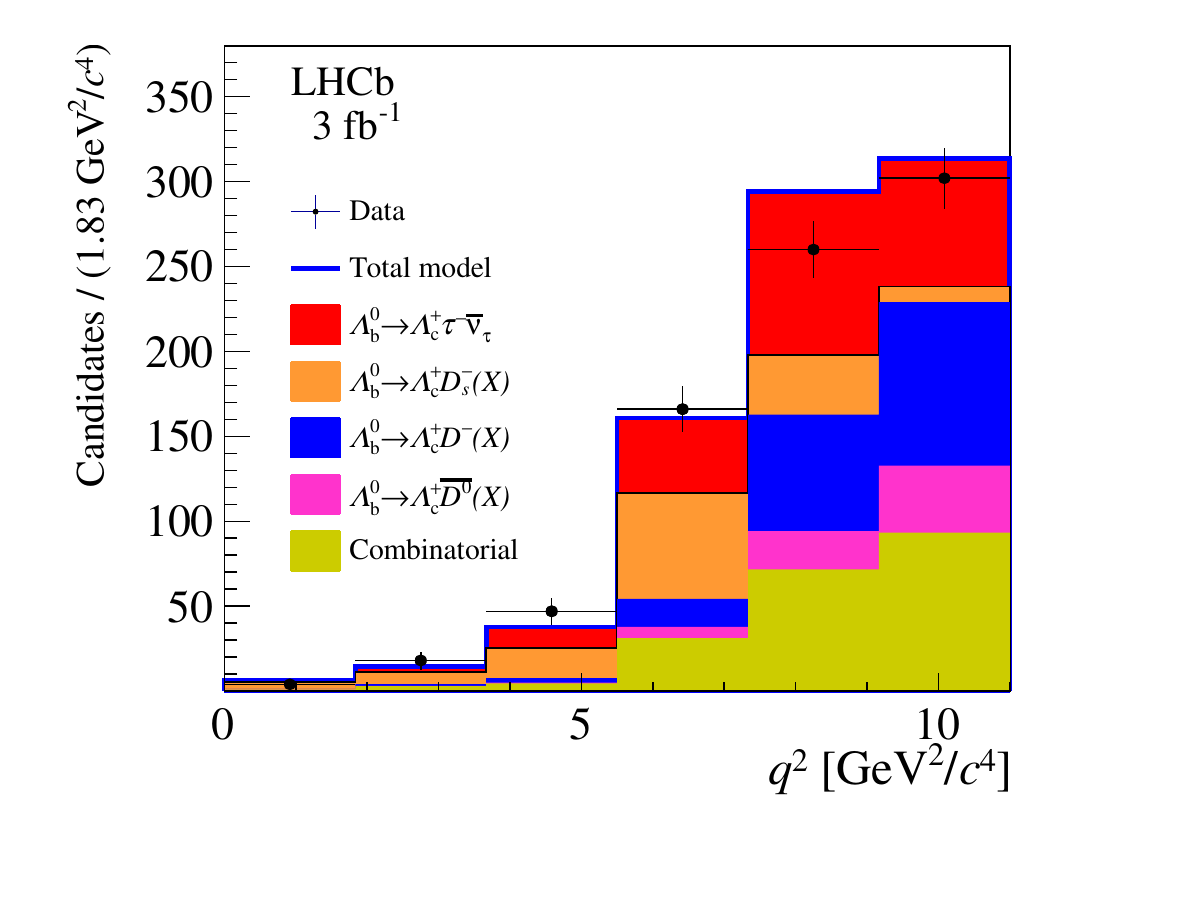}
    \vspace*{-5mm}
   \caption{Distributions of $q^2$ for $\Lb\to\Lc\tau^-\nu_\tau$ candidates having a BDT output value (top) below and (bottom) above $0.66$. 
   Projections of the three-dimensional fit ar overlaid. The various fit components are described in the legend. 
  }
   \label{fig:fitq2_results}
 %\captionsetup[figure]{font=small,skip=0pt}
 \end{figure}

\begin{figure}[htb]
        \includegraphics[width=0.40\textwidth]{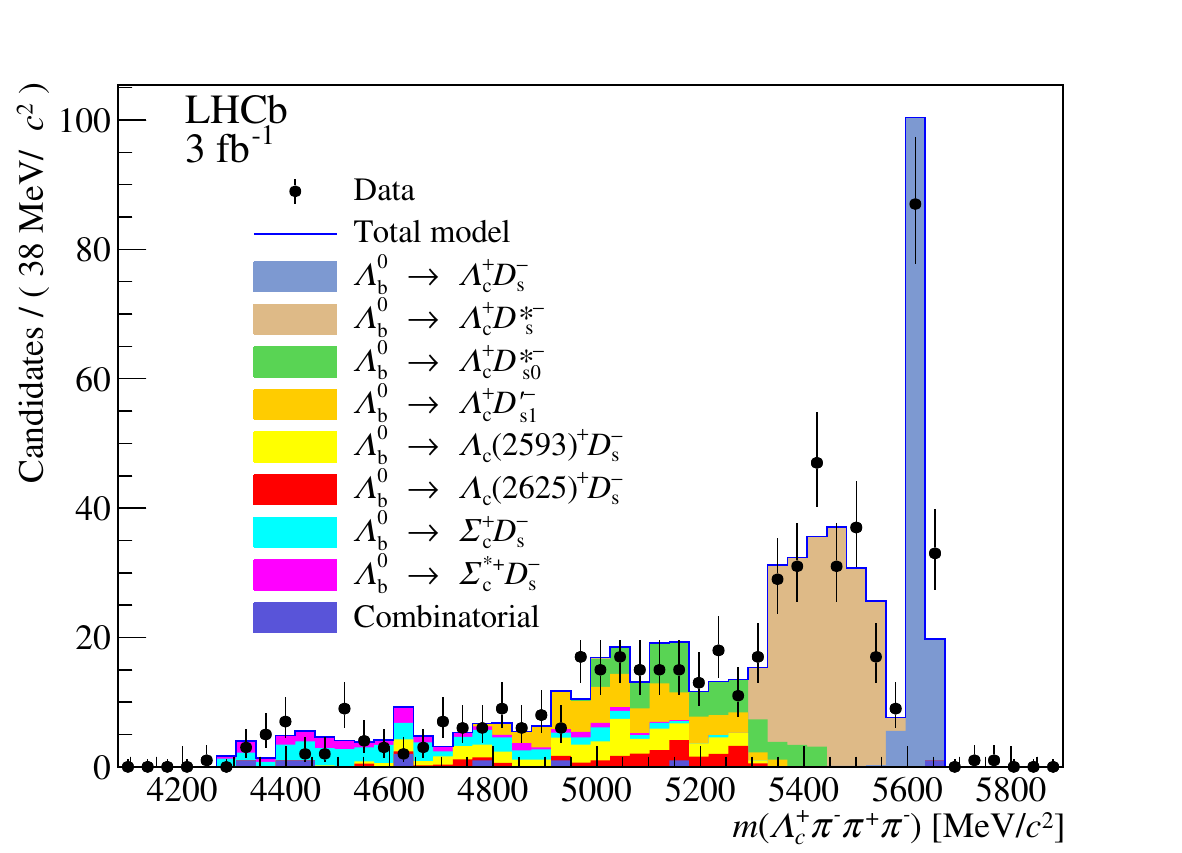}
         \vspace*{5mm}
        \caption{ Distribution of the $\Lc\pim\pip\pim$ invariant mass for the $\Lb\to\Lc\Dsm (X)$ control sample, with $\Dsm\to\pim\pip\pim$.
	The components contributing to the fit model are indicated in the legend.}
  \label{fig:Ds_control_Fit}
\end{figure}

The signal yield is measured using a three-dimensional binned maximum-likelihood fit
to $t_{\tau}$, the BDT output, and $q^2$. 
Their distributions are shown in Figs~\ref{fig:fit_results} and~\ref{fig:fitq2_results}. 
The fit model includes: a signal component; background components due to $\B\to\Lc\Dsm (X)$,
$\Lb\to\Lc\Dm (X)$ and $\Lb\to\Lc\Dzb (X)$ decays; background due to misreconstructed $\Lc$ candidates; and combinatorial background. 
Template distributions for signal and background are obtained from simulation, 
except for the latter two contributions, which are constructed from data-based control samples, using events where the $\Lc$ candidate 
has a mass that falls outside a window of $15$\,MeV around the known $\Lc$ mass central value~\cite{Zyla:2020zbs}, and where  the $\Lc$ and the $3\pi$ system have the same charge, respectively.
The \Dsm decay model used in the simulation is taken from the higher precision $\Bzb\to\Dstarp\Dsm$ sample~\cite{ourPRD}
and applied to each \Dsm branching fraction. The background originating from \Lb\to$\Lc\Dsm(X)$ is divided
into contributions from \Dsm, \Dssm, and higher-order excited \Dsm states, using events where the $3\pi$ invariant mass is selected to be within the  $\Dsm$ mass peak, as shown in Fig.~\ref{fig:Ds_control_Fit}. 
A contribution from $\Lb\to\Lcstar\tau^-\nu_\tau$ decays, where $\Lcstar$ denotes  the $\Lcstar(2595)$ and $\Lcstar(2625)$ excited charm baryons, 
forms a feed-down to the signal. Its yield fraction is constrained to be  (10 $\pm$ 5)\% of the signal.

The signal yield is  $N_{\text{sig}}=349 \pm 40$. 
The $\chisq$ variation derived from the change of the fit maximum likelihood  between the nominal fit and the one where the signal yield is forced to $0$ corresponds to an increase of $6.1\,\sigma$.
A clear separation between signal and the main background originating from $\Lb\to\Lc\Dsm(X)$ decays is obtained, as demonstrated in the BDT distribution of Fig.~\ref{fig:fit_results}.
Figure~\ref{fig:fitq2_results} shows that the $\Lb\to\Lc\Dsm(X)$ background is dominant at low BDT values, while a good signal-to-background ratio is observed at high BDT output.

\begin{figure}[htb]
        \includegraphics[width=0.40\textwidth]{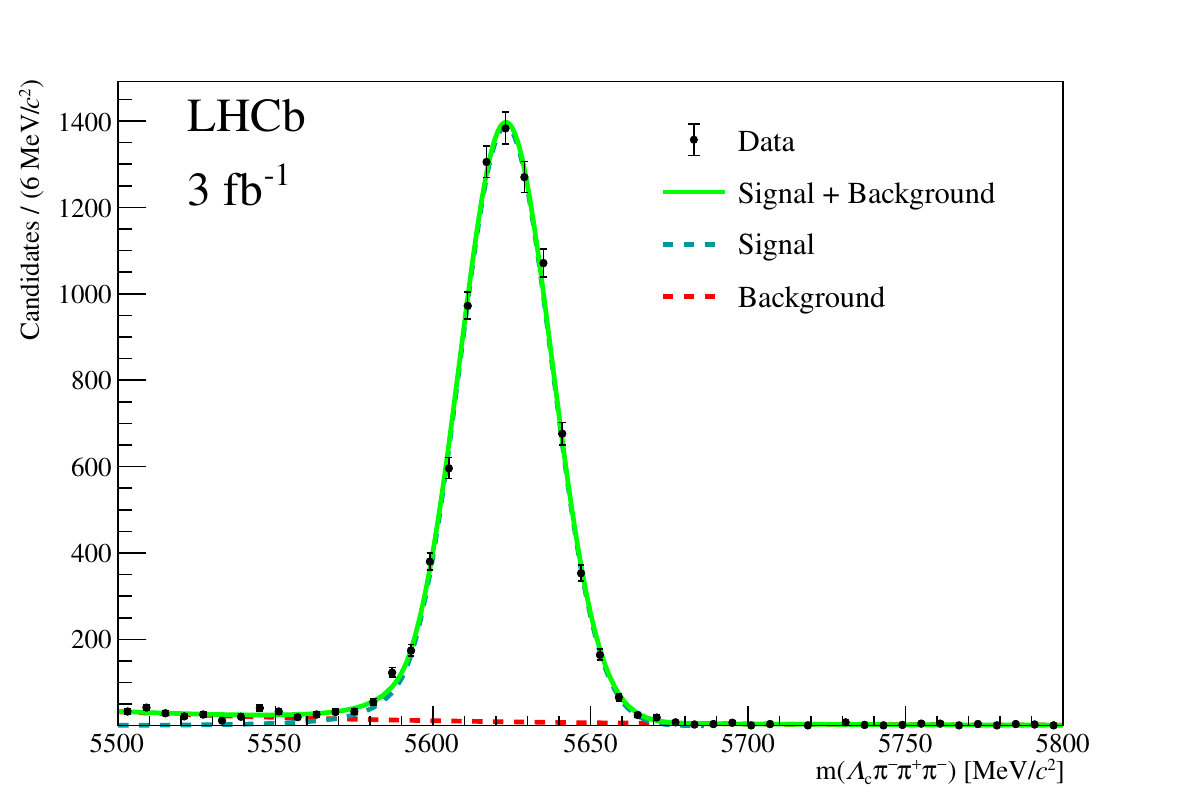}
        \caption{ Distribution of the $\Lc\pim\pip\pim$ invariant mass for all candidates in the normalisation channel, 
        after removal of the $\Lcstar$ contributions. 
        The fit components are indicated in the legend. 
        The signal is described by a Crystal Ball function, and the background by an exponential term.}
  \label{figsupp:normalisation}
\end{figure}

The $\Lb\to\Lc3\pi$ decay is chosen as a normalisation channel,  leading to the measurement of the ratio  
\begin{equation}
{\cal{K}}(\Lc)\equiv\frac{{\cal{B}}(\Lb \to \Lc \tau^-\nu_\tau)}{{\cal{B}}(\Lb \to \Lc3\pi)} \,.
\label{eqn:kappa}
\end{equation}
 The  contribution  from excited baryons which decay to $\Lc\pip\pim$, $\Lc\pip$, or $\Lc\pim$  is explicitely vetoed from the normalisation channel. 
 This ensures that the $3\pi$ dynamics resembles that of the signal, and leads to a reduced systematic uncertainty.
Fig.~\ref{figsupp:normalisation}  shows the invariant-mass distribution of selected $\Lc3\pi$ candidates, with a  yield of  $N_{\mathrm{norm}} = 8584~\pm$ 102, 
after subtraction of a small contribution of $168\pm 20$ $\Lb\to\Lc\Dsm (\to 3\pi)$ decays.

The ratio of branching fractions in Eq.~\eqref{eqn:kappa} is determined to be ${\cal{K}}(\Lc)= 2.46 \pm 0.27\pm 0.40$, 
where the first uncertainty is statistical and the second systematic.
The  sources of systematic uncertainty of  ${\cal{K}}(\Lc)$  are reported in Table~\ref{tab:systematics}. 
For ${\cal{B}}(\Lb\to\Lc\taum\nu_\tau)$ and $\RLc$, the systematic uncertainties related to the external branching fractions are added in quadrature.

\begin{table}
  \renewcommand*{\arraystretch}{1.1}
  \newcolumntype{C}{ >{\centering\arraybackslash } m{3cm} <{}}
  \newcolumntype{R}{ >{\raggedright\arraybackslash } m{5cm} <{}}
  \centering
  \caption{Relative systematic uncertainties in ${\cal{K}}(\Lc)$.}
  \label{tab:systematics}
    \begin{tabular}{R C}
      \hline
      \hline
      Contribution                         & $\delta \mathcal{K}(\Lc) /
                                        \mathcal{K}(\Lc) [\%]$ \\
      \hline
      Simulated sample size     & 3.8 \\ % 4.1 in quadrature with 1.7 and 1.6
      Fit bias                            & 3.9 \\
      Signal  modelling               &  2.0 \\ % 0.7 in quadrature with 1.0, 0.4, 1 and 0.7 
      $\Lb\to\Lcstar\tau^-\nu_{\tau}$ feed-down & 2.5 \\ %1.5 in quadrature with 2.3
      $D_s^- \to 3\pi Y$ decay model         & 2.5 \\
      $\Lb\to  \Lc D_s^- X$, $\Lb\to \Lc D^- X$, & \\ \quad $\Lb\to\Lc \Dzb X$ bkg. & 4.7 \\ % 2.9 in quadrature with 2.6
      Combinatorial bkg.             & 0.5 \\
      Particle identification & \\ \quad and trigger corrections & 1.5 \\
      Isolation BDT classifier and & \\ \quad vertex selection requirements & 4.5 \\
      \Dsm, \Dm, & \\ \quad \Dzb template shapes &13.0 \\
      Efficiency ratio                  & 2.8 \\
      Normalisation channel efficiency & \\ \quad (modelling of $\Lb\to\Lc 3\pi$) & 3.0 \\ 
        \hline
      \textbf{Total uncertainty}                  & \textbf{$16.5$} \\
      \hline
       \hline	
    \end{tabular}
\end{table}
 
Using ${\cal{B}}(\Lb\to\Lc 3\pi)= (6.14 \pm 0.94) \times 10^{-3}$~\cite{Zyla:2020zbs} 
corresponding to an average of  measurements by the CDF~\cite{cdf} and LHCb~\cite{LHCB-PAPER-2011-016} experiments, 
the signal branching fraction is determined to be
${\cal{B}}(\Lb\to\Lc\taum\nu_\tau)= (1.50 \pm 0.16 \pm 0.25 \pm 0.23)\%$, where the third uncertainty is due to the  external branching fraction measurement.  
The branching fraction  ${\cal{B}}(\Lb\to\Lc\mu^-\nu_\mu) = (6.2 \pm 1.40)\%$ from the DELPHI experiment~\cite{delphi}, updated in~\cite{Zyla:2020zbs}, 
is used to obtain the ratio of semileptonic branching fractions (cf.~\cite{Bernlochner:2022hyz})
\begin{equation}
	\RLc = 0.242 \pm 0.026\pm 0.040 \pm 0.059\,.
\end{equation}	
This value is below but in good agreement with the SM prediction in Eq.~\eqref{eqn:th:RLcpred}.
It tends to rule out some fraction of parameter space for models such as found in~\cite{Datta}, where \RLc is predicted to be far above the SM value.
See also the discussion in~\cite{Blanke:2018yud,Blanke:2019qrx,Fedele:2022iib}.

%%%%%%%%%%%%%%%%%%%%%%%%%%%%%% LHCB R(J/Psi) %%%%%%%%%%%%%%%%%%%%%%%%%%%%%%
\subsubsection{\RJ with \taumunu}
\label{sec:lhcb_rjpsi}
%%%%%%%%%%%%%%%%%%%%%%%%%%%%%% LHCB R(J/Psi) %%%%%%%%%%%%%%%%%%%%%%%%%%%%%%

%% Introduction and selection
The ratio \RJ was measured for the first time in 2018 by the LHCb experiment~\cite{Aaij:2017tyk}, thus opening
the possibility for the exploration of LFUV in decays subject to very different sources of both experimental and theoretical uncertainties 
compared to those in \RDx.  This measurement leverages two of the key techniques developed for the muonic \RDsp analysis
described in Sec.~\ref{sec:lhcb_muonic_rds}: the isolation BDT and the rest frame approximation. Just as for
the \RDsp measurement, the $\tau$ lepton is reconstructed via \taumunu, 
so that signal \BJtaunu and normalization \BJmunu decays share the same final state. The event is
selected if the only additional tracks close to the muon coming from the $\tau$ decay are a pair of oppositely charged muons
that form a vertex separated from the PV and whose invariant mass is compatible with the $\jpsi\to\mu\mu$ decay.

%% Signal Fit
The signal and normalization yields are extracted from a four-dimensional binned maximum likelihood fit to
$q^2$, \mmiss, \Esl, and the proper time elapsed between the production and decay of the $B_c$ meson: the decay time. 
The first three variables are calculated with the same techniques as used in the muonic \RDsp analysis
(Sec.~\ref{sec:lhcb_muonic_rds}). The inclusion of the decay time among the fit variables improves the
separation of $B_c$ decays from $B_{u,d,s}$ decays, because the $B_c$ lifetime is almost
three times shorter than that of $B_{u,d,s}$ mesons.

%% Fit to LHCb B->JPsitaunu muonic signal sample
\begin{figure*}
  \includegraphics[width=0.98\textwidth]{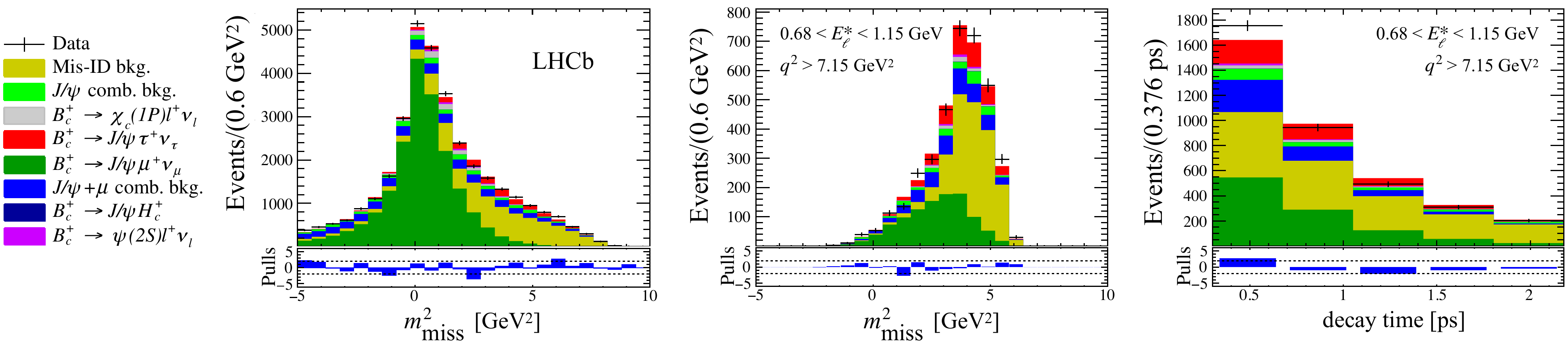} \caption{Projections of the signal
    fit for the LHCb muonic measurement of \RJ~\cite{Aaij:2017tyk}. Left: Full \mmiss projection;
   Middle: \mmiss projection in the highest $q^2$ and lowest \Esl bins; and Right: decay time projection in the
    highest $q^2$ and lowest \Esl bins.}
\label{fig:lhcb_rjpsi_signal_fit.pdf}
\end{figure*}

%% Backgrounds
A key difference with respect to the \RDx measurements is that background contributions from partially
reconstructed $B_c$ decays are significantly reduced thanks to the narrow invariant mass of the \jpsi meson
and its clean dimuon final state. As a result of this reduction and the overall small $B_c$ production rate,
the main sources of background in the \RJ analysis are misidentified $H_b\to\jpsi h^+$ decays,
where $H_b$ is a more abundant $b$-hadron and $h^+$ is a hadron incorrectly identified as a muon,
as well as random combinations of muons.

The template for the $\jpsi h^+$ contribution is estimated by applying the 
misidentification probabilities for different hadron species, as determined in high-purity samples of
identified hadrons, to a control sample with a \jpsi and an additional track that fails the muon identification. This
template is treated as free-floating in the signal fit. The combinatorial backgrounds are estimated in the sidebands
of the $B_c$ mass and the \jpsi masses, $m\left(\jpsi\mu\right)>6.4$~GeV and
$3150<m\left(\mu^+\mu^-\right)<3190$~MeV, respectively. The small contributions from higher-mass
$B^-_c \rightarrow \psi(2S) \,\ell^-\overline{\nu}_\ell$ and
$B^-_c \rightarrow \chi_c(1P) \,\ell^-\overline{\nu}_\ell$ are extracted from the fit with templates taken
from MC simulation.

%% Fit result
Figure~\ref{fig:lhcb_rjpsi_signal_fit.pdf} shows the fit projections for \mmiss over the full range, as well as
\mmiss and the $B_c$ decay time in the \Esl and $q^2$ ranges with the highest signal-to-background
ratio. The agreement is good overall and a small but significant signal contribution at high \mmiss and low
decay times can be observed.

%% Systematics table
\begin{table}
  \renewcommand*{\arraystretch}{1.1}
  \newcolumntype{C}{ >{\centering\arraybackslash } m{1.25cm} <{}}
  \centering
  \caption{Summary of the relative uncertainties for the LHCb muonic measurement of \RJ~\cite{Aaij:2017tyk}. }
  \label{tab:lhcb_rjpsi_errors}
  \vspace{1ex}
  \begin{tabular}{lCC}
    \hline\hline
\multirow{2}{*}{Contribution}              & \multicolumn{2}{c}{Uncertainty [\%]} \\
                                           & Sys. & Stat.                         \\
\hline
Signal/norm. form factors                         & 17.0                                 \\
Simulated sample size                      & 11.3                                 \\
Fit model                                  & 11.2                                 \\
Misidentified $\mu$ bkg.                   & 7.9                                  \\
Partial $B_c$ bkg.                         & 6.9                                  \\
Combinatorial bkg.                         & 6.5                                  \\
$\epsilon_\text{sig}/\epsilon_\text{norm}$ & 0.9                                  \\ 
\vspace{1mm}
\textbf{Total systematic}                  & \textbf{25.4}                        \\
\textbf{Total statistical}                 &      & \textbf{23.9}                 \\
\hline
\textbf{Tota}l                             & \multicolumn{2}{c}{\textbf{34.9}}     \\
\hline\hline
\end{tabular}
\end{table}

%% Systematics and result
Table~\ref{tab:lhcb_rjpsi_errors} summarizes the sources of uncertainty in this measurement. The leading
contribution comes from the \BJlnu decay form factors, which have not been measured yet and had to be determined
in the signal fit itself. As discussed in
Sec.~\ref{sec:th:other_states}, HQET cannot be used to describe a decay with a heavy spectator quark. As a result, at
the time of publication of this measurement only quark model predictions, untested by experiment, were available. The recent
results of lattice calculations will reduce this uncertainty substantially. Sizeable uncertainties also
arise due to the limited size of the simulated samples and the fit model. These are also expected to be reduced
in future measurements.

The result of this measurement is 
\begin{equation}
	\RJ = 0.71 \pm 0.17\,\text{(stat)} \pm 0.18\,\text{(syst)}\,, 
\end{equation}
which lies within 2 standard deviations of the SM prediction in Eq.~\eqref{eqn:th:RJPsi}.

%%%%%%%%%%%%%%%%%%%%%%%%%%%%%%%%%%%%%%%%%%%%%%%%%%%%%%%%%%%%%%%%%%%%%%%%%%%%%
%%%%%%%%%%%%%%%%%%%%%%%%%%%% BELLE POLARIZATION %%%%%%%%%%%%%%%%%%%%%%%%%%%%
\subsection{Belle polarization measurements}
\label{sec:belle_polarization}
%%%%%%%%%%%%%%%%%%%%%%%%%%%% BELLE POLARIZATION %%%%%%%%%%%%%%%%%%%%%%%%%%%%

\subsubsection{$\tau$ polarization with \taupinu and \taurhonu}
\label{sec:bfactories_hadtag_taupol}
The Belle experiment measured in~\cite{Hirose:2016wfn,Hirose:2017dxl} the $\tau$ polarization fraction $P_\tau(D^*)$ introduced in Sec.~\ref{sec:th:longpol}. 
The analysis strategy is similar to that of the hadronic tag measurements of \BDstaunu decays~\cite{Lees:2012xj, Lees:2013uzd,Huschle:2015rga}, 
but reconstructs the $\tau$ lepton in the hadronic one-prong \taupinu and \taurhonu modes. 
For these final states, the helicity angle $\cos \theta_{h}$ can be explicitly reconstructed
by taking advantage of the fully reconstructed tag-side $B$ meson to boost the visible $\tau$ daughter particles
into the center-of-mass frame of the $\tau\nutb$ lepton pair whose 4-momentum
\begin{align}
 	q = p_{e^+ e^-} - p_{\Btag} - p_{D^*} \,.
\end{align}
The terms on the right-hand side are the momenta of the colliding $e^+ e^-$ pair, the reconstructed tag-side $B$ meson, and the reconstructed $D^*$ candidate, respectively. 
In the $\tau\nutb$ center-of-mass frame, the $\tau$ energy and momentum magnitude are fully determined by $q^2$ and the $\tau$ lepton mass $m_\tau$,
as follows
\begin{align}
	E_\tau =  \frac{q^2 + m_\tau^2}{2 \sqrt{q^2}}\,, \qquad |\vec p_\tau| = \frac{q^2 - m_\tau^2}{2 \sqrt{q^2}}\,.
\end{align} 
In this frame, the cosine of the angle between the spatial momenta of the $\tau$ lepton and its daughter meson, $h$, is
\begin{align}
 \cos \theta_{\tau h} = \frac{2 E_\tau E_h - m_\tau^2 - m_h^2}{2 |\vec p_\tau| |\vec p_h|} \,,
\end{align}
in which $E_h$ and $|\vec p_h|$ are the daughter meson energy and absolute spatial momentum, respectively.
By applying a boost into the $\tau$ rest frame, one can then express the cosine of the helicity angle as
\begin{align}
 \cos \theta_h = \frac{1}{|\vec p_h^{ \, \, \tau} |} \big( \gamma |\vec p_h|   \cos \theta_{\tau h}  - \gamma \beta E_h \big) \, .
\end{align}
Here, $\gamma = E_\tau / m_\tau$, $\beta = |\vec p_\tau| / E_\tau$, and $|\vec p_h^{\,\,\tau}| = {( m_\tau^2 - m_h^2)} / (2 m_\tau)$
denotes the absolute daughter meson spatial momentum in the $\tau$ rest frame. 

To reduce backgrounds, only candidates with \mbox{$q^2 > 4 \, \text{GeV}^2$} and with a physical value of $ \cos \theta_h  \in [-1,1]$ are retained. 
Unassigned neutral energy depositions fulfilling photon-energy reconstruction criteria are summed to reconstruct \ECL 
and only candidates with  $\ECL < 1.5 \, \text{GeV}$ are retained. 
In order to not be dependent on the \Btag reconstruction, whose efficiency likely differs between data and simulation, 
the measured signal event yields are normalized to \BDsellnu events. 
These can be identified and separated from background processes using $\mmiss$ (cf. Sec.~\ref{sec:bframe}). 
For both signal and normalization candidates, events with additional charged tracks or $\pi^0$ candidates are rejected. 

\begin{figure}
 \centering
 \includegraphics[angle=90,width=0.3\textwidth]{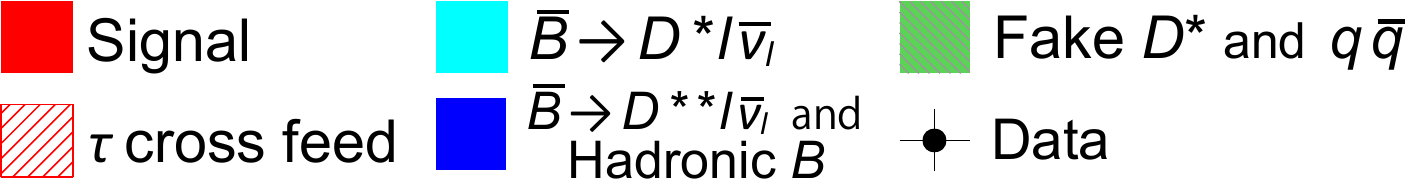} 
 \includegraphics[width=0.38\textwidth]{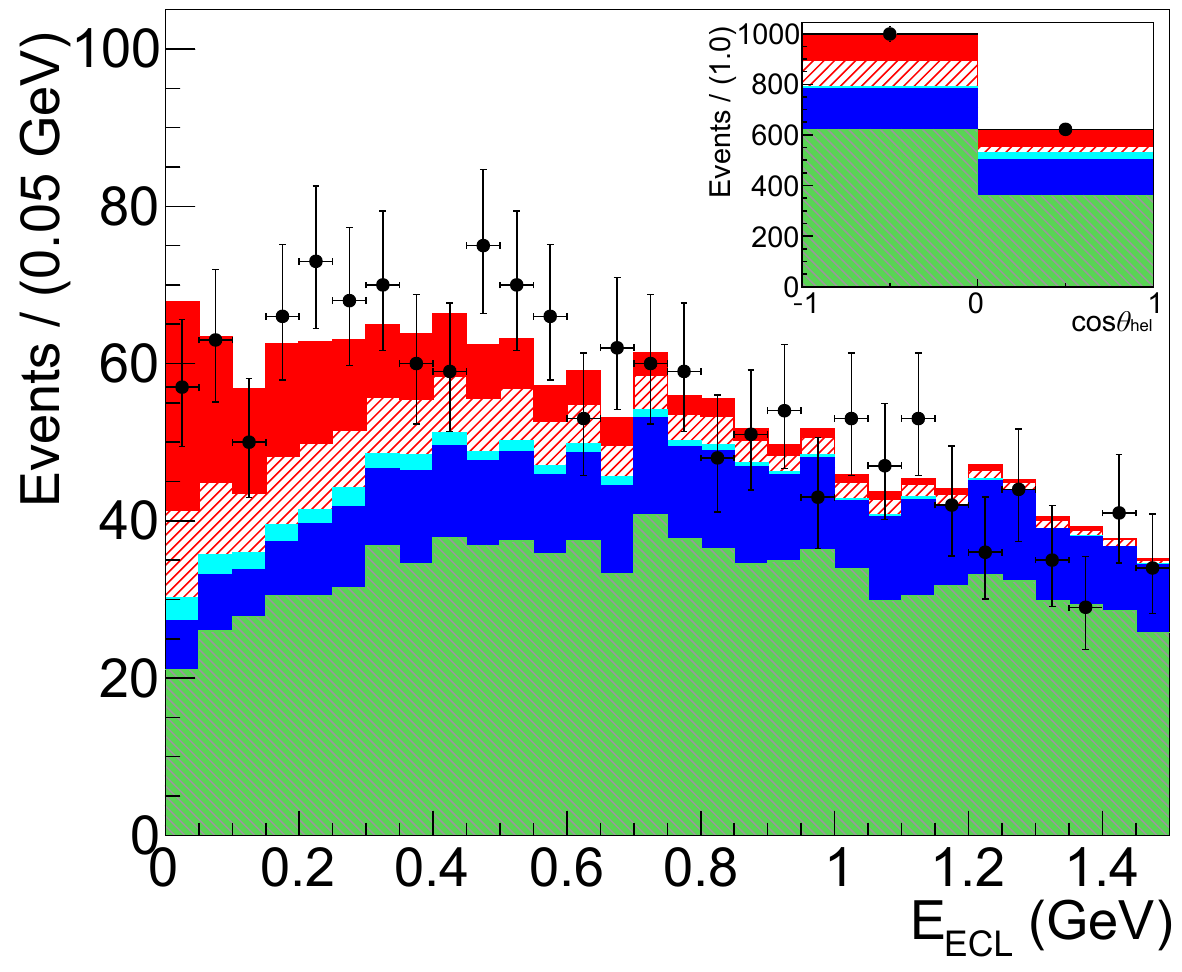} 
  \caption{Signal fit for the measurement of the $\tau$ polarization fraction
   $P_\tau(D^*)$ by Belle~\cite{Hirose:2016wfn}. The fits to the neutral and charged $B$ candidates as well as
   the \taupinu and \taurhonu decay modes and the two $\cos \theta_h$ bins are combined.}
   \label{fig:Belle_E_ECL_B0}
\end{figure}

\begin{figure}
  \includegraphics[width=0.5\textwidth]{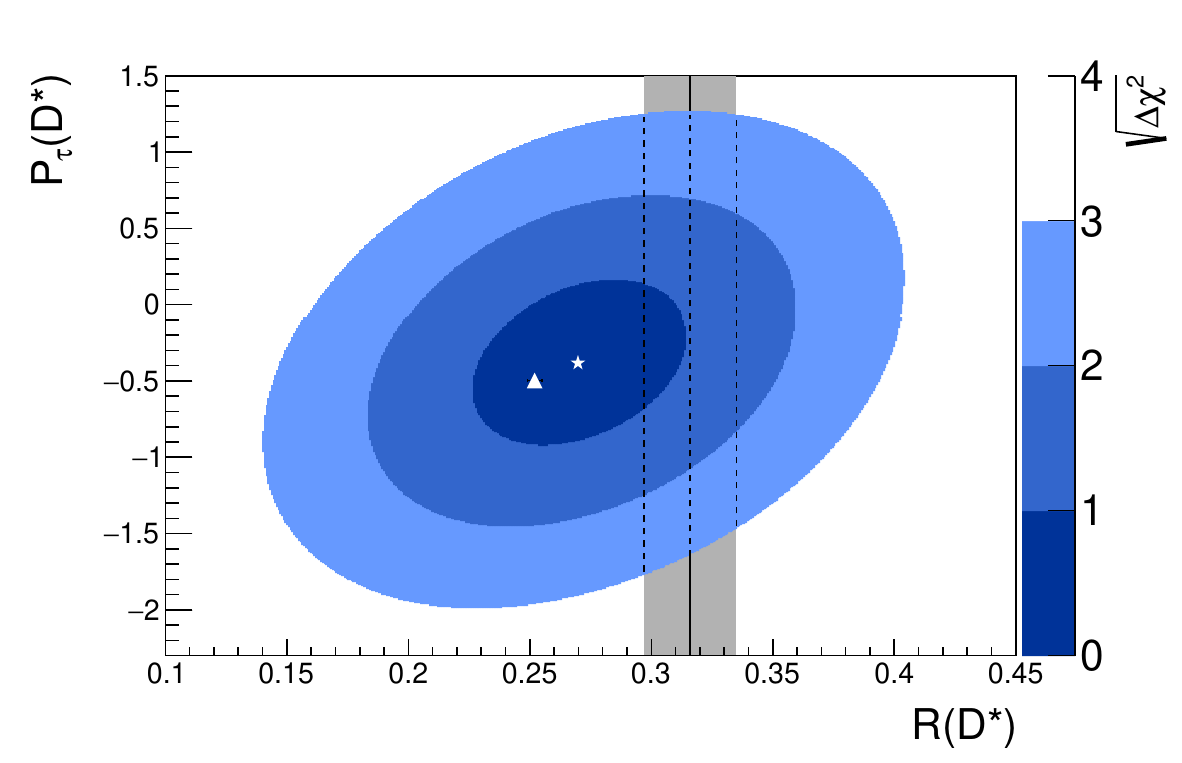} 
  \caption{
   The values of $\RDs$ and $P_\tau(D^*)$ (white star) and the $1\sigma$, $2\sigma$, and $3\sigma$ contours as measured by Belle~\cite{Hirose:2016wfn}.
   The SM expectations~\cite{Amhis:2019ckw,PhysRevD.87.034028} are shown by the white triangle. The gray band shows the then-world-average measurement of $\RDs$.
   }
\label{fig:Belle_Pol_2D}
\end{figure}

The observables $\RDs$ and $P_\tau(D^*)$ are extracted from a fit to the \ECL distribution in two bins of $\cos \theta_h$: \mbox{$[-1, 0]$} and \mbox{$[0, 1]$}. This fit is
performed simultaneously to the two $\tau$ decay samples, \taupinu and \taurhonu. 
The free parameters in the fit include the yields for the \BDstaunu, \BDsellnu, 
\BDsslnu, continuum, and fake $D^*$ contributions, among others. 
Figure~\ref{fig:Belle_E_ECL_B0} shows the fitted \ECL distribution for all the reconstructed modes combined together. 
The fitted signal yields are then converted into measurements of $\RDs$ and $P_\tau(D^*)$ with
\begin{align}
 	\RDs & = \frac{1}{ \BR(\tau \to h \nu) } \times \frac{\epsilon_\text{norm}}{\epsilon_\text{sig}}  \times \frac{N_{\mathrm{sig}}}{N_{\mathrm{norm}}} \, , \\
 	P_\tau(D^*) & = \frac{2}{ \alpha} \frac{ N_{\mathrm{sig}}^{\cos \theta_h > 0}  - N_{\mathrm{sig}}^{\cos \theta_h < 0}  }{  N_{\mathrm{sig}}^{\cos \theta_h > 0}  + N_{\mathrm{sig}}^{\cos \theta_h < 0}  } \, , 
\end{align}
with $\alpha$ being a factor that accounts for the sensitivity on the polarization and efficiency differences of both channels. The obtained values are
\begin{align}
  R(D^*) = 0.270 \pm 0.035 (\mathrm{stat}) {}^{+0.028}_{-0.025} (\mathrm{syst}) \, , \\
  P_\tau(D^*) = -0.38 \pm 0.51 (\mathrm{stat}) {}^{+0.21}_{-0.16} (\mathrm{syst}) \, , 
\end{align}
with a total correlation including systematic uncertainties of $\rho=0.33$. These results are in good agreement with the SM expectations, as shown in  
Fig.~\ref{fig:Belle_Pol_2D}. 
A summary of the uncertainties on these measurements can be found in Table~\ref{tab:had_tag_had_tau_errors}. 
The largest systematic uncertainties stem from the composition of the hadronic $B$ meson background and the limited size of the simulated samples
used to determine the fit PDFs. 

\begin{table}
  \renewcommand*{\arraystretch}{1.1}
  \newcolumntype{C}{ >{\centering\arraybackslash } m{1.25cm} <{}}
  \newcolumntype{R}{ >{\raggedright\arraybackslash } m{3cm} <{}}
  \centering
  \caption{Summary of the relative uncertainties for Belle's hadronic tag measurement of $\RDs$ and $P_\tau(D^*)$~\cite{Hirose:2016wfn,Hirose:2017dxl}. }
  \label{tab:had_tag_had_tau_errors}
  \vspace{1ex}
  \begin{tabular}{cRCC}
    \hline\hline
    \multirow{2}{*}{Result }       & \multirow{2}{*}{Contribution}              & \multicolumn{2}{c}{Uncertainty [\%]} \\
    				   &                                            & sys.         & stat.                 \\
    \hline 
    \multirow{7}{*}{\RDs}          & $B \to D^{**} \ell \bar \nu_\ell$          & 2.4          &                       \\
                                   & PDF modeling                               & 3.4          &                       \\
                                   & Other bkg.                                 & 8.4          &                       \\
                                   & $\epsilon_\text{sig}/\epsilon_\text{norm}$ & 3.2          &                       \\ 
                                   \vspace{1mm}
                                   & \textbf{Total systematic}                  & \textbf{9.9} &                       \\
                                   & \textbf{Total statistical}                 &              & \textbf{12.9}         \\ \cline{2-4}
                                   & \textbf{Total}                             & \multicolumn{2}{c}{\textbf{16.3}}    \\  \hline
    \multirow{5}{*}{$P_\tau(D^*)$}  & PDF modeling                               & 33          &                       \\
                                   & Other bkg.                                 & 31          &                       \\
                                   \vspace{1mm}
                                   & \textbf{Total systematic}                  & \textbf{48} &                       \\
                                   & \textbf{Total statistical}                 &              & \textbf{134}         \\ \cline{2-4}
                                   & \textbf{Total}                             & \multicolumn{2}{c}{\textbf{143}}    \\                           
    \hline\hline
  \end{tabular}
\end{table}

%%%%%%%%%%%%%% D* polarization
\subsubsection{$D^*$ polarization with inclusive tagging}
\label{sec:exp:Dspol}
The Belle experiment reported in~\cite{Abdesselam:2019wbt} a first, preliminary, measurement of the longitudinal $D^*$ polarization fraction $F_{L,l}(D^*)$ (see Sec.~\ref{sec:th:longpol})
based on inclusively tagged events (Sec.~\ref{sec:tagging}). 
First, a viable $B^0 \to \Dstarm \tau^+ \nu_\tau$ signal candidate with \tauellnu or \taupinu and $\Dstarm \to \Dbar^0 \pi^-$ is reconstructed. 
The $\Dbar^0$ meson is reconstructed in $\Dbar^0 \to K^+ \pi^-$,  $\Dbar^0 \to K^+ \pi^- \pi^0$, and  $\Dbar^0 \to K^+ \pi^+ \pi^- \pi^-$ modes. 
Thereafter, no explicit reconstruction is attempted of the other (tag) $B$ meson produced in the $e^+ e^-$ collision. 
Instead, an inclusive reconstruction approach that sums over all unassigned charged particles and neutral energy depositions 
above a certain energy threshold in the calorimeter is employed. 
Compared to hadronic or semileptonic tagging, this approach has the benefit of a higher reconstruction efficiency,
as it does not rely on the correct identification of the decay cascades, but results in a poorer $B$ momentum resolution. 

The tag side is required to be compatible with a well-reconstructed $B$ meson by requiring
\begin{equation}
M_{\mathrm{tag}} = \sqrt{E_{\mathrm{beam}}^2 - | \mathbf{p}_{\mathrm{tag}}|^2}>5.2\text{ GeV}\,,
\end{equation}
and $-0.30 < E_{\rm tag}-E_{\rm beam}<0.05\text{ GeV}$,
where $E_{\mathrm{beam}}  = \sqrt{s}/2$ is the energy of each of the colliding \epem beams in the CM frame.

The sizeable background contributions are suppressed with the signal-side normalized variable
\begin{equation}
X_{\mathrm{miss}} = \frac{E_{\mathrm{miss}} - | \mathbf{p}_{D^*} + \mathbf{p}_{d_\tau} |}{\sqrt{E^2_{\rm beam}-m^2_{B^0}}}\,, 
\end{equation}
where $E_{\mathrm{miss}} = E_{\mathrm{beam}} - (E_{D^{*}} + E_{d_\tau})$ and $d_\tau$ refers to the visible $\tau$ daughter.
Events with one neutrino have values of $X_{\mathrm{miss}}$ in the range \mbox{$[-1, 1]$}, while events with multiple
undetected particles tend to take larger values. The analysis optimizes the signal significance by requiring
$X_{\mathrm{miss}}$ to be larger than 1.5 or 1 for the \tauellnu and \taupinu decay modes, respectively.

\begin{figure}
  \includegraphics[width=0.33\textwidth]{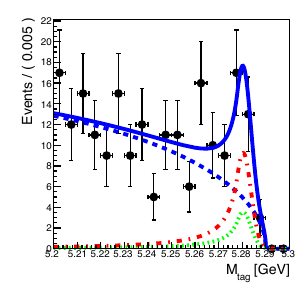}
  
  \caption{Signal fit to the lowest \cosHel bin, $[-1,-0.67]$, in the $\Dbar^0 \to K^+ \pi^- \pi^0$ channel for
  the measurement of the longitudinal $D^*$ polarization fraction by Belle~\cite{Abdesselam:2019wbt}. The red (gray dash-dot) curve
  corresponds to the signal contribution, and the blue (dark gray dashed) and green (light gray dotted) curves display the non-resonant and resonant background
  contributions, respectively.}
  
\label{fig:Belle_Mtag_cosv}
\end{figure}

\begin{figure}
  \includegraphics[width=0.3\textwidth]{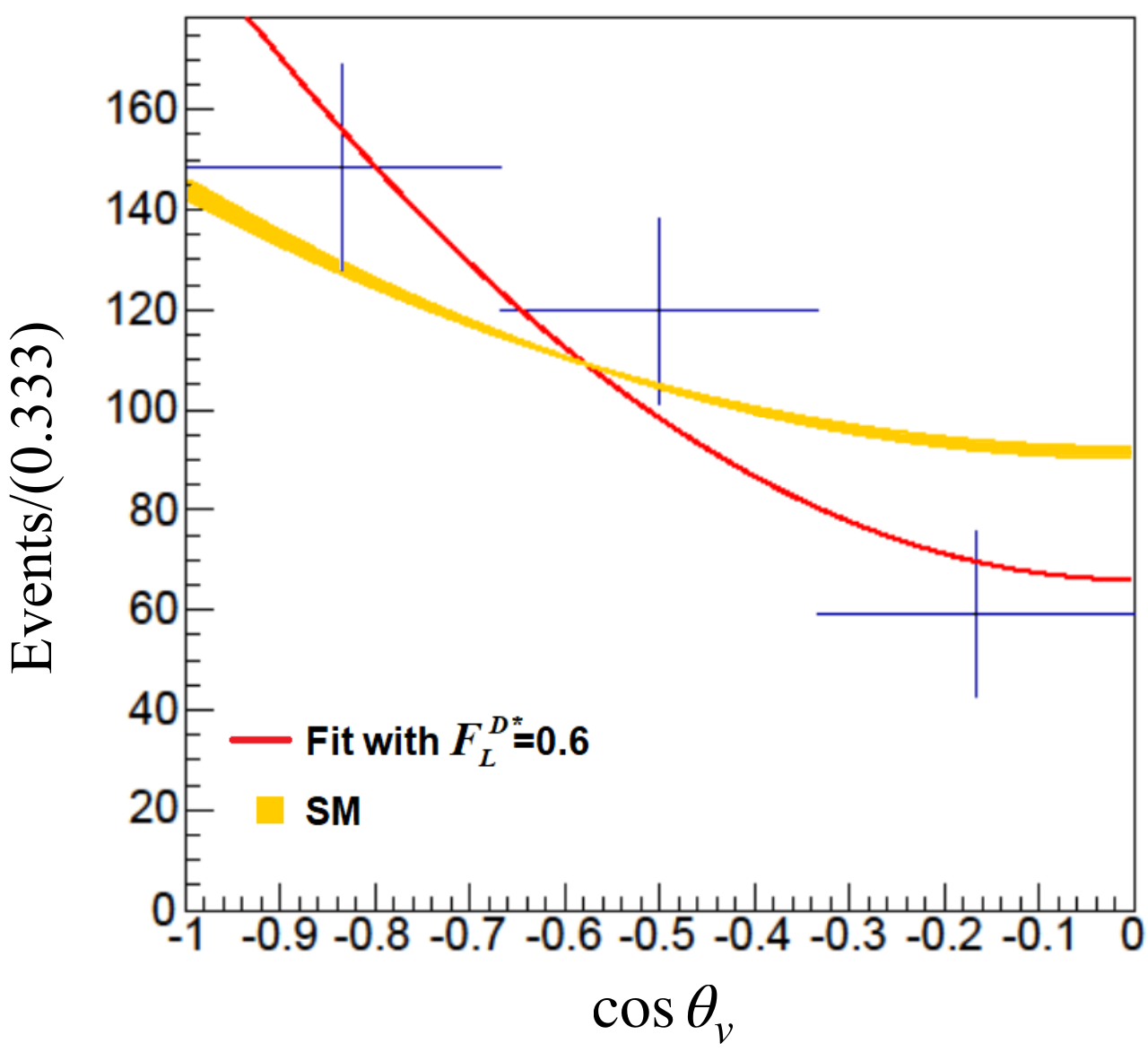}
  \caption{Measured \cosHel distribution in $B^0 \to \Dstarm \tau^+ \nu_\tau$ decays for the determination of
  the longitudinal $D^*$ polarization fraction by Belle, adapted from~\cite{Abdesselam:2019wbt}.  The red (dark gray) solid curve shows the
  best fit of the longitudinal polarization fraction and the yellow (light gray) band corresponds to the SM
  expectation~\cite{Huang:2018nnq}.}
\label{fig:Belle_cosv}
\end{figure}

The helicity angle $\theta_{v}$ is defined as the angle between the reconstructed $\Dbar^0$
and the direction opposite to the $B^0$ meson in the $D^{*-}$ frame (see the definition in Fig.~\ref{fig:polar_def}; 
the Belle analysis uses the notation $\theta_{\text{hel}}$).
Because of the low $D^*$ reconstruction efficiency
for $\cosHel>0$, the analysis focuses on the $-1 \leq \cosHel \leq 0$ range.
The signal yields are extracted in three bins of \cosHel from fits to the $M_{\mathrm{tag}}$ distribution,
see Fig.~\ref{fig:Belle_Mtag_cosv} for an example.
Most backgrounds do not peak in this variable, with the exception of semileptonic decays into light leptons.
The yields for these peaking contributions are determined in the sidebands of kinematic variables.
The $D^*$ polarization fraction is determined by a fit to the signal yields as a function of \cosHel.
Given the size of the \cosHel bins, resolution effects are assumed to be negligible.
Figure~\ref{fig:Belle_cosv} shows the measured helicity angle distribution, corrected for acceptance effects. 
The resulting fitted value for the longitudinal $D^*$ polarization fraction is 
\begin{equation}
	F_{L,\tau}(D^*) = 0.60 \pm 0.08\text{(stat)} \pm 0.04\text{(sys)}   \, ,
\end{equation}
with its uncertainty dominated by the limited size of the data sample. 
The largest systematic uncertainty in this measurement stems from the signal and non-resonant background shapes used in the $M_{\mathrm{tag}}$ fits, 
followed by the uncertainty on the modeling of \BDsstaunu decays.

This result agrees with the SM prediction of $F_{L,\tau}(D^*)_{\rm SM} = 0.455(6)$ 
(Sec.~\ref{sec:th:longpol}, from an arithmetic average of the various SM predictions) 
at the $1.6\sigma$ level.
An important control measurement is the $D^*$ polarization of the light-lepton states, $F_{L,\ell}(D^*) = 0.56\pm 0.02$ (statistical uncertainty only),
which is in agreement with the prediction of $F_{L,\ell}(D^*)_{\rm SM}^{\rm BLPR} =0.517(5)$ within 2.1 standard deviations.

\section{Common Systematic Uncertainties and Future Prospects}
\label{sec:systematics}

%% RD(*) systematic uncertainties
\begin{table*} 
  \newcolumntype{R}{ >{\centering\arraybackslash\bf } r <{}}
  \caption{Summary of the uncertainties on the $\RDx$ measurements. The ``Other bkg.'' column includes primarily
    contributions from $DD$ and combinatorial backgrounds. The ``Other sources'' column is dominated by particle identification
  and external branching fraction uncertainties.}
  \renewcommand{\arraystretch}{1.15}
  \label{tab:summary_uncertainties} \vspace{0.1in}
\begin{threeparttable}
  \begin{tabular}{llll  |ccccc  |RR|R} \hline\hline
    &&&& \multicolumn{5}{c|}{Systematic uncertainty [\%]} & \multicolumn{3}{c}{{\bf Total uncert. [\%]}} \\
    Result                & Experiment      & $\tau$ decay   & Tag  & ~MC stats~ & ~$\Dx l\nu$~ & ~$D^{**}l\nu$~ & ~Other bkg.~ & ~Other sources~ & ~Syst.~ & ~Stat.~ & ~Total \\ \hline
    \multirow{3}{*}{\RD}  & \babar\tnote{a} & $\ell\nu\nu$    & Had. & 5.7        & 2.5          & 5.8                      & 3.9          & 0.9             & 9.6     & 13.1    & 16.2   \\ 
                          & Belle\tnote{b}  & $\ell\nu\nu$    & Semil.  & 4.4        & 0.7          & 0.8                     & 1.7          & 3.4             & 5.2     & 12.1    & 13.1   \\                                                                                           
                          & Belle\tnote{c}  & $\ell\nu\nu$    & Had. & 4.4        & 3.3          & 4.4                      & 0.7          & 0.5             & 7.1     & 17.1    & 18.5   \\\hline
                                                                                                                           
    \multirow{6}{*}{\RDs} & \babar\tnote{a} & $\ell\nu\nu$    & Had. & 2.8        & 1.0          & 3.7                      & 2.3          & 0.9             & 5.6     & 7.1     & 9.0    \\ 
                          & Belle\tnote{b}  & $\ell\nu\nu$    & Semil. & 2.3        & 0.3          & 1.4                      & 0.5          & 4.7             & 4.9     & 6.4    & 8.1   \\              
                          & Belle\tnote{c}  & $\ell\nu\nu$    & Had. & 3.6        & 1.3          & 3.4                      & 0.7          & 0.5             & 5.2     & 13.0    & 14.0   \\
                          & Belle\tnote{d}  & $\pi\nu$,  $\rho\nu$       & Had.      & 3.5        & 2.3         & 2.4                      & 8.1         & 2.9             & 9.9     & 13.0    & 16.3   \\              
                          & LHCb\tnote{e}   & $\pi\pi\pi(\pi^0)\nu$ & ---   & 4.9        & 4.0          & 2.7                      & 5.4          & 4.8             & 10.2    & 6.5     & 12.0   \\
                          & LHCb\tnote{f}   & $\mu\nu\nu$    & ---   & 6.3        & 2.2          & 2.1                      & 5.1          & 2.0             & 8.9     & 8.0     & 12.0   \\ \hline

    \hline\hline
  \end{tabular}
  \begin{tablenotes}[para]
  	\item [a] \cite{Lees:2012xj, Lees:2013uzd}
	\item [b] \cite{Belle:2019rba}
	\item [c] \cite{Huschle:2015rga}
	\item [d] \cite{Hirose:2017dxl}
	\item [e] \cite{Aaij:2015yra}
	\item [f] \cite{Aaij:2017deq}
  \end{tablenotes}
\end{threeparttable}
\end{table*}

The different measurements of \RDx so far are fairly independent of each other because their uncertainties are
dominated by the limited size of the data and the simulation samples. However, over the next decade and half,
Belle~II and LHCb will collect data samples 50 to 200 times larger than those used for the present
measurements of $\RDx$ (Table~\ref{tab:production_comp}), so the relative impact of other systematic uncertainties will increase. Some
of these uncertainties are due to aspects of the experimental analysis that are shared among all measurements
and can therefore lead to common systematic uncertainties. As a result, the combination of the measurements
will entail a more complex treatment of these uncertainties. Table~\ref{tab:summary_uncertainties} and the
following subsections describe the main sources of systematic uncertainty in the measurement of \RDx,
and the level of commonality among the various approaches.\footnote{It is worth noting that, while some
  uncertainties are \emph{multiplicative}, i.e. they scale with the resulting central value (e.g.,
  uncertainties on the signal efficiency), the majority of the uncertainty is \emph{additive} (e.g., uncertainties associated with the background subtraction or signal shapes). As a result,
  changes in the central values would alter the value of the uncertainty when expressed as a
  percentage. However, given that the overall uncertainty has become smaller than 20\% and the central values
  are starting to converge (see Fig.~\ref{fig:rdx_time.pdf}), the presentation of uncertainties as percentages
  should give a broadly accurate representation of the uncertainties and allow for comparisons across
  different measurements.}
  
We also discuss in the following subsections the future prospects to reduce the total uncertainty in $\RDx$, 
as well as on LFUV ratios in many other decay modes, down to a few percent or less. 
In particular, reducing the systematic uncertainties commensurately with the
statistical uncertainties will require meeting key challenges in computation, the modeling of $b$-hadron
semileptonic decays, and background estimation in the years to come.

\subsection{Monte Carlo simulation samples}
\label{sec:sys:MC}
Table~\ref{tab:summary_uncertainties} shows that one of the principal sources of uncertainty in the $\RDx$
measurements arises from the limited size of the simulation samples. 
This limitation results in large uncertainties through two different, but parallel, considerations:
First, $\BDxltnu$ decays have some of the largest $B$ branching fractions,
necessitating very large simulation samples to acceptably model the data. 
Such uncertainties, however, are statistical in nature and thus independent among different experimental analyses.

Second, semitauonic decays involve final states with multiple neutrinos, which escape detection.
As a result, the reconstructed kinematic distributions employed to separate signal from background events are
broad and difficult to describe analytically. Instead, experiments rely upon Monte Carlo simulation to derive the
templates that are used in the signal extraction fit. Because of the broad nature of these distributions,
multiple dimensions are necessary to disentangle the various contributions, which results in the simulated
events being widely distributed among the numerous bins in the templates. 

Of course, Monte Carlo-based uncertainties can be reduced simply by producing more simulated events. 
However, given the size of future data samples, it will be both a time and cost challenge
to continue producing simulated events in sufficient numbers such that these uncertainties remain controlled.
Thus, different solutions will need to be considered. At present the most promising approaches are:
\begin{enumerate}[wide, labelwidth=!, labelindent=0pt, label = \quad \textbf{(\roman*)}, noitemsep, topsep =0pt]
  \item \emph{Hardware}: The High Energy Physics (HEP) community has historically relied upon the exponential increase in computing throughput for
  relatively stable investments. As this exponential growth slows, either greater funding will have to be found or new
  avenues will need to be explored to keep up. Monte Carlo simulations are highly parallelizable, which makes
  them a favorable target for graphics processing unit (GPU) computation.  Efforts to make increasing use of
  GPUs are underway, and expertise and appropriate tools will have to be further developed by the HEP
  community to ensure the widespread adoption of GPUs and reap their benefits.

  \item \emph{Fast simulation (FastSim)}: the most resource intensive step in the generation of simulated events is
  the simulation of the detector response. Several procedures have been developed and are already in use that
  accelerate this step by simulating only parts of the detectors, or parameterizing its response (see examples at
  \cite{Muller:2018vny,deFavereau:2013fsa}). New machine
  learning techniques such as generative adversarial networks may be able to further optimize this aspect of event simulation. 
  See e.g. \cite{Vallecorsa_2018,Erdmann:2018jxd} for proof-of-concept studies of this. 

  \item \emph{Aggressive generator-level selections}: these can help reduce the number of events that need
  to be fully simulated. Fiducial selections are already widely applied, but as data become abundant and the
  computing resources are stretched thin, analyses may have to start focusing on reduced regions of phase
  space with an even better signal-to-noise ratio. The generator-level selections would then have to be
  adjusted as closely as possible to these reduced areas to maximize the physics output of the simulation.
  For Belle~II an attractive option to increase the size of simulated samples in analyses that use hadronic tagging 
  would be to generate only the low branching fraction modes actually targeted by the tagging algorithms. 
  See e.g.~\cite{ediss24013} for a proof-of-concept implementation using generative adversarial networks.
 \end{enumerate}
 
It is important to note that none of these approaches alone will be sufficient to cover all future
needs. For instance, the FastSim implementations currently employed at LHCb allow for simulated events
to be produced with about ten times fewer resources than with full simulation. However, this order of
magnitude improvement only covers the increased needs from Run 1 (3.1~\invfb) to Run 2 (6~\invfb, twice
the $\bbbar$ cross section, and higher efficiency than in Run 1). Meeting the needs for the $50\invab$ that will be collected by Belle~II, or the
$300\invfb$ by LHCb, will probably involve the combined use of the approaches listed above and perhaps others. 

\subsection{Modeling of $\BDxltnu$}
\label{sec:syst:dxlnu}
As discussed at length in Sec.~\ref{sec:theory}, the predominant theory uncertainties in the modeling of $b \to c\tau\nu$ decays 
arise in the description of their hadronic matrix elements.
Precision parametrizations of these matrix elements are currently achieved by either data-driven model-independent approaches, 
such as fits to HQET-based parametrizations (Sec.~\ref{sec:th:hqet}), or by lattice QCD results (Sec.~\ref{sec:th:LQCD}), or a combination of both.
This applies to predictions both for the ground states as well as the excited states (Sec.~\ref{sec:th:other_states}) that often dominate background contributions.
In the case of \BDxellnu, these approaches have led to form factor determinations whose uncertainties only contribute at the $1$--$2\%$ level in the measurements of $\RDx$.

Especially for semitauonic analyses using the electronic or muonic $\tau$ decay channels, 
a reliable description of \BDxellnu semileptonic decays is a critical input, in order to control lepton cross-feed backgrounds.
The hadronic $\tau$ decay analyses also rely on these light semileptonic inputs, but to a lesser extent.
Finally, there is some additional uncertainty in the modeling of the detector resolution for the kinematic variables
that these analyses depend upon, that can be shared across results from the same experiment.

\subsection{$\BDsstaunu$ and other $B\to D^{**}$ backgrounds}
\label{sec:syst:dssbg}
\subsubsection{Systematic uncertainty evaluation and control}
\label{sec:syst:dss_ff}
Excited $\Ddouble$ states decay to $\Dstar$, $\Dz$, or $\Dpm$ mesons plus additional photons or pions,
which can escape detection. 
As a result, both $\BDssellnu$ and $\BDsstaunu$ decays can easily lead to extraneous
candidates in $\RDx$ analyses, though the former contributes only to measurements that employ the leptonic decays of the $\tau$ lepton. 
In hadronic-$\tau$ analyses, another background source associated with \Ddouble production is formed by $B\to\Ddouble D_s^{(*,**)}$ decays with \Ds\to\pip\pim\pip X.
While all analyses exploit dedicated $\Ddouble$ control samples where some of the
parameters describing these contributions are measured, a number of assumptions are shared among the various
measurements, namely the form factor parameterization of the $\BDssltnu$ decays (Sec.~\ref{sec:th:other_states})
and the $\Ddouble$ decay branching fractions. 

First data-driven fits of the $B \to D^{**}$ form factors have been performed~\cite{Bernlochner:2016bci,Bernlochner:2017jxt}, but the resulting parameters---especially 
for the broad states---are not yet well constrained. The chosen approach, however, will be improvable with future data.
Just as for the $B \to \Dx$ modes, data-driven predictions for $B \to D^{**}$ (Eq.~\eqref{eqn:th:RDss})
are thus likely to improve in precision until they reach the naive order of $1/m_c^2$ contributions---i.e. a few percent---beyond 
which the number of parameters required to describe higher-order effects becomes too large to be effectively constrained. 
Combination with future LQCD results (see e.g. \cite{Bailas:2020cho}), however, may permit even more precise predictions.
Additionally, the $\calR(\Ddouble)$ ratios have not yet been measured, 
so the various experiments have relied on theoretical predictions, assigning a relatively large uncertainty. 
The size of this uncertainty is however arbitrary and could lead to a common underestimate of the systematic uncertainty from the \Ddouble feed-down (see Sec.~\ref{sec:syst:dss_bf}).
With the latest theoretical predictions (Eq.~\eqref{eqn:th:RDss}), this uncertainty should be reduced in the future.

Dedicated experimental efforts are also presently ongoing to further address these issues. In particular:
\begin{enumerate}[wide, labelwidth=!, labelindent=0pt, label = \quad \textbf{(\roman*)}, noitemsep, topsep =0pt]
	\item Improved measurements are anticipated for the $\BDssellnu$ relative branching fractions and kinematic distributions such as the four-momentum 
	transfer squared or further angular relations. This is especially important for the broad $D_1'$ and $D_0^*$ states, which are still poorly known compared to the narrow $D_1$ and $D_2^*$ states.
	Such measurements can in principle already be carried out with currently available data sets. 
  	\item Measurements involving a hadronized $W \to D_s^+$, i.e. $B \to (\Ddouble \to \Dx\pi)\Ds$~\cite{PhysRevD.101.032005,lhcb19-026}.
	This approach offers much better sensitivity to decays involving the wide $\Ddouble$ states because the $\Dx\pi$ 
	spectrum can be cleanly measured via the sideband subtraction on the narrow $B$ mass peak. 
	Additionally, the presence of a $\Ds$ meson in the final state offers two unique features: 
		(a) in contrast to decays where the virtual $W$ produces a single pion, 
		the $q^2$ range for production of a $\Ds$ meson is in the range of interest for semitauonic decays; 
		and (b) the relative rates of the various $\Ddouble$ states can be measured when associated with both spin-$0$ ($D_s$) and spin-$1$ ($\Dsstar$) states.
	\item The direct measurement of $\BDsstaunu$ decays for the narrow states $\Ddouble = D_1$ or $D_2^*$.
	When combined with the estimated branching fractions for the narrow $\Ddouble$ versus the total $\Ddouble$ rate,
	and expectations from isospin symmetry (the feed-down is dominated by $D^{**\pm}$ states while much better experimental precision will be achieved for $D^{**0}$),
	these $\BDsstaunu$ results might be used to control the $\Ddouble$ feed-down rate into the $\RDx$ signal regions.
\end{enumerate}

Significant progress can therefore be expected in the control of this important common systematic uncertainty in the near term, 
such that the systematic uncertainty due to $\BDssltnu$ decays is likely to be reduced to the percent level or less.

\subsubsection{\Ddouble branching fraction assumptions in \RDx analyses}
\label{sec:syst:dss_bf}

While the estimation of the normalization of the contributions from background \BDsslnu decays is largely data-driven, a number of
assumptions in the various branching fractions involved can have a significant impact in the measurement of
\RDx. These are:
\begin{enumerate}[wide, labelwidth=!, labelindent=0pt, label = \quad \textbf{(\roman*)}, noitemsep, topsep =0pt]
  \item $\Br(D^{**} \to \Dx \pi (\pi))$: These branching fractions are primary inputs to all the \BDsslnu
    templates employed in the signal extraction fits. Using the approach of \cite{Bernlochner:2016bci},
    $\BR(\Ddouble \to \Dx\pi)$ can be estimated by combining data for the ratios $\BR(\Ddouble \to
    \Dstar\pi)/\BR(\Ddouble \to D\pi)$~\cite{Zyla:2020zbs}, isospin relations, and measurements of ratios of
    non-$D^*$-resonant three-body $D_1^0$ and $D_2^{*0}$ decays to $D^0\pi^+\pi^-$ versus two-body decays to
    $D^{*+} \pi^-$~\cite{Aaij:2011rj}.  The latter are used to estimate the total non-$D^*$-resonant
    branching fraction to all possible $D\pi\pi$ final states with an isospin correction factor $\simeq
    2$.  The resulting estimates for exclusive two-body decays, and sum of non-$D^*$-resonant three-body
    decays, are shown in Table~\ref{tab:DssBRs}.  The experimental analyses, however, have used various other
    sets of different numbers, which is worth revisiting.

\begin{table}[t]
	  \renewcommand*{\arraystretch}{1.25}
 	 \newcolumntype{C}{ >{\centering\arraybackslash $} m{1.25cm} <{$}}
	\caption{Estimates for $D^{**}$ strong decay branching fractions to exclusive two-body decays, 
	  and the sum of non-$D^*$-resonant three-body decays, $\sum D\pi\pi$. Based on the approach of~\cite{Bernlochner:2016bci} and
            measurements from~\cite{Aaij:2011rj,Zyla:2020zbs}.}
	\label{tab:DssBRs}
	\begin{tabular*}{\linewidth}{@{\extracolsep{\fill}}cCCCCC}
	\hline\hline
	\multirow{2}{*}{Parent} & \multicolumn{5}{c}{Final state} \\
					 & D^{*} \pip & D^{*}  \piz &  D \pip & D  \piz & \sum D\pi\pi\\
	\hline
	$D_2^*$ 		& 0.26 	& 0.13 	& 0.40 		& 0.20 		& \text{---} \\
	$D_1$		& 0.42 	& 0.21 	& \text{---}		& \text{---} 	& 0.36\\
	$D_1^{\prime}$	& 0.67 	& 0.33 	& \text{---}		& \text{---} 	&  \text{---}\\
	$D_0$	&  \text{---}	&  \text{---}	 & 0.67 	& 0.33 	&  \text{---}\\
	\hline\hline
	\end{tabular*}
\end{table}

  \item $\Br(\BDssellnu)$: As mentioned above, the hadronic-$\tau$ measurements are not sensitive to this
    contribution. The leptonic-$\tau$ analyses have some sensitivity to these branching fractions, but it is
    small because the total contribution from \BDssellnu decays for the four \Ddouble states is floated in the
    various fits. Since the four contributions are combined together in the same fit template, the relative
    \BDssellnu branching fractions---typically taken from~\cite{Zyla:2020zbs}---impact the measured \RDx values
    at the 0.3--0.8\% level~\cite{Lees:2013uzd}.

  \item $\Br(\BDsstaunu)$: All \RDx measurements are rather sensitive to this contribution because the
    kinematics of the final state particles in these decays are similar to those in signal decays.  Some
    leptonic-$\tau$ measurements tie this contribution to the fitted \BDssellnu yields via \RDdouble or merge it
    with other background contributions.  The \babar
    analysis~\cite{Lees:2013uzd} assumes $\RDdouble=0.18$ for all \Ddouble states. 
    Investigation of the numerical simulation inputs used by Belle
    analyses~\cite{Huschle:2015rga,Belle:2019rba} suggests they assumed an average of $\RDdouble=0.15$,
    while the LHCb result~\cite{Aaij:2015yra} uses $\RDdouble=0.12$. 
    The hadronic-$\tau$ \RDs measurement from LHCb~\cite{Aaij:2017deq} ties the \BDsstaunu
    yield to be 11\% of the fitted \BDstaunu yield, and further decreases the value of \RDs by 3\% to take into
    account an additional contribution from $\Bs \to D_{s1}^{\prime}\tau\nu$ decays.
    Notably, all these assumed values for $R(\Ddouble)$ are significantly above the predicted central
    values (Eq.~\eqref{eqn:th:RDss}), by about $50\%$. The impact on the measured values can be estimated from
    the \RDdouble systematic uncertainty estimated in~\cite{Lees:2013uzd}. A 50\% downwards variation of the
    assumed $\RDdouble=0.18$ value results in \RDx increasing by $1.7$--$1.8$\%. A shift of this magnitude would
    result in an increase of the tension of the \RDx world average with the SM predictions by more than
    $0.5\sigma$.   For future measurements, we therefore advocate that experiments revisit their assumptions regarding the \Ddouble
    feed-down in light of available data-driven predictions.

  \item $\Br(B_s \to D_s^{**} X)$: Additional feed-down contributions to the LHCb measurements of \RDx come from
    decays involving partially reconstructed heavy $D_s^{**}$ mesons, namely, $B_s \rightarrow D_{s1}^*$
    and $B_s \rightarrow D_{s2}^*$. The $D_{s1}^*$ and $D_{s2}^*$ mesons are heavy enough so that they
    decay primarily as $D_s^{**}\rightarrow D^{(*)} K$. Given that the \BsDssslnu branching fractions have not
    yet been measured and the considerable $B_s$ meson production at the LHC (see
    Table~\ref{tab:production_comp}), these decays can lead to sizable uncertainties on \RDx. In a similar
    fashion to the \BDsslnu decays, \BsDsssellnu decays contribute only to measurements that employ the
    leptonic decays of the $\tau$ lepton, while \BsDssstaunu decays contribute to both leptonic- and
    hadronic-$\tau$ measurements, and $\Bs\to D_s^{**}\Ds X$ decays with $\Ds\to\pip\pim\pip X$ to
    hadronic-$\tau$ results.  As an example of the potential size of these contributions, the present
    correction due to the \BsDssstaunu feed-down in the hadronic-$\tau$ measurement of \RDs by LHCb is 3\%, with a
    relative uncertainty of 50\%. Future measurements of the $B_s \to D_s^{**}$ branching fractions
    will thus be very important to be able to reach percent-level uncertainties on the LHCb measurements of
    \RDx---as well as on $\calR(D^{(*)}_s)$.

\end{enumerate}
   
\subsection{Modeling other signal modes}
Some insight into the precision of future form factor predictions, and their role in LFUV analyses, can be obtained from considering the case of $B_c \to \Jpsi \tau \nu$. 
As can be seen in Table~\ref{tab:lhcb_rjpsi_errors}, a dominant systematic uncertainty---$17\%$---in the 2018 LHCb analysis~\cite{Aaij:2017tyk} 
arose from the poorly-known description of the $B_c \to \Jpsi$ form factors.  
At the time, the prediction for $\RJ$ was known only at the $10\%$ level, or worse. 
However, recent LQCD results for the $B_c \to \Jpsi$ form factors~\eqref{eqn:th:RJPsi} now permit percent-level predictions, 
such that one might expect the corresponding systematic uncertainty to similarly drop by an order of magnitude in a future analysis.

With regard to $\Lambda_b \to \Lambda^{(*)}_c$ decays, while the ground state form factors are known to high precision already, 
a combination of anticipated LQCD results and future data may similarly permit the excited state form factors to be constrained at or beyond the $1/m_c^2$ level.
Finally, future LQCD studies may be expected to improve predictions for $B_s \to D_s^{(*,**)}$ form factors to a level comparable to that for $B \to D^{(*,**)}$, 
well beyond the $\sim 20\%$ uncertainties from flavor symmetry arguments.

\subsection{Other background contributions}
\label{sec:overview:other_bkg}

Double charm decays of the form $B\to D^{(*,**)}D_s^{(*,**)}$ and $B\to D^{(*,**)}D^{(*,**)}K^{(*)}$ can lead
to final state topologies very similar to those of semitauonic processes, whenever the decay of one of the charm
mesons mimics that of a $\tau$ lepton. Examples are $D_s^{(*,**)} \to X\tau\nu,\,X\pi^+\pi^-\pi^+$ or $D^{(*,**)} \to X \ell \nu$
with $X$ referring to unreconstructed particles. 
Such processes are very significant background modes for \RDx measurements at LHCb, and to a somewhat lesser extent, for $B$-factory measurements.
While several of these analyses estimate the overall double-charm contribution using data control samples, all
measurements rely on averages of previously measured branching fractions of $B$ and $D$ decays from the Particle Data Group compilation~\cite{Zyla:2020zbs}. 
These averages are used as an input to produce the right mixture of decay modes for background templates. 
Additionally, the extrapolations into the signal regions often rely on simulations whose
models for the decay dynamics might not reflect the full resonance structure of such transitions. This set of
assumptions can be common to several experiments.

Although a wealth of branching fraction determinations regarding these and other relevant decays have been
accumulated by BESIII~\cite{Ablikim:2009aa}, \babar, Belle, and LHCb, there are significant areas where measurements that are in principle feasible 
have not been carried out or are not precise enough to provide useful constraints. Instances of these are
double charm decays with excited kaons in the final state or hadronic and double charm processes involving
$\Ddouble$ states. These are especially important because they cover the high $q^2$
range that has the highest signal purity in \RDx measurements. In the near future, Belle~II and LHCb will 
provide new results of branching fractions for such decays that will alleviate the reliance on common assumptions for
the various double-charm decay modes. Additionally, more precise information about the semileptonic
and $\pi^+\pi^-\pi^+$ decays of charm mesons, which can be provided by BESIII in the near future, will be needed.

\subsection{Other systematic uncertainties}
The remaining uncertainties in Table~\ref{tab:summary_uncertainties} are dominated by particle identification
and external branching fraction uncertainties. The latter are especially relevant for measurements that
utilize the hadronic decays of the $\tau$ lepton. The final state for the signal decays in these measurements
does not correspond to that of the \BDxellnu decays needed for the \RDx denominator and, as a result,
intermediate normalization modes are employed. For instance, the current precision on the
normalization decays for the \taupipipi analysis from LHCb~\cite{Aaij:2017deq},
$B\to D^*\pi^+\pi^-\pi^+$ and $\BDsmunu$ as shown in Eq.~\eqref{eq:rd_hadronic}, is limited to $3$--$4\%$,
so new measurements of these branching fractions are necessary to reduce the overall uncertainty beyond that
level. In fact, what is required is the ratio of these two quantities. 
This can be measured more precisely than each branching ratio separately: 
a measurement that Belle~II may be able to perform relatively easily.

Radiative contributions from $B \to \Dx l \gamma \nu$ decays reconstructed as $B \to \Dx l \nu$ are further
sources of common systematic uncertainties. These may arise at approximately the few percent level, and are
thought to be well-approximated in experimental simulations by \texttt{PHOTOS}~\cite{Barberio:1993qi}, although
Coulomb-term corrections may eventually also become
important~\cite{deBoer:2018ipi,Cali:2019nwp,Klaver:2019wgf}.

%% RD(*) results

\section{Combination and Interpretation of the Results}
\label{sec:interpretation}
The semitauonic measurements described in Sec.~\ref{sec:measurements} exhibit various levels of disagreement with the SM predictions.
In this section, we further examine these results and explore these tensions.
To briefly resummarize, the following recent measurements are currently available 
(see also Table~\ref{tab:overview} and references therein): 
\begin{enumerate}
  \item In \BDxtaunu decays
    \begin{enumerate}[(a)]
    \item Six measurements of \RDs and three of \RD. For convenience we resummarize here these results in Table~\ref{tab:summary_results}. 
    \item One measurement of the $\tau$ polarization fraction, $P_\tau(D^*)=-0.38\pm0.51^{+0.21}_{-0.16}$.
    \item One measurement of the $D^*$ longitudinal polarization fraction, $F_{L,\tau}(D^*) = 0.60 \pm 0.08 \pm 0.04$.
    \item Two measurements of the efficiency-corrected $q^2$ distributions shown in Fig.~\ref{fig:had_tag_q2.pdf}.
    \end{enumerate}    
  \item One measurement of a \bctaunu transition using $B_{c}$ decays, $\RJ = 0.71 \pm 0.17 \pm 0.18$.
  \item One measurement of a \butaunu transition, $\calR(\pi)  = 1.05 \pm 0.51$.
\end{enumerate}

\begin{table}[t]
  \renewcommand*{\arraystretch}{1.5}
  \newcolumntype{C}{ >{\centering\arraybackslash } m{2.5cm} <{}}
  \caption{Summary of $\RDx$ measurements and world averages. The hadronic-$\tau$ LHCb
    result~\cite{Aaij:2017deq} has been updated taking into account the latest HFLAV average of $\Br(B^0 \to
    D^{*+} \ell\nu) = (5.08 \pm 0.02 \pm 0.12)$\%.
    The values for ``Average ($\hat \rho_{D^{**}}$)'' are calculated by profiling the unknown
    \BDsslnu correlation and obtaining $\hat \rho_{D^{**}} = -0.88$ as described in Sec.~\ref{sec:averages}.}
  \label{tab:summary_results}
  \vspace{1ex}
\begin{threeparttable}
\scalebox{0.78}{
\parbox{1.2\linewidth}{
   \begin{tabular*}{0.62\textwidth}{@{\extracolsep{\fill}}lll CCc} \hline\hline
    Experiment\quad                           & $\tau$ decay                        & Tag    & \RD                        
                                         & \RDs                                & $\rho_{\rm tot}$     \\ \hline
    \babar\!\tnote{a}                      & $\mu\nu\nu$                         & Had.   & $0.440(58)(42)$ 
                                         & $0.332(24)(18)$         & $-0.31$              \\ \hline
    Belle\tnote{b}                       & $\mu\nu\nu$                         & Semil. & $0.307(37)(16)$ 
                                         & $0.283(18)(14)$         & $-0.52$              \\
    Belle\tnote{c}                       & $\mu\nu\nu$                         & Had.   & $0.375(64)(26)$ 
                                         & $0.293(38)(15)$         & $-0.50$              \\
    Belle\tnote{d}                       & $\pi\nu$,  $\rho\nu$                & Had.   & 
                                         & $0.270(35)^{(+28)}_{(-25)}$ & --                   \\ \hline
    LHCb\tnote{e}                        & $\pi\pi\pi(\pi^0)\nu$               & --     & --
                                         & $0.280(18)(25)(13)$         & --                   \\
    LHCb\tnote{f}                        & $\mu\nu\nu$                         & --     & -- 
                                         & $0.336(27)(30)$         & --                   \\ \hline
    \multicolumn{2}{l}{\textbf{\!Average ($\hat \rho_{D^{**}}$)}}                               &        & $\boldsymbol{0.337(30)}$ 
                                         & $\boldsymbol{0.298(14)}$      & $\boldsymbol{-0.42}$ \\
    \multicolumn{2}{l}{HFLAV Avg.\tnote{g}}                                                   &        & $0.340(30)$ 
                                         & $0.295(14) $                  & $-0.38$              \\
    \hline\hline
  \end{tabular*}
  \begin{tablenotes}[para]
  	\item [a] \cite{Lees:2012xj, Lees:2013uzd}
	\item [b] \cite{Belle:2019rba}
	\item [c] \cite{Huschle:2015rga}
	\item [d] \cite{Hirose:2017dxl}
	\item [e] \cite{Aaij:2017deq}
	\item [f] \cite{Aaij:2015yra}
	\item [g] \cite{Amhis:2019ckw}
  \end{tablenotes}
  }} %% resizebox
\end{threeparttable}
\end{table}

In Sec.~\ref{sec:analysis_rdx}, we inspect the measurements of \RDx in terms of the light-lepton normalization modes, the isospin-conjugated modes,
and their measured values as a function of time.
Thereafter we revisit in Sec.~\ref{sec:averages} the combination of the measured \RDx values. 
In particular, we discuss the role of non-trivial correlation effects on such averages and point out that with more precise measurements on the horizon 
these effects will need to be revisited.
In Sec.~\ref{sec:incl_excl.pdf} we discuss the saturation of the measured inclusive rate by exclusive contributions as implied by the current world averages of \RDs and \RD 
together with the expected \BDsstaunu rates. 
Finally, Secs.~\ref{sec:np_interpretations} and~\ref{sec:fcnc} discuss the challenges in developing self-consistent new physics interpretations of the 
observed tensions with the SM and possible connections to the present-day flavor-changing neutral-current (FCNC) anomalies, respectively.  

\subsection{Dissection of \RDx results and SM tensions}
\label{sec:analysis_rdx}

The current status of LFUV measurements versus SM predictions, and the significance of their respective tensions or agreements, is summarized in Table~\ref{tab:int:tensions},
and includes the current HFLAV combination of the $\RDx$ data. For the SM predictions the arithmetic averages discussed in Sec.~\ref{sec:theory} are quoted.  
The individual tensions of all LFUV measurements with the SM expectations range from $0.2\sigma$ to $2.5\sigma$. The combined value of $\RD$ and $\RDs$ is in tension with the SM expectation by $3.1\sigma$ 
because of their anticorrelation. 
Also note that the value of $P_\tau(D^*)$ is slightly correlated with both averages. 

\begin{table}[t]
	\renewcommand*{\arraystretch}{1.75}
	\newcolumntype{C}{ >{\centering\arraybackslash $} c <{$}}
 	\caption{Current status of LFUV measurements (see Sec.~\ref{sec:measurements}) versus SM predictions in Sec.~\ref{sec:theory},
	and their respective agreements or tensions (negative when below the predictions). 
	For $P_{\tau}(\Dstar)$ and $F_{L,\tau}(\Dstar)$ we show a na\"ive arithmetic average of the SM predictions (Table~\ref{tab:th:FLPt}) as done for $\RDx$.
	For $\RDx$ we show the world average from the HFLAV combination~\cite{Amhis:2019ckw};
	below the line we show for comparison the results of the $\RDx$ world average obtained in this work (see Sec.~\ref{sec:averages}).}  
 	\label{tab:int:tensions}
        \vspace{1ex}
    	\begin{tabular*}{0.98\linewidth}{@{\extracolsep{\fill}}l C C C C}
      		\hline\hline
     		 Obs.	&  \shortstack{\\ Current \\ World AvData} & \shortstack{\\ Current \\SM Prediction} & \multicolumn{2}{c}{Significance} \\
      		\hline
      		$\RD$ 				& 0.340\pm0.030 		& 0.299 \pm 0.003 	& 1.2\sigma & \multirow{2}{*}{$\Bigg\} 3.1\sigma$}\\
      		$\RDs$ 				& 0.295\pm0.014 		& 0.258 \pm 0.005 	& 2.5\sigma & \\
		$P_{\tau}(\Dstar)$	 	& -0.38 \pm 0.51^{+0.21}_{-0.16} 	& -0.501 \pm 0.011 & 0.2\sigma & \\
		$F_{L,\tau}(\Dstar)$	 	& 0.60 \pm 0.08 \pm 0.04 	& 0.455 \pm 0.006 	& 1.6\sigma & \\
		$\RJ$ 				& 0.71\pm0.17\pm0.18 	& 0.2582\pm0.0038  & 1.8\sigma & \\
		$\calR(\pi)$ 			& 1.05 \pm0.51 		& 0.641 \pm0.016  	& 0.8\sigma & \\
		 $\RLc$       			& 0.242 \!\pm\! 0.026 \!\pm\! 0.071		& 0.324 \pm 0.004 	& -1.1\sigma & \\                  
		\hline
		$\RD$ 				& \bm{0.337\pm0.030} 	& 0.299 \pm 0.003 	& 1.3\sigma & \multirow{2}{*}{$\Bigg\} \bm{3.6\sigma}$}\\
      		$\RDs$ 				& \bm{0.298\pm0.014} 	& 0.258 \pm 0.005 	& 2.5\sigma & \\
		\hline\hline
    	\end{tabular*} 
\end{table}

\begin{figure}[t]
  \includegraphics[width=0.47\textwidth]{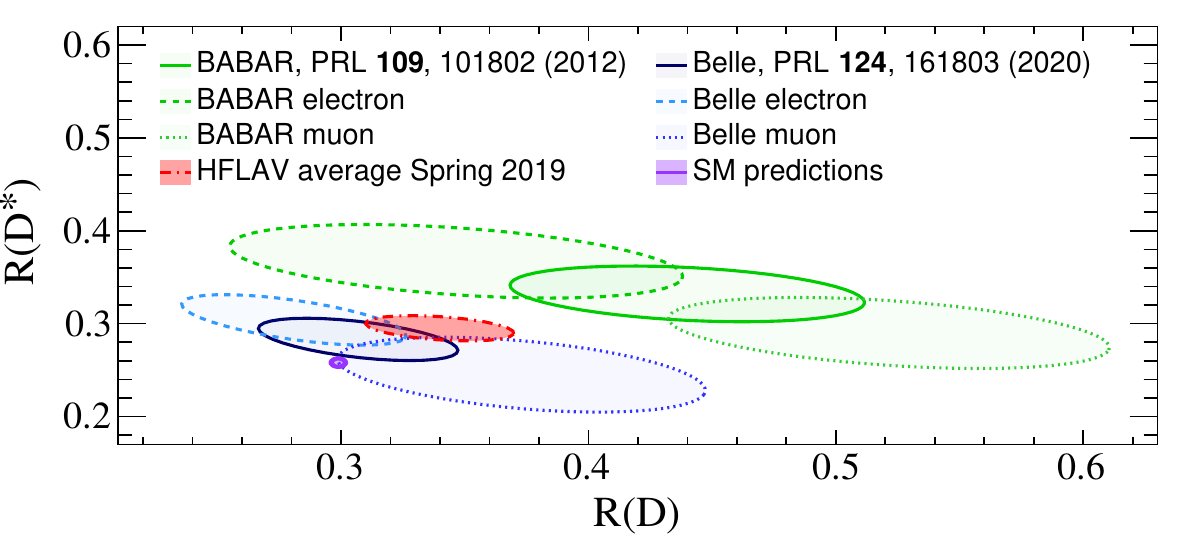}
  \caption{Measurements of \RDx, $\calR(D^{(*)})_e$, and $\calR(D^{(*)})_\mu$ from
    \babar~\cite{FrancoSevilla:2012thesis} and
    Belle~\cite{Caria:2019thesis}. }
\label{fig:rdx_emu.pdf}
\end{figure}

A subset of the existing measurements provide values of \RDx normalized to either electron or muon final states. 
These results present an important check because the values reported for the semitauonic ratios are typically an average 
for the electron and muon normalizations, assuming 
\begin{equation}
  \RDx = \calR(D^{(*)})_e = \calR(D^{(*)})_\mu,
\end{equation}
with
\begin{align}
	\RDx_e &\equiv \frac{\BR\left(\Bbar\rightarrow \Dx \tau^-\nutb \right)}{\BR\left(\Bbar\rightarrow \Dx e^-\nueb\right)} \, , \\
	\RDx_\mu &\equiv \frac{\BR\left(\Bbar\rightarrow \Dx \tau^-\nutb \right)}{\BR\left(\Bbar\rightarrow \Dx \mu^-\numb\right)}.
\end{align}
LHCb measures only $\calR(D^{(*)})_\mu$, but the $B$-factories have access to the electron normalization as
well.  Figure~\ref{fig:rdx_emu.pdf} compares $\RDx_e$ and $\RDx_\mu$ and no systematic deviation between the two ratios is observed.
Note that these results were released as stability checks that compare the compatibility of the electron and
muon channels, not as optimized measurements of $\calR(D^{(*)})_{e/\mu}$. For instance, \cite{FrancoSevilla:2012thesis}
did not include the full systematic uncertainties and correlation for the electron and muon \RDx, so the
values from the full \RDx results are used in Fig.~\ref{fig:rdx_emu.pdf}, increasing the correlation to account for the larger statistical
uncertainty of the $\calR(D^{(*)})_e$ and $\calR(D^{(*)})_\mu$ results. Additionally, the double ratio
\begin{equation}
  \RDx_\text{light} = \frac{\calR(D^{(*)})_\mu}{\calR(D^{(*)})_e }
  = \frac{\BR\left(\Bbar\rightarrow D^* e^-\nueb \right)}{\BR\left(\Bbar\rightarrow D^* \mu^-\numb\right)}\,
\end{equation}
that would be obtained from dividing these results would have unnecessarily large uncertainties because the
common $\BR\left(\BDxtaunu \right)$ factor is obtained with \tauenu decays
in the case of $\calR(D^{(*)})_e$ and \taumunu decays for $\calR(D^{(*)})_\mu$. A high-precision
measurement of $\RDx_\text{light}$ was recently released by the Belle collaboration~\cite{Waheed:2018djm}
\begin{equation}
	\label{eqn:Remu}
 	 \RDx_\text{light} = 1.01\pm0.01\pm0.03
\end{equation}
and is compatible with unity.

%% RD(*) measurements for electron and muon

%% Isospin-unconstrained RD(*) results
\begin{table}[t] 
 \renewcommand*{\arraystretch}{1.2}
  \caption{Results of the isospin-unconstrained fits for the \babar analysis~\cite{Lees:2012xj, Lees:2013uzd}. The first uncertainty is
    statistical and the second systematic.}
  \label{tab:iso_unconstrained} \vspace{0.1in}
  \begin{tabular*}{0.75\linewidth}{@{\extracolsep{\fill}} l c} \hline\hline
    Result & \babar                      \\ \hline
    \RDz   & $0.429 \pm 0.082 \pm 0.052$ \\
    \RDp   & $0.469 \pm 0.084 \pm 0.053$ \\
    \RDsz  & $0.322 \pm 0.032 \pm 0.022$ \\
    \RDsp  & $0.355 \pm 0.039 \pm 0.021$ \\ 
    \hline\hline
  \end{tabular*}
\end{table}

Table~\ref{tab:iso_unconstrained} shows the results of the isospin-unconstrained fits of the \babar \RDx analysis, 
exhibiting good compatibility between charged and neutral $D$ and $D^*$ modes. 
Such measurements might be particularly interesting in the context of obtaining data-driven insight into
the size of semiclassical radiative corrections, expected to enter at the subpercent level.

\begin{figure}[t]
  \includegraphics[width=0.47\textwidth]{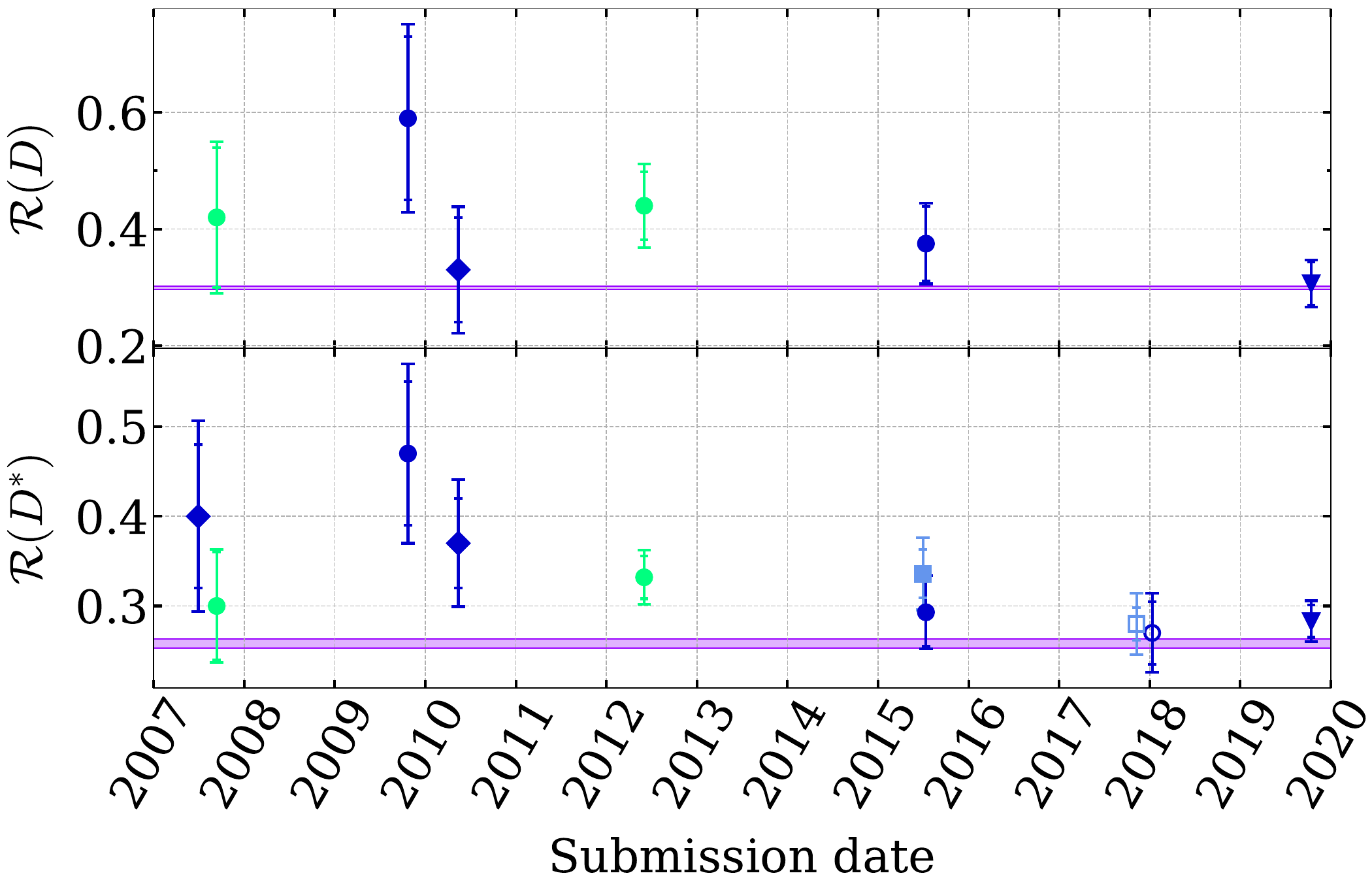} \caption{Measurements of \RDx as a function of
  paper submission time. Green (light gray) refers to \babar, dark blue (dark gray) to Belle, light blue (medium gray) to LHCb, and violet to the SM
  predictions. Circular markers refer to hadronic tagging, triangles to semileptonic tagging, diamonds to
  inclusive tagging, and squares to untagged measurements. Filled markers refer to measurements using muonic
  decays of the $\tau$ lepton while hollow to hadronic decays. Some of the earlier results measured
  $\Br(B\to\Dx\tau\nu)$ instead of \RDx. In those cases, the values for \RDx were obtained by normalizing the
  $\tau$ branching fraction with the latest world averages for $\Br(B\to\Dx\ell\nu)$~\cite{Zyla:2020zbs}.}
\label{fig:rdx_time.pdf}
\end{figure}

Another interesting comparison is to examine the measurements of \RDx as a function of time: more precise knowledge of normalization and 
background processes can lead to shifts in the central values. Figure~\ref{fig:rdx_time.pdf} displays the measured value as a function of paper submission time,
illustrating the improving precision with time.
Notably, the most recent measurements tend to display better agreement with the SM expectations. It is not clear, however, whether this is a systematic
shift or a statistical fluctuation as there have not been meaningful changes in
the procedures that determine the background, normalization, and signal components.
Note also that all measurements are compatible among themselves, with a $\chi^2$ probability of 27\%.

%% RD(*) measurements as a function of time

\subsection{Revisiting of \RDx world averages via $D^{**}$ correlations}
\label{sec:averages}

\begin{figure*}[tb]
		\includegraphics[width=0.47\textwidth]{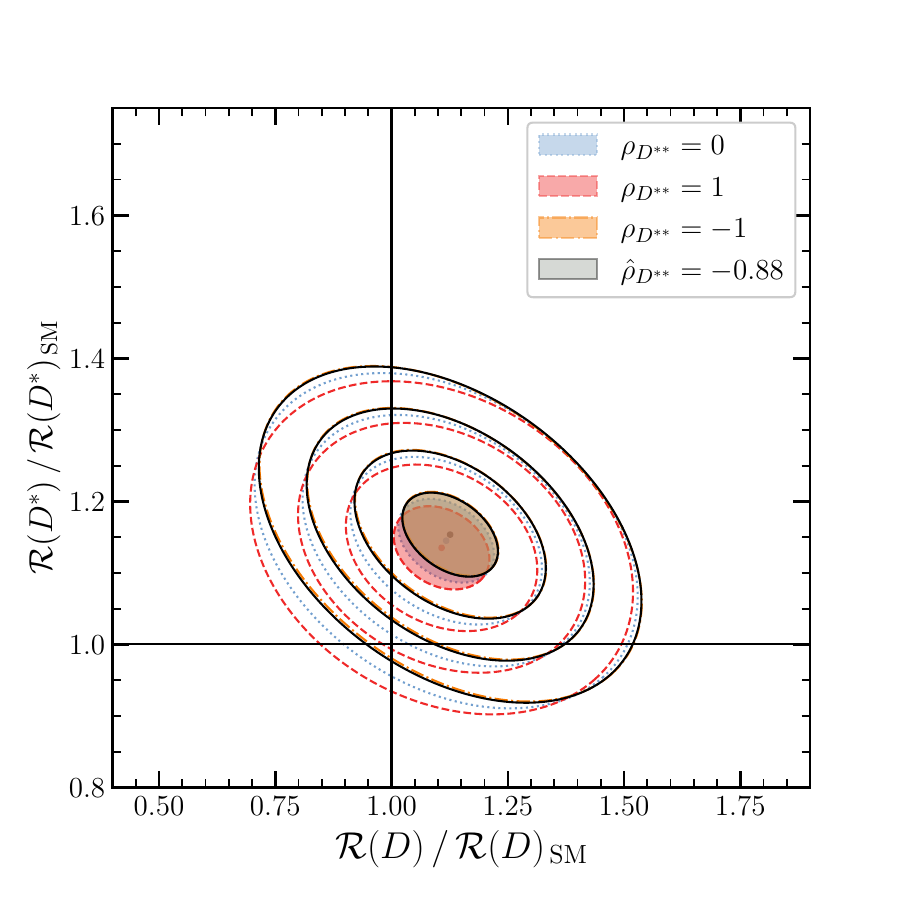}
		\includegraphics[width=0.47\textwidth]{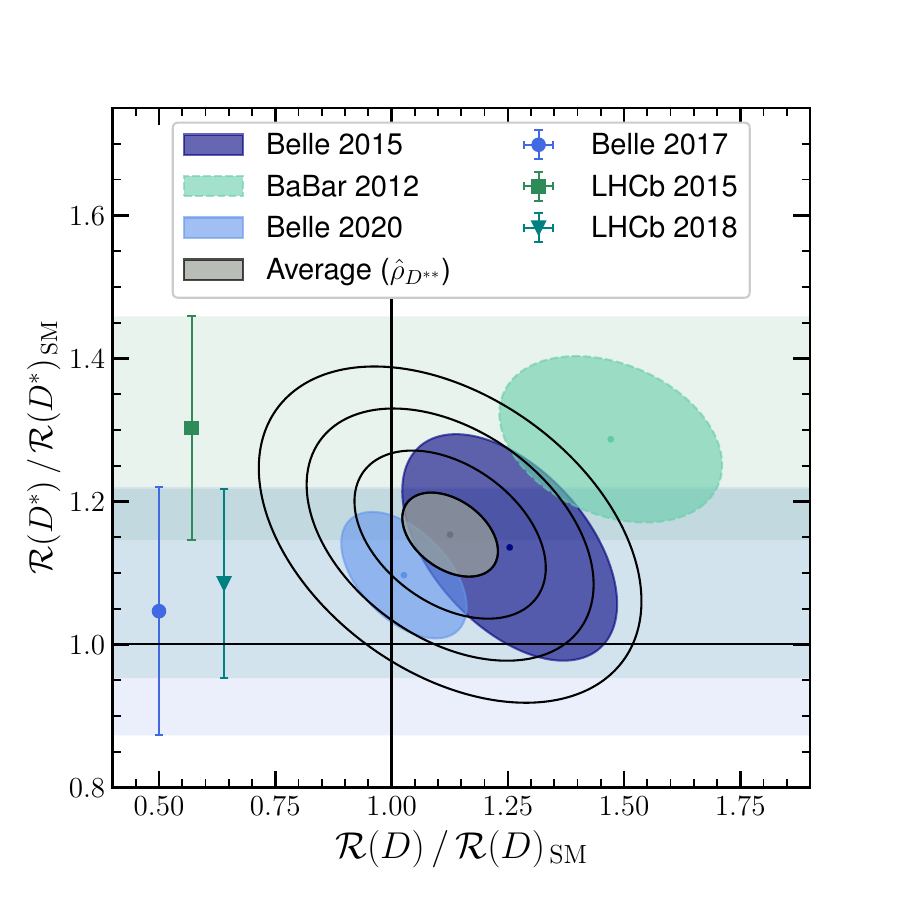}
  \caption{Left panel: \RDx world averages with different assumptions for the unknown correlation $\rho_{D^{**}}$:  
  The average with $\rho_{D^{**}} = 0$ (light blue or light gray dotted curves) is based on similar assumptions as \cite{Amhis:2019ckw} and shows a compatibility with the SM expectation of 3.2 standard deviations
  taking into account the small uncertainties of the theoretical predictions;
   the scenarios $\rho_{D^{**}} = 1$ (red or medium gray dashed) and $\rho_{D^{**}} = - 1$ (orange or light gray dash-dotted curves) agree with the SM expectation within $2.9$ and $3.7$ standard deviations, respectively. 
   In our quoted average we profile the unknown correlation and obtain $\hat \rho_{D^{**}} = -0.88$ (heather gray solid curves) 
   with a compatibility with the SM of 3.6 standard deviations.  
   Right panel: our world average of $\mathcal{R}(D)$ and $\mathcal{R}(D^*)$ (black solid curves), compared to the various measurements of \RDx.  
  The unknown correlation $\rho_{D^{**}}$ is treated as a free, but constrained, parameter of the average (see the main text for more details). 
   }
\label{fig:averages_2}
\end{figure*}

To further investigate the tension of the measured values of \RDx with the SM, we examine and update their averages. 
We note that the systematic uncertainties of all measurements have significant  
correlations (see Sec.~\ref{sec:systematics}) that need to be properly taken into account.
The most important ones stem from the modeling of the \BDsslnu processes, which 
comprise a significant background source in all measurements to date. The manner in which the uncertainties of these background contributions
are estimated varies considerably. As discussed in Sec.~\ref{sec:syst:dss_ff}, the normalization or shape uncertainties from the hadronic
form factors are, in some measurements, validated or constrained by control regions. Thus, a simple correlation model will not be able to properly 
quantify such correlations. 

One particularly important point here is the treatment of the correlations of these systematics between the \RDs and \RD measurements. 
In individual measurements that measure both quantities simultaneously, this treatment is straightforward. 
However, it becomes unclear how to relate systematic uncertainties between e.g. \RD and \RDs in two separate measurements. 
To provide a concrete example, consider the \babar measurement of \RD (in the context of the 
combined \RDx determination of \cite{Lees:2012xj, Lees:2013uzd}) and the Belle measurement of \RDs (in the combined \RDx analysis of \cite{Huschle:2015rga}). 
In the individual measurements, the systematic uncertainty associated with $B \to D^{**} \ell \bar \nu_\ell$ is $45\%$ and $-15\%$ correlated between \RD and \RDs, respectively\footnote{Both measurements provide the systematic uncertainties associated with $D^{**}$ in a different granularity. The quoted correlations are obtained by summing for  \cite{Lees:2012xj, Lees:2013uzd} the resulting covariance matrices for the $D^{**}$ form factor and the various branching fraction uncertainties. For \cite{Huschle:2015rga} the covariance for the  $B \to D^{**} \ell \bar \nu_\ell$ shape and the $D^{**}$ are summed. }. 
From this information alone it is impossible to derive the correct correlation structure between \RD and \RDs across the measurements. 

We further investigate the dependence of the world average on the $B \to D^{**} \ell \bar \nu_\ell$ correlation structure across the \RD and \RDs measurements by parametrizing them with a single factor $\rho_{D^{**}}$. 
In Fig.~\ref{fig:averages_2} (left panel) we show the world average assuming that such correlation effects are negligible (labeled as $\rho_{D^{**}} = 0$) 
and we reproduce a world average very similar to HFLAV~\cite{Amhis:2019ckw}. 
The numerical values, normalized to the arithmetic average of the SM predictions (cf. Table~\ref{tab:th:RDDs} in Sec.~\ref{subsec:SM_pred}), are
\begin{align}
  \RD / \RD_{\mathrm{SM}} & = 1.12 \pm 0.10 \, , \\
  \RDs / \RDs_{\mathrm{SM}} & = 1.15 \pm 0.06 \, ,
\end{align} 
with an overall correlation of $\rho = -0.33$. In addition to the $B \to D^{**} \ell \bar \nu_\ell$ uncertainties, 
the uncertainties in the leptonic $\tau$ branching fractions and the \BDxlnu FFs are fully correlated across measurements. 
The compatibility with the SM expectation is within $3.2$ standard deviations (close to the value quoted by \cite{Amhis:2019ckw} of $3.1\sigma$). 
Figure~\ref{fig:averages_2} (left panel) also shows the impact of setting this unknown correlation to either $\rho_{D^{**}} = 1$ or $\rho_{D^{**}} = -1$, 
resulting in compatibilities with the SM predictions of $2.9$ or $3.7$ standard deviations, respectively. 

A possible way to deal with an unknown parameter such as $\rho_{D^{**}}$ in this type of problem was outlined in~\cite{Cowan:2018lhq}.
Instead of neglecting the value, we can incorporate it as a free parameter of the problem and constrain it within its probable range. 
A possible choice that limits this missing correlation 
to fall between $[-1,1]$ is to assign it a double Fermi-Dirac distribution\footnote{$f(x,w) = 1 / \left( 2 ( 1 + \exp(w(x-1)) )( 1 + \exp(-w(x-1)) \right) $}
with a large shape parameter, e.g. $w = 50$. Carrying out our average with such a setup results in
\begin{align}
  \RD / \RD_{\mathrm{SM}} & = 1.13 \pm 0.10 \, , \\
  \RDs / \RDs_{\mathrm{SM}} & = 1.15 \pm 0.06 \, ,
\end{align} 
with $\hat \rho_{D^{**}} = -0.88$ and an overall correlation of $\rho = -0.40$.  
This results in an increased tension of about $3.6\sigma$ with respect to the SM. 

Although neither of these world averages are based on completely correct assumptions, 
they illustrate the need for future \RDx measurements to provide more detailed breakdowns of their uncertainties. 
It is intriguing that introducing an additional correlation structure of a systematic uncertainty can shift the agreement with the SM expectation over a range of $0.8$ standard deviations. 
Table~\ref{tab:summary_results} lists the numerical values of this average---denoted as ``Average ($\hat \rho_{D^{**}}$)''---and the HFLAV average~\cite{Amhis:2019ckw}; 
see also Table~\ref{tab:int:tensions}.
We show this world average for $\RDx$ compared to the various measurements in Fig~\ref{fig:averages_2} (right panel).

\subsection{Exclusive saturation of the inclusive rate}
\label{sec:incl_excl.pdf}

\begin{figure}  
 \centering
  \includegraphics[width=0.95\linewidth]{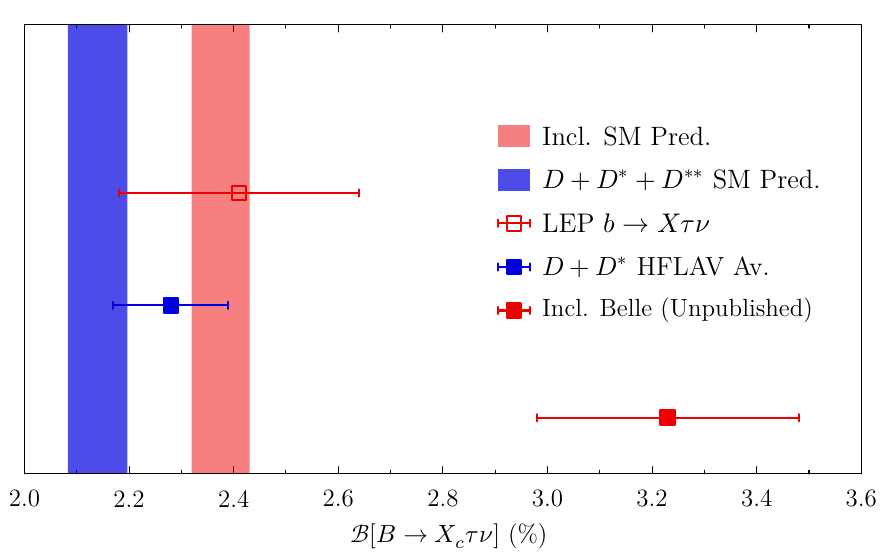} 
  \caption{Saturation of the inclusive SM prediction (red or medium gray band) for $\BR(B \to X_c \tau \nu)$ 
  by the sum of the measured exclusive branching fractions that are implied by the $\RD$ and $\RDx$ world averages (blue or dark gray square).
  By comparison, the SM prediction for the sum of $B \to D^{(*,**)} \tau \nu$ exclusive branching fractions (blue or dark gray band), is compatible with, and does not saturate, the inclusive prediction.
  Also shown are (i) the measured inclusive branching fraction measurements for $b \to X \tau \nu$ from LEP~\cite{Zyla:2020zbs} (open square),
  which is normalized against the total number of tagged $b\bbar$ events.
  Assuming that hadronization effects cancel, it can be interpreted as $\BR(B \to X \tau \nu)$; and (ii)
  the unpublished inclusive measurement of~\cite{handle:20.500.11811/7578} using Belle data (red or medium gray filled square), which shows a large excess.
  }
  \label{fig:saturation}
\end{figure}

The SM prediction for the semitauonic inclusive branching ratio is
\begin{equation}
	\label{eqn:incl_excl:inclSM}
	\BR(B \to X_c \tau\nu) = 2.37(6)\times 10^{-2}\,,
\end{equation}
obtained by combining the SM prediction in Eq.~\eqref{eqn:th:Rincl} with the data for the flavor-averaged light-lepton branching ratio $\BR(B \to X_c \ell \nu)$~\cite{Zyla:2020zbs}.
This value of the inclusive branching fraction should correspond to the sum of branching fractions of all possible exclusive final states, i.e. the sum of decay rates of exclusive states should saturate the inclusive rate. The degree of this saturation can 
be explored by comparing the inclusive branching ratio to that for the sum of $\Dx$ and $D^{**}$.
For simplicity, in the following we treat the uncertainties for each mode as independent.
Using the HFLAV-averaged SM prediction for $\RDx$ (Table~\ref{tab:th:RDDs})
together with the average branching ratio for $\BR(B^0 \to \Dx \ell \nu)$ and  $\BR(B^- \to \Dx \ell \nu)$, one finds
\begin{subequations}
\begin{align}
	\BR(B \to D \tau\nu) & = 0.72(4)\times 10^{-2}\,, \\
	 \BR(B \to D^* \tau\nu) & =1.28(4) \times 10^{-2}\,, 
\end{align}
\end{subequations}
and similarly one may use the combined $D^{**}$ SM prediction in Eq.~\eqref{eqn:th:RbarDss} with world averages for $\BR(B^- \to D^{**} \ell \nu)$~\cite{Bernlochner:2016bci}, 
yielding
\begin{equation}
	\sum_{X_c \in D^{**}} \BR(B \to X_c \tau\nu) =  0.14(2) \times 10^{-2}\,.
\end{equation}
Adding these contributions, one obtains the SM prediction $\sum_{X_c \in D^{(*,**)}} \BR(B \to X_c \tau\nu) = 2.14(6)\times 10^{-2}$, 
which is compatible with, and does not saturate, the inclusive SM prediction in Eq.~\eqref{eqn:incl_excl:inclSM}, as shown in Fig.~\ref{fig:saturation}.

One can characterize the degree of LFUV in the semitauonic system by comparing the inclusive SM prediction with the sum of measured branching ratios for $\BR(B \to \Dx \tau \nu)$.
In this case the SM prediction in Eq.~\eqref{eqn:incl_excl:inclSM} arises from theory inputs, and features theory uncertainties, 
that are independent of the inputs used for predictions of $\RDx$ (see Sec.~\ref{sec:th:incl}).
Figure~\ref{fig:saturation} compares the inclusive SM prediction to the sum of the \BDxtaunu branching fractions arising from the $\RDx$ world averages,
as well as to the measured inclusive $b \to X \tau \nu$ branching fraction from LEP~\cite{Zyla:2020zbs}, 
and the result for $B \to X \tau \nu$ from the PhD thesis \cite{handle:20.500.11811/7578} using Belle data.
One sees that the $\RDx$ world averages already imply near saturation of the inclusive SM prediction, 
while the unpublished result from the Belle data is more than $3\sigma$ in tension with it.

\subsection{New physics interpretations}
\label{sec:np_interpretations}
\begin{figure*}[t]
	\includegraphics[width = 0.245\textwidth]{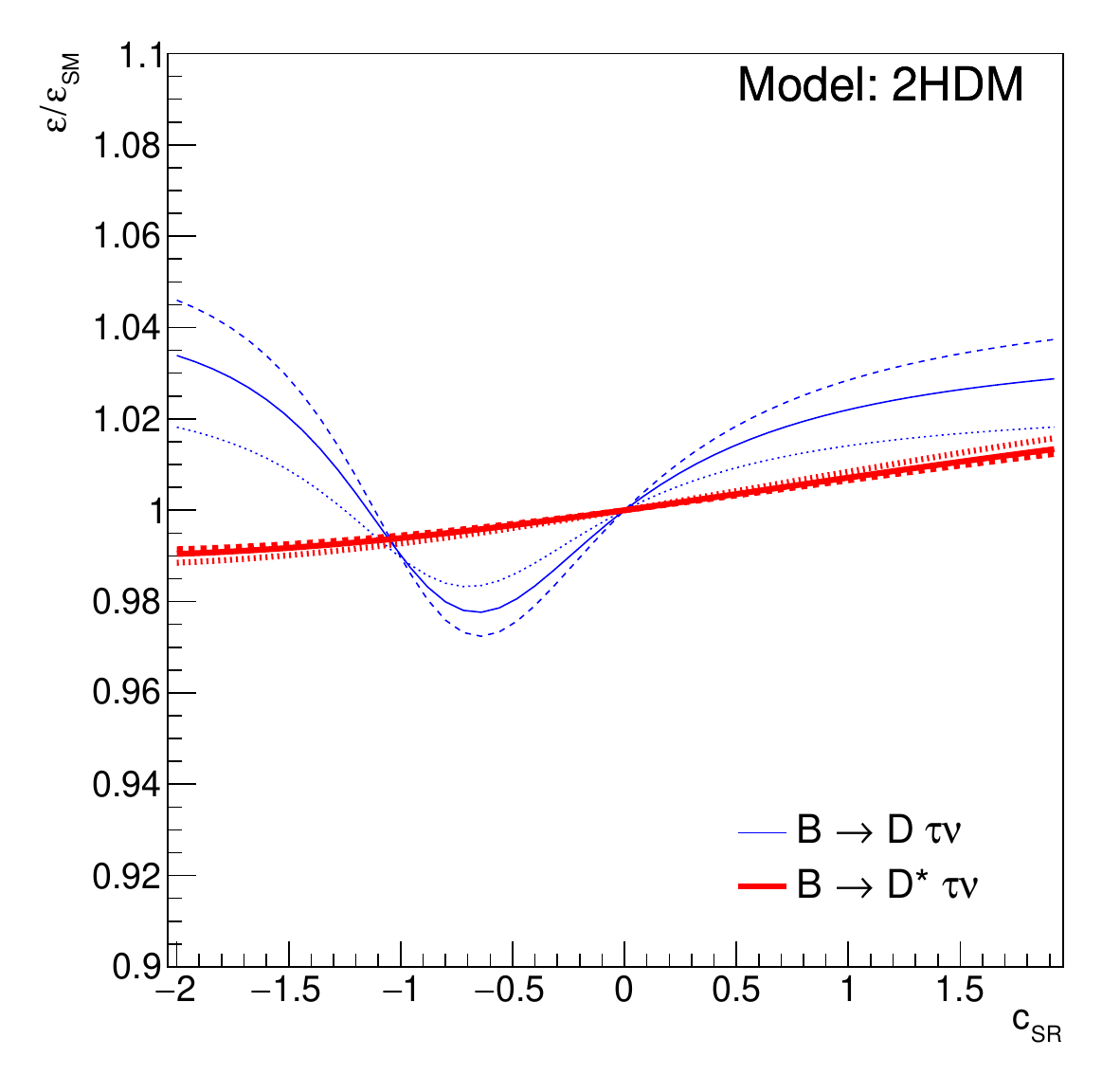}
	\includegraphics[width = 0.245\textwidth]{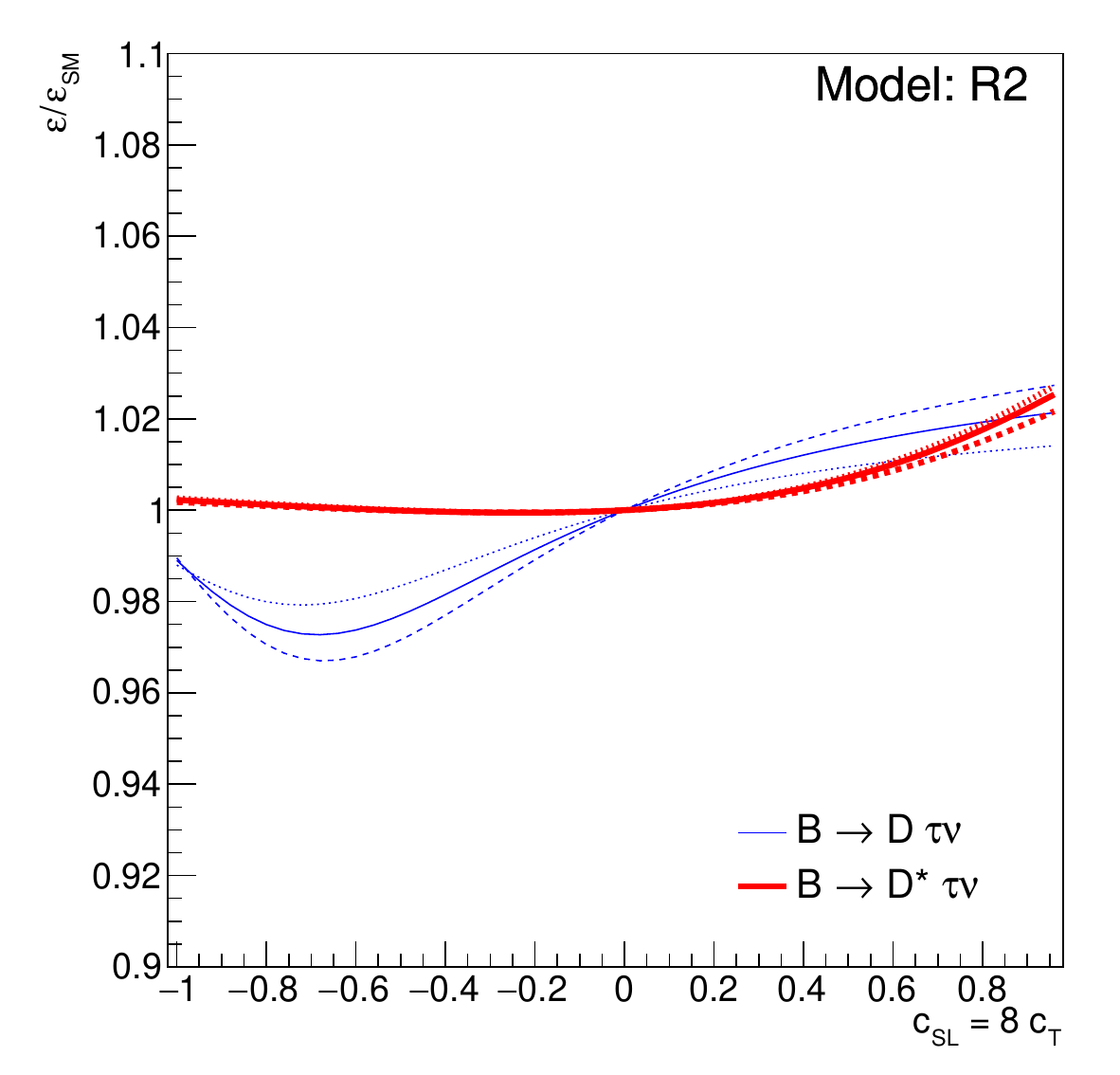}
	\includegraphics[width = 0.245\textwidth]{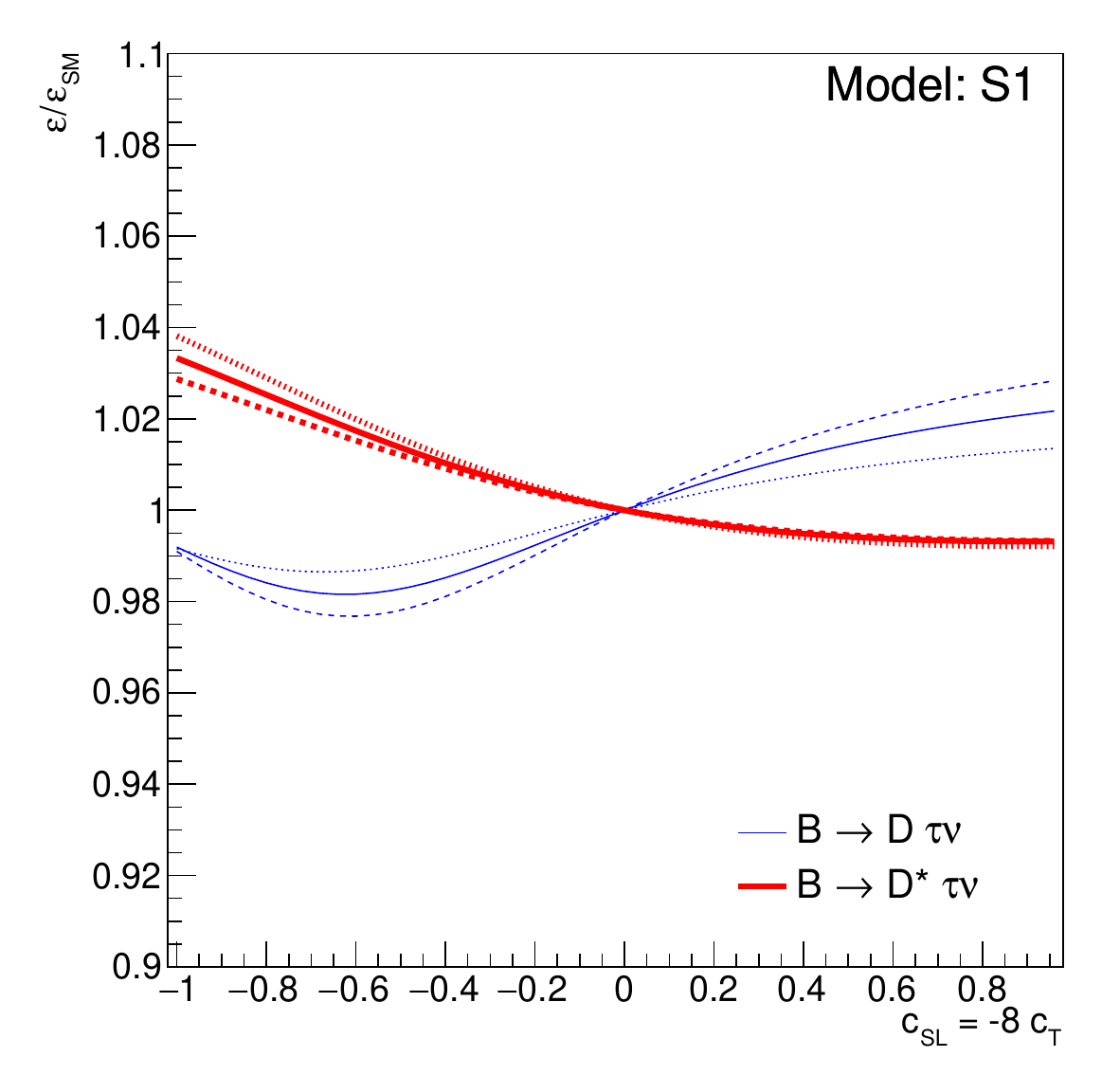}
	\includegraphics[width = 0.245\textwidth]{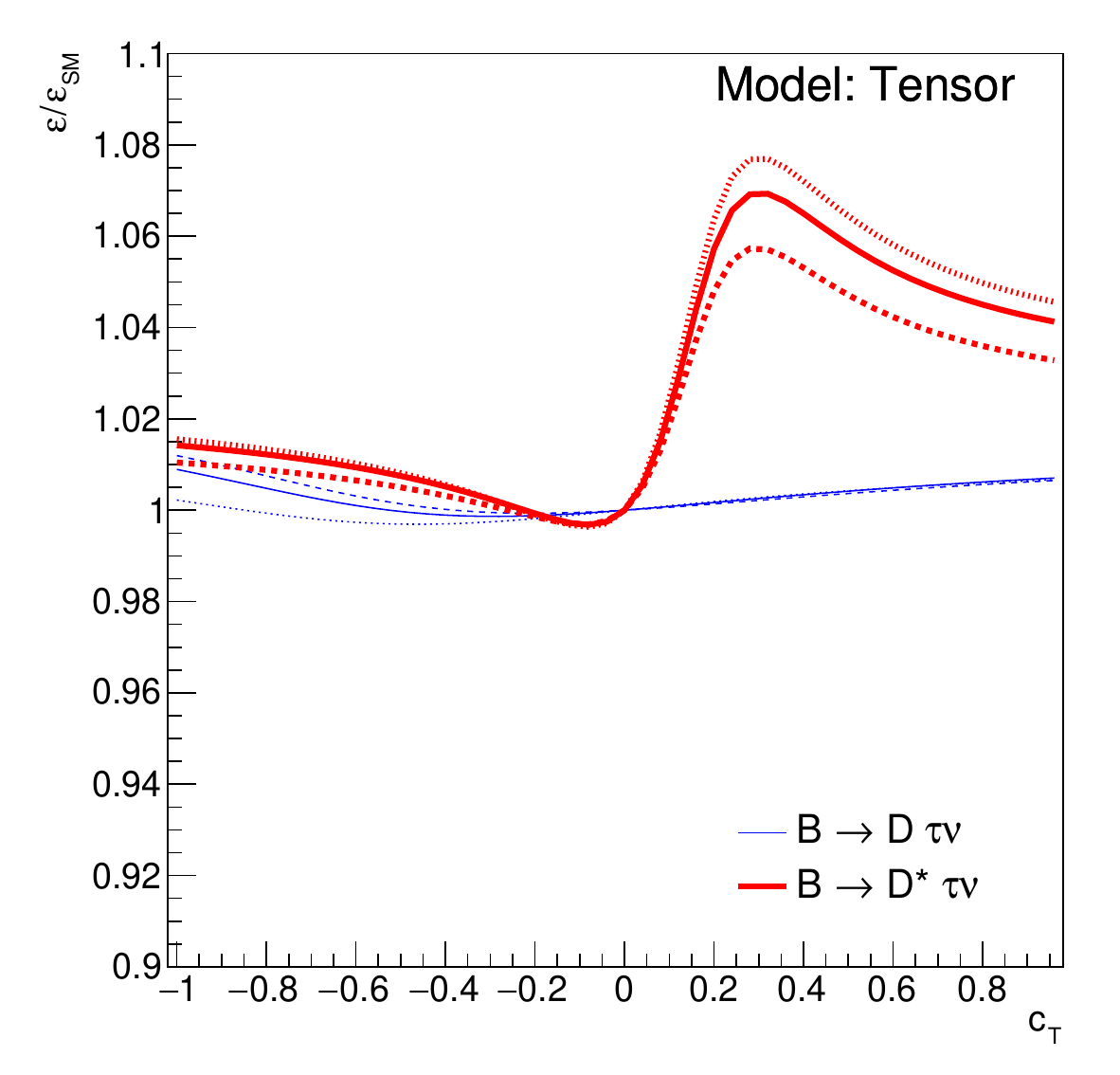}\\
	\includegraphics[width = 0.245\textwidth]{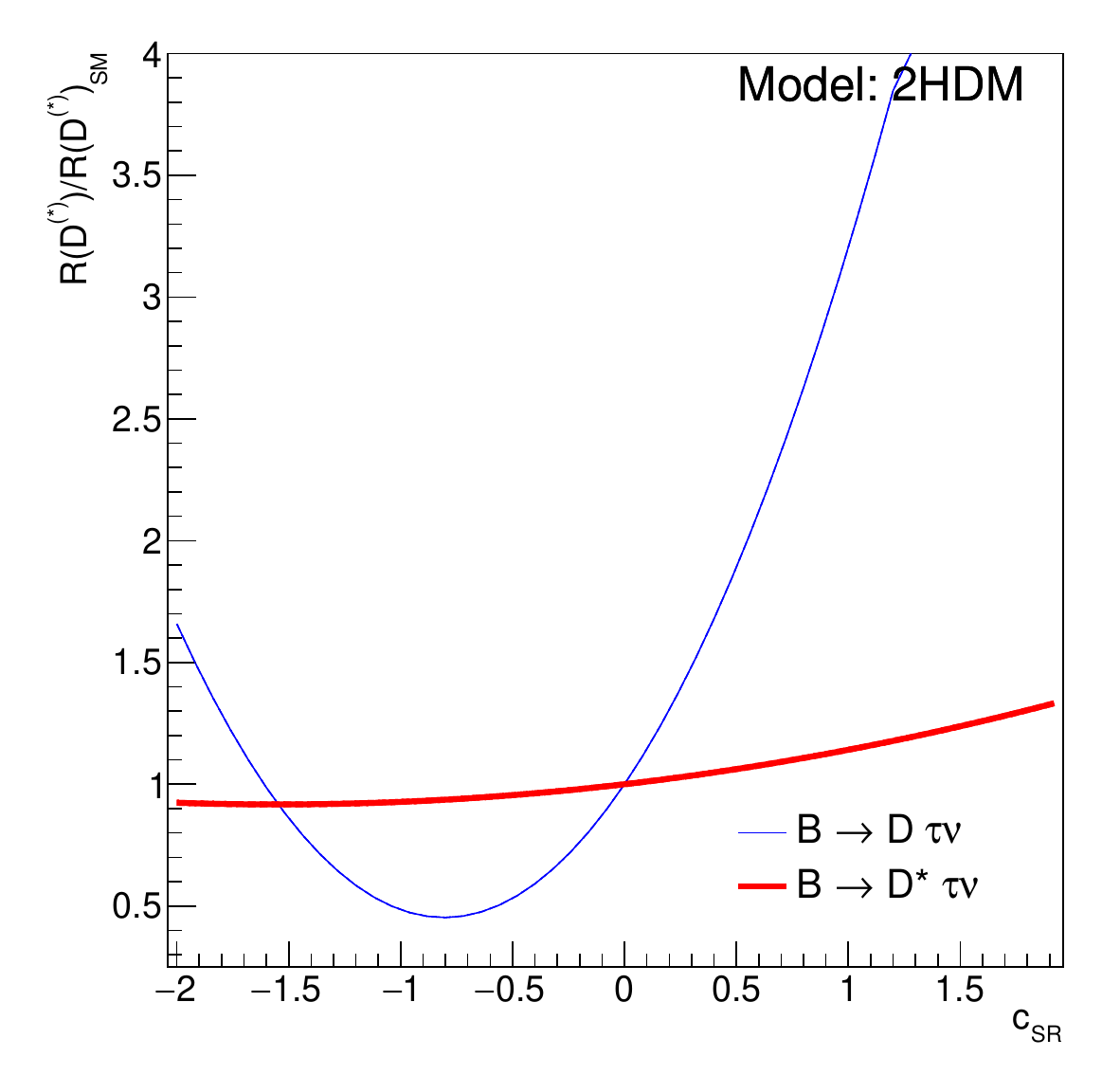}
	\includegraphics[width = 0.245\textwidth]{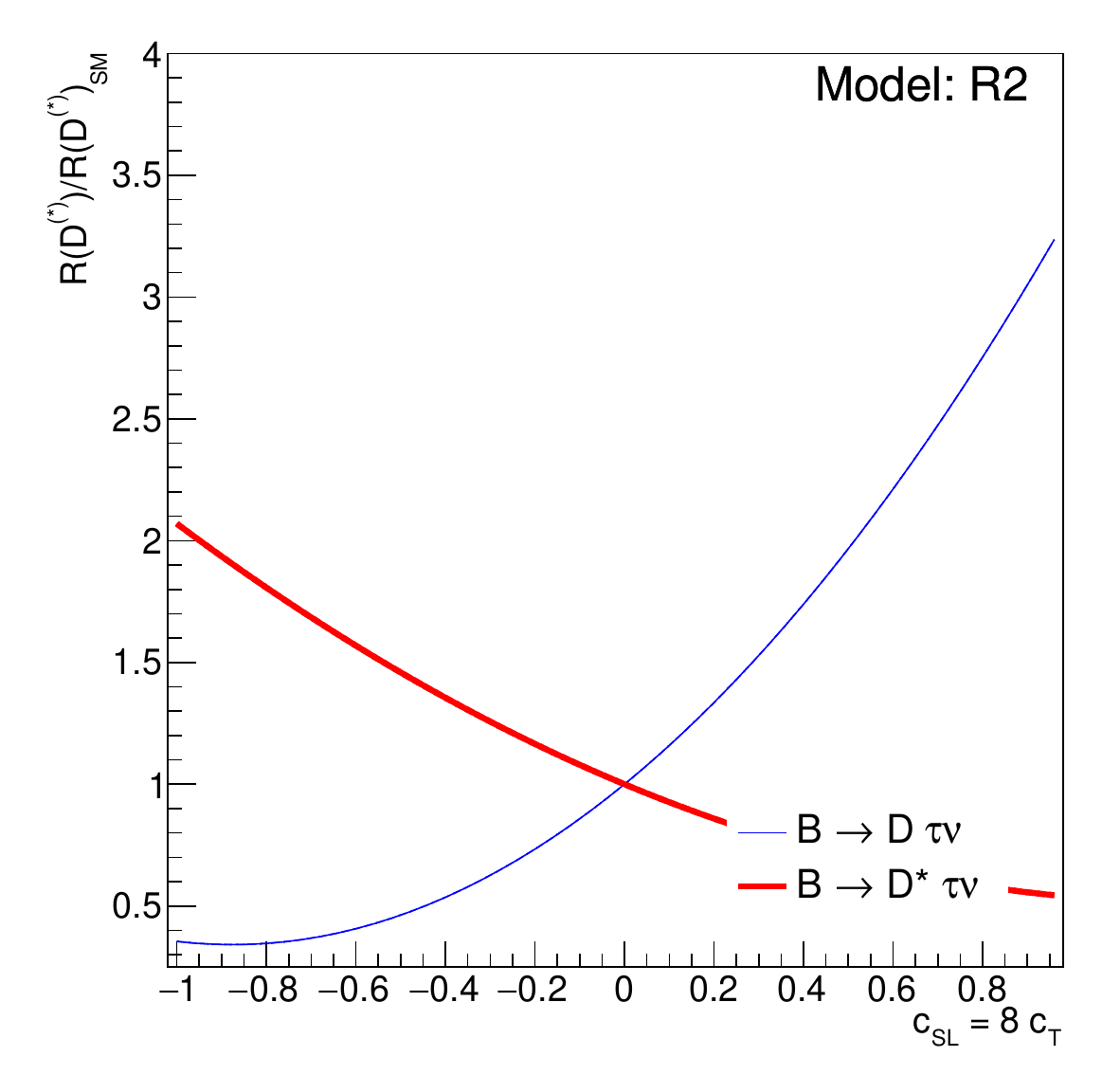}
	\includegraphics[width = 0.245\textwidth]{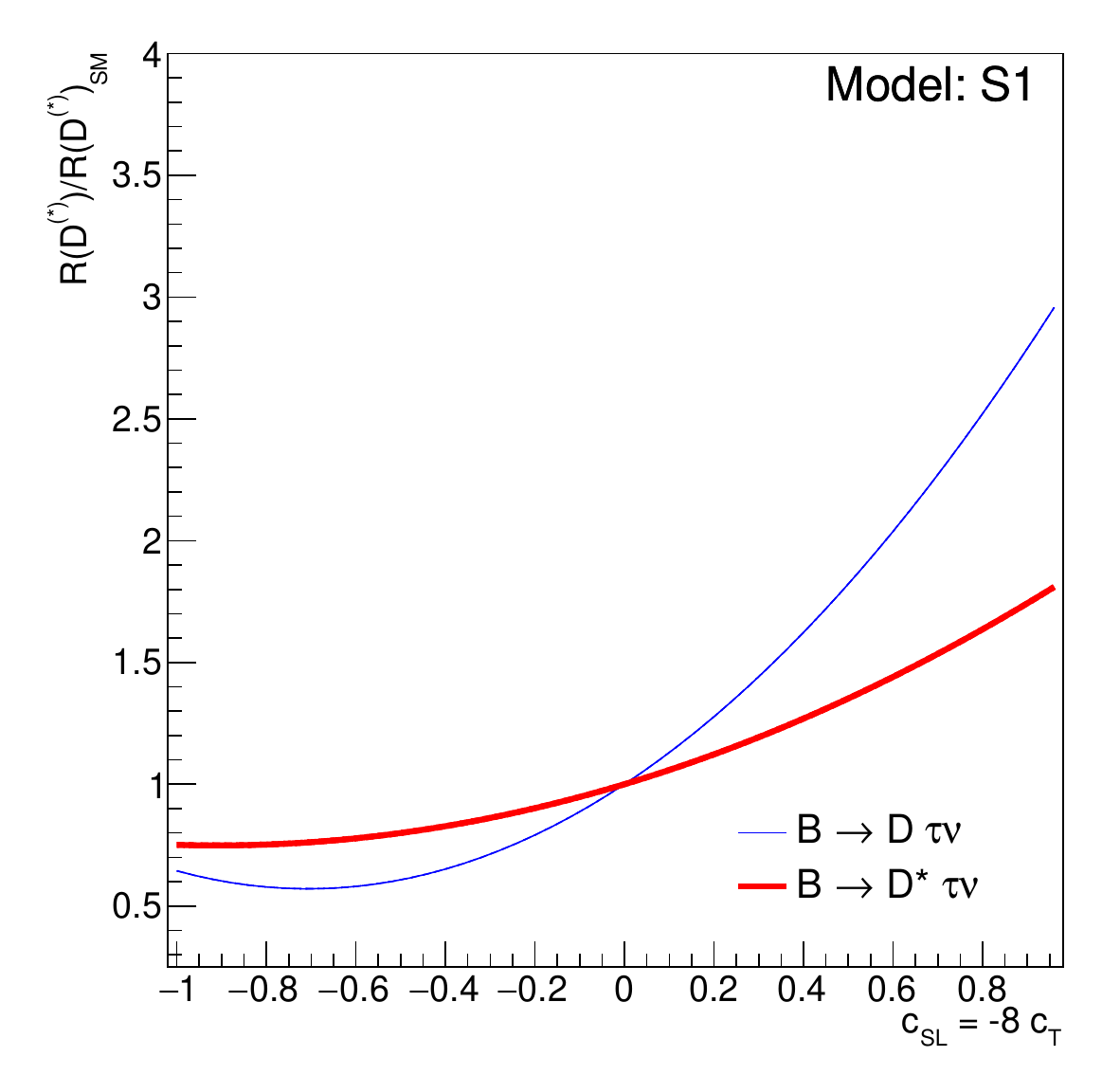}
	\includegraphics[width = 0.245\textwidth]{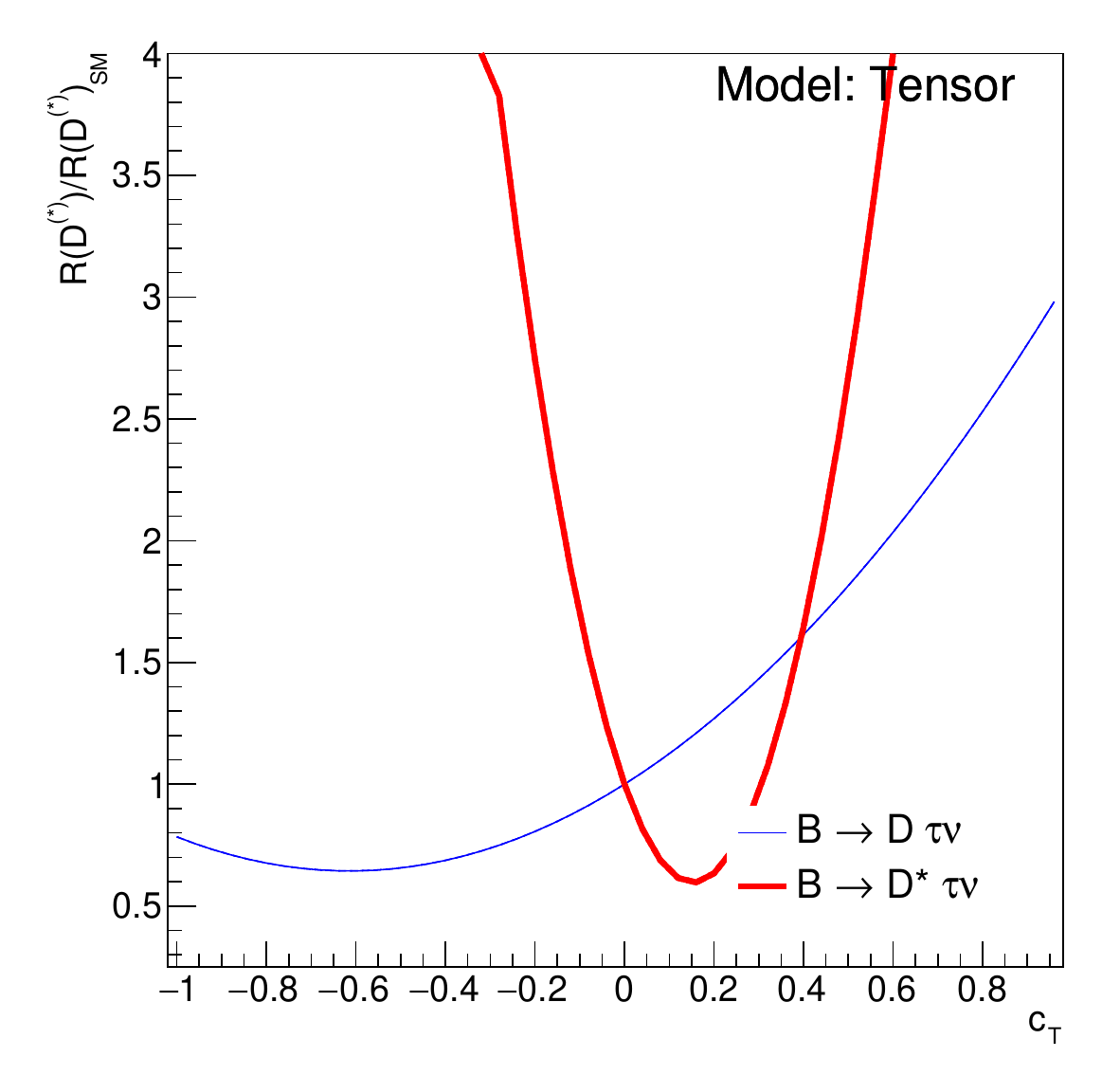}
	\caption{Top row: typical variation of experimental acceptances for the 2HDM, the leptoquark models $R_2$ and $S_1$, and a pure tensor current, 
	normalized with respect to the SM acceptance $\varepsilon_{\text{SM}}$, for $B \to D \tau \nu$ (thin blue lines) and $B \to (D^* \to D\pi) \tau \nu$ (thick red lines), with $\tau \to e \nu\nu$.
	The dotted, solid and dashed lines show the resulting acceptances for $q^2$ resolutions (see text) of $0.8$, $1.2$ and $1.6$\,GeV$^2$, respectively.
	Bottom row: variation in $\RDx/{\RDx}_{\text{SM}}$ for the same models.
	}
	\label{fig:NPint:acc}
\end{figure*}

\subsubsection{Parametrization of SM tensions}
The measured lepton universality ratios $\RDx$ naively express tensions with respect to SM predictions in terms of the overall decay rates or branching ratios.
As such, typically many phenomenological interpretations of these results simply require that any New Physics (NP) accounts for the measured ratios 
(or other observables such as polarization fractions) within quoted uncertainties. 
However, this naive approach may lead to biases in NP interpretations.

The reason for this is that in practice, as discussed in Sec.~\ref{sec:measurements}, the $\RDx$ ratios are recovered from fits in multiple reconstructed observables. 
In these fits, the signal \BDxtaunu decay distributions (as well as backgrounds) are assumed to have SM shapes---their reconstructed observables 
are assumed to have an \emph{SM template}---while their normalization is allowed to float independently. 
In the SM, the ratio of $\RD$ to $\RDs$ is itself tightly predicted up to small form factor uncertainties.
Thus, the current experimental approach can be thought of introducing a \emph{NP fit template}, 
that is parametrized by variation in the double ratio $\RD/\RDs$ as well as, say, the overall size of $\RDs$.

Variation of $\RDs$, while keeping $\RD/\RDs$ fixed to its SM prediction, is consistent with NP contributions from the $c_{\VL}$ Wilson coefficient. 
This Wilson coefficient by definition still generates SM-like distributions,
so incorporating $c_{\VL}$ contributions is self-consistent with the fit template assumptions from which the measured $\RDx$ values were recovered.

However, to explain the variation in $\RD/\RDs$ from the SM prediction requires further NP contributions that 
generically also alter the $B \to \Dx \tau \nu$ signal (and some background) decay distributions and acceptances. 
(It is possible that there exist NP contributions that modify only the neutrino distributions. Because the experiments marginalize over missing energy, 
this particular NP could permit $\RD/\RDs$ to simultaneously float from the SM prediction while preserving the SM template for reconstructed observables.)
These NP contributions are thus generically inconsistent with the assumed SM template in the current measurement and fit, 
and may affect the recovered values of $\RDx$ themselves. 
As a result, while the current world average for $\RD$--$\RDs$ unambiguously indicates a tension with the SM, 
it does not \emph{a priori} allow for a self-consistent NP interpretation or explanation.
A self-consistent BSM measurement of any recovered observable instead requires e.g. dedicated fit templates for each BSM point of interest, which we discuss further below.

\begin{figure*}[t]
	\includegraphics[width = 0.245\textwidth]{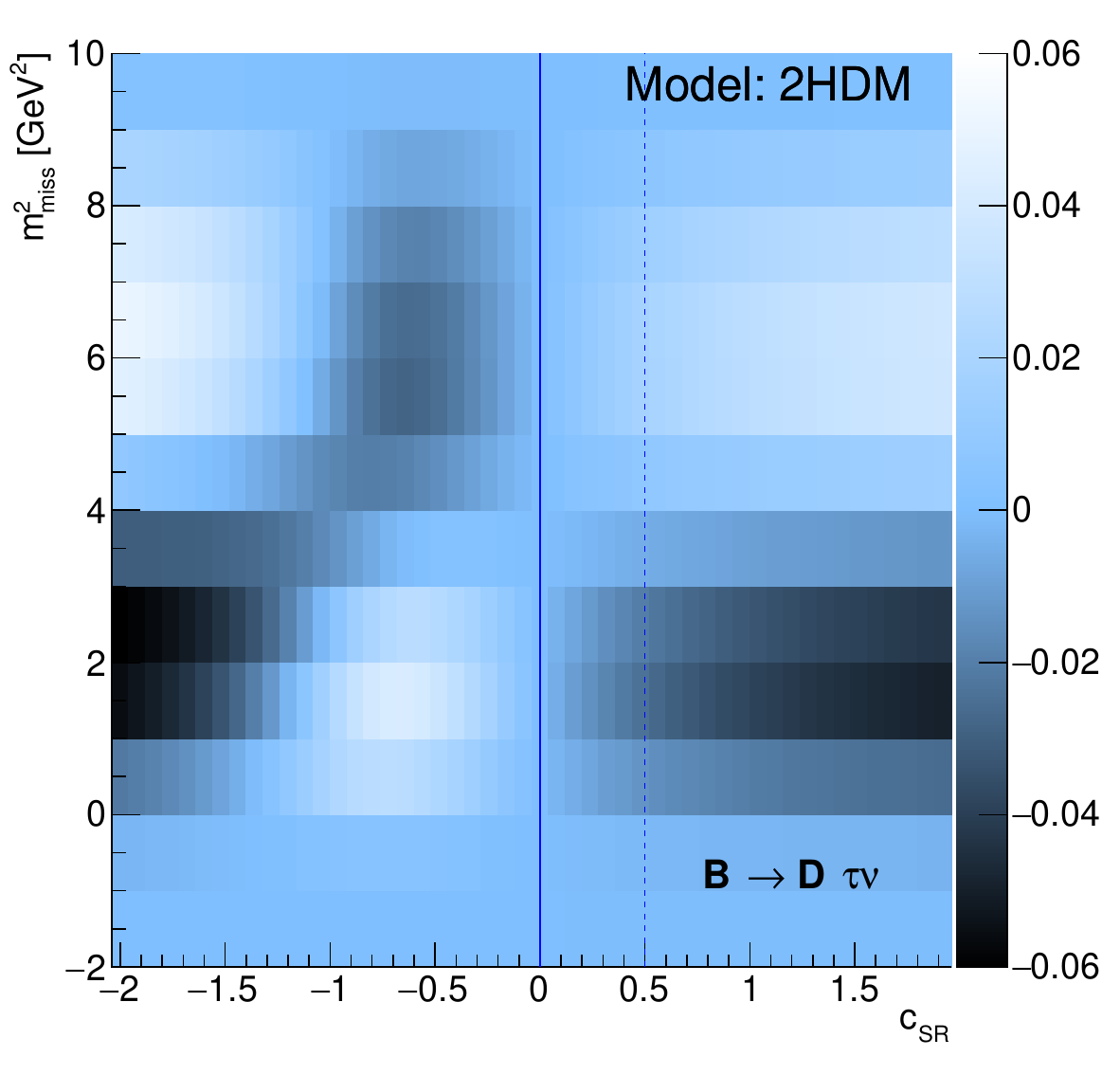}
	\includegraphics[width = 0.245\textwidth]{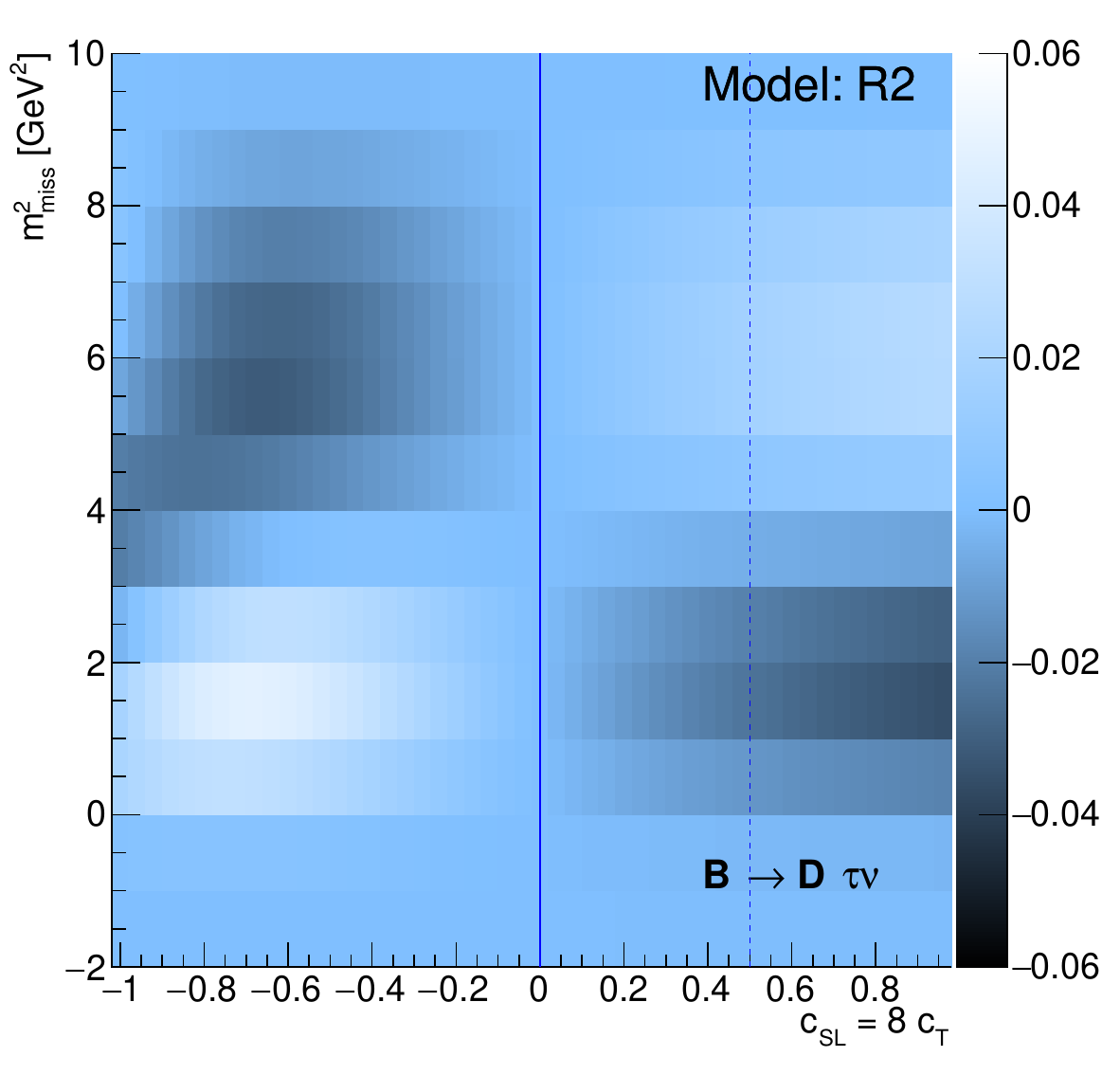}
	\includegraphics[width = 0.245\textwidth]{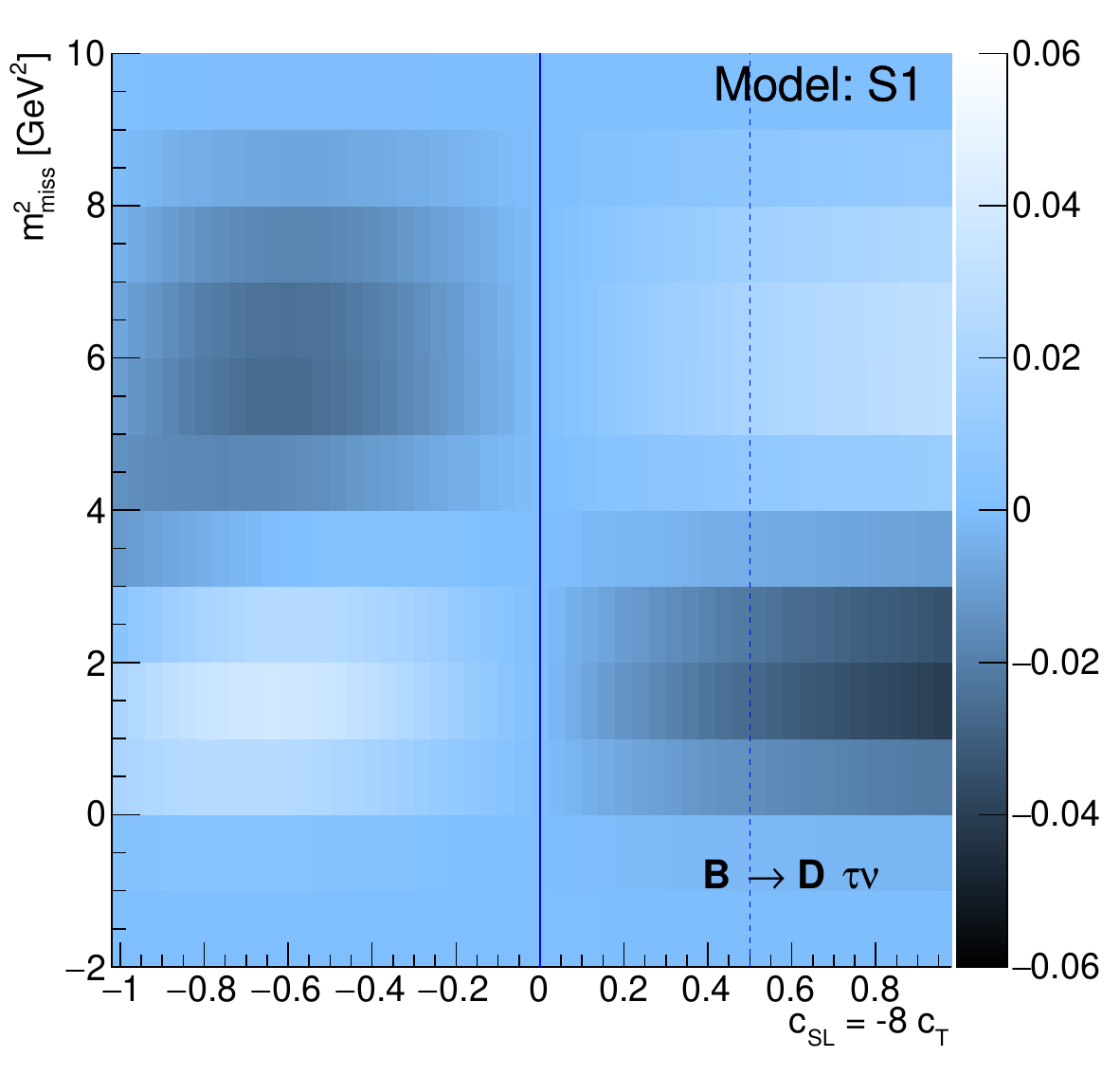}
	\includegraphics[width = 0.245\textwidth]{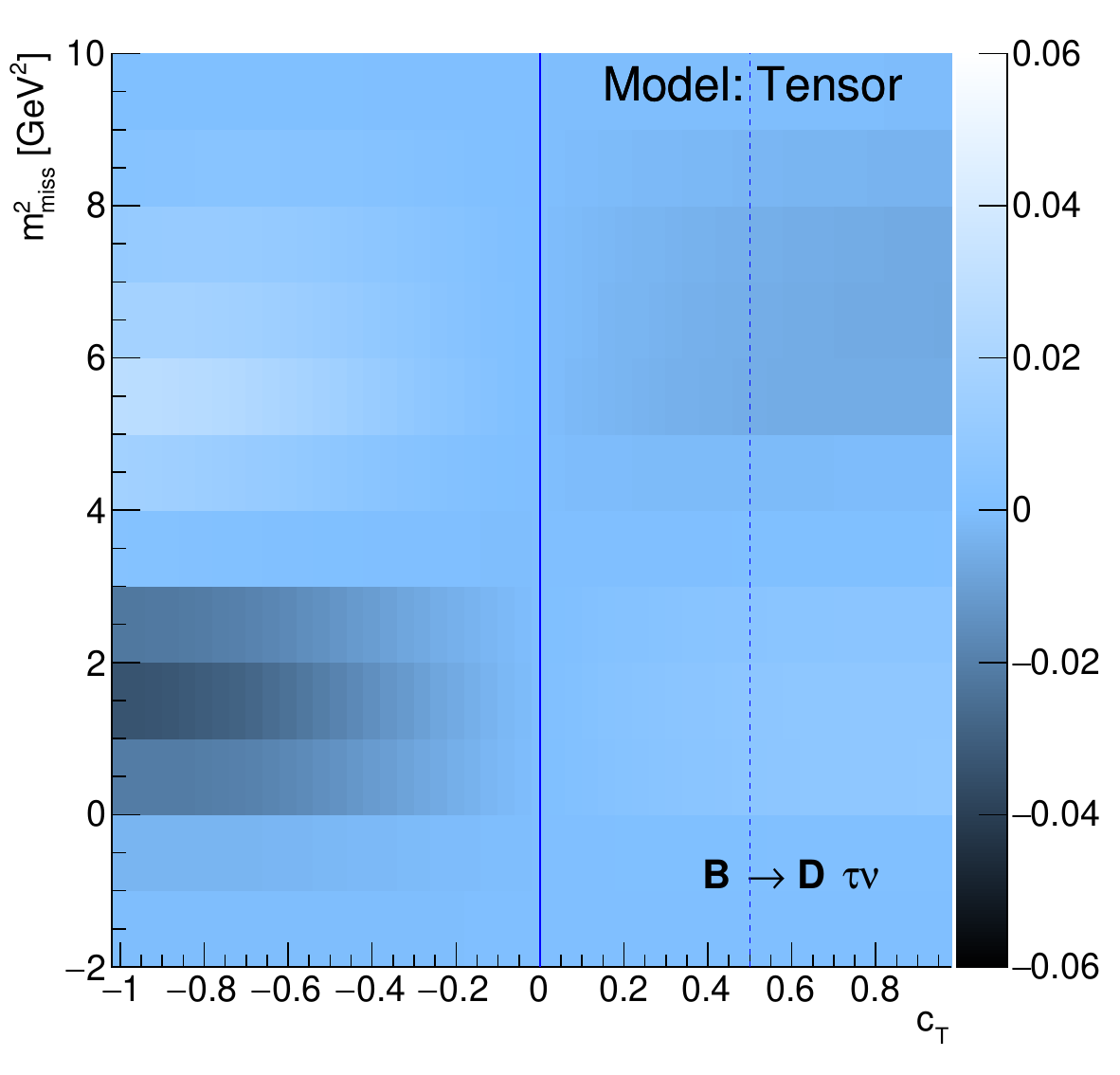}\\
	\includegraphics[width = 0.245\textwidth]{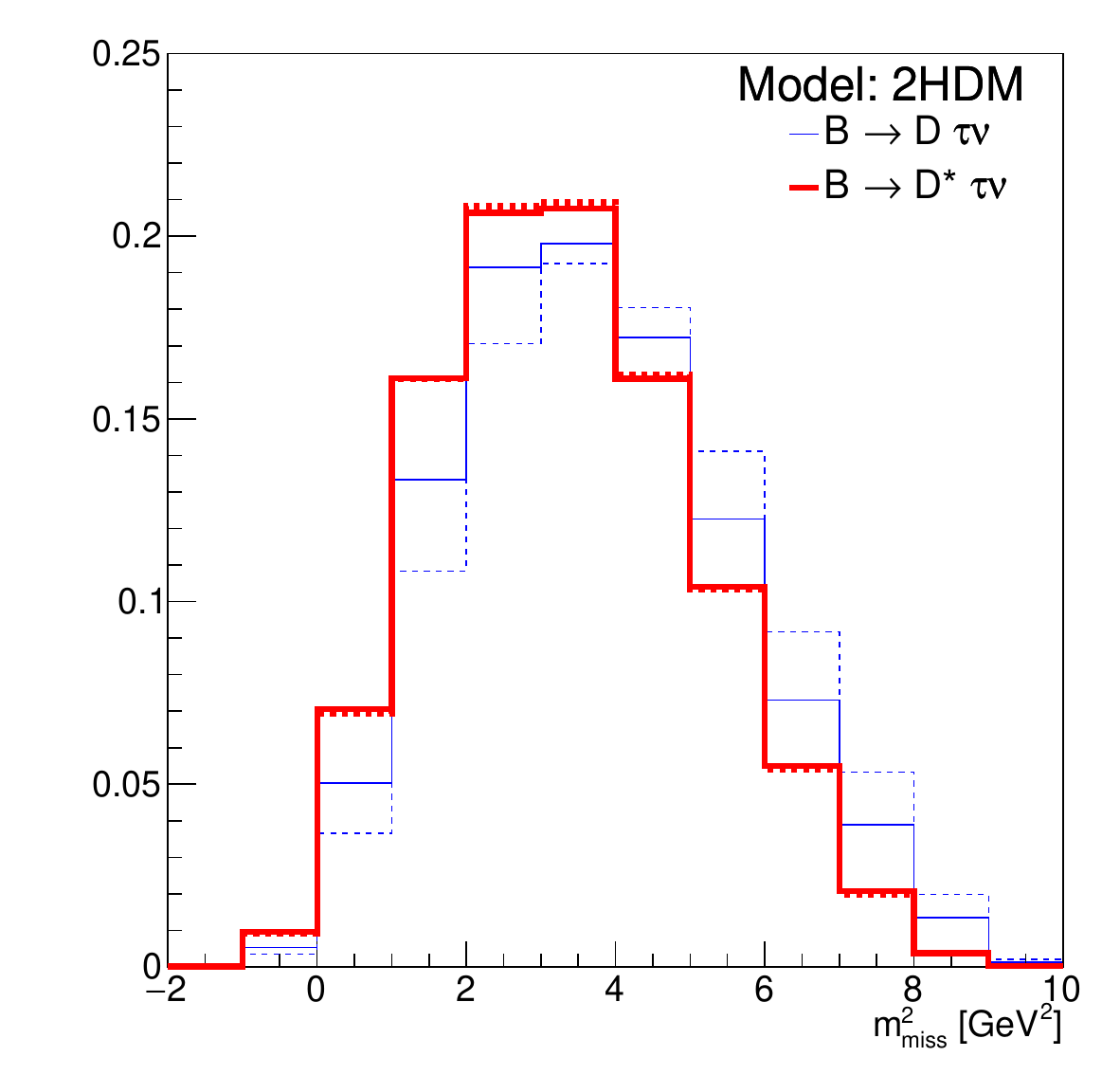}
	\includegraphics[width = 0.245\textwidth]{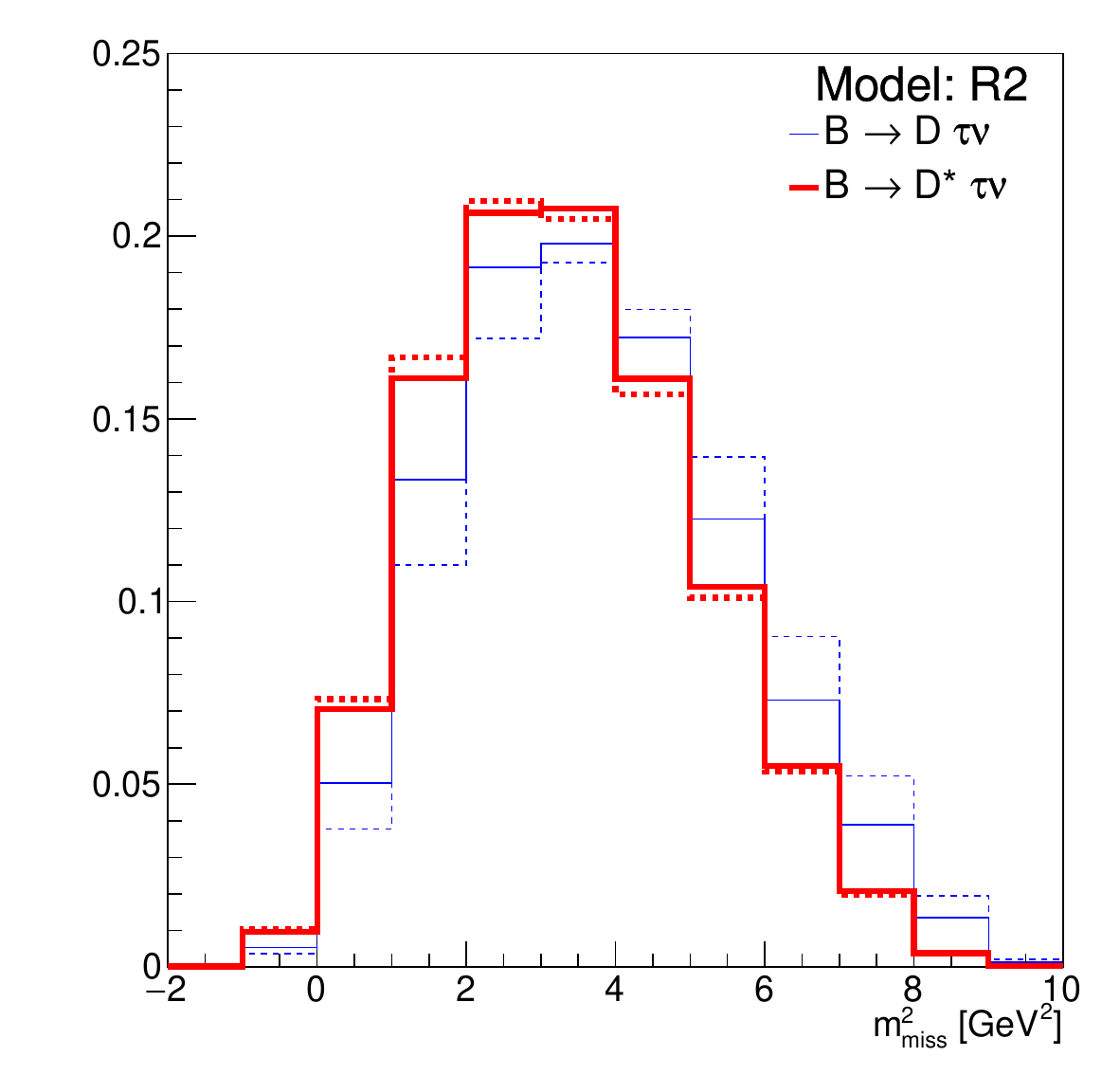}
	\includegraphics[width = 0.245\textwidth]{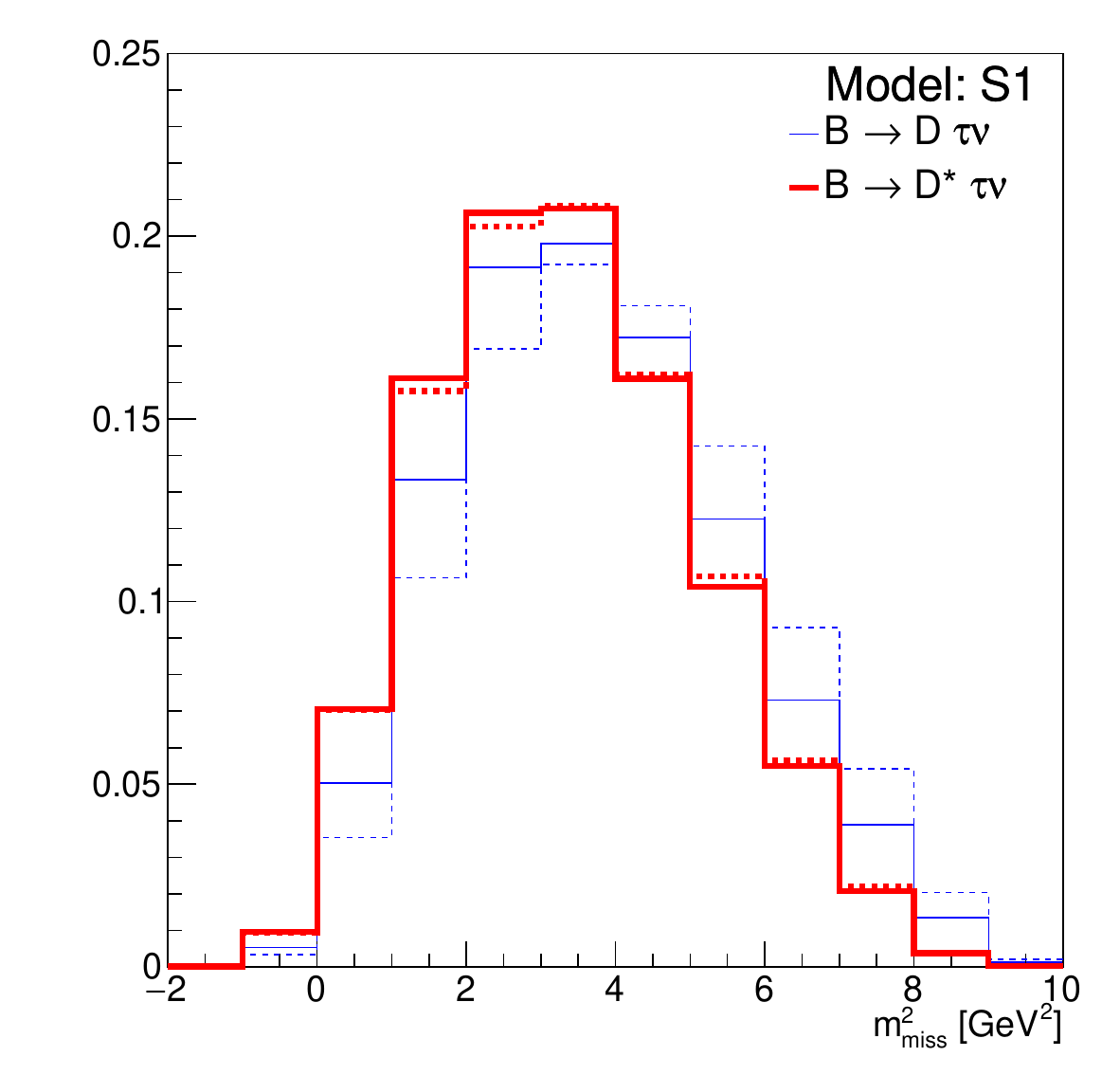}
	\includegraphics[width = 0.245\textwidth]{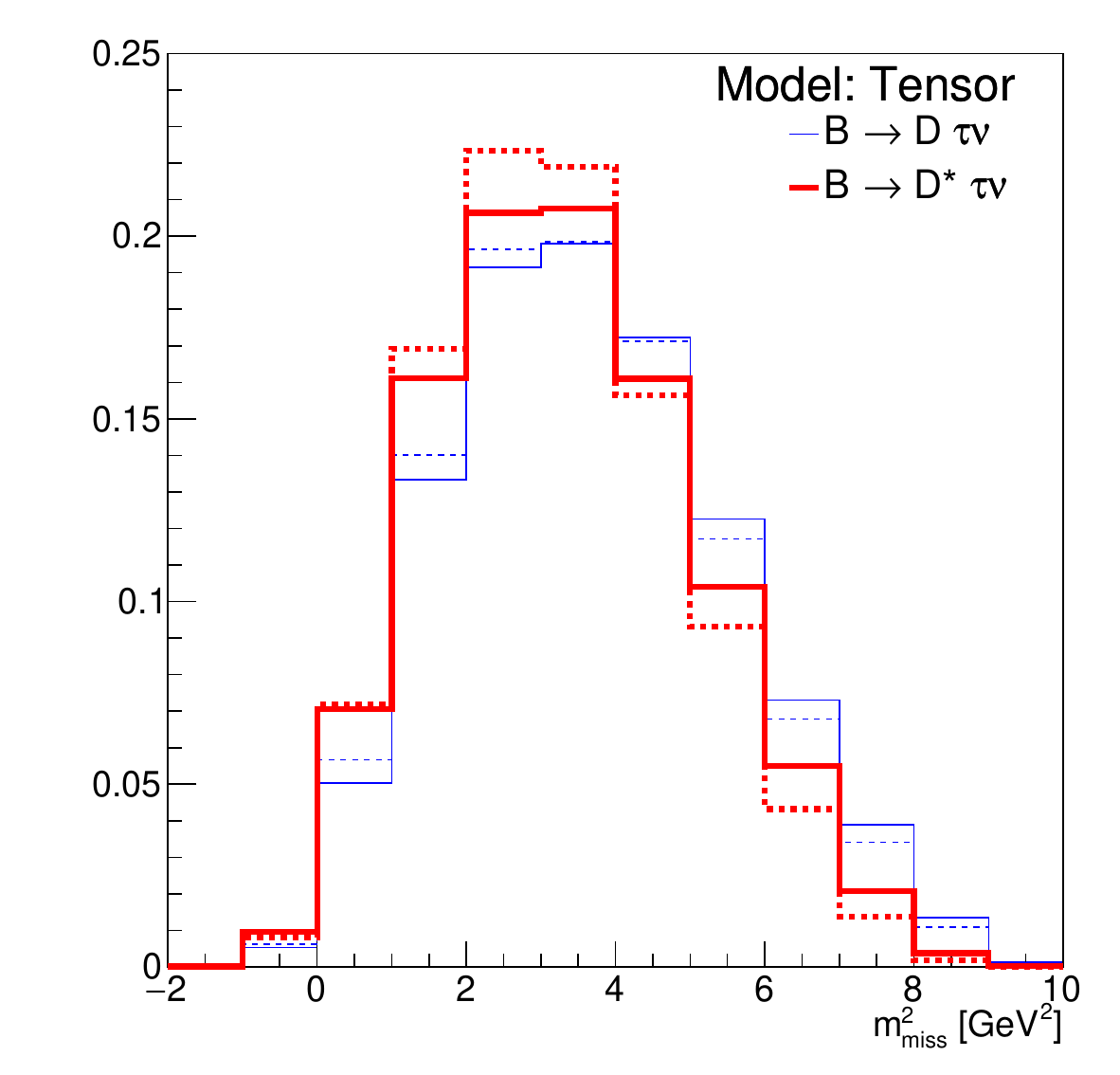}
	\caption{Top row: color map of the percent variation per bin in the reconstructed $m^2_{\text{miss}}$ normalized distribution for $\Bbar \to D\tau\nu$, 
	comparing the SM to a range of couplings for the 2HDM, 
	the leptoquark models $R_2$ and $S_1$, and a pure tensor current. 
	The variations for $B \to D^*\tau\nu$ are similar but somewhat smaller, ranging up to the $1$-$2\%$ level.
	Bottom row: examples of normalized $m^2_{\text{miss}}$ distributions for the SM (solid lines) versus NP (dashed lines) for $\Bbar \to D\tau\nu$ (thin blue lines) and $\Bbar \to D^*\tau\nu$ (thick red lines). 
	The chosen NP coupling for each model is shown as a dashed line in the corresponding top row panels.
	}
	\label{fig:NPint:obs}
\end{figure*}

A similar tension with the SM can be established when additional observables such as asymmetries, longitudinal fractions, 
and polarization fractions are compared to SM predictions (see Sec.~\ref{sec:th:longpol}), 
and there is much literature studying their in-principle NP discrimination power.
However, the same caveat with regard to NP interpretations applies: NP contributions may alter the recovered values of these parameters.

\subsubsection{Sensitivity and biases in recovered observables} 
\label{sec:int:rec}
To gain a sense of the size of these effects, we consider an approximate mock-up of an $e^+e^-$ experimental environment 
and examine the variation in acceptances $\varepsilon$ for $B \to D \tau \nu$ and $B \to (D^* \to D\pi) \tau \nu$, with $\tau \to e \nu\nu$ in the presence of NP.
In this mock-up, the beam energies are fixed to $7$ and 4~GeV, and we require visible final state particles to fall within an angular acceptance of $20^\circ$ to $150^\circ$. 
We impose a minimum electron energy threshold of $E_e > 300$~MeV, 
and an approximate turn-on efficiency is included to account for the slow pion reconstruction efficiencies in $D^* \to D\pi$ decays.
We further include a Gaussian smearing added to the truth level $q^2$ with a width of 1.2~GeV$^2$, in order to account for detector resolution and tag-$B$ reconstruction, 
and require the reconstructed $q^2 > 4$~GeV$^2$.

For this mock-up, we show in Fig.~\ref{fig:NPint:acc} the ratio of the NP experimental acceptance compared to the SM, $\varepsilon/\varepsilon_{\text{SM}}$, 
for several different simplified models (cf.~\cite{Lees:2013uzd} which studied this effect for the Type-II 2HDM).
To characterize the sensitivity to the $q^2$ cut and smearing, we also show acceptances for better and poorer $q^2$ resolutions, 
with widths of $0.8$ and 1.6~GeV$^2$ for the Gaussian smearing, respectively.
To provide further insight into the NP variability of the differential distributions, in Fig.~\ref{fig:NPint:obs} 
we show the percent variation per bin in the reconstructed $m^2_{\text{miss}}$ normalized 
distribution for $\Bbar \to D \tau\nu$ for the same set of simplified models, over the identical range of NP couplings, 
as well as examples of $\Bbar \to \Dx \tau\nu$ distributions in the reconstructed $m^2_{\text{miss}}$ for particular NP coupling values.

One typically sees a few percent variation in the acceptances as well as in the differential $m^2_{\text{miss}}$ distribution, with up to $5\%$ or so variations in some cases. 
Although this might seem small in comparison to the typical $15$\%--$20$\% size of currently measured LFUV in $\RDx$, 
such variations are already comparable to the typical size of systematic uncertainties in current analyses, such as those shown in Table~\ref{tab:ls_tag_errors}.
It is not surprising then that mismatches between the SM and NP signal templates can introduce significant biases into the analyses.
This was observed in the \babar analysis~\cite{Lees:2013uzd}. 
A similar but more detailed mock-up analysis in an $e^+e^-$ collider environment suggests biases at greater than the $4\sigma$ level may be expected to typically arise 
with $5$\,ab$^{-1}$ of data~\cite{Bernlochner:2020tfi}.
This effect may also be important in the extraction of the CKM parameter $|V_{cb}|$, which is sensitive to the assumed form factor parametrization used to generate the fit templates.

Future semileptonic analyses may address these biases through a variety of approaches. 
We discuss these below in Sec.~\ref{sec:outlook:distributions}.

\subsection{Connection to FCNCs}
\label{sec:fcnc}
Measurements of the $b \to s \ell \ell$ ratios $R_{K^{(*)}}$ (Eq.~\eqref{eqn:th:RK}) in various ranges of the dilepton invariant mass have produced an indication of lepton flavor universality violation.
For instance, the most precise measurements of these ratios in the range $q^2 = m^2(\ell\ell) \in [1.1, 6.0]$\,GeV$^2$ currently are~\cite{Aaij:2017vbb,Aaij:2021vac}
\begin{equation}
	\calR_{K^+}  = 0.846^{+0.044}_{-0.041}\,,\qquad \calR_{K^{*0}} = 0.69^{+0.12}_{-0.09}\,,
\end{equation}
but are expected to be unity to the subpercent level. 
Angular analyses of $B \to K^* \mu\mu$ decays exhibit components that are in similar tension with theoretical predictions, but subject to potentially large theory uncertainties.
However, various other less precise measurements of $\RKx$ from Belle and \babar are consistent with unity~\cite{Amhis:2019ckw}; 
see also the recent $\Lb \to p K \ell \ell$ analysis by LHCb~\cite{Aaij:2019bzx}.
As discussed in Sec.~\ref{sec:th:conn}, because the neutrino belongs to an electroweak doublet, 
nontrivial (model-dependent) connections may arise between $b \to c \ell \nu$ and $b \to s \ell \ell$ or $b \to s \nu \nu$ operators.
Studies of possible connections between the $\RDx$ and $\RKx$ anomalies thus explore common origins of NP in $b \to c \tau \nu$ versus $b \to s \ell \ell$,
such as various leptoquark mediators and flavor models, that are not also excluded by other precision measurements (see Sec.~\ref{sec:th:conn}).
See e.g~\cite{Bhattacharya:2014wla, Calibbi:2015kma, Buttazzo:2017ixm, Kumar:2018kmr}, for some representative works in an extensive literature.

In light of these results, it is also interesting to consider how much LFUV can be tolerated in the electron versus muon couplings from $b \to c l \nu$ measurements alone.
As above, the Belle direct measurement (Eq.~\eqref{eqn:Remu}) constrains LFUV to no more than percent-level deviations between the electron and muon semileptonic modes.

An additional constraint arises from exclusive measurements of $|V_{cb}|$ and associated $q^2$ distributions from \BDxellnu decays.
Although they are not a focus of this review, these measurements are presently quite sensitive to the $B \to \Dx$ form factor parametrization:
Precision fits leave little room for the presence of additional form factors beyond those of the $V-A$ interactions,
because introducing such form factors would significantly distort the well-measured $q^2$ distributions for these decays.
Moreover, shape fits to the electron and muon modes separately are in good agreement (see e.g.~\cite{Aubert:2008yv,Glattauer:2015teq,Waheed:2018djm}).
These results suggest that in the $b \to c e \nu$ and $b \to c \mu \nu$ systems, one can plausibly introduce NP only via $V-A$ NP currents,
and one can plausibly produce electron-muon LFUV at most at the percent level.
Based on this qualitative discussion we eagerly anticipate further quantitative studies of bounds on LFUV in \BDxellnu.

\section{Prospects and Outlook}
\label{sec:outlook}

As detailed in Sec.~\ref{sec:interpretation}, the world averages for \RD and \RDs currently exceed their SM
predictions by about 14\% each. While the theory uncertainties on the \RDx SM predictions are already 1\% to 2\%
(see Table~\ref{tab:th:RDDs}), the uncertainties on the corresponding measurements are $5$--$10$ times larger.  
If key challenges in computation, the modeling of $b$-hadron semileptonic decays, and background estimation are met
in the years to come, as discussed in Sec.~\ref{sec:systematics}, the large amount of data that LHCb and Belle~II
will collect over the next two decades will bring down the experimental uncertainties to the 1\% level. 
At the present level of discrepancy with the SM, this degree of precision would nominally be sufficient to either establish an observation of LFUV 
or resolve the present anomalies.

However, highly significant but isolated results will arguably not be sufficient to fully establish the presence of NP in this manner,
given the vast number of experimental and theoretical effects that can influence the interpretation of these indirect searches for BSM physics. 
Spurred on by the \RDx anomalies, a wide program of LFUV measurements and calculations,
that encompasses several experimental and theoretical communities across particle physics, 
will likely be the key to disentangling potential BSM signals from sources of uncertainty that may not be fully understood. 

To this end, in this last section we discuss various aspects of this program, including:
efforts underway to measure other important ratios such as \RJ, $\calR(\pi)$, $\calR(D_{(s)}^{(*)})$ and $\calR(\Lambda_c)$ (Sec.~\ref{sec:outlook:ratios});
analyses that exploit the fully differential information measured in semitauonic $b$-hadron decays
to complement and enhance the sensitivity to NP (Sec.~\ref{sec:outlook:distributions}); and
should these indirect searches end up establishing the presence of NP, 
the role of proposed future colliders, that may be able to either directly observe NP mediators, 
or further characterize established anomalies with related measurements (Sec.~\ref{sec:outlook:future_colliders}).

%%%%%%%%%%%%%%%%%%%%%%%%%%%%%%%%%%%%%%%%%%%%%%%%%%%%%%%%%%%%%%%%%%%%
%%%%%%%%%%%%%%%%%%%%%%%%%%%%%%%%% RATIOS %%%%%%%%%%%%%%%%%%%%%%%%%%
\subsection{Measurement of the ratios ${\cal R}(H_{c,u})$}
\label{sec:outlook:ratios}
%%%% Ratios R(Hcu)

As described throughout this review, the ratios $\calR(H_{c,u})$ defined in Eq.~\eqref{eqn:defRHc}
are powerful probes of LFUV and NP, in part because of the significant cancellation of theoretical and experimental uncertainties in the ratios. 
The SM predictions for $\calR(D_{(s)}^{(*)})$, $\RJ$, and $\calR\left(\Lambda_c\right)$ now have uncertainties in the
1--3\% range (see Sec.~\ref{sec:theory}), and improvements in lattice QCD together with new experimental
measurements are expected to bring these down further. Over the next two decades, LHCb and Belle~II will
collect enough data to reduce the statistical uncertainty on the $\calR(H_{c,u})$ measurements down to a
few percent or less. However, the systematic uncertainties on the best known ratios, \RDx, are currently
significantly higher than that, as shown in Table~\ref{tab:summary_uncertainties}. Thus, quantifying the achievable
precision on $\calR(H_{c,u})$ as a probe of NP after LHCb and Belle~II complete their data taking rests
primarily on estimating the extent to which the associated experimental systematic uncertainties can be
reduced.

As detailed in Sec.~\ref{sec:systematics}, if already ongoing theoretical and experimental efforts are
sustained in the following years, the majority of the systematic uncertainty on $\calR(H_{c,u})$ is
expected to decrease commensurately with the increasing size of the data samples being collected. For
instance, the uncertainty from the background contributions will decrease as the data control samples grow, and
the size of the simulated data samples will continue increasing proportionately if the power of GPUs and fast
simulation algorithms is appropriately harnessed. Of course, these improvements are likely to have their own
limitations, and a certain level of irreducible systematic uncertainty will be reached. Based on the
considerations described in Sec.~\ref{sec:systematics}, one may estimate that floors of $\sim1$--$2\%$ uncertainty
in \RDx are achievable, while a floor of $\sim3$--$4\%$ is plausible for other $\calR(H_{c,u})$ ratios, in which the form factor parametrization
cannot be measured as precisely. 
To illustrate the variability of these estimations, we present extrapolations for the anticipated $\calR(H_{c,u})$ precision 
that LHCb and Belle~II are likely to reach under two scenarios: 
(i) a \emph{pessimistic} scenario, with irreducible systematic uncertainties of 2\% for \RDx and
5\% for the other $\calR(H_{c,u})$ ratios; 
and (ii) an \emph{optimistic} scenario, with uncertainty floors of 0.5\% for \RDx 
and 3\% for the other $\calR(H_{c,u})$ ratios. 
Further assumptions included in these extrapolations are detailed below.

%%%%%% LHCb
\subsubsection{Prospects for $\calR(H_{c,u})$ at LHCb}

%% R(Hc) prospects for LHCb
\begin{figure*}
  \includegraphics[width=0.49\textwidth]{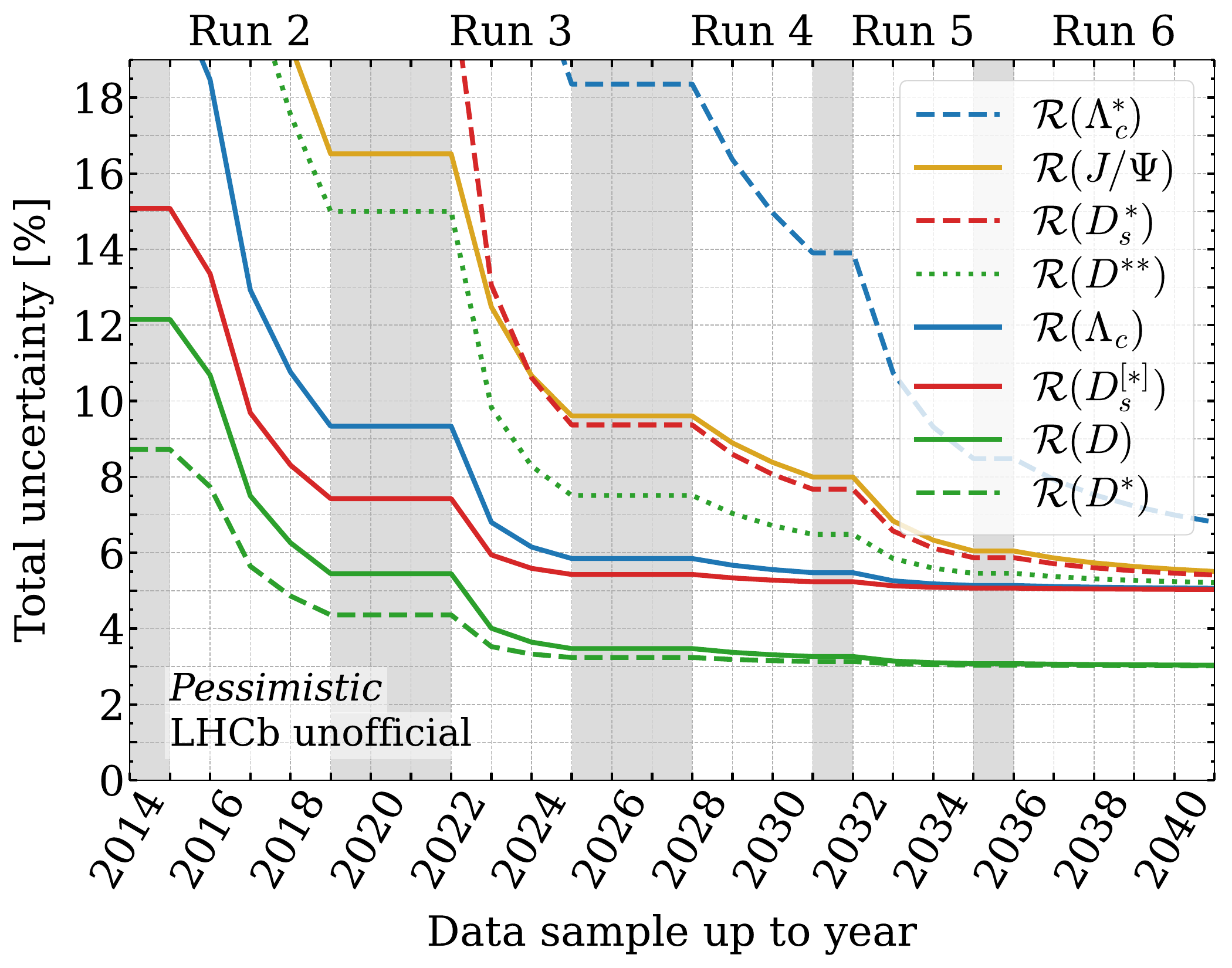}
  \includegraphics[width=0.49\textwidth]{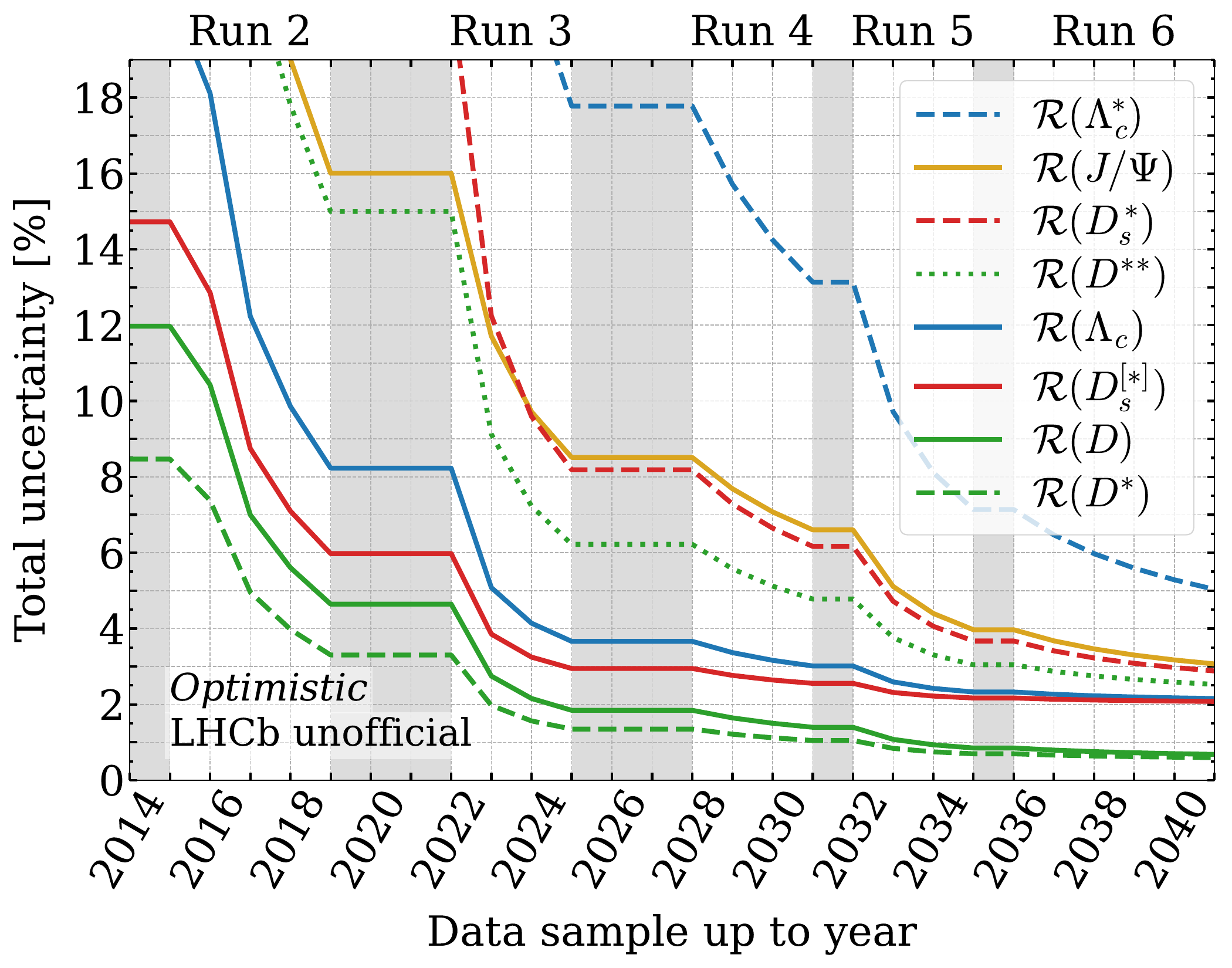}
  \caption{Projections for the expected precision on the measurement of selected $\calR(H_{c})$ ratios at
    LHCb as a function of the year in which the corresponding data sample becomes available. The order of the curves
    in the legend corresponds to the order of the curves on the plot for the year 2026.
    Left: pessimistic scenario for an irreducible systematic uncertainty of 3\% on \RDx and 5\% on the other ratios.
    Right: optimistic scenario for an irreducible systematic uncertainty of 0.5\% on \RDx and 2\% on the other ratios.
    These extrapolations are based on the current muonic-$\tau$ measurements of \RDx and \RJ,
    as well as the forthcoming hadronic-$\tau$ measurement of $\calR(D^{0}_1)$ for the $\calR(D^{**})$ curve.
    The symbol $\calR(D_s^{[*]})$ refers to the sum of the $D_s$ and $D_s^*$ yields, as described in the text.
    The $\calR(\Lambda_c^*)$ entry in the legend refers to $\calR(\Lambda_c^*(2625))$.
    The shaded regions correspond to the long shutdowns during which there is no data taking at the LHC and have been updated
    including the latest estimates~\cite{hl_lhc:2020tdr}.}
\label{fig:ratios:lhcb}
\end{figure*}

As described in Sec.~\ref{sec:production}, the high center-of-mass energy at the LHC gives LHCb access to
large samples of many $b$-hadron species. Thus far, LHCb has published results on \RDs and \RJ
(see Sec.~\ref{sec:measurements}), and measurements of \RD, $\calR(D^{**})$, $\calR(D_s)$, $\calR(D^{*}_s)$,
$\calR\left(\Lambda_c\right)$, $\calR(\Lambda_c^{*})$ as well as the non-semitauonic ratios 
$\calR(D^{(*)})_\text{light}$ are underway. We can project the sensitivity to some of these ratios based on 
the $b$-hadron samples expected in the next two decades (Table~\ref{tab:production_comp}), the reduction of
the systematic uncertainty described above, and the following broad assumptions:\footnote{These projections are
  for the measurements that employ the muonic decays of the $\tau$ lepton. The projections for the hadronic
  measurements would be similar except that the irreducible systematic uncertainty would be asymptotically higher because of the
  external branching fractions used to normalize the result.}
\begin{enumerate}[wide, labelwidth=!, labelindent=0pt, label = \quad \textbf{(\roman*)}, noitemsep, topsep =0pt]
  \item \RDs: The current Run 1 results for $\calR(D^{*+})$ have a total uncertainty of 12\%, but this value
    should be reduced by about $\sqrt{2}$ when $\calR(D^{*0})$ is also included in the measurement. This can
    be done by inclusively reconstructing \BDsztaunu decays via their feed-down to $D^0\mu^-$ samples in combined \RD--\RDs measurements.
    Starting in Run 2, a dedicated trigger achieved 50\% higher efficiency and the \bbbar cross
    section increased by a factor of around two.  Another factor of two will be gained when the hardware
    trigger is replaced by a software-only trigger~\cite{CERN-LHCC-2014-016} starting in the next data taking period (Run 3).
    
  \item \RD: The same assumptions apply as for the measurement of \RDs in terms of triggers and the combination of $D^0$ and $D^+$, 
  but data samples are expected to be about 50\% smaller due to the difference in branching fractions and \RD.

  \item $\calR(D^{**})$: The projections are specifically for $\calR(D^{0}_1)$ which provides the most accessible final state.
  The projections are based on the expected uncertainty of about 15\% for a combined analysis of Run 1 and 2 data, and include
  a factor of two efficiency increase starting in Run 3 thanks to the software-only trigger.
  
  \item $\calR(D_s^{(*)})$: At LHCb, the reconstruction of neutral particles is challenging (see
  Sec.~\ref{sec:reco:neutral}). As a result, the reconstructed number of signal events for $\calR(D_s^{*})$ is
  expected to be about 40 times smaller than for \RDs, due to both the smaller $B_s$ production fraction
  and the requirement to reconstruct a photon in the $D_s^{*+} \to D_s^{+}\gamma$ decay (resulting in
  about a factor of ten lower efficiency), although these are partially compensated for by the larger reconstructed
  branching fractions of the $D_s^{*+}$ decay chain. Given the limitations associated with the reconstruction
  of neutral particles, another possibility is the measurement of
  $\calR(D_s^{[*]})=[\Br(B_s \rightarrow D_s \tau\nu) + \Br(B_s \rightarrow D_s^* \tau\nu)]/[\Br(B_s \rightarrow D_s \mu\nu) + \Br(B_s \rightarrow D_s^* \mu\nu)]$
  which avoids the explicit reconstruction of the photon. The data samples
  for this measurement are expected to be about 3 times smaller than those for \RDs.

  \item $\calR\left(\Lambda_c\right)$: Data samples are expected to be six times smaller than for \RDs, according to the
  smaller $\Lambda_b$ production fraction, as well as the requirement to reconstruct an additional track in
  the $\Lambda_c^+ \to p K^- \pi^+$ decay (which results in a factor of two lower efficiency due primarily to
  the limited LHCb acceptance as well as the PID and tracking efficiencies).

  \item $\calR(\Lambda_c^*)$: A preliminary study by LHCb~\cite{Lupato:2315592} using the muonic decays of the
  $\tau$ finds a factor of 45 smaller data samples for $\calR(\Lambda_c^*(2625))$ than those expected
  for \RDs. This study, however, is not able to constrain the unmeasured $\Lambda_b \to \Lambda_c^* D_s^{(*)}$
  background. Instead, we project $\calR(\Lambda_c^*(2625))$ based on the same assumptions as for
  $\calR\left(\Lambda_c\right)$ but with 33 times smaller data samples due to the smaller $\Lambda_b \to \Lambda^*_c l \nu$
  branching fraction and the efficiency of the $\Lambda_c^* \to \Lambda_c \pi\pi$ reconstruction. This
  is estimated in a preliminary LHCb study of $\Lambda_b \to \Lambda_c^{(*)} \pi \pi \pi$ events under the assumption
  that the ratio of the $\Lambda_b \to \Lambda_c^{(*)} \pi \pi \pi$ branching fractions is the same as that for
  $\Lambda_b \to \Lambda_c^{(*)} \tau \nu$.  The projections for $\calR(\Lambda_c^*(2595))$ would be similar
  but with data samples a factor of two smaller than those for $\calR(\Lambda_c^*(2625))$.
  
  \item \RJ: We scale the 2018 result based on the expected data samples.
\end{enumerate}

Figure~\ref{fig:ratios:lhcb} shows the results of these projections. The years on the horizontal axis refer to the
dates at which data samples became or will become available, which will eventually result in the plotted total uncertainties once analyses are completed. 
For instance, the 8.5\% uncertainty on \RDs shown at the beginning of 2015 corresponds to the eventual precision achievable for
the combined measurement of $\calR(D^{*+})$ and $\calR(D^{*0})$ with the Run~1 data sample, but the analysis is not expected to be completed until 2021.
These projections illustrate the enormous benefit that the data samples collected after the ongoing
LHCb \emph{Upgrade~I} will have on the measurement of $\calR(H_{c})$. The proposed LHCb \emph{Upgrade~II}, 
which would take place in 2031, 
would allow LHCb to further improve the precision on these ratios down to the $0.5$--$2\%$ level, 
if the irreducible systematic uncertainties can be reduced accordingly.

Finally, \butaunu transitions are especially interesting because their potential NP couplings could be quite
different from those potentially involved in \bctaunu transitions. The most direct way to access these
transitions at LHCb could be through $B\to\ppbar\tauon\neu$ decays, for which the normalization
$B\to\ppbar\ell\neu$ channel was recently observed~\cite{Tien:2013nga,Aaij:2019bdu} and is quite clean. A
measurement of $\calR(\ppbar)$ is currently underway. Additionally, LHCb also has plans to measure $\Lb\to
p\tauon\neu$, although this process is more challenging due to the lack of a $\Lambda_b$ decay vertex and
sizable feed-down backgrounds from $\Lambda_b \to \Lambda_c$ processes.

%%%%%% Belle II
\subsubsection{Prospects for $\calR(H_{c,u})$ at Belle~II}
\label{sec:out:belle}

%% R(Hc) prospects for Belle II
Belle~II will profit from the much cleaner environment of $B$ meson pair production in electron-positron annihilations, i.e. even with its smaller data samples 
with respect to LHCb, highly competitive results will emerge. One of the major challenges will be to retain this clean environment at high luminosities and reduce 
the impact of beam and other backgrounds as much as possible. In addition, several orthogonal datasets can be obtained by leveraging different analysis or tagging approaches
(see Sec.~\ref{sec:tagging}). The most important results will be:
\begin{enumerate}[wide, labelwidth=!, labelindent=0pt, label = \quad \textbf{(\roman*)}, noitemsep, topsep =0pt]
 	\item \RDx with exclusive tagging: In principle four statistically independent measurements can be carried out this way, 
 namely either with hadronic or semileptonic tagging and with the focus on either leptonic or hadronic $\tau$-lepton decays. 
 The results with the best control of the systematic uncertainty will be obtained from the combination of hadronic tagging and leptonic or hadronic $\tau$ decays. 
 For these, the $B$ rest frame will be accessible and, in the case of hadronic single-prong $\tau$ decays, the $\tau$ polarization will also be accessible. 
 These results will suffer, however, from the low overall efficiency of hadronic tagging caused by the small branching fractions of such processes. 
 
 Semileptonically tagged events will retain much higher numbers of semitauonic decays, but these will in principle suffer from higher systematic uncertainties.
 Nonetheless, all reconstructed particles in such signatures can still be assigned to either the signal or the tag side, which will allow for reliable measurements.
 It is worth noting that additional energy depositions from beam-background processes will lead to conditions  more challenging than those that impacted the present-day results.
 Further, only measurements with leptonic $\tau$ decays have been realized to date, so it will be an exciting challenge for Belle~II 
 to establish measurements with hadronic $\tau$ decays using this technique. 

%% R(Dx) and R(pi) prospects for Belle II (placed here so that figure is not too far)
\begin{figure*}
  \includegraphics[width=0.49\textwidth]{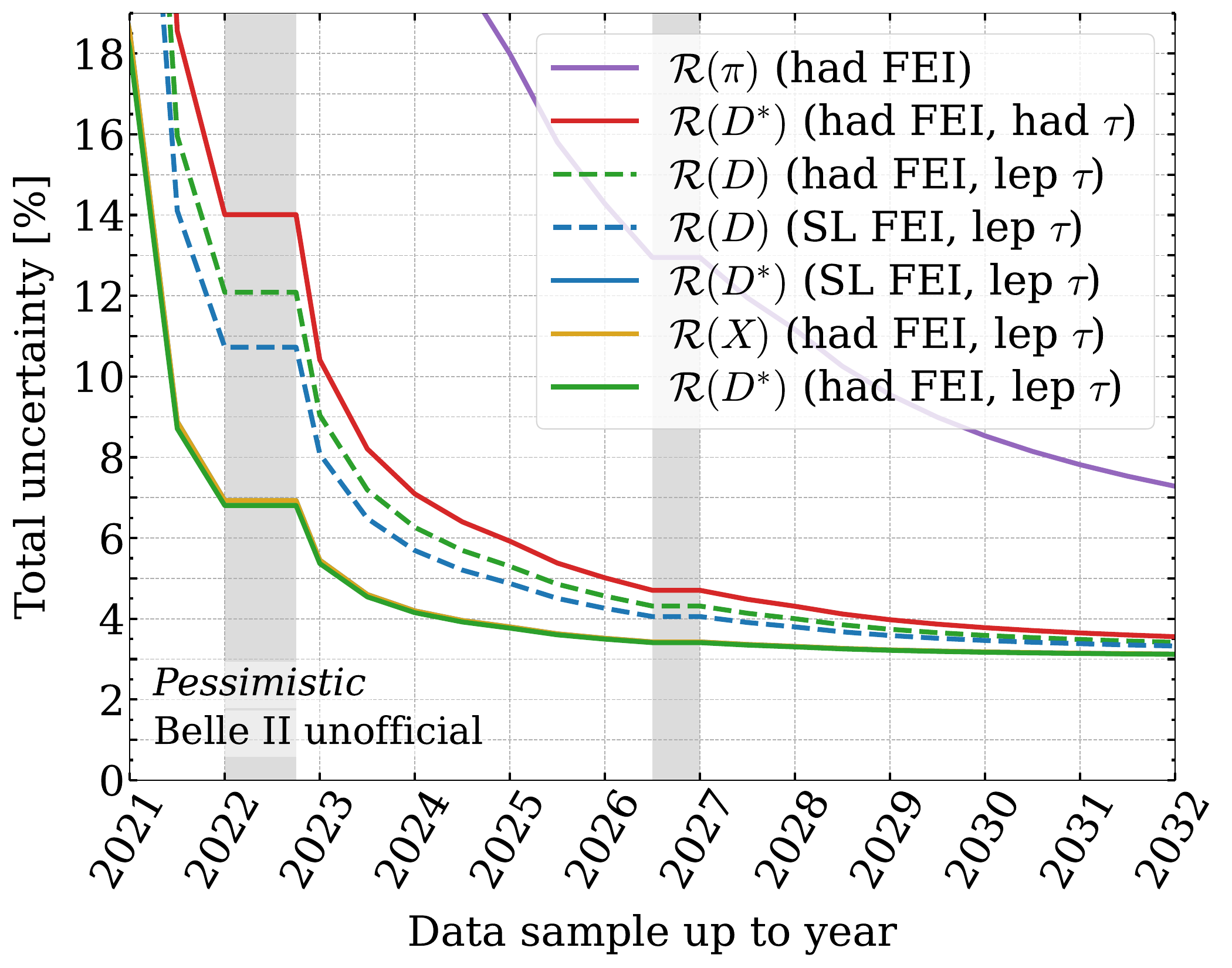}
  \includegraphics[width=0.49\textwidth]{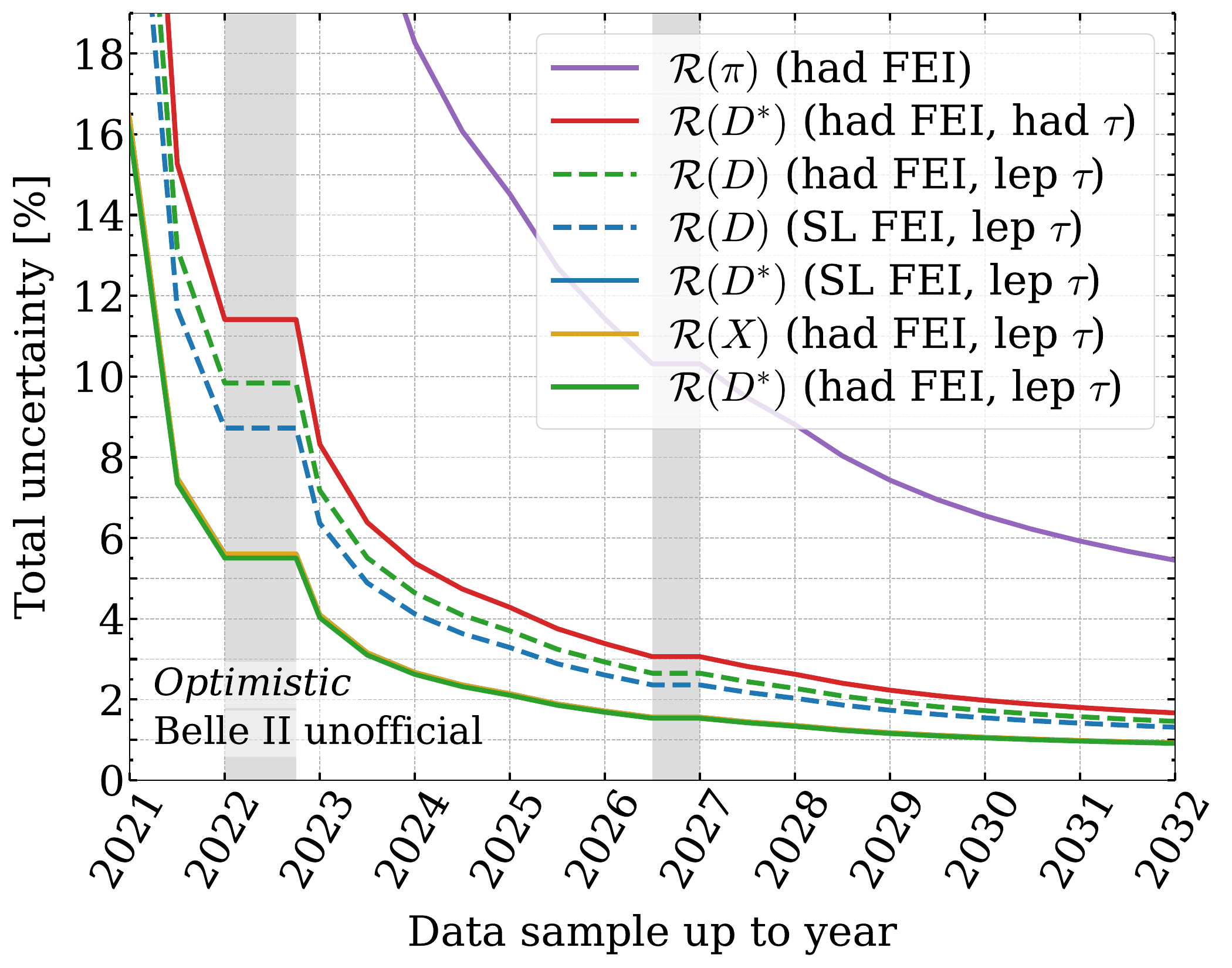}
  \caption{Projections for the expected precision on the measurements of \RDx,  $\calR(X)$, and $\calR(\pi)$  at
    Belle~II as a function of the year in which the corresponding data sample becomes available. The order of the curves
    in the legend corresponds to the order of the curves on the plot for the year 2022. The ``\RDx (SL FEI, lep $\tau$)'' curve
    sits under the ``$\calR(X)$ (had FEI, lep $\tau$)'' curve because their projected uncertainties are very similar.
    An irreducible systematic uncertainty of (left panel) 3\% for the pessimistic scenario and (right panel) 0.5\% for the optimistic one is assumed.
    The optimistic scenario also assumes a 50\% increase in the reconstruction efficiency of the exclusive tagging algorithms.
    The shaded regions indicate years in which significant downtime is expected due to upgrades of the detector and/or the accelerator.}
\label{fig:ratios:belleii}
\end{figure*}

 \item \RDx with inclusive or semi-inclusive tagging: Compared to hadronic or semileptonic tagging, inclusive tagging offers much higher reconstruction efficiency at the cost
 of higher backgrounds and lower precision in the reconstruction of $B$-frame kinematic variables. 
 Nonetheless, such measurements will offer additional orthogonal data sets that can be analyzed. 
 A particularly interesting option might involve the use of semi-inclusive tagging via a charmed seed meson ($D$, $D^*$, $J/\psi$, $D_s$, or $D_s^{*}$). 
 Such an approach could offer more experimental control than purely inclusive tagging, while still retaining a high reconstruction efficiency. 
 It is unclear at present how precise such measurements will be, as no detailed studies have been carried out, and we therefore do not include these in our projections. 
 \item $\calR(\pi / \rho  / \omega)$: Belle~II will have a unique opportunity to further investigate semitauonic processes involving $b \to u$ transitions. 
 The existing search (detailed in. Sec.~\ref{sec:belle_pitaunu}) focused on charged pion final states. 
 Interesting additional channels with higher branching fractions are decays to $\rho$ and $\omega$ mesons, 
 although the large width of the $\rho$ meson is a challenge. 
 Nonetheless, Belle~II will improve the existing limits and with a substantial dataset of 10--15 ab$^{-1}$
 the discovery of these decays, assuming that their branching fraction is of the size of the SM expectation, is feasible. 
 \item $\calR(D_s^{(*)})$: Belle~II anticipates collecting a clean sample of $\epem \to \Upsilon(5S) \to B_s^{(*)} \, \overline{B}_s^{(*)}$ events. 
 The experimental methodology applied to the study of semitauonic $B$ meson decays can also be applied to these datasets.
 For instance, future measurements of $\calR(D_s^{(*)})$ based on hadronic or semileptonic tagging can be done in a fashion similar to the \RDx measurements. 
 It is unclear, however, whether a precision can be reached that would rival LHCb, because of the much smaller number of produced $B_s$ mesons. 
 \item $\calR(X_{(c)})$ with hadronic tagging: Belle~II  will further be able to produce measurements of fully inclusive or semi-inclusive semitauonic final states. 
 These will allow measurements of $\calR(X_{(c)})$. 
 We use the preliminary measurement of \cite{handle:20.500.11811/7578} to estimate the sensitivity for $\calR(X)$, but caution the reader that Belle II will need to demonstrate the feasibility of such measurements. 
\end{enumerate}

Figure~\ref{fig:ratios:belleii} displays the expected sensitivity as a function over time. 
The left panel displays our pessimistic scenario based on the statistical and systematic uncertainties of existing measurements and an irreducible systematic uncertainty of 3\%, as described above.
The right panel shows the same progression for the optimistic scenario, which includes an irreducible systematic uncertainty of 0.5\%
and an increase in the efficiency of the exclusive tagging algorithms of $50\%$. 
Such an improvement is not completely unexpected since novel ideas, such as the use of deep learning concepts and attention maps, 
have already shown promising efficiency gains in simulated events~\cite{Tsaklidis:2122}. 
However, it remains to be seen whether such efficiency gains are also retained in the analysis of actual collision events, 
and whether the identified events are clean enough to provide an actual gain in sensitivity. 
In both scenarios the uncertainties are expected to decrease with luminosity until the systematic uncertainty floor is reached. 

The gray bands indicate years in which significant down-time is expected due to upgrades of the detector and/or the accelerator.
In 2022, the Belle~II pixel detector will be replaced with its final version, and more radiation-hard photomultipliers for the time-of-propagation-detector will be integrated as well. 
In 2026, the Belle~II interaction region will be upgraded to allow for the increase of the instantaneous luminosity to its design value: 
The superconducting magnets that perform the final focusing will be placed farther away from the beam crossing point to reduce the chance of quenches. 
Measurements of \RDs will be somewhat more precise because of their cleaner signature and lack of feed-down contributions, compared to \RD measurements, 
but in both cases a precision of 4--5\% and about 3\% will be reached by 2026 in the pessimistic and optimistic scenarios, respectively.
Inclusive \RDx measurements and measurements of \RDs with hadronic $\tau$ final states will reach 3.5\% precision in the pessimistic scenario and below 2\% in the optimistic case.
All measurements, except for the ones explicitly probing $b \to u$ transitions, will reach precisions close to their irreducible systematic uncertainties by 2031.

%%%%%%%%%%%%%%%%%%%%%%%%%%%%%%%%%%%%%%%%%%%%%%%%%%%%%%%%%%%%%%%%%%%%%%%%%%%%%%
%%%%%%%%%%%%%%%%%%%%%%%%%%%%%%%%% DISTRIBUTIONS %%%%%%%%%%%%%%%%%%%%%%%%%%
\subsection{Exploiting full differential information}
\label{sec:outlook:distributions}

\subsubsection{Angular analyses and recovered observables}
A $2$--$3$\% systematic floor for LFUV ratio measurements might be reached quickly 
given the high statistical power provided by the LHCb and Belle II experiments together. 
Combined with the fact that the ratios $\calR(H_{c,u})$ are recovered observables from template fits to differential distributions, 
this suggests that attention might increasingly turn towards other measurable properties.
These include angular correlations, longitudinal and polarization fractions of the $D^*$ and $\tau$
(see Sec.~\ref{sec:th:longpol}), and asymmetries and so on.

Many such observables using angular correlations have been put forward in a wide range of
phenomenological studies, in particular as a means to distinguish SM from NP interactions in \bctaunu transitions. 
On the experimental side, the most accessible of these is the $D^*$ longitudinal fraction, $F_{L,\tau}(D^*)$, which can easily be reconstructed. 
As discussed in Sec.~\ref{sec:exp:Dspol}, Belle has already provided a preliminary measurement for this variable based on \BDstaunu decays.
This result is compatible with the SM expectations within $2\sigma$. 
LHCb is expected to soon publish a similar analysis with slightly improved sensitivity.

The $\tau$ polarization (Sec.~\ref{sec:bfactories_hadtag_taupol}) was also measured for the first time by Belle, 
which used the \taupinu single-prong decay channel, although with limited precision. 
Preliminary studies in LHCb have demonstrated that the measurement of the $\tau$ polarization is possible 
using the \taupipipi decay mode, 
recycling techniques developed at LEP involving optimized variables~\cite{Davier:1992nw}. 
This analysis is much more complex than the single-prong mode, in which the pion momentum in the $\tau$ rest frame acts as an in-principle perfect polarizer, 
because the analyzing power of the $\pipipi$ final state is comparatively small (see Eq.~\eqref{eqn:th:ptanz}):
The analyzing power of the dominant $\aone$ resonance in \taupipipi features a numerical cancellation on-shell, 
$\alpha_{\aone} = (1-2m^2_{\aone}/m_\tau^2)/(1+2m^2_{\aone}/m_\tau^2) \simeq 0.02$.
The expected LHCb sensitivity to $P_{\tau}(\Dx)$ in the three-prong mode is not yet known.

A recent study~\cite{Hill_2019} showed that LHCb may be able to reliably recover the angular coefficients describing the $B \to (\Dstar \to D\pi)(\tau \to h \nu)\nu$ decay,
assuming a sample size of around $10^5$ signal events.
A dataset of this size is expected to be available at the end of Run 3 of the LHC; first attempts along these lines may be performed using the full Run 2 dataset.

\subsubsection{Future strategies}
\label{sec:out:fut}
However, as discussed in Sec.~\ref{sec:int:rec}, mismatches between SM and NP signal templates 
can introduce significant biases into analyses that consider recovered observables, 
 such that one cannot consistently determine the compatibility of the data with any particular NP model.
Future semileptonic analyses may address these biases through a variety of approaches: 
One possibility is to attempt to carefully control the size of these biases when experiments quote their results. 
A different, more robust, approach is for experiments to adapt their analyses such that instead of reporting recovered observables, 
they perform fits directly in the multidimensional space of the NP couplings---the Wilson coefficients---themselves. 
This approach has the additional advantage of making it more straightforward to combine results from different experiments.

The latter approach is sometimes referred to as ``forward folding''. 
A key obstacle is that generating sufficient simulated data for the SM analysis alone is challenging (see Sec.~\ref{sec:sys:MC}); 
generating enough data to study a space of NP models is naively computationally prohibitive. 
This difficulty can be resolved, however, with matrix element reweighting, which allows for large MC samples to be converted from the SM to any desired NP template, 
or to any description of the hadronic matrix elements, without regenerating the underlying MC data.
In recent years, new software tools, such as the \texttt{Hammer} library~\cite{bernlochner_florian_urs_2020_3993770}, 
have been developed by experimental-theory collaborations to permit fast and efficient MC reweighting of this type.

\begin{figure}[t]
\centering
  \includegraphics[width=0.45\textwidth]{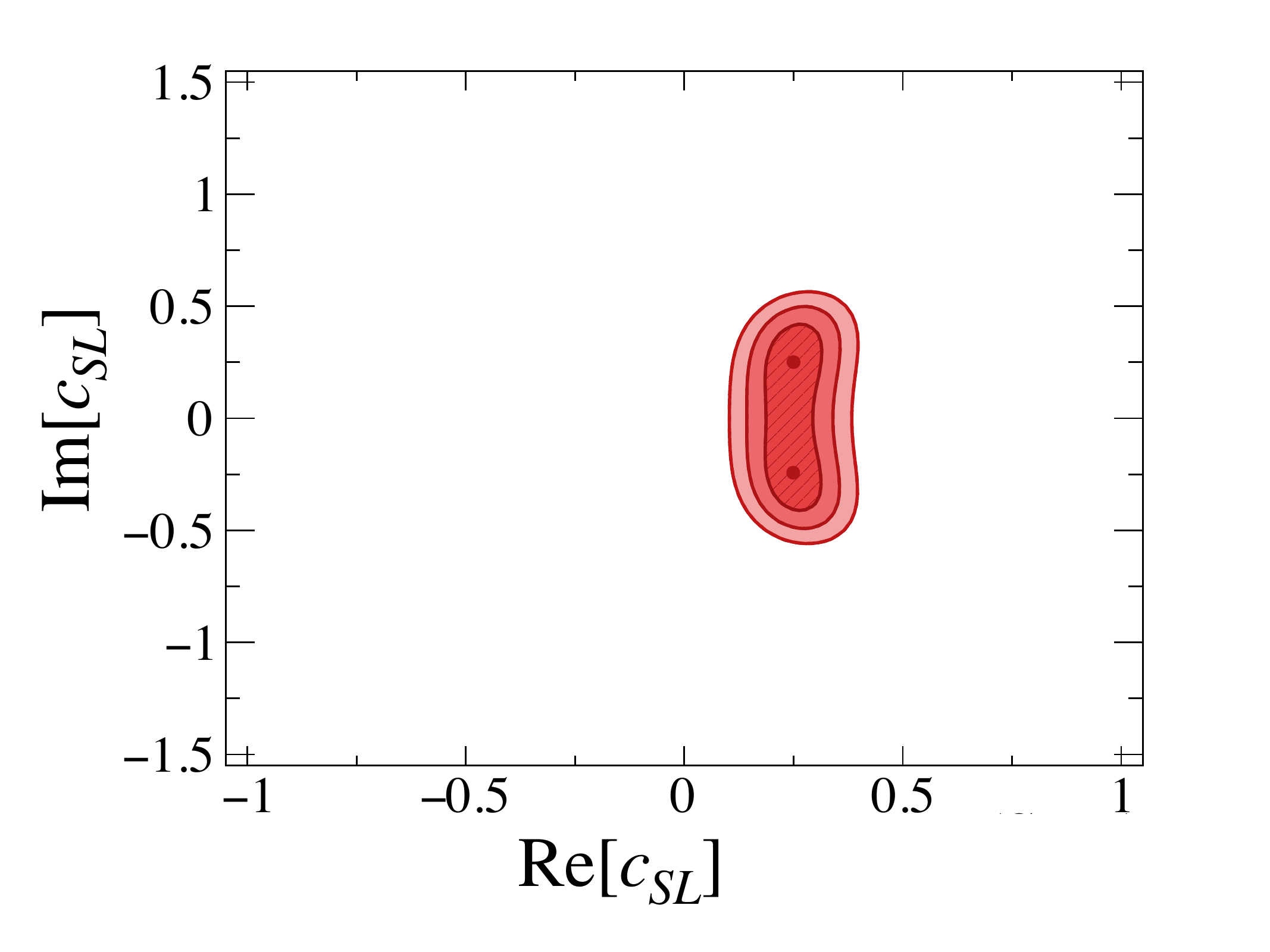}
\caption{The 68\%, 95\% and 99\% CL allowed regions for the $R_2$ simplified model coupling $c_{SL} = 8c_{T}$ fitting to an Asimov dataset with $c_{SL} = 8c_{T} = 0.25(1+i)$. 
The best fit recovered points are shown by gray dots.} 
\label{fig:NP_Rec}
\end{figure}

As an example, one can consider the mock-up reweighting analysis of~\cite{Bernlochner:2020tfi}, 
which uses the differential information in the missing invariant mass $m^2_{\text{miss}}$ and lepton momentum $|\bm{p}_{\ell}|$, 
including an approximation of the effects of various backgrounds and reconstruction effects.
In Fig.~\ref{fig:NP_Rec} we show the potential recovered CLs from this analysis for the (complex) NP Wilson coefficients of the $R_2$ simplified model, 
defined by $c_{SL} \simeq 8c_{T}$, compared to the ``truth'' value $c_{SL} (= 8c_{T}) = 0.25(1+i)$.
This mock-up forecasts that with $5$\,ab$^{-1}$ of future data, one would be able to not only exclude the SM,
but also recover the ``true'' NP Wilson coefficient up to a mild twofold degeneracy in its imaginary part.
Because the forward-folding approach can use all differential information by construction, it may supersede approaches based on measuring recovered observables.

%%%%%%%%%%%%%%%%%%%%%%%%%%%%%%%%%%%%%%%%%%%%%%%%%%%%%%%%%%%%%%%%%%%%%%%%%%%%%%
%%%%%%%%%%%%%%%%%%%%%%%%%%%%%%%%% FUTURE COLLIDERS %%%%%%%%%%%%%%%%%%%%%%%%%%
\subsection{Outlook for future colliders}
\label{sec:outlook:future_colliders}

If NP were to be discovered through indirect LFUV searches, 
future colliders could be instrumental in further characterizing the nature of the new interactions.
In some scenarios, NP mediators can escape the discovery reach of the High Luminosity (HL)-LHC while still giving rise to the observation of LFUV in semitauonic $b$-hadron decays.
Future hadron machines such as the FCC-$hh$ collider~\cite{Mangano:2651294}, presently under study at CERN, 
would extend the reach for direct observation of NP mediators into the multi-TeV range, covering most of these scenarios.
An indirect NP observation could also be possible at FCC-$hh$ by, e.g., detecting deviations from the predicted inclusive $\tau\tau$ production rate in the SM~\cite{Mangano:2651294}. 

High-luminosity \epem colliders may also play a crucial role because the characteristics of $b$-hadron production on the $Z$ pole 
combine several of the advantages enjoyed by $B$-factory experiments with those of hadron colliders. 
In particular, the advantages of the former include: a very favorable ratio of $B$ production divided by total cross section (22\%); 
a low-multiplicity environment (perfect separation of the two $B$ mesons); 
and good knowledge of the $B$ center-of-mass frame, achieved by exploiting jet direction measurements and the peaked fragmentation function.
The advantages of the latter include large production of all $b$-hadron species, and the large boost of the hadrons themselves, 
which allows one to more easily separate their decay products from primary fragments, and to fully reconstruct secondary and tertiary vertices.

The ``TeraZ'' class of proposed \epem colliders---either FCC-$ee$~\cite{abada:hal-02277870} or CEPC~\cite{thecepcstudygroup2018cepc1, thecepcstudygroup2018cepc2}---could provide enough $B$ mesons
produced in this very favorable $Z$-pole environment to measure very complex decays such as $\Bp\to\Kp\taup\taum$, 
that are very difficult to probe otherwise~\cite{Kamenik_2017}. 
A precise measurement of this branching ratio and its angular distributions 
would provide a critical test of LFUV in the neutral-current decays involving the $\tau$ lepton. 
This might in turn provide evidence of a link between the LFUV hints from $\calR(H_{c,u})$, which involve charged-current decays to $\tau$ leptons, and those of $\RKx$, 
which involve neutral current decays to the first two lepton families only (see Sec.~\ref{sec:th:conn}).
In a similar vein, rare \Bc decays such as $\Bc\to\tau\nu$ could also be studied at a TeraZ factory~\cite{zheng2020analysis}. 
A precision of 1\% of this branching fraction could be reached, thereby providing strong constraints on many NP models.

\subsection{Parting thoughts}
In this review we have provided an in-depth look into the theoretical and experimental foundations for semitauonic LFUV measurements. 
This comprised a detailed overview of the theoretical state-of-the-art and an extensive survey of the experimental environments 
and measurement methodologies at the \bfacs and LHCb. 
We further reexamined the current combinations and NP interpretations of the data as well as their limitations, 
and the future prospects to control systematic uncertainties, 
all of which will be crucial not only for establishing a tension with the SM, should one exist, but also for understanding the nature of the New Physics responsible for it.

Driven by the intriguing and persistent anomalies in $\RDx$, 
the host of planned and ongoing measurements of lepton flavor universality violation in semitauonic $b$-hadron decays
will provide new data-driven insights into, if not resolutions for, the current LFUV puzzles.
A golden era in flavor physics is just ahead of us.

%\appendix

\makeatletter
	\let\section\t@section
\makeatother

\acknowledgements{
We thank Hassan Jawahery and Zoltan Ligeti for their comments on the manuscript. 
We also are grateful to Maria R\'o$\dot{\text{z}}$a\'nska for her input on the Belle measurements of $R(\Dx)$ and to Marcello Rotondo for his expertise on the LHCb projections for ${\cal R}(H_{u})$, the $B_s\to D^{**}_s$ contributions, and other matters. 
We thank Ana Ovcharova for her help with the formatting of several plots.
We thank Patrick Owen for sharing his work on the LHCb projections for ${\cal R}(H_{c})$ and subsequent discussions.
We thank CERN for its hospitality during the initial preparation of this work. 
FB is supported by DFG Emmy-Noether Grant No. BE 6075/1-1 and BMBF Grant No. 05H19PDKB1.
MFS is supported by the National Science Foundation under contract PHY-2012793.
DJR is supported in part by the Office of High Energy Physics of the U.S. Department of Energy under contract DE-AC02-05CH11231.
}

\fancyhf{} 
	%\fancyhead[L]{\textsl{\rightmark}}
	\fancyhead[L]{}
	\fancyhead[R]{\textsl{References} \quad \thepage}

\onecolumngrid

\end{document}